\documentclass[11pt,a4paper]{article}

\usepackage[T1]{fontenc}
\usepackage[utf8]{inputenc}
\usepackage{lmodern}
\usepackage[margin=1in]{geometry}
\usepackage{amsmath,amssymb,mathtools}
\usepackage{graphicx}
\usepackage{microtype}
\usepackage{array,booktabs,longtable}
\usepackage[font=small,labelfont=bf,labelsep=colon]{caption}
\usepackage{placeins}
\usepackage{cite}
\usepackage{xcolor}
\usepackage{xurl}
\usepackage[unicode,colorlinks=true,linkcolor=red!65!black,citecolor=green!45!black,urlcolor=blue!55!black]{hyperref}
\hypersetup{
  pdftitle={Turbulence simulation on quantum circuits without linear solves},
  pdfauthor={Byeonggeon Kim and Donghyun You}}

\DeclareUnicodeCharacter{02E2}{\textsuperscript{s}}
\DeclareUnicodeCharacter{0393}{\ensuremath{\Gamma}}
\DeclareUnicodeCharacter{0394}{\ensuremath{\Delta}}
\DeclareUnicodeCharacter{0398}{\ensuremath{\Theta}}
\DeclareUnicodeCharacter{039E}{\ensuremath{\Xi}}
\DeclareUnicodeCharacter{03A0}{\ensuremath{\Pi}}
\DeclareUnicodeCharacter{03A3}{\ensuremath{\Sigma}}
\DeclareUnicodeCharacter{03A5}{\ensuremath{\Upsilon}}
\DeclareUnicodeCharacter{03A6}{\ensuremath{\Phi}}
\DeclareUnicodeCharacter{03A9}{\ensuremath{\Omega}}
\DeclareUnicodeCharacter{03B1}{\ensuremath{\alpha}}
\DeclareUnicodeCharacter{03B2}{\ensuremath{\beta}}
\DeclareUnicodeCharacter{03B3}{\ensuremath{\gamma}}
\DeclareUnicodeCharacter{03B4}{\ensuremath{\delta}}
\DeclareUnicodeCharacter{03B5}{\ensuremath{\varepsilon}}
\DeclareUnicodeCharacter{03B7}{\ensuremath{\eta}}
\DeclareUnicodeCharacter{03B8}{\ensuremath{\theta}}
\DeclareUnicodeCharacter{03BA}{\ensuremath{\kappa}}
\DeclareUnicodeCharacter{03BB}{\ensuremath{\lambda}}
\DeclareUnicodeCharacter{03BD}{\ensuremath{\nu}}
\DeclareUnicodeCharacter{03C0}{\ensuremath{\pi}}
\DeclareUnicodeCharacter{03C1}{\ensuremath{\rho}}
\DeclareUnicodeCharacter{03C3}{\ensuremath{\sigma}}
\DeclareUnicodeCharacter{03C4}{\ensuremath{\tau}}
\DeclareUnicodeCharacter{03C6}{\ensuremath{\varphi}}
\DeclareUnicodeCharacter{03C7}{\ensuremath{\chi}}
\DeclareUnicodeCharacter{03C8}{\ensuremath{\psi}}
\DeclareUnicodeCharacter{03C9}{\ensuremath{\omega}}
\DeclareUnicodeCharacter{1D34}{\textsuperscript{H}}
\DeclareUnicodeCharacter{1D40}{\textsuperscript{T}}
\DeclareUnicodeCharacter{1D57}{\textsuperscript{t}}
\DeclareUnicodeCharacter{2016}{\ensuremath{\|}}
\DeclareUnicodeCharacter{2020}{\ensuremath{\dagger}}
\DeclareUnicodeCharacter{2032}{\ensuremath{'}}
\DeclareUnicodeCharacter{2033}{\ensuremath{''}}
\DeclareUnicodeCharacter{2070}{\textsuperscript{0}}
\DeclareUnicodeCharacter{2074}{\textsuperscript{4}}
\DeclareUnicodeCharacter{2075}{\textsuperscript{5}}
\DeclareUnicodeCharacter{2076}{\textsuperscript{6}}
\DeclareUnicodeCharacter{2077}{\textsuperscript{7}}
\DeclareUnicodeCharacter{2078}{\textsuperscript{8}}
\DeclareUnicodeCharacter{2079}{\textsuperscript{9}}
\DeclareUnicodeCharacter{207A}{\textsuperscript{+}}
\DeclareUnicodeCharacter{207B}{\textsuperscript{-}}
\DeclareUnicodeCharacter{207F}{\textsuperscript{n}}
\DeclareUnicodeCharacter{2080}{\textsubscript{0}}
\DeclareUnicodeCharacter{2081}{\textsubscript{1}}
\DeclareUnicodeCharacter{2082}{\textsubscript{2}}
\DeclareUnicodeCharacter{2083}{\textsubscript{3}}
\DeclareUnicodeCharacter{2084}{\textsubscript{4}}
\DeclareUnicodeCharacter{2088}{\textsubscript{8}}
\DeclareUnicodeCharacter{2095}{\textsubscript{h}}
\DeclareUnicodeCharacter{2113}{\ensuremath{\ell}}
\DeclareUnicodeCharacter{2192}{\ensuremath{\to}}
\DeclareUnicodeCharacter{2194}{\ensuremath{\leftrightarrow}}
\DeclareUnicodeCharacter{21D2}{\ensuremath{\Rightarrow}}
\DeclareUnicodeCharacter{2202}{\ensuremath{\partial}}
\DeclareUnicodeCharacter{2207}{\ensuremath{\nabla}}
\DeclareUnicodeCharacter{2208}{\ensuremath{\in}}
\DeclareUnicodeCharacter{220E}{\ensuremath{\blacksquare}}
\DeclareUnicodeCharacter{2212}{\ensuremath{-}}
\DeclareUnicodeCharacter{221A}{\ensuremath{\surd}}
\DeclareUnicodeCharacter{221D}{\ensuremath{\propto}}
\DeclareUnicodeCharacter{221E}{\ensuremath{\infty}}
\DeclareUnicodeCharacter{222B}{\ensuremath{\int}}
\DeclareUnicodeCharacter{223C}{\ensuremath{\sim}}
\DeclareUnicodeCharacter{2248}{\ensuremath{\approx}}
\DeclareUnicodeCharacter{2260}{\ensuremath{\neq}}
\DeclareUnicodeCharacter{2261}{\ensuremath{\equiv}}
\DeclareUnicodeCharacter{2264}{\ensuremath{\leq}}
\DeclareUnicodeCharacter{2265}{\ensuremath{\geq}}
\DeclareUnicodeCharacter{226A}{\ensuremath{\ll}}
\DeclareUnicodeCharacter{226B}{\ensuremath{\gg}}
\DeclareUnicodeCharacter{2272}{\ensuremath{\lesssim}}
\DeclareUnicodeCharacter{2273}{\ensuremath{\gtrsim}}
\DeclareUnicodeCharacter{2282}{\ensuremath{\subset}}
\DeclareUnicodeCharacter{2297}{\ensuremath{\otimes}}
\DeclareUnicodeCharacter{22A5}{\ensuremath{\perp}}
\DeclareUnicodeCharacter{230A}{\ensuremath{\lfloor}}
\DeclareUnicodeCharacter{230B}{\ensuremath{\rfloor}}
\DeclareUnicodeCharacter{27E8}{\ensuremath{\langle}}
\DeclareUnicodeCharacter{27E9}{\ensuremath{\rangle}}
\DeclareUnicodeCharacter{2AB0}{\ensuremath{\succeq}}
\DeclareUnicodeCharacter{0304}{\ensuremath{{}^{\scriptscriptstyle -}}}
\DeclareUnicodeCharacter{0302}{\ensuremath{{}^{\scriptscriptstyle\wedge}}}
\DeclareUnicodeCharacter{0303}{\ensuremath{{}^{\scriptscriptstyle\sim}}}

\newcommand{\disp}[2]{\par\smallskip\noindent
  \begin{minipage}{\dimexpr\linewidth-4em\relax}\centering #1\end{minipage}\hfill\makebox[3.6em][r]{(#2)}\par\smallskip}
\newcommand{\displ}[1]{\par\smallskip\noindent\begin{minipage}{\linewidth}\centering #1\end{minipage}\par\smallskip}

\makeatletter
\newenvironment{SIpart}{%
  \setcounter{section}{0}%
  \renewcommand{\@seccntformat}[1]{\ifnum\pdfstrcmp{##1}{section}=0 Supplementary Note~\fi\csname the##1\endcsname\quad}%
}{}
\newenvironment{NoteCnumbering}{%
  \setcounter{subsection}{-1}%
}{}
\makeatother

\renewenvironment{abstract}{\begin{quotation}\noindent\ignorespaces}{\end{quotation}}

\title{Turbulence simulation on quantum circuits without linear solves}
\author{Byeonggeon Kim and Donghyun You\thanks{Corresponding author: \href{mailto:dhyou@postech.ac.kr}{dhyou@postech.ac.kr}}\\[6pt]
  {\normalsize\itshape Department of Mechanical Engineering,}\\
  {\normalsize\itshape Pohang University of Science and Technology (POSTECH), Pohang, Republic of Korea}}
\date{}

\begin{document}
\maketitle

\begin{abstract}
Quantum algorithms evolve linear, unitary dynamics efficiently, but incompressible flow is nonlinear, dissipative and
divergence-constrained~\cite{Liu2021,Fang2023,Dalzell2025}. We present a hybrid quantum--classical Navier--Stokes
integrator in which pressure evolves by artificial compressibility and a classical midpoint predictor re-linearizes the
flow at every step. On a staggered grid, symmetric splitting turns each linearized step into a unitary factor between two
contractions, applied as polynomials of self-inverse block encodings by generalized quantum signal processing. Time
marching therefore needs no linear solve, no condition number enters the query count or success probability, and the
exact factors stay bounded for any step size: stiff pressure waves impose no stability limit, and our $256^3$ turbulence
run needed 1.9--3.3 times fewer steps than explicit integration of the same system at its stability limit. Counted with
reference polynomial degrees, transport and pressure coupling also need about a hundred times fewer block-encoding calls
per step than a projection method in geometries without a fast diagonalizing transform. Gate-level simulations on small
grids verify the encodings, and the predictor restores second-order convergence in a $16^3$ test. In exact classical
emulation the integrator carries three-dimensional decaying turbulence through the dissipation peak, within 3.25\% of an
in-house pseudo-spectral reference in kinetic energy and 1.6\% in band-averaged large-scale spectra. With errors injected
at every step, the part fed back into the operator grows several-fold more than a passive part, while the total remains
below or comparable to matched accumulation. This provides a structure-preserving route to quantum-compatible
integration of turbulent flows.

\end{abstract}

\section{Introduction}\label{sec:introduction}

Quantum algorithms for linear differential equations are now well developed: unitary dynamics can be simulated with
near-optimal query complexity~\cite{LowChuang2019,Gilyen2019}, and non-unitary linear evolution can be treated by time
marching, linear combinations of Hamiltonian simulations or Schr\"odingerization~\cite{Fang2023,AnLiuLin2023,%
AnChildsLin2026,JinLiuYu2024}. Incompressible fluid flow fits these frameworks poorly. The Navier--Stokes equations are
nonlinear, whereas quantum evolution is linear; viscous dissipation is not unitary, so it must be realized by
post-selection or dilation, whose success can decay over repeated steps~\cite{Fang2023}; incompressibility is a global
constraint that projection methods enforce by solving a pressure-Poisson equation at every step; and a solution held in
the amplitudes of a quantum state is expensive to read out~\cite{Liu2021,Dalzell2025}. Turbulence adds chaos, in which
small differences between trajectories grow~\cite{Boffetta2017}.

Existing approaches resolve different subsets of these mismatches~\cite{Cecile2026}. Carleman and Koopman--von
Neumann~\cite{Joseph2020} embeddings keep the evolution coherent in a larger linear space and are efficient when
dissipation dominates the nonlinearity~\cite{Liu2021,Krovi2023}, a regime that turbulence lies outside; for chaotic
dynamics, lower bounds exist~\cite{Lewis2024,Ameri2026}. Linear-solver approaches place an implicit step or the pressure
projection in a quantum linear-systems algorithm~\cite{Harrow2009,Costa2022,Lapworth2022}, whose cost grows with a
condition number; for the pressure-Poisson operator this number grows with the square of the resolution, and within a
projection method the solve can be replaced by a direct transform only where a fast diagonalizing transform is
available. Lattice-kinetic formulations replace the pressure solve by a local collision
step~\cite{Wang2025,Jennings2026}. Hybrid loops measure the state at every step and rebuild the operator
classically~\cite{Griffin2019,Lapworth2022,Bharadwaj2023}; such loops have been run on noisy quantum
hardware~\cite{Song2025}, and when the full state is measured at every step their measurement and re-preparation cost
grows with the number of grid points~\cite{Bharadwaj2023}. In such a loop the cost of the quantum--classical
interface, where the classical data enter the operator and the state is read out and re-prepared, can be considered
separately from the query count of the evolution between two readouts, which is decided by how the time step is
constructed. From these studies we take the following design requirements for the time-marching part of a quantum-compatible fluid
integrator: the nonlinearity should be handled without enlarging the state, the pressure without a linear solve whose
cost grows with a condition number, the dissipation by operations whose success is fixed by their normalization, and
the whole step with the accuracy of a standard time integrator. Readout and state preparation, which the loop presented
here shares with earlier per-step hybrids, belong to the interface and are treated separately
(Section~\ref{sec:cost}; Methods).

Here we introduce a hybrid quantum--classical integrator designed around these requirements, whose quantum-implementable
part has a guaranteed structure in every step once the advecting velocity is frozen. A classical midpoint predictor re-linearizes the flow at every
step, so the nonlinearity never enters the quantum part and the state keeps the size of the discretized field. Pressure
is advanced together with velocity in an entropically damped artificial-compressibility (EDAC)
formulation~\cite{Clausen2013a}, so no pressure-Poisson equation is solved after the initial condition. On a staggered
grid with skew-symmetric advection~\cite{Harlow1965,Morinishi1998}, the frozen convection--acoustic generator is
anti-Hermitian for any real advecting velocity, and diffusion is a constant negative semidefinite operator. A symmetric
splitting~\cite{Strang1968} separates the two structures, leaving a unitary convection--acoustic factor between two
contractive diffusion factors, each a function of a normal operator with a known spectral range. Each factor is applied as a
polynomial of a block-encoded operator, synthesized by generalized quantum signal processing
(GQSP)~\cite{LowChuang2019,Gilyen2019,Motlagh2024}, with a success probability set by the normalization of the
synthesis: one quarter for the unitary factor and, depending on the state, close to one quarter for diffusion.

The contribution of this work is a time-marching architecture for nonlinear, three-dimensional incompressible flow in
which these elements act together, and its verification. The pressure--velocity coupling and the pressure advection sit
inside the unitary factor. The stiff acoustic coupling is kept with convection and, because diffusion commutes with it
on uniform periodic grids with equal diffusivities, enters the splitting error only through a cross term with
advection, so that the splitting is exact for a uniform advecting velocity. A second-order midpoint predictor supplies
the coefficients of every frozen step. As a result, a step requires no linear solve, its query count is
governed by a normalized evolution time (the time step multiplied by the normalization of the block-encoded generator),
which includes the acoustic stiffness of the formulation, rather than by a condition number, and its exact factors are
bounded for any time step. Individual ingredients have precedents: unitary
advection has been split from post-selected diffusion for
passive scalars~\cite{Pfeffer2025}; the same diffusion--advection--diffusion composition appears in concurrent work on
variable-coefficient transport~\cite{Akash2026}, and artificial compressibility in concurrent work on Stokes
flow~\cite{Jin2026}; and classical refresh of an operator inside a hybrid loop has been proposed and
used~\cite{Griffin2019,Lapworth2022}. In the literature we searched, no quantum or hybrid time stepper for the
macroscopic flow equations combines artificial-compressibility pressure evolution with nonlinear convection, and none
applies a Strang split into a separately exponentiated dissipative factor and a unitary conservative factor to a coupled
incompressible-flow system; lattice-kinetic schemes avoid the pressure solve in a different, weakly compressible
formulation~\cite{Wang2025,Jennings2026}. We construct the block encodings as self-inverse circuits and verify them at
the gate level on small grids, show that the predictor restores second-order temporal convergence on a smooth test
state, and apply the integrator in exact classical emulation to three-dimensional decaying isotropic turbulence at
$256^3$. We then establish its computational structure and resource scaling, and we estimate, in block-encoding calls,
what the replacement of the pressure projection saves in geometries without a fast diagonalizing transform.

Finally, because every step of a hybrid loop crosses a quantum--classical interface where errors enter, we use a
property of the integrator, namely that it rebuilds its operator from the real part of its state only, to separate how
a repeated per-step error is fed back into the dynamics from how it is merely carried along.

\section{Results}\label{sec:results}

\subsection{A structure-preserving hybrid integrator}\label{sec:framework}

\begin{figure}[tbp]
\centering
\includegraphics[width=\linewidth]{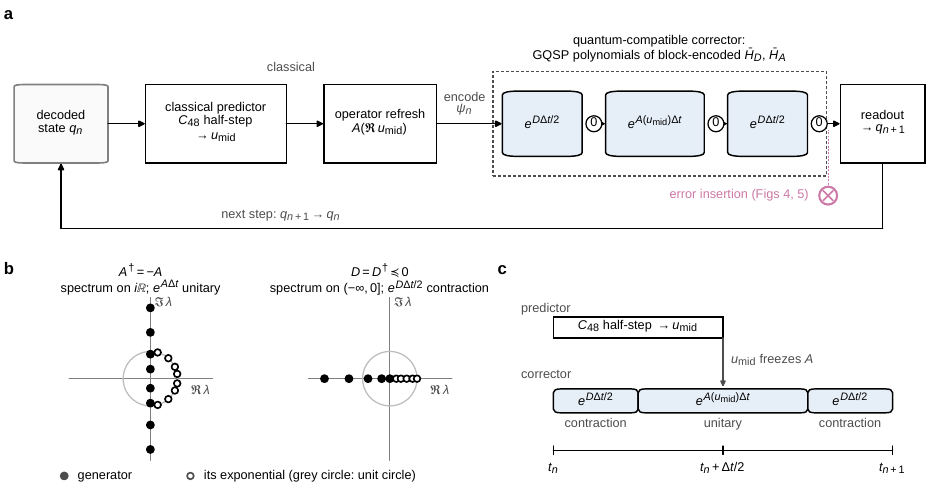}
\caption[{The hybrid quantum--classical integrator.}]{\textbf{The hybrid quantum--classical integrator.} \textbf{a},
One time step. A classical predictor supplies the midpoint velocity $u_{\mathrm{mid}}$ that re-linearizes the flow; the
convection--acoustic operator is refreshed from its real part; the quantum-compatible corrector applies $e^{D\Delta
t/2}$, $e^{A(u_{\mathrm{mid}})\Delta t}$ and $e^{D\Delta t/2}$, each as a GQSP polynomial of a block-encoded operator
followed by post-selection of the ancillas (circled 0); the output is read out as the next state. Crossed circle, where
the per-step error of Figs~\ref{fig:transfer} and~\ref{fig:feedback} is inserted. \textbf{b}, Why the step is split
(schematic spectra): the anti-Hermitian convection--acoustic generator $A$ has its spectrum on the imaginary axis, so
$e^{A\Delta t}$ is unitary; the negative semidefinite diffusion operator $D$ has its spectrum on the negative real axis,
so $e^{D\Delta t/2}$ is a contraction. Each factor is therefore a bounded function of one normal operator, which GQSP
can apply. \textbf{c}, How a step is completed: the classical half-step supplies $u_{\mathrm{mid}}$, which freezes $A$
for the symmetric composition of two contractions around one unitary factor.}\label{fig:framework}
\end{figure}

The integrator is built around one requirement: every operation that is delegated to a quantum processor should be a
bounded matrix function of an operator whose structure is known before the step, so that its cost and success do not
depend on a condition number and no step-size restriction is needed for its stability. Four properties of incompressible flow stand in the way,
and each is met by one design choice (Fig.~\ref{fig:framework}). The nonlinearity is kept out of the quantum part by
freezing the advecting velocity within a step and updating it classically between steps. The incompressibility
constraint, which projection methods enforce by solving a pressure-Poisson equation at every step, is replaced by
pressure evolution: we integrate the EDAC equations~\cite{Clausen2013a,Clausen2013b}, in which a hyperbolic pressure equation with pressure diffusion replaces the
constraint and approaches the incompressible equations with $O(\mathrm{Ma}^2)$ errors (Methods). Convection and the
acoustic pressure--velocity coupling are discretized so that together they generate a unitary evolution, and viscous
diffusion so that it generates a contraction. Operator splitting then separates the two, so that each factor can be
applied as a polynomial of a block-encoded operator.

On a periodic staggered marker-and-cell grid~\cite{Harlow1965} of $N^3$ cells, with skew-symmetric advection of all four
fields~\cite{Morinishi1998}, freezing the advecting velocity at a real field $\bar u$ turns the semi-discrete equations
into the linear system
\begin{equation}
  \frac{dq}{dt} = A(\bar u)\,q + D\,q,\qquad A(\bar u)^{\dagger} = -A(\bar u),\qquad D^{\dagger} = D \preceq 0,
  \label{eq:frozen-main}
\end{equation}
for $q = (u, v, w, p_{\mathrm{s}})$, with $u = (u, v, w)$ the velocity and $p_{\mathrm{s}} = \mathrm{Ma}\,p$ the scaled
pressure (Fig.~\ref{fig:framework}b). The convection--acoustic generator $A(\bar u)$ is anti-Hermitian for every real
$\bar u$, without a divergence-free condition: each advection block is real skew-symmetric, and on the staggered grid
the discrete divergence is the negative transpose of the gradient, which keeps the acoustic coupling
skew~\cite{Bosch2025}. The diffusion operator $D = \nu L_h$, with $L_h$ the seven-point Laplacian, is real symmetric,
negative semidefinite and independent of the flow, because EDAC diffuses pressure with the momentum viscosity
(Supplementary Note~\ref{note:A}). Hence $e^{A\Delta t}$ is unitary, as in Hamiltonian simulation of
advection~\cite{Brearley2024}, and $e^{D\Delta t/2}$ is a contraction.

One step combines a classical predictor with a quantum-compatible corrector (Fig.~\ref{fig:framework}a,c):
\begin{equation}
  \begin{gathered}
    \tilde q = C_{48}\!\left[e^{D\Delta t/2}\right] C_{48}\!\left[e^{A(u_n)\Delta t/2}\right] q_n,\qquad
    u_{\mathrm{mid}} = \Re\,\tilde u,\\
    q_{n+1} = e^{D\Delta t/2}\, e^{A(u_{\mathrm{mid}})\Delta t}\, e^{D\Delta t/2}\, q_n .
  \end{gathered}
  \label{eq:step}
\end{equation}
The predictor evaluates degree-48 Chebyshev series $C_{48}$ classically and keeps only the midpoint velocity; the
corrector is a symmetric (Strang) splitting~\cite{Strang1968} of the system frozen at $u_{\mathrm{mid}}$.
Equation~\eqref{eq:step} shows the exact factors; the implementation replaces each corrector factor by a degree-48
polynomial approximation (equation~\eqref{eq:gqsp}). The operator $A$ is rebuilt classically twice per step, from $u_n$
and from $u_{\mathrm{mid}}$, using real velocities only; $D$ is built once, and no pressure-Poisson equation is solved
after the initial condition.

The splitting is what gives the frozen step a structure that quantum signal processing can use. The unsplit generator
$L = A + D$ is normal only if advection commutes with diffusion, since $LL^{\dagger} - L^{\dagger}L = 2[A, D]$
(Supplementary Note~\ref{note:C}, §9.1); its exponential is therefore not a function of the eigenvalues of a Hermitian
or unitary operator, nor of the singular values of $L$, and cannot be applied as one polynomial of a block encoding of
$L$ by quantum signal processing or singular value transformation. After splitting, the step consists of one unitary
factor and two contractions, each a function of a normal operator whose spectrum lies on a known axis
(Fig.~\ref{fig:framework}b,c). Each factor then has a bounded target and a success probability that does not depend on
a condition number. A linear combination of Hamiltonian simulations~\cite{AnLiuLin2023,AnChildsLin2026} applies the
unsplit step at comparable query cost and without splitting error, but with a quadrature register; the split step needs
neither that register nor the history registers of solvers based on linear systems. A series encoding of the unsplit
step that ignores this structure would, without amplification, succeed with a probability of at most about $10^{-9}$
per step at the production parameters, against $1/64$ per step for the synthesis used here
(Section~\ref{sec:implementation}); this is a weak baseline, because uniform singular-value
amplification~\cite{Fang2023} restores a success of order one at additional query cost (Supplementary
Note~\ref{note:C}, §9.4).

The price of the splitting is a commutator error, whose structure the discretization controls. On the uniform periodic
staggered grid with equal momentum and pressure diffusivities, diffusion commutes exactly with the acoustic part of the
generator, $[D, A_{\mathrm{ac}}] = 0$ (with walls this commutation is not established), so the splitting is exact for a
uniform advecting velocity, and its error is bounded by nested commutators with explicit constants and no growth
factor, for any $\Delta t$ (Supplementary Note~\ref{note:C}, Lemma S2 and Proposition S3). The stiff acoustic coupling
therefore enters the error only through a cross term with the commutator of advection and diffusion; this term
dominates the error, scales as $1/\mathrm{Ma}$ and has no counterpart in a projected formulation. On three saved
production states the realized splitting error of a frozen step is $1.9$--$6.7\times10^{-6}$ of the state norm, 255--286
times smaller than if the acoustic coupling were also split from advection, which is why the coupling is kept inside
the unitary factor.

The predictor deals with the nonlinearity. Freezing the advecting velocity at the start of a step commits a local error
of second order in $\Delta t$, so the frozen-coefficient step is only first-order accurate however the frozen system is
split. Freezing it at a midpoint velocity predicted by a classical half-step, an exponential-midpoint step of Magnus
type~\cite{Hochbruck2010}, removes this error at leading order and restores second-order convergence
(Section~\ref{sec:verification}). The predictor and the coefficient refresh are classical and are the only nonlinear
operations of the step. Re-linearizing at every step also keeps the quantum state at the size of the discretized field,
with no embedding into a larger linear space and no truncation order (Supplementary Note~\ref{note:C}, §10).

Within each step the convection--acoustic factor conserves the discrete energy $\|u\|^2 + \|p_{\mathrm{s}}\|^2$ of the
frozen transport and the diffusion factors can only remove it, so every exact factor of the corrector has norm at most one
for any $\Delta t$ and introduces no linear stability restriction; the polynomial approximations exceed this bound by at
most their approximation error (Supplementary Note~\ref{note:C}, §7). These are properties of each frozen step, not a
proof of stability or conservation for the nonlinear integration.

\subsection{Quantum implementation of the split factors}\label{sec:implementation}

Each factor acts through a Hermitian operator of norm at most one, $\bar H_A = -iA(u_{\mathrm{mid}})/\lambda_A$ or
$\bar H_D = D/\lambda_D$, accessed by a self-inverse block encoding $B$ ($B^2 = I$, $\langle 0|B|0\rangle = \bar H$).
The diffusion encoding is a linear combination of shift unitaries that does not depend on the flow. The convection
encoding is a six-branch linear combination of unitaries on the staggered layout whose top-left block equals
$-iA(\bar u)/\lambda_A$ by construction, with exactly the normalization $\lambda_A$ used in the simulations
(Supplementary Note~\ref{note:B}); it carries the whole dependence of the step on the flow. A
generalized-quantum-signal-processing sequence $W_f$ of controlled qubitization walks of
$B$~\cite{LowChuang2019,Motlagh2024} then implements
\begin{equation}
  \bigl(\langle 0| \otimes I\bigr)\, W_f \,\bigl(|0\rangle \otimes I\bigr) = \tfrac12 P_f(\bar H),\qquad
  P_f \approx f,\qquad f_A(x) = e^{i\tau_A x},\quad f_D(x) = e^{\tau_D x},
  \label{eq:gqsp}
\end{equation}
with the normalized evolution times $\tau_A = \lambda_A\Delta t$ and $\tau_D = \lambda_D\Delta t/2$ and with $P_f$ the
Chebyshev truncation of $f$ at degree 48, so that post-selecting the ancillas on $|0\rangle$ applies the corresponding
factor of equation~\eqref{eq:step} up to the constant $1/2$. The success probability is $\|P_f(\bar H)\psi\|^2/4$ for a
normalized input $\psi$. For the unitary convection factor it is $1/4$ for every input, exactly for the target and to
within the polynomial approximation error for $P_f$; for diffusion it is $1/4$ times the fraction of the norm that the
contraction retains, which depends on the state and was 0.2497--0.2500 in the production run. The design success of a
step, $(1/4)^3 = 1/64$, is therefore set by the normalization alone. The factor $1/2$ belongs to the synthesis used
here, not to the factors: for unit-normalized syntheses of the same factors the success probability of a step equals,
up to the synthesis margins, the fraction of the norm that the frozen step retains (Supplementary Note~\ref{note:C},
Propositions 1--3). These constructions are proved but not implemented; the diffusion factor would need a modified
encoding, and with the present synthesis routines the success of a step would be about 0.96. The three factors use 53,
79 or 78, and 53 queries, 184--185 per step: the Laurent synthesis adds inverse walks, so degree and query count differ
(Methods).

Simulating these circuits at $256^3$ requires a representation of the full ancilla--system state. Because a
self-inverse block encoding confines the GQSP sequence to two-dimensional invariant subspaces of
qubitization~\cite{LowChuang2019}, the state of the whole sequence is represented exactly by four system-size vectors,
for every encoding with $B^2 = I$ and $\langle 0|B|0\rangle = \bar H$. This qubitized-pair representation yields the
post-selected output, the success probability and the ancilla-sector populations of the circuit without storing the
ancillas, which makes exact classical emulation of the corrector at $256^3$ ($n = 26$ system qubits) possible.

\subsection{Verification of the quantum-compatible step}\label{sec:verification}

\begin{figure}[tbp]
\centering
\includegraphics[width=\linewidth]{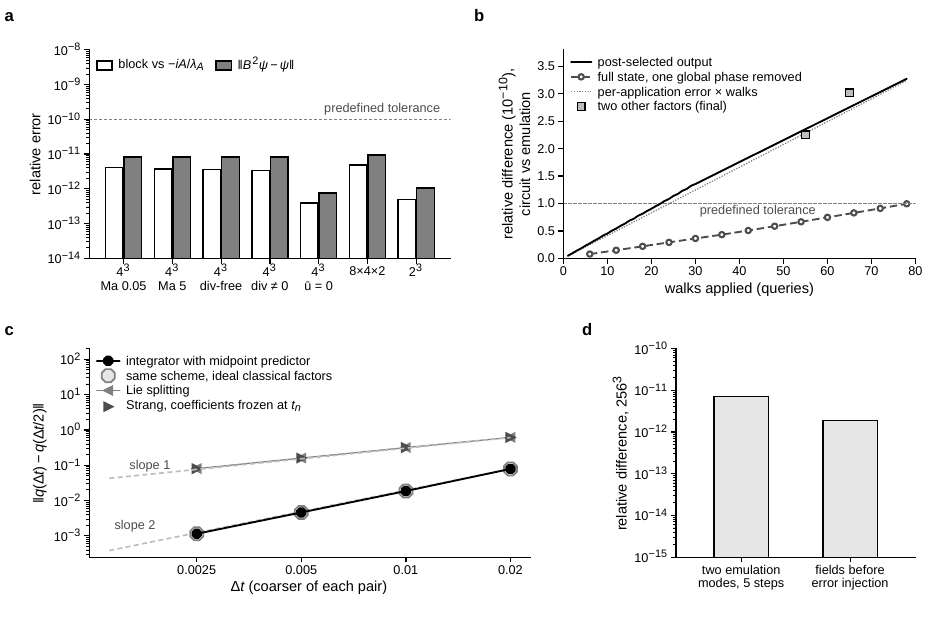}
\caption[{Verification of the staggered quantum-compatible step.}]{\textbf{Verification of the staggered
quantum-compatible step.} \textbf{a}, Operator representation: for the compiled staggered convection block encoding,
relative Frobenius difference between its top-left block and the production operator $-iA(\bar u)/\lambda_A$ on all
basis states (white), and the largest $\|B^2\psi - \psi\|$ over random full states (grey), for seven test cases (Ma,
Mach number; div-free and div $\neq 0$, solenoidal and non-solenoidal advecting fields; $\bar u = 0$, purely acoustic);
dashed, predefined tolerance. \textbf{b}, Circuit action: its GQSP factor at the production normalized evolution time on
a $4^3$
grid, relative difference between the circuit and the qubitized-pair emulation after each walk, for the post-selected
output (solid) and for the full ancilla--system state after removal of one global phase (circles, a later diagnostic);
dotted, the independently measured round-off of one application of the compiled encoding times the number of walks;
squares, final values of two other factors; dashed, predefined tolerance, which the raw difference exceeds.
\textbf{c}, Temporal accuracy on a $16^3$ staggered grid: difference between solutions with steps $\Delta t$ and
$\Delta t/2$; slopes 1 and 2 dashed. \textbf{d}, Consistency of the two emulation modes at $256^3$: relative difference
of scalar diagnostics over five steps and of the fields before error injection.}\label{fig:verification}
\end{figure}

We tested the step at three levels (Fig.~\ref{fig:verification}): whether the block encoding represents the intended
operator, whether the GQSP circuit reproduces the intended action of a factor, and whether the nonlinear integrator
reaches its design order in time.

\emph{Operator representation.} The staggered convection encoding used by the integrator was built as a circuit and
checked on every basis state in seven test cases on $4^3$, $8\times4\times2$ and $2^3$ grids, covering acoustically and
advectively dominated, solenoidal and non-solenoidal advecting fields and the purely acoustic case. Its top-left block
reproduces the production operator $-iA(\bar u)/\lambda_A$ to at most $4.8\times10^{-12}$ (relative Frobenius norm), and
the compiled circuit is self-inverse to at most $9.4\times10^{-12}$, both within the predefined tolerance of $10^{-10}$
(Fig.~\ref{fig:verification}a).

\emph{Circuit action.} The GQSP convection factor at the production normalized evolution time (78 queries on a $4^3$
grid) agrees
with the qubitized-pair emulation used for the production runs to $3.3\times10^{-10}$ (Fig.~\ref{fig:verification}b).
This exceeds the predefined tolerance of $10^{-10}$, and the check is reported as failed. The difference grows linearly
with the number of walks, at the rate of the round-off of one application of the compiled encoding, and is mostly a
global phase; the comparison therefore confirms the construction to the floating-point floor of a simulation
of about $3\times10^{7}$ gates rather than to the predefined tolerance (Methods). The diffusion encoding and a collocated
analogue of the convection encoding agree with Qiskit statevector evolution~\cite{Qiskit2024} within their predefined
tolerance of $10^{-9}$ (Extended Data Fig.~\ref{edfig:collocated}). These are small-grid checks: at $8^3$ only sampled
checks were made, and the $256^3$ runs evaluate the corrector in exact classical emulation without executing a circuit.
The two emulation modes used for those runs agree to $7.0\times10^{-12}$ in scalar diagnostics over five steps and to
$1.9\times10^{-12}$ in the fields before error injection (Fig.~\ref{fig:verification}d), a consistency check of the
emulation, not of a circuit.

\emph{Temporal accuracy.} On a $16^3$ staggered grid, Strang splitting with coefficients frozen at the start of the
step converges at first order (observed order 0.98), as the freezing error implies, whereas the midpoint predictor
restores second order (2.08, 2.02 and 2.00 over successive refinements; Fig.~\ref{fig:verification}c). This holds for a
smooth test state at acoustic Courant numbers up to 1.02; the production runs use 1.63, where the corrector factors
remain bounded but the order of accuracy has not been measured.

\subsection{Three-dimensional decaying turbulence}\label{sec:turbulence}

We applied the integrator to decaying homogeneous isotropic turbulence in a periodic box on a $256^3$ staggered grid at
$\mathrm{Re} = 800$ and $\mathrm{Ma} = 0.05$, for 1,650 steps to $t = 3.3$, with the corrector factors evaluated in
exact classical emulation. The initial velocity has a Passot--Pouquet-type spectrum~\cite{Passot1987} and the initial
pressure is balanced so that the discrete divergence does not change at $t = 0$ (Methods). Over the run the kinetic
energy falls from 1.43 at $t = 0.5$ to 0.21 at $t = 3.0$, and the Taylor-microscale Reynolds number from about 200 to
41, through a dissipation peak near $t = 1.5$ (Extended Data Table~\ref{edtab:stats}).

\begin{figure}[tbp]
\centering
\includegraphics[width=\linewidth]{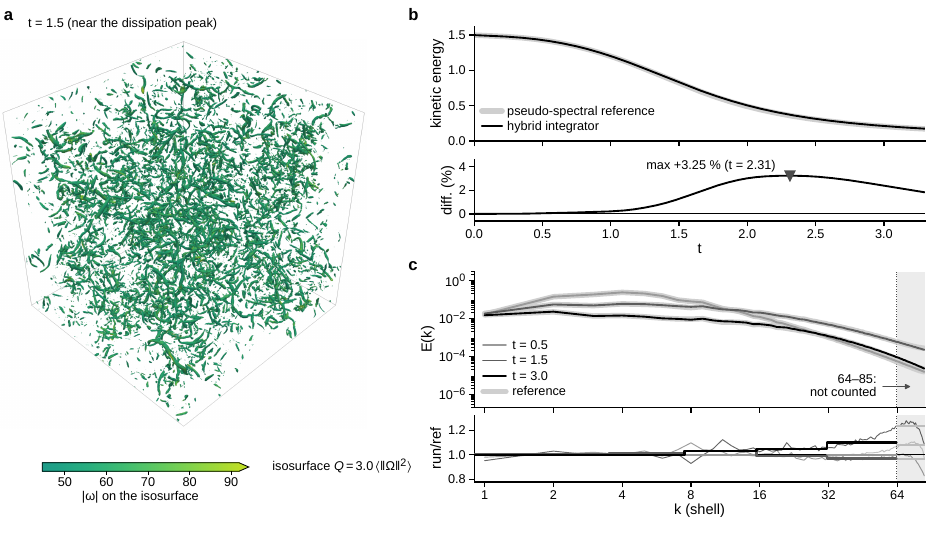}
\caption[{Three-dimensional decaying turbulence integrated by the hybrid integrator.}]{\textbf{Three-dimensional
decaying turbulence integrated by the hybrid integrator.} $256^3$ staggered grid, $\mathrm{Re} = 800$,
$\mathrm{Ma} = 0.05$, corrector factors in exact classical emulation. \textbf{a}, Vortex tubes at $t = 1.5$, near the
dissipation peak: isosurface $Q = 3.0\langle\|\Omega\|_F^2\rangle$ (the 98th percentile of positive $Q$), coloured by the
vorticity magnitude $|\omega|$, computed from the production velocity field (Methods); an illustration, not a validation.
\textbf{b}, Kinetic energy of the run and of an in-house pseudo-spectral reference from the same initial condition (top)
and their relative difference (bottom; maximum $+3.25\%$ at $t = 2.31$). \textbf{c}, Shell energy spectra at $t = 0.5$,
1.5 and 3.0 (top) and run-to-reference ratios (bottom; steps, band averages; faint lines, single shells). Shading,
shells 64--85 next to the reference's dealiasing limit, shown but not counted in the comparison.}\label{fig:turbulence}
\end{figure}

Near the dissipation peak, isosurfaces of $Q$, the second invariant of the velocity-gradient tensor, show thin,
elongated vortex tubes throughout the domain (Fig.~\ref{fig:turbulence}a). The level is three times the volume mean of
$\|\Omega\|_F^2$, the squared norm of the rotation-rate tensor, and is exceeded by the most strongly rotation-dominated
2\% of the points with $Q > 0$ (Methods). The visualization illustrates the three-dimensional vortical structure of the
integrated field; agreement with a reference solution is assessed through the energy and its spectra.

The quantitative test is a comparison with an in-house pseudo-spectral solution from the same initial condition. The
kinetic energy agrees within 1.3\% up to the dissipation peak and within 3.25\% over the whole run, with the largest
difference at $t = 2.31$ (Fig.~\ref{fig:turbulence}b). At large scales, in the two lowest wavenumber bands (shells
1--7), which contain the spectral peak, the band-averaged energy spectra agree within 1.6\% at the six compared times,
$t = 0.5$, 1.0, \ldots, 3.0; at smaller scales the band ratios depart further from one (Fig.~\ref{fig:turbulence}c;
Methods and Supplementary Note~\ref{note:D}). The dissipation rate of the run, evaluated
with the scheme's own Laplacian, differs from the shell-convention value of the reference by $-2.4\%$ to $+5.5\%$; with
the shell convention on both sides the difference is $+0.7\%$ to $+11.0\%$ (Extended Data Table~\ref{edtab:stats}). The
integrator, with its quantum-compatible factors evaluated exactly, therefore follows the decay of the energy and of its
large-scale spectrum through the dissipation peak and the subsequent decay. The discrete divergence stays
bounded (Extended Data Fig.~\ref{edfig:divergence}). The staggered layout matters here: at $t = 3.13$ a collocated
$256^3$ run had 49 and 60 times the divergence maxima of two staggered runs, an excess that we attribute to the seven
additional Nyquist null modes of the collocated operators.

\subsection{Computational structure and resources of a step}\label{sec:cost}

The formulation fixes the computational character of a step before any flow is specified (Extended Data
Table~\ref{edtab:comparisons}; Methods). Time marching contains no linear solve: the pressure is advanced by the same
unitary and contractive factors as the velocity, so no condition number enters the query count or the success
probability of a step. Each factor is a function of a normal operator whose spectrum lies on a known axis and within a
known bound, the normalization of its block encoding, so the polynomial that implements it is determined by a
normalized evolution time and a tolerance alone. Every exact factor is bounded for any $\Delta t$: the acoustic
stiffness of the artificial-compressibility system, which limits the time step of explicit schemes, appears instead in
the normalized evolution time of the unitary factor ($\tau_A = 10.2$--$10.9$ in production, of which the acoustic part
contributes 9.8).

The quantum resources follow from this structure. A grid of $N^3$ points with four fields occupies
$n = 3\log_2 N + 2$ system qubits, 26 at $256^3$ and three more for each doubling of the resolution. The encodings add
a number of ancillas that is independent of $N$, and at $256^3$ the whole corrector acts on 32 qubits when ancillas are
reused (Supplementary Note~\ref{note:C}, §5.4); the advecting field and the field between steps remain classical data.
For black-box access at the encoding normalization, the unitary factor can be
implemented with
\begin{equation}
  Q_A = \Theta\!\left(\tau_A + \frac{\log(1/\eta)}{\log\bigl(e + \log(1/\eta)/\tau_A\bigr)}\right)
  \label{eq:queries}
\end{equation}
block-encoding queries for an approximation error $\eta$, the optimal scaling in that model~\cite{Gilyen2019};
unit-normalized GQSP constructions need $O(\tau_A + \log(1/\eta))$, and the diffusion factors ($\tau_D = 0.025$) only
the tolerance-dependent part (Supplementary Note~\ref{note:C}, §4). The
synthesis used here takes 78 or 79 queries for the unitary factor, against a reference degree of 31--32 (the degree at
which the Chebyshev truncation of $e^{i\tau_A x}$ reaches $\eta = 10^{-12}$), and 184--185 for a step, per attempt at
the design success of $1/64$ (Section~\ref{sec:implementation}). In the query model, in which the block encodings are
given and every call is counted once, the frozen step has the cost profile of quantum simulation of sparse
operators~\cite{LowChuang2019}. At fixed Courant numbers, $\tau_A$ and, with it, the query count of the unitary factor
per step do not depend on the grid, while the number of steps grows in proportion to $N$. A classical evaluation of the
same polynomials~\cite{Hochbruck2010} needs the same order of operator applications, each a stencil product of
$\Theta(\mathcal{N})$ operations, $\mathcal{N} = N^3$, where the corrector makes a query on the $n$-qubit register; the
two are different units, and what a query costs depends on the access to the advecting field (below). The flow enters
the unitary factor only: the diffusion operator, its block encoding and its GQSP phases are the same in every step of a
run. The classical refresh changes the angles of the circuits that load the advecting field and, for the unitary
factor, the normalization $\lambda_A$ of its block encoding, the GQSP phases and the query count (78 or 79); the
structure of the encodings does not change.

The removal of the pressure solve can be quantified in calls to the block encodings. A projection method needs a
linear solve at every step, whose quantum cost grows with a condition number~\cite{Harrow2009,Costa2022}: where no fast
diagonalizing transform is available, a projector applied by quantum singular value transformation needs
$\Theta(\kappa_G\log(1/\eta))$ calls~\cite{Gilyen2019}, with the condition number $\kappa_G$ of the discrete gradient
proportional to $N$. Pressure evolution replaces this multiplicative dependence on a condition number by an additive
dependence on an evolution time: the acoustic coupling adds its stiffness to the normalized evolution time of the
unitary factor, an addition that does not grow with the grid at fixed Courant numbers. With reference degrees at equal
$\Delta t$ and $\eta$, the unitary factor takes 31--32 calls per step at $256^3$, about 20 of them for the acoustic
part, whereas a projected step built from the same machinery takes 9--12 for its transport factor plus about
$3.9\times10^3$ for its projector. For transport and pressure coupling, the projection-free step thus needs about a
hundred times fewer calls per step (122--126), a ratio that grows in proportion to $N$ at fixed Courant numbers
(Supplementary Note~\ref{note:C}, Proposition E1; Methods; Extended Data Table~\ref{edtab:comparisons}).

The bounded factors also change how stiffness enters time marching. Explicit integration of the same weakly
compressible system is bound by the acoustic Courant number: at their stability limits, explicit schemes would need
$3.2$--$5.4\times10^3$ steps for this run, 1.9--3.3 times the 1,650 steps taken here at an acoustic Courant number of
1.63 (Supplementary Note~\ref{note:C}, Proposition E3). The corrector has no such stability limit: the stiffness is paid in queries inside a step rather than in additional
steps, each of which would add a classical refresh and a crossing of the quantum--classical interface.

What these query counts cost in elementary operations depends on the interface between the quantum register and the
classical data, that is, on how the advecting field is loaded into the operator and how the state is prepared and read
out, as for quantum linear-algebra algorithms generally~\cite{Dalzell2025}. A diffusion query is a fixed circuit on the
index registers ($6.2\times10^3$ and $1.8\times10^4$ CNOT gates on $8^3$ and $16^3$ grids). A convection query costs
what access to the advecting field costs: polylogarithmic in $\mathcal{N}$ for compactly described fields, logarithmic
in depth with a quantum random-access memory of $\Theta(\mathcal{N})$ components~\cite{Giovannetti2008}, and
proportional to $\mathcal{N}$, up to a logarithm, when an arbitrary field is compiled into gates, in which case the
frozen step has the same scaling as its classical evaluation (Supplementary Note~\ref{note:C}, Proposition 5). The loop
analysed here is of the last kind, because neither a compact description of a turbulent field nor a quantum memory is
assumed, and by design the full field is read out, processed classically and re-prepared at every step. These
interface operations, rather than the polynomial synthesis of the factors, dominate the cost of the complete loop, whose accounting, including the comparison with a classical step, is given in Methods.
The structure of the step, its query counts and the boundedness of its factors do not depend on the interface and carry
over to any loop that uses the same step with a different interface. Access to the advecting field and the readout of
the solution are therefore the parts of the architecture with most to gain (Discussion).

\subsection{Propagation of a repeated per-step error}\label{sec:transfer}

Every step of the integrator ends at a quantum--classical interface, where the output is read out and re-encoded and
where errors from sampling, imperfect gates or approximate preparation would enter. Repeated perturbations are known to
grow in chaotic flows~\cite{Senoner2008,Karp2025,Ma2024} and in noise models of quantum flow
solvers~\cite{Yang2026,Wang2025}, and sampling noise in a hybrid scheme can drive a chaotic system arbitrarily far from
its exact solution~\cite{Andress2025}. Growth alone, however, does not show whether the dynamics merely accumulate the
injected errors or transform them. We therefore repeated the turbulence run with an error of exactly known size injected
at the interface (Fig.~\ref{fig:framework}a, crossed circle): once per step, before decoding, the post-selected state
$\psi$ is replaced by a norm-preserving rotation $\psi'$ with $\|\psi' - \psi\|/\|\psi\| = \varepsilon$, a per-step
relative state error. Its direction combines a white complex Gaussian component (weight 0.91) with a structured
component (weight 0.09) whose per-qubit weights follow the two-qubit gate census of a compiled convection circuit
(Methods). Errors are injected from $t = 1.41$, and the noisy fields $V$ are compared with the noise-free run $U$ through
\begin{equation}
  \delta = \frac{\|\Re\,\Delta\|}{\|U\|},\qquad \Delta = V - U,
  \label{eq:delta}
\end{equation}
over velocity and scaled pressure. The decoded fields are complex because the injected directions are, and $\delta$
measures the real part, the only part that the refresh reads.

\begin{figure}[tbp]
\centering
\includegraphics[width=\linewidth]{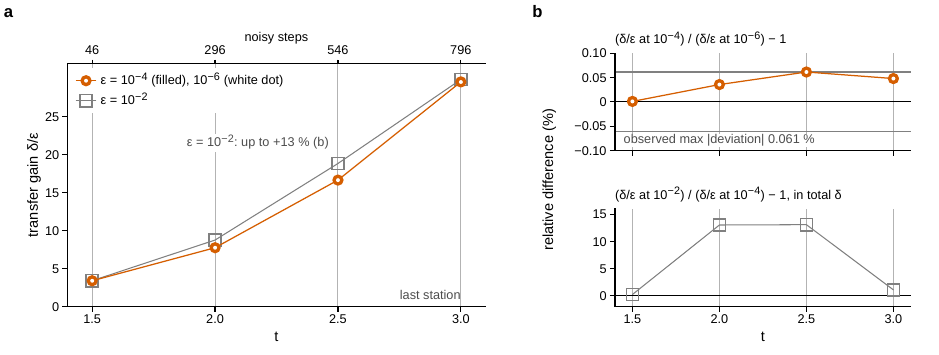}
\caption[{Transfer of a repeated per-step error.}]{\textbf{Transfer of a repeated per-step error.} The $256^3$ run of
Fig.~\ref{fig:turbulence} with a norm-preserving error of relative size $\varepsilon$ inserted into the post-selected
state at every step from $t = 1.41$ (Fig.~\ref{fig:framework}a); one noise realization, with the same random draws at
every amplitude. \textbf{a}, Transfer gain $\delta/\varepsilon$ (real part, full state) at $\varepsilon = 10^{-4}$
(filled circles) and $10^{-6}$ (white dots; the two coincide at this scale) and at $10^{-2}$ (squares); top axis, number
of steps with error injection; vertical lines, measurement stations. \textbf{b}, Relative difference of the gain between
$\varepsilon = 10^{-4}$ and $10^{-6}$ (top; horizontal lines, observed maximum deviation) and between $10^{-2}$ and
$10^{-4}$ (bottom).}\label{fig:transfer}
\end{figure}

The transfer gain $\delta/\varepsilon$ rose from 3.4 at $t = 1.5$ to 7.8, 16.7 and 29.6 at $t = 2.0$, 2.5 and 3.0, after
796 injections (Fig.~\ref{fig:transfer}a). It is a cumulative ratio, not the amplification of a single error, and its
magnitude depends on the configuration and the error direction: it is lower on $64^3$ and $128^3$ grids, an ordering
across co-varying configurations rather than a trend, and 14--47\% above that of a purely white direction (Extended
Data Table~\ref{edtab:grids}). At $\varepsilon = 10^{-6}$ and $10^{-4}$, run with the same random draws, the two gains
agree within 0.061\% (Fig.~\ref{fig:transfer}b), consistent with proportional scaling at these amplitudes. At
$\varepsilon = 10^{-2}$ the gain departs from this scaling by up to 13\% at $t = 2.0$--$2.5$, mainly through the
contraction that each rotation imposes on the state (Methods; Extended Data Fig.~\ref{edfig:amplitude}). These results
come from one noise realization and one initial condition.

\subsection{Feedback versus accumulation}\label{sec:feedback}

The integrator supplies a control for this question (Extended Data Fig.~\ref{edfig:structure}a). Given the refreshed
coefficients, a step is a real linear map, and the refresh reads only the real part of the decoded fields. Both parts of
the contamination are therefore advected, coupled acoustically and diffused by the same operator, but only the real part
changes that operator at the next refresh. To first order the two parts differ only through the linearized advective
coupling of the velocity contamination $u'$ with the noise-free flow, $(u'\cdot\nabla)U$, the standard coupling term of
the Navier--Stokes equations linearized about a base state~\cite{Moulin2019}, plus $\frac12(\nabla\cdot u')U$ in the
skew-symmetric form, which vanishes for solenoidal $u'$. We call the real part the fed-back part and the imaginary part
the passive part. The control is approximate: the two parts are different random fields with equal statistics, so each
is normalized by its share of the injected energy; the injection exchanges a fraction of order $0.17\varepsilon$ between
them, and its contraction acts on the real part only. Comparing $\varepsilon = 10^{-4}$ with $10^{-6}$ bounds these two
effects below $10^{-3}$ of the result in total $\delta$ (Methods).

\begin{figure}[tbp]
\centering
\includegraphics[width=\linewidth]{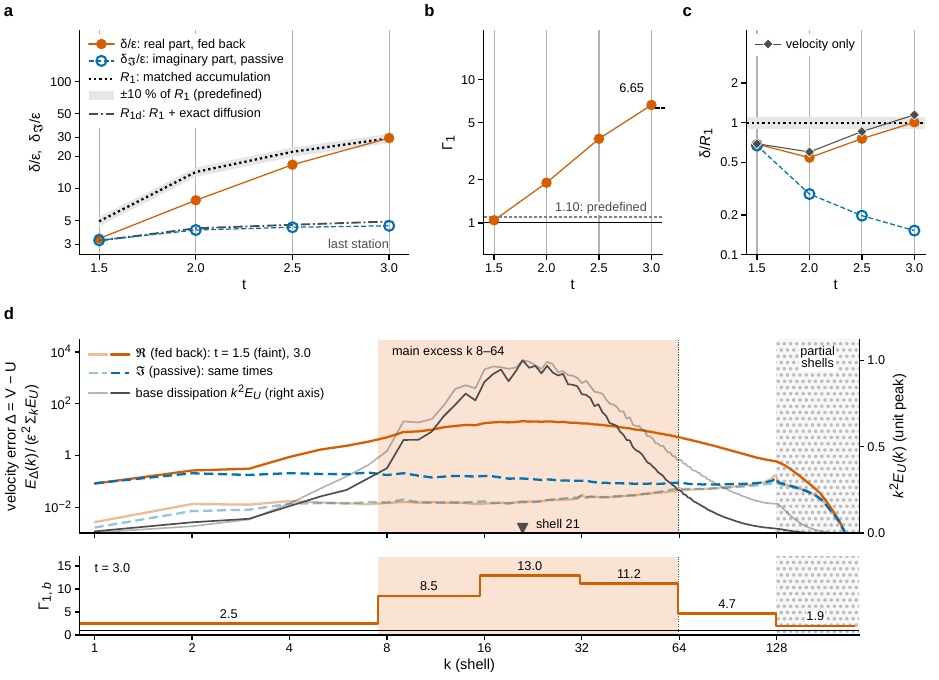}
\caption[{Feedback versus accumulation.}]{\textbf{Feedback versus accumulation.} Same run as Fig.~\ref{fig:transfer},
$\varepsilon = 10^{-4}$. \textbf{a}, Contamination of the fed-back (real; vermillion) and passive (imaginary; blue) parts,
$\delta/\varepsilon$ and $\delta_{\Im}/\varepsilon$; dotted, matched accumulation $R_1$ of the fed-back part with the
predefined $\pm10\%$ tolerance (grey band); dash-dotted, accumulation with exact diffusion $R_{1\mathrm{d}}$ of the
passive part (Methods). \textbf{b}, Growth of the fed-back part relative to the passive part, $\Gamma_1$; ticks,
alternative share normalizations (6.36 and 6.38); dashed, predefined guide value. \textbf{c}, Each part against its own
matched accumulation, $\delta/R_1$ (filled) and $\delta_{\Im}/R_{1,\Im}$ (open); grey band, predefined $\pm10\%$
tolerance; diamonds, velocity only. \textbf{d}, Velocity-error spectra $E_\Delta(k)$ of the fed-back (solid) and passive
(dashed) parts at $t = 1.5$ (faint) and 3.0, divided by $\varepsilon^2$ times the base-flow kinetic energy, with the
base dissipation spectrum $k^2E_U$ (grey, right axis, unit peak). Shading, shells 8--64, where the excess is
concentrated; dotted line, $k = 64$, the lower edge of the shells not counted in the comparison of
Fig.~\ref{fig:turbulence}c; triangle, shell 21,
where the base dissipation peaks; dotted hatching, partially filled shells beyond $k = 128$. Lower strip, band excess
$\Gamma_{1,b}$ of the fed-back part at $t = 3.0$.}\label{fig:feedback}
\end{figure}

The passive part grew far less than the fed-back part (Fig.~\ref{fig:feedback}a). The growth of the fed-back part
relative to the passive control, $\Gamma_1 = (\delta/\sqrt{s_{\Re}})/(\delta_{\Im}/\sqrt{s_{\Im}})$ with
$\delta_{\Im} = \|\Im\,\Delta\|/\|U\|$ and $s$ the injected energy shares, rose from 1.04 at $t = 1.5$ to 6.4--6.7 at
$t = 3.0$, the range spanning alternative share normalizations (Fig.~\ref{fig:feedback}b), and exceeded a predefined
guide value of 1.10 from $t = 2.0$. A weakly chaotic $64^3$ flow at $\mathrm{Re} = 120$ with eight noise realizations
gives a much weaker excess at $t = 3.0$, 1.15--1.37, which is context rather than an uncertainty for the $256^3$ value
(Extended Data Fig.~\ref{edfig:passive}).

This relative enhancement does not make the total exceed matched accumulation $R_1$, the contamination that the same
injections would give if each kept its norm. The ratio $\delta/R_1$ was 0.69, 0.54, 0.76 and 1.01 at the four stations
(Fig.~\ref{fig:feedback}c): below accumulation up to $t = 2.5$ and comparable to it at $t = 3.0$, within a predefined
$\pm10\%$ tolerance for the full state (for the velocity alone the value, 1.15, lies above that tolerance). The
near-equality at $t = 3.0$ is a crossing at the last measured station, not a balance. By an exact identity,
$\delta/R_1 = \Gamma_1 \times (\delta_{\Im}/R_{1,\Im}) \times f$ with $f = 0.98$--$0.99$, so at $t = 3.0$ a passive part
that retains only 15\% of non-dissipative accumulation, multiplied by the 6.7-fold excess of the fed-back part, gives
the observed total. In this run the total alone does not reveal the feedback.

The two parts also differ in structure (Fig.~\ref{fig:feedback}d; Extended Data Fig.~\ref{edfig:structure}b). Injected
mostly above $k = 64$, the fed-back velocity error comes to peak at shell 21 by $t = 3.0$, where the dissipation of the
base flow also peaks, whereas the passive part stays weighted towards the grid scale; the band excess relative to the
passive control is 8.5--13.0 in $k = 8$--64 against 1.9--4.7 in the other bands. The fed-back error also becomes almost
purely solenoidal (dilatational energy fraction 0.29 at $t = 1.5$ and 0.005 at $t = 3.0$), whereas the passive part
stays near the white-noise value of 1/3, and the pressure contamination shows no excess comparable to that of the
velocity. These differences are consistent with the feedback but are not attributed to it alone, because the two parts
are different random fields weakly coupled by the injection. At the largest scales, $k = 1$--4, no excess is resolved;
there, the structured component of the error model is the majority origin of the velocity-error energy under the
as-compiled register order on three grids, an attribution that depends on the register order (Extended Data
Fig.~\ref{edfig:origin}).

\section{Discussion}\label{sec:discussion}

We have presented a hybrid quantum--classical integrator for incompressible flow in which the nonlinearity is handled by
a classical midpoint predictor and everything delegated to the quantum side is a bounded matrix function of an operator
with a guaranteed structure. Its central idea is to let the structure of the discretized equations decide how each part
of a step is executed: the pressure is evolved rather than projected, convection and acoustic coupling form an
anti-Hermitian generator, diffusion forms a constant negative semidefinite one, and a symmetric splitting turns each
frozen step into a unitary factor between two contractions.

This structure is what makes a step of a nonlinear flow solver compatible with quantum signal processing. The unsplit
frozen generator is non-normal whenever advection and diffusion do not commute, and its exponential then cannot be
applied as one polynomial of a block encoding of the generator by quantum signal processing or singular value
transformation. After splitting, each factor is a function of a normal operator whose spectral range is known before
the step, its post-selection success follows from the normalization (one quarter for the unitary factor and close to
one quarter for diffusion in the synthesis used here), and its exact form has norm at most one for any time step. Skew-symmetric
advection on a staggered grid keeps the generator anti-Hermitian for any real advecting velocity, without requiring a
divergence-free field. Because diffusion commutes with the acoustic coupling on uniform periodic grids with equal
diffusivities, keeping that coupling with convection confines its contribution to the splitting error to a cross term,
and the splitting is exact for a uniform advecting velocity. Artificial compressibility replaces the pressure-Poisson
solve, whose quantum cost grows with a condition number where no fast diagonalizing transform exists, by acoustic
stiffness that is paid in block-encoding queries rather than in a stability limit on the time step. In exchange, the
formulation carries an $O(\mathrm{Ma}^2)$ modelling error, a splitting error dominated by an acoustic cross term, and
query counts that grow with the acoustic Courant number. These properties hold for each frozen step; the classical
midpoint predictor connects the steps, re-linearizing the flow without enlarging the state and restoring second-order
temporal convergence on a smooth test state.

The construction has been realized as circuits on small grids and exercised on turbulence in exact classical emulation.
The block encodings exist as self-inverse circuits whose blocks match the production operators on small grids, and the
GQSP factor reproduces the emulated action to the floating-point floor of a simulation of $3\times10^{7}$ gates, though
not to its predefined tolerance. An exact representation of the full ancilla--system state by four system-size vectors
then allows the corrector to be emulated at scale: the integrator carries three-dimensional decaying turbulence at
$256^3$ resolution over 1,650 steps through the dissipation peak, with the kinetic energy and the band-averaged
large-scale spectra close to an in-house pseudo-spectral reference. The
structure-preserving step thus remains usable for a strongly nonlinear, multi-scale flow.

The computational significance of the formulation lies in what it removes from the quantum part of a step and in what
it leaves there. It removes the linear solve and, with it, any dependence of the query count or of the success
probability on a condition number. What remains are polynomials of two block-encoded operators with known spectral
bounds, one of them identical in every step, on a register whose size grows logarithmically with the number of grid
points, a property of the register rather than of the storage of the flow data. Two consequences can be quantified
(Section~\ref{sec:cost}; Extended Data Table~\ref{edtab:comparisons}). First, the pressure coupling adds an increment
to the normalized evolution time, where a projector applied by quantum singular value transformation needs a number of
calls proportional to a condition number. In geometries without a fast diagonalizing transform, the projection-free
step therefore needs about a hundred times fewer block-encoding calls per step at $256^3$ for transport and pressure
coupling, a benefit that grows with the resolution. Second, because the factors are bounded for any time step, the run
took 1.9--3.3 times fewer steps than explicit integration of the same weakly compressible system would need at its
stability limit; each step saved removes one coefficient refresh and one crossing of the quantum--classical interface.

How these query-level properties translate into elementary operations is decided by the interface between the quantum
register and the classical data. In the loop analysed here an arbitrary advecting field is compiled into every query of
the unitary factor and, as in other hybrids with per-step readout~\cite{Bharadwaj2023}, the full field is read out and
re-prepared at every step; these operations dominate the cost of the loop (Methods). The costs of this interface are
separate, and each at present grows with the number of grid points. Access to the advecting field and preparation of
the state would need quantum memory or, where a field admits a compact description, structured loading. Readout would
have to deliver observables in place of full fields, and the predictor and the refresh of the operator would have to
work without processing the full field classically at every step.

Because every step crosses a quantum--classical interface, the integrator also exposes how repeated errors enter the
dynamics. Its refresh reads only the real part of the state, which turns the imaginary part of any injected error into an
in-run passive control. In the run analysed here, the fed-back part of the error grew 6.4--6.7-fold more than this
control, concentrated at the most dissipative scales and became solenoidal, whereas the total remained below or
comparable to matched accumulation; the total error alone would therefore have suggested simple accumulation. The
magnitude of the gain is specific to this configuration: it rests on one noise realization and one initial condition,
it differs across the grids and error directions tested, error growth depends on the resolved
scales~\cite{Budanur2022}, and decaying turbulence is more predictable than stationary turbulence~\cite{Metais1986}.
The error model is an effective end-of-step perturbation; noise on present processors is local and structured and need
not scramble into white noise~\cite{vandenBerg2023,Foldager2024}, and a stochastic effective channel represents it on
average only under randomized compiling~\cite{Wallman2016}. What carries over is the measurement design: a shared-draw
amplitude sweep that isolates the dependence on the error size, and a passive control wherever a step propagates a
component that its refresh never reads.

The design principles of the integrator can be stated independently of the configuration studied: a constraint is
evolved instead of being enforced by a solve, the terms of the discretized equations are grouped by the algebraic
structure of their generators rather than by their physical origin, and the nonlinearity is confined to a classical
update of coefficients. The same separation of anti-Hermitian transport from Hermitian dissipation could be applied to
other transport equations, including nonlinear ones such as those of flows with active scalars. Several further steps
follow from the present structure. Unit-normalized syntheses would raise the success probability of a step
from $1/64$ towards the fraction of the norm that the step retains. In geometries without a fast diagonalizing
transform, where the query-level benefit of pressure evolution would apply, the splitting error remains to be
established for discretizations in which diffusion no longer commutes exactly with the acoustic coupling. A quantum
realization of the predictor or less frequent coefficient updates could reduce the classical part of the loop and the
number of interface crossings. Execution of the factors on hardware, together with in-circuit noise models and multiple
noise realizations, would turn the error-transfer measurements into statistics of reliability. By matching the
structure of the discretized equations to bounded linear operations on a quantum register, the framework provides a
structure-preserving route to quantum-compatible integration of nonlinear flows.

\FloatBarrier
\section{Methods}\label{sec:methods}

\subsection{Equations and discretization}\label{sec:m-equations}

The integrator advances the EDAC Navier--Stokes equations~\cite{Clausen2013a,Clausen2013b} in code units, with the scaled
pressure $p_{\mathrm{s}} = \mathrm{Ma}\,p$:
\begin{equation}
  \begin{aligned}
    \frac{\partial u}{\partial t} &= -(u\cdot\nabla)u - \frac{1}{\mathrm{Ma}}\nabla p_{\mathrm{s}} + \nu\nabla^{2}u,\\
    \frac{\partial p_{\mathrm{s}}}{\partial t} &= -(u\cdot\nabla)p_{\mathrm{s}} - \frac{1}{\mathrm{Ma}}\nabla\cdot u
    + \nu\nabla^{2}p_{\mathrm{s}},
  \end{aligned}
  \label{eq:edac}
\end{equation}
where $\nu = 1/\mathrm{Re}$ and Ma is based on the initial root-mean-square velocity; the system approaches the
incompressible equations with $O(\mathrm{Ma}^{2})$ errors~\cite{Clausen2013b}. On a periodic staggered marker-and-cell
(MAC) grid~\cite{Harlow1965} of $N^{3}$ cells ($h = 2\pi/N$), $p_{\mathrm{s}}$ sits at cell centres and $u$, $v$, $w$ on
faces, and the forward-difference gradient $G$ and backward-difference divergence Div satisfy $\mathrm{Div} = -G^{T}$
exactly. Advection takes the skew-symmetric form~\cite{Morinishi1998} $\tfrac{1}{2}[(\bar{u}\cdot\nabla)f +
\nabla\cdot(\bar{u}f)]$ on all four fields, including $p_{\mathrm{s}}$. The operators of equation~\eqref{eq:frozen-main}
are
\begin{equation}
  A(\bar{u}) = -\left[A_{\mathrm{adv}}(\bar{u}) + \frac{1}{\mathrm{Ma}}\begin{bmatrix} 0 & G\\ \mathrm{Div} & 0\end{bmatrix}\right],
  \qquad D = \nu L_{h},
  \label{eq:frozen}
\end{equation}
with $L_{h}$ the periodic seven-point Laplacian on all four fields. Each advection block is real skew-symmetric for any
real coefficient field, without a divergence-free condition, and $\mathrm{Div} = -G^{T}$ makes the acoustic block skew,
so $A(\bar{u})$ is anti-Hermitian for every real $\bar{u}$; $D$ is Hermitian, negative semidefinite and constant
(Supplementary Notes~\ref{note:A} and~\ref{note:D}; Note~\ref{note:A} also treats the pressure-advection form).

\subsection{Initial condition}\label{sec:m-initial-condition}

The initial velocity has complex Gaussian modes with the energy spectrum
$E(k) = 16(2/\pi)^{1/2}(u_{\mathrm{rms}}^{2}/k_{p})(k/k_{p})^{4}\exp[-2(k/k_{p})^{2}]$, $k_{p} = 4$, $u_{\mathrm{rms}} =
1$, usually referred to as the Passot--Pouquet spectrum~\cite{Passot1987}, projected to be discretely solenoidal and
rescaled to the exact $u_{\mathrm{rms}}$. The pressure is EDAC-balanced, $\hat{p}_{\mathrm{s}} = \mathrm{Ma}\,
\hat{S}/|m|^{2}$, with $S$ the MAC divergence of the advection term and $m$ the MAC divergence symbol, so that the
discrete $d(\mathrm{Div}\,u)/dt$ vanishes at $t = 0$ (static check at $256^{3}$: maximum 241 with a uniform pressure,
$2.3\times10^{-11}$ balanced). The projection and pressure solve, both by FFT at $t = 0$, are the only field-size linear
solves of a run. The balanced pressure suppresses the acoustic start-up mode that artificial-compressibility initial data
excite; in the classical artificial-compressibility method even incompressible-consistent initial data excite an acoustic
mode with an $O(\mathrm{Ma}^{2})$ velocity amplitude~\cite{Ohwada2010}. Its dynamic effect was measured only in a $64^{3}$
collocated test; at $256^{3}$ only the static check exists (Extended Data Fig.~\ref{edfig:divergence};
Supplementary Note~\ref{note:D}).

\subsection{Time step and interface}\label{sec:m-step}

\emph{Encoding}: $\psi_{n} = q_{n}/\|q_{n}\|$ (state preparation is not modelled). \emph{Predictor} (classical):
$\psi_{\mathrm{mid}} = C_{48}[e^{D\Delta t/2}]\,C_{48}[e^{A(u_{n})\Delta t/2}]\,\psi_{n}$, with $C_{48}$ a degree-48
Chebyshev series applied directly to the state, without ancillas or post-selection; only
$u_{\mathrm{mid}} = \Re(\psi_{\mathrm{mid}}^{u})\,\|q_{n}\|$ (likewise $v$, $w$) is kept. \emph{Refresh} (classical):
$A$ and its normalization $\lambda_{A}$ are built from $u_{n}$ for the predictor and from $u_{\mathrm{mid}}$ for the
corrector, with real weights, so only $\Re(u)$ enters; $D$ is never rebuilt. \emph{Corrector}: the Strang
splitting~\cite{Strang1968} $e^{D\Delta t/2}e^{A(u_{\mathrm{mid}})\Delta t}e^{D\Delta t/2}$, each factor a GQSP
polynomial~\cite{Motlagh2024}, with phases synthesized classically every step, of $\bar{H}_{A} =
-iA(u_{\mathrm{mid}})/\lambda_{A}$ or $\bar{H}_{D} = D/\lambda_{D}$, targeting $e^{i\tau_{A}x}/2$ or $e^{\tau_{D}x}/2$.
The targets are truncated at Chebyshev degree 48 and converted into GQSP sequences by a Laurent shift with 48 inverse
controlled walks, trimming positive powers below $10^{-12}$; the factors then use 53, 79 or 78, and 53 queries (the
convection count uses $\lambda_{A}$, which the staggered encoding attains exactly). Post-selection succeeds with
probability $\|P_{f}(\bar{H})\psi\|^{2}/4$, 1/4 for convection (to the polynomial approximation error) and
0.249745--0.249987 for diffusion, computed from exact amplitudes and never sampled. \emph{Readout}: $q_{n+1} = \psi_{\mathrm{out}}\,\|q_{n}\|\cdot 8$, with the
deterministic design gain $8 = (1/(1/2))^{3}$; the decoded fields are complex. Readout and post-selection are logical
steps of the method: the emulation decodes the state exactly and performs no quantum measurement.

\subsection{Block encodings, emulation and verification}\label{sec:m-verification}

The staggered convection block encoding is a six-branch linear combination of unitaries whose top-left block is
$-iA(\bar{u})/\lambda_{A}$ by construction, with the production normalization $\lambda_{A}$, and which is self-inverse
(Supplementary Note~\ref{note:B}); it loads the advecting field with a number of gates proportional to the number of grid
points per query. The circuits were built in Qiskit~\cite{Qiskit2024}, compiled and simulated as statevectors. All
tolerances below were fixed before the checks were run and are reported unchanged. \emph{Block and self-inverse
checks.} Seven cases cover $4^{3}$, $8\times4\times2$ and $2^{3}$ grids, the production Mach number and an advectively
dominated field, divergence-free and non-solenoidal advecting fields and $\bar u = 0$. The relative Frobenius
difference between the circuit's top-left block, taken over all basis states, and $-iA(\bar u)/\lambda_{A}$ is at most
$4.8\times10^{-12}$, and $\|B^{2}\psi - \psi\|$ over random normalized full states is at most $9.4\times10^{-12}$ for the
compiled circuit ($2.2\times10^{-11}$ for the uncompiled circuit), against a tolerance of $10^{-10}$ for each.
\emph{GQSP checks.} On one $4^{3}$ field at the production Mach number, the convection factor at the production
normalized evolution time $\tau_{A} = 10.5$ (78 queries) was executed as a circuit in segments of forward and inverse
walks;
segmented execution reproduces whole-circuit execution on a four-query sequence to $4.6\times10^{-11}$ (tolerance
$10^{-10}$). The error measure is the relative 2-norm difference $\|\psi_{\mathrm{c}} - \psi_{\mathrm{e}}\|/
\|\psi_{\mathrm{e}}\|$ between the circuit output $\psi_{\mathrm{c}}$ and the qubitized-pair emulation
$\psi_{\mathrm{e}}$, for the post-selected system state and for the full ancilla--system state, with a tolerance of
$10^{-10}$ for each; the success probability and the ancilla-sector populations have absolute tolerances of $10^{-10}$,
and the comparison with the direct polynomial and with the exact exponential a tolerance of $10^{-9}$. The post-selected
difference is $3.3\times10^{-10}$ (full state $3.3\times10^{-10}$), alone and within the three-factor corrector
($3.3\times10^{-10}$ for two diffusion settings); the largest sector-population difference is $1.1\times10^{-10}$. Both
exceed their tolerance, and these checks are recorded as failed. The success probability agrees to
$4.4\times10^{-11}$ and the output with the exact exponential to $3.3\times10^{-10}$, within their tolerances. Two
further factors, at $\tau_{A} = 0.15$ (55 queries) and at $\tau_{A} = 3.0$ on an advectively dominated field
(65 queries), give $2.3\times10^{-10}$ and $3.0\times10^{-10}$, consistent with a difference set by the number of walks
rather than by $\tau_{A}$. \emph{Origin of the excess.} A diagnostic run after these checks recorded the difference after every walk.
The full-state difference grows linearly, by $4.2\times10^{-12}$ per walk, which matches the error of one application of
the compiled encoding measured independently ($4.15\times10^{-12}$). At the end of the sequence the output carries a
global phase of $-3.1\times10^{-10}$\,rad relative to the emulation; after removal of that phase the difference is
$9.4\times10^{-11}$ for the post-selected state and $9.9\times10^{-11}$ for the full state, of the same size as the norm
drift of the simulated state, $8.6\times10^{-11}$. The excess over the tolerance is therefore accumulated
double-precision round-off of a simulation of about $3\times10^{7}$ gates, mostly a global phase; the phase
removal is a diagnostic and does not change the recorded outcome. The checks cover one field per setting on $4^{3}$
grids; at $8^{3}$ only sampled block and self-inverse checks were made, without a GQSP check, and the $256^{3}$ runs do
not execute the circuit (Supplementary Note~\ref{note:D}). The GQSP checks were not repeated with a simulation of
lower round-off.

The production emulation (qubitized-pair mode) represents the full ancilla--system state of a GQSP sequence exactly by
four system-size vectors, using the invariant subspaces of qubitization~\cite{LowChuang2019}; its post-selected output,
success probability and sector populations are identical for every self-inverse block encoding $B$ with
$\langle 0|B|0\rangle = \bar{H}$ (Supplementary Note~\ref{note:B}). A system-only mode applies each factor's
post-selected polynomial directly to the system state. These are the two emulation modes (backends in the Supplementary
Information); the system-only mode ran the three amplitude-sweep runs, and the qubitized-pair mode ran the base run and
all other noisy runs (Extended Data Table~\ref{edtab:runs}). Before noise, the system-only $\varepsilon = 10^{-4}$ run
differs from the base run by $\delta = 1.87\times10^{-12}$. The diffusion encoding and a collocated analogue of the
convection encoding were compared with Qiskit statevector evolution on a non-degenerate $4^{3}$ grid at Chebyshev degree
6 (Extended Data Fig.~\ref{edfig:collocated}a); the residual, below the predefined $10^{-9}$, is mostly norm loss in
Qiskit's circuit synthesis. The temporal-order test uses a smooth $16^{3}$ test state that is neither solenoidal nor pressure-balanced,
4--64 steps and acoustic Courant numbers up to 1.02; frozen-coefficient Strang splitting gives order 0.98 and the
predictor restores 2.081, 2.016 and 2.003 through the production path, 2.03 with ideal classical factors
(Supplementary Note~\ref{note:D}).

\subsection{Turbulence run and reference}\label{sec:m-turbulence}

The production run has $N = 256$, $\mathrm{Re} = 800$, $\mathrm{Ma} = 0.05$ and $\Delta t = 0.002$ for 1,650 steps to
$t = 3.3$, an acoustic Courant number $\Delta t/(\mathrm{Ma}\,h) = 1.63$; the local Mach number reaches at least about 0.26. The
reference is an in-house pseudo-spectral incompressible solution (Fourier--Galerkin, SSP-RK3 time
stepping~\cite{ShuOsher1988}, 2/3 dealiasing~\cite{Orszag1971} with the cube mask $|k_{i}| < 85.3$) with the same $N$,
Re, $\Delta t$ and initial Gaussian draws; it has not been validated independently, and because the two codes discretize
differently, their differences combine spatial and temporal error. Spectra use shells $\mathrm{rint}(|k|)$; band ratios
compare band sums over the shell ranges 1--4, 4--8, 8--16, 16--32, 32--64 and 64--85. The large scales are the two
lowest bands (shells 1--7), which contain the spectral peak at all compared times; the 64--85 band lies next to the
reference's dealiasing limit and is not counted in the comparison. The dissipation rate $\varepsilon_{\mathrm{d}}$
uses either the shell convention $2\nu\sum_{k}k^{2}E(k)$ or the scheme's seven-point Laplacian; $k_{\max}\eta$ uses
$k_{\max} = N/2$ for both codes, and $\mathrm{Re}_{\lambda} = u_{\mathrm{rms}}\lambda/\nu$ with
$u_{\mathrm{rms}} = (2E/3)^{1/2}$ and $\lambda = (15\nu u_{\mathrm{rms}}^{2}/\varepsilon_{\mathrm{d}})^{1/2}$ (Extended
Data Table~\ref{edtab:stats}). No time-step study was made for the turbulent run.

\emph{Vortex visualization} (Fig.~\ref{fig:turbulence}a). The field is the saved production snapshot at $t = 1.5$ (step
750); no simulation was run for it. The face-centred MAC velocities are averaged to the cell centres,
$u_{c}[i] = \tfrac12(u[i] + u[i-1])$ along each velocity's own axis, and the velocity-gradient tensor
$G_{ij} = \partial u_{i}/\partial x_{j}$ is formed by periodic second-order centred differences; with
$S = \tfrac12(G + G^{\mathsf{T}})$ and $\Omega = \tfrac12(G - G^{\mathsf{T}})$, the second invariant is
$Q = \tfrac12(\|\Omega\|_{F}^{2} - \|S\|_{F}^{2})$, positive where rotation dominates strain. The isosurface level,
$Q = 3.0\langle\|\Omega\|_{F}^{2}\rangle$, is the 98th percentile of the positive values of $Q$; it selects the most
strongly rotation-dominated 2\% of the rotation-dominated cells, and the major tubes persist at the 97th and 99th
percentiles. The surface is coloured by $|\omega|$ over a fixed range. The pseudo-spectral reference did not save velocity
fields, so the visualization has no reference counterpart. Procedure, checks and rendering settings are given in
Supplementary Note~\ref{note:D}, \ref{si:D.7}.

\subsection{Effective per-step error model}\label{sec:m-error-model}

Once per step, after the third corrector factor and before decoding, the post-selected system state $\psi$ is replaced
by
\begin{equation}
  \psi' = \Bigl(1 - \frac{\varepsilon^{2}}{2}\Bigr)\psi + \varepsilon\Bigl(1 - \frac{\varepsilon^{2}}{4}\Bigr)^{1/2}
  \|\psi\|\, n_{\perp},
  \label{eq:rotation}
\end{equation}
with $n_{\perp}$ a unit direction orthogonal to $\psi$ in the complex inner product, so that $\|\psi'\| = \|\psi\|$ and
$\|\psi' - \psi\| = \varepsilon\|\psi\|$ exactly. The amplitude $\varepsilon$ is a per-step relative state error, not a
gate error rate or an infidelity (Supplementary Note~\ref{note:E}). The direction is $\sqrt{0.91}\,\hat{g} +
\sqrt{0.09}\,\hat{d}_{\mathrm{sys}}$, projected off $\psi$ and normalized: $\hat{g}$ is a normalized white complex
Gaussian vector, and $\hat{d}_{\mathrm{sys}}$ applies one random traceless complex $2\times2$ generator to each system
qubit, weighted by a per-qubit ladder. The two-qubit gate census of a compiled collocated $16^{3}$ convection walk (91\%
of gate touches on ancillas, 9\% on system qubits) sets the split and the ladder; at $256^{3}$, 12 of the 24 address bits
are extrapolated. Random draws follow a fixed order per step, so runs of the same noise realization share their white
directions. Noise acts from $t = 1.41$ to $t = 3.3$ (946 steps); contamination is stored at $t = 1.5$, 2.0, 2.5 and 3.0.
The model is a lumped end-of-step perturbation of the post-selected state: it has no ancilla dynamics, readout, shot or
measurement noise, its census split and ladder are untested at $256^{3}$ and for the staggered layout, and its
equivalence to in-circuit errors is untested (runs: Extended Data Table~\ref{edtab:runs}; construction: Supplementary
Note~\ref{note:E}).

\subsection{Contamination measures and references}\label{sec:m-measures}

For a noisy run $V$ and the noise-free run $U$ at equal step, $\Delta = V - U$ over the decoded fields
$(u, v, w, p_{\mathrm{s}})$; $\delta = \|\Re\,\Delta\|/\|U\|$ and $\delta_{\Im} = \|\Im\,\Delta\|/\|U\|$, and
velocity-only versions use $(u, v, w)$. The noise-free run is real up to a uniform global phase below $10^{-12}$.
Contamination spectra use the bands $k < 8$, 8--16, 16--32, 32--64, 64--128 and $\ge 128$, with band amplitudes
$(E_{X}(b)/E_{U}(b))^{1/2}$. Dilatational fractions use the MAC projection, the energy
$\sum_{k}|m(k)\cdot F(k)|^{2}/|m(k)|^{2}$ of the velocity transform $F$ over its total, an exact orthogonal discrete
Helmholtz split because $\mathrm{Div} = -G^{T}$ (white noise gives 1/3).

Let $S_{x}(k; j)$ be the per-mode injection spectrum of component $x$ (real or imaginary), averaged over the draws near
noisy step $j$ and normalized so that $\sum_{x}\sum_{k}S_{x} = 1$; $s_{x}(j) = \sum_{k}S_{x}(k; j)$ and $s_{x}$ is its
mean. With $\lambda(k) = -\nu(4/h^{2})\sum_{a}\sin^{2}(\pi k_{a}/N)$ the decay rate of mode $k$ under the seven-point
Laplacian, the references per unit $\varepsilon$ at observation step $n$ are
\begin{equation}
  R_{1} = \frac{\bigl[\sum_{j}\|U(j)\|^{2}s_{x}(j)\bigr]^{1/2}}{\|U(n)\|},\qquad
  R_{1\mathrm{d}} = \frac{\bigl[\sum_{k}\sum_{j}\|U(j)\|^{2}S_{x}(k; j)\,e^{2\lambda(k)(n - j)\Delta t}\bigr]^{1/2}}{\|U(n)\|}.
  \label{eq:refs}
\end{equation}
$R_{1}$ is matched non-dissipative accumulation (each injection keeps its norm); $R_{1\mathrm{d}}$ adds exact diffusion
but no advection, acoustic exchange or feedback, a reference rather than a prediction. A predefined $\pm10\%$ tolerance
classifies all-band ratios $\delta/R_{1}$ as below, consistent with or above accumulation; $B_{2}$ and $B_{2\mathrm{d}}$
are $R_{1}$ and $R_{1\mathrm{d}}$ per velocity band. The relative growth
$\Gamma_{1} = (\delta/\sqrt{s_{\Re}})/(\delta_{\Im}/\sqrt{s_{\Im}})$ is 1 for identical dynamics; its band version
$\Gamma_{1,b}$ uses the mean band fractions of the injected real and imaginary velocity. At $t = 3.0$ alternative share
normalizations give 6.36 and 6.38 against 6.65, a normalization range rather than a statistical spread. By construction
$\delta/R_{1} = \Gamma_{1}\times(\delta_{\Im}/R_{1,\Im})\times f$, with $f = (s_{\Re}/s_{\Im})^{1/2}R_{1,\Im}/R_{1,\Re}
= 0.984$--$0.994$ (Supplementary Note~\ref{note:F}).

\subsection{Passive control}\label{sec:m-passive}

Given the refreshed coefficients the step is a real linear map (GQSP polynomials real-matrix-valued to
$2.5\times10^{-15}$), and the refresh reads only real parts: a purely imaginary perturbation of relative size
$10^{-4}$--$10^{-2}$ ($32^{3}$, production initial condition) changed the real part after ten steps by at most
$1.6\times10^{-15}$. The imaginary part, propagated like the real part but never entering the refresh, is therefore an
approximately passive control. It is approximate in three respects: the injection's real and imaginary parts are
different random fields with equal statistics (energy shares 0.498 and 0.502), handled by share normalization; the
injection exchanges a fraction of order $\kappa\varepsilon$ ($\kappa = 0.172$) between them, the only route from the
imaginary to the real part; and the contraction of the injection acts on the essentially real state. The runs at
$\varepsilon = 10^{-4}$ and $10^{-6}$ give the same $\Gamma_{1}$ within $6.6\times10^{-4}$, which bounds the last two
effects below $10^{-3}$ of the result in total $\delta$ (Supplementary Note~\ref{note:F}).

\subsection{Contraction at large amplitude}\label{sec:m-contraction}

Each injection shortens the state's component along its pre-injection direction by exactly $1 - \varepsilon^{2}/2$,
compounding to $\approx\exp(-N_{n}\varepsilon^{2})$ in energy after $N_{n}$ injections, which accounts for 84--91\% of
the kinetic-energy deficit at $\varepsilon = 10^{-2}$. The random part of the field error, orthogonal to $U$, exceeds
the contraction part 3.7--524-fold (Extended Data Fig.~\ref{edfig:amplitude}d). A run at $\varepsilon = 10^{-2}$ with
per-step renormalization by $1/(1 - \varepsilon^{2}/2)$ and a real filter brings the persisting excess at $k < 8$ to
within 1.5\% of linear, so the departure at $\varepsilon = 10^{-2}$ is attributed mainly to the built-in contraction; the
real filter was excluded as a cause only at $64^{3}$ (Supplementary Note~\ref{note:E}).

\subsection{Origin attribution}\label{sec:m-origin}

In the linear regime the contamination of the census-weighted run superposes the responses to the white and structured
parts of the injection; with a white-only run sharing the white draws, they give per velocity band the energies $E_{W}$
and $E_{\mathrm{sp}}$ and their cross term, from which follows the structured share at any structured weight
(Supplementary Note~\ref{note:G}). The decomposition was validated against direct structured-only runs at $64^{3}$ only,
and applied with one realization at $128^{3}$, where linearity in $\varepsilon$ was not tested, and at $256^{3}$, where
the two runs used different emulation modes. The as-compiled register order places the $z$-index bits first; under an
interleaved order ($z_0, y_0, x_0, z_1, \ldots$; an in-model permutation, not a compiled circuit) the $k = 1$--4
majority becomes marginal or disappears. Read literally, the predefined placement rule therefore grades the attribution
as not robust, and restricting it to the as-compiled order is a scope decision made after the fact. The attribution
identifies an origin, not a route; it does not explain the gain and implies no choice of register layout.

\subsection{Computational complexity}\label{sec:m-complexity}

Computational statements fall into three categories: (I) properties of the structure of a step, proved for the frozen
operators; (II) costs of subproblems, compared with other methods for the same task under a stated input, output and
access model; and (III) the cost of the complete hybrid loop. Extended Data
Table~\ref{edtab:comparisons} lists the comparisons with their settings; derivations, assumptions and published
comparators are given in Supplementary Note~\ref{note:C}, whose §14 states the setting of each comparison. Here
$\mathcal{N} = N^{3}$ is the number of grid points and $\eta$ the polynomial tolerance of a factor.

\emph{Cost models.} In the query model the unit of cost is one call to a block encoding of a frozen operator, assumed
available, and inputs and outputs are quantum states. In the compiled-circuit model one- and two-qubit gates are
counted, classical data enter only through the circuit description and no quantum memory persists between calls. The
complete loop, as designed here, outputs a classical description of the full field at every step and re-prepares the
state from classical data; this interface is a property of the design, not of the flow problem.

\emph{Structure (I).} For every real frozen velocity $A$ is anti-Hermitian and $D$ is Hermitian, negative semidefinite
and constant, so every exact corrector factor is a contraction for any $\Delta t$; this is a stability statement for
the frozen factors and does not by itself establish the accuracy limit of the step. Because every exact factor is a
contraction, the errors of the polynomial approximations accumulate at most linearly over a sequence of frozen steps
as long as their sum is small (Supplementary Note~\ref{note:C}, Proposition 8). The step contains no projector or
linear solve. Unit-normalized syntheses exist for which the convection factor succeeds with probability close to one
for every input and a step succeeds with the fraction of the norm that it retains, up to the synthesis margins
(Supplementary Note~\ref{note:C}, Propositions 1--3); these are proved constructions that are not implemented. The
splitting error is bounded by nested commutators (Proposition S3). On the uniform periodic MAC grid with equal
diffusivities $[D, A_{\mathrm{ac}}] = 0$ (Lemma S2), and the acoustic cross term
$[A_{\mathrm{ac}}, [A_{\mathrm{adv}}, D]]$, which scales as $1/\mathrm{Ma}$, carries at least 99.8\% of the leading term
on the production states; a projected formulation has no such term, and with walls the commutation requires a
compatible vector Laplacian and is not shown here. The encodings add five ancillas and a signal qubit for convection
and four qubits for each diffusion factor, independent of $N$; the corrector acts on 32 qubits with ancilla reuse, 40
with disjoint registers.

\emph{Frozen linear step (II).} The production configuration uses 184--185 corrector queries per step at a design
success of 1/64. If post-selection is measured after each factor and a step is aborted at the first failure, this gives
4,852--4,868 expected corrector queries per accepted step (state preparations and readout repetitions excluded). Because
$A(\bar u)$ is anti-Hermitian, the convection factor is unitary and its query count can scale as
equation~\eqref{eq:queries}, the optimal black-box scaling at the encoding
normalization~\cite{LowChuang2019,Gilyen2019}; this is not a lower bound for these structured operators, and the
production synthesis does not reach it (78--79 queries against a reference degree of 31--32). At fixed Courant
numbers, $\tau_{A}$ does not depend on the grid; at fixed $\Delta t$ it grows in proportion to $N$, mainly through its
acoustic part. With the unit-normalized constructions a step would need 70--72 queries, and 43--45 if the convection factor used a published
directional-control construction~\cite{Berry2024}, at a step success of about 0.96 with the synthesis margin that the
present routines support; these configurations are prospective and not implemented (Supplementary Note~\ref{note:C},
§§4.5 and 4.6). Because the convection factor is unitary, one round of oblivious amplitude
amplification~\cite{BerryChildsCleve2014,Gilyen2019} would also make it deterministic with three uses of the factor,
which would reduce the expected number of trials per step of the production design from 64 to 16; it does not apply to
the diffusion factors and is not implemented (Supplementary Note~\ref{note:C}, §5.3). The classical counterpart is the
evaluation of the same polynomials by stencil products of $\Theta(\mathcal{N})$ operations each; the emulation applies
81 or 80 products of $A$ and $2\times55$ of $D$ for the corrector (286--287 stencil applications per step with the
predictor). In gates,
the flow-independent diffusion walk compiles to 6,193 and
17,641 CNOT gates on $8^{3}$ and $16^{3}$ grids. For the convection encoding, block-encoding an arbitrary advecting
field needs $\Omega(\mathcal{N}/\log\mathcal{N})$ gates per call for Lipschitz-parameterized gates in the
compiled-circuit model (Supplementary Note~\ref{note:C}, Proposition 5): the staggered encoding compiles, unoptimized,
to $1.0\times10^{5}$ and $8.0\times10^{5}$ CNOT gates per application on $4^{3}$ and $8^{3}$ grids, in proportion to
$\mathcal{N}$, and, extrapolated, a $256^{3}$ convection factor would need about $2.7\times10^{12}$ CNOT gates
(Supplementary Note~\ref{note:C}, §5.5). For arbitrary fields in this model the frozen step therefore has the same
$\Theta(Q\mathcal{N})$ scaling as its classical evaluation, up to a logarithm, with unmatched outputs (a state against a
field); a query and a stencil product are different units, and no ratio between them is formed. The bound does not
apply to fields with a compact description, which a turbulent field is not assumed to have, or when a quantum
random-access memory is counted as an oracle, in which case its $\Theta(\mathcal{N})$ hardware
remains~\cite{Giovannetti2008,Dalzell2025}. Among quantum algorithms for the same frozen step, a linear combination of
Hamiltonian simulations (LCHS)~\cite{AnLiuLin2023,AnChildsLin2026} has a comparable query cost and no splitting error:
with the truncation that its kernel requires at $\eta = 10^{-12}$, the reference degree of one LCHS run is 40--41
(48--49 if the identity branch of the diffusion encoding is not factored out), against 39--42 for the three factors of
the split (Supplementary Note~\ref{note:C}, §9.4).

\emph{Pressure coupling (II).} Quantum linear-system algorithms~\cite{Harrow2009,Costa2022} solve a different task from
the exponential factors of the integrator, whose step contains no linear system; they enter the comparison only through
formulations that contain a linear solve, here the pressure projection. Where no fast diagonalizing transform exists, a
projector takes $\Theta(\kappa_{G}\log(1/\eta))$ block-encoding calls with quantum singular value transformation
($\kappa_{G}$, the condition number of the discrete gradient) or $\Theta(\kappa_{\Delta}\log(1/\eta))$ with optimal
linear-system solvers ($\kappa_{\Delta}$, that of the Laplacian)~\cite{Costa2022,Gilyen2019}. The comparison of Results
is made in the query model, per step, at equal $\Delta t$ and $\eta = 10^{-12}$. It counts the reference polynomial
degree of the unitary factor of the integrator (31--32, of which about 20 are due to the acoustic part) against those
of the transport factor and the projector of a projected step built from the same machinery (9--12 for transport plus
about $3.9\times10^{3}$ for a once-per-step projector). The constant of the projector degree is set to one, and the
condition number of the periodic grid, $\kappa_{G} = 141$ at $256^{3}$, is used as a proxy for geometries without a
fast transform. The projection-free step then needs, for transport and pressure coupling, 122--126 times fewer calls
per step and about 30--110 times fewer per trajectory; with the two diffusion factors, which both formulations contain, included on both sides the
per-step ratio is about 93--100. The reference degree of a factor is the degree at which the Chebyshev truncation of
its target reaches $\eta$; it is a property of the target function and differs from the number of queries of a
particular synthesis. Both sides of the ratio are counted with reference degrees, and the production synthesis of the
integrator, with 184--185 queries per step at a design success of 1/64, is not its basis. The result holds under four
conditions (Supplementary Note~\ref{note:C}, Proposition E1). The first two are that every call is counted once and
that no fast diagonalizing transform is available, which excludes periodic and rectangular boxes. The other two are
that the splitting error of the integrator stays acceptable in geometries without such a transform, where
$[D, A_{\mathrm{ac}}] = 0$ is not established and cut-cell or immersed-boundary treatments break it, so that the
acoustic part may enter the splitting error directly, and that the accuracy target does not lie below the
$O(\mathrm{Ma}^{2})$ modelling error. It compares calls, not cost at matched accuracy, because the
two formulations solve different models, each with its own splitting error. Two other settings
give the opposite ordering. In periodic and rectangular boxes a fast transform applies the projector, and a projected
step needs 2.6--3.6 times fewer velocity-dependent calls (9--12 against 31--32). Counted in gates in the
compiled-circuit model, the acoustic increment of about 20 calls is paid on calls that load the field, whereas the
projector calls contain no flow data, and the projected step needs fewer gates at $256^{3}$; the ordering would change
only for loaders below about $1.2\times10^{7}$ CNOT gates per call, fewer than one two-qubit gate per loader angle
(Supplementary Note~\ref{note:C}, Proposition E2 and §§8.4, 13 and 14.3).

\emph{Time step (II).} The corrector imposes no stability limit on $\Delta t$. The production run took 1,650 steps at an
acoustic Courant number of 1.63 and an advective Courant number of at most 0.89. At their stability limits, explicit
acoustically limited schemes would need 1.9--3.3 times as many steps ($3.2$--$5.4\times10^{3}$, depending on the scheme
and on the definition of the Courant number), whereas a projection scheme at unit advective Courant number would need
1.1--3.7 times fewer (450--1,470). These counts compare stability limits and measure refreshes and interface crossings,
not total work or equal accuracy: as a classical exponential integrator, the scheme applies about 18 times more stencil
products as run, and about 3.5 times more with polynomial degrees trimmed to the tolerance and a reduced predictor,
than explicit fourth-order Runge--Kutta integration of the same system at its stability limit, with about half as many
steps (estimates). Second-order convergence was measured at acoustic Courant numbers up to 1.02;
whether the production step, or a larger one, is accurate enough is not established, because no time-step study was
made for the turbulent run (Supplementary Note~\ref{note:C}, Proposition E3 and §8.3).

\emph{Complete loop (III).} The classical side of a step is dominated by the predictor (96 stencil applications,
reducible to 17--27) and costs about 1.3--15 steps of a classical pseudo-spectral solver (an estimate). With a full
readout at every step, each copy of the $n = 26$-qubit output carries at most $n$ bits of accessible
information~\cite{Holevo1973,Haah2017}, so a $b$-bit readout needs at least about
$(1 - P_{\mathrm{err}})\,4\mathcal{N}(b - 1)/n$ copies per step ($\ge 5.3\times10^{7}$ at $b = 32$,
$P_{\mathrm{err}} = 1/3$) and, in the compiled-circuit model, $\Omega(\mathcal{N}^{2}/\log^{2}\mathcal{N})$ gates per
step. With exact preparation of generic states~\cite{Plesch2011} this floor is about $1.8\times10^{15}$ CNOT gates at
$256^{3}$ at leading order, against about $2.7\times10^{10}$ floating-point operations for a pseudo-spectral step of the
same problem (an estimate; Supplementary Note~\ref{note:C}, §§6.2 and 6.3). Under these assumptions the loop costs more than
a classical computation of the same steps, and no end-to-end speedup is claimed. These floors belong to the interface
of the loop as designed, that is, to per-step full-field readout and to loading in the compiled-circuit model, not to
the problem: with observable outputs the readout floor does not apply, and with compactly described fields the loading
bound does not, and both changes are needed together. If every call and every state preparation took
$O(\mathrm{polylog}\,\mathcal{N})$ time on quantum hardware of $\Theta(\mathcal{N})$ components while the full field
were still read out, at most a logarithmic factor would remain in favour of the loop against a sequential classical
step, and none against classical hardware of the same size. Lower bounds for coherent simulation of chaotic dynamics~\cite{Lewis2024,Ameri2026} do not
cover an interface that re-prepares the state from classical data and reads it out every step.

\subsection{Statistics and reproducibility}\label{sec:m-statistics}

No statistical tests were performed. Every $256^{3}$ noise result rests on one noise realization and one initial
condition; the amplitude sweep shares one sequence of random draws and tests proportionality, not spread. Eight noise
realizations of a $64^{3}$ flow at $\mathrm{Re} = 120$ from one initial condition (Extended Data
Table~\ref{edtab:grids}; Extended Data Fig.~\ref{edfig:passive}e) indicate the spread at that resolution only. The guide
value $\Gamma_{1} > 1.10$ and the $\pm10\%$ tolerance were fixed before $\Gamma_{1}$ or any reference value was
computed, but not blind, because the running output had already shown the raw real-to-imaginary ratio
(Supplementary Note~\ref{note:F}). No error bars are plotted; bands and lines mark predefined tolerances, observed
maximum deviations, alternative normalizations or examined parameter ranges, as named in each legend.


\clearpage
\section*{Extended Data}\label{sec:extended-data}
\addcontentsline{toc}{section}{Extended Data}
\setcounter{figure}{0}\renewcommand{\figurename}{Extended Data Fig.}\renewcommand{\theHfigure}{ED.\arabic{figure}}
\setcounter{table}{0}\renewcommand{\tablename}{Extended Data Table}\renewcommand{\theHtable}{ED.\arabic{table}}
\captionsetup{hypcap=false}
\newcommand{\edgraphic}[1]{\makebox[\linewidth][c]{\includegraphics[width=183mm]{#1}}}

\begin{center}
\edgraphic{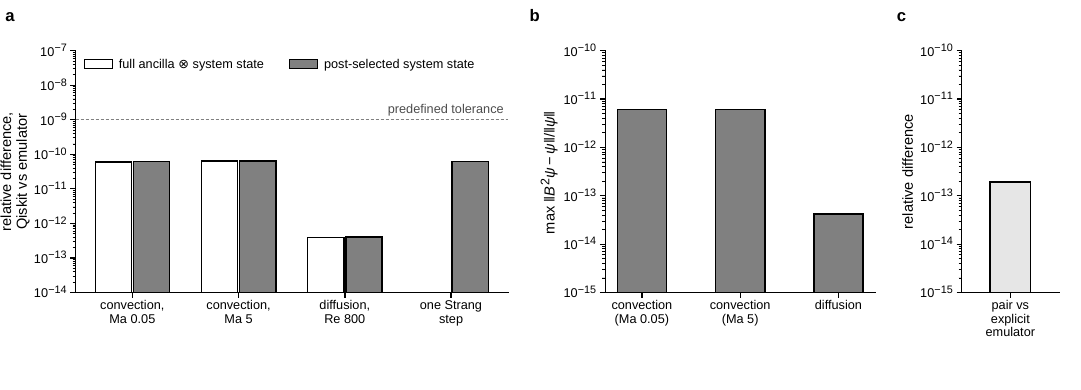}
\captionof{figure}[{Gate-level checks of the collocated circuits.}]{\textbf{Gate-level checks of the collocated
circuits.} These circuits use the collocated layout of earlier tests; the staggered convection encoding used by the
integrator is verified in Fig.~\ref{fig:verification}. \textbf{a}, Relative $L^2$ difference between Qiskit statevector
evolution and the emulator on a non-degenerate $4^3$ grid (Chebyshev degree 6) for each block encoding and for one
block-sequential Strang step: full ancilla $\otimes$ system state (white) and post-selected system state (grey); dashed,
predefined tolerance $10^{-9}$. Ma, Mach number; Re, Reynolds number. \textbf{b}, Self-inversion $B^2 = I$ checked on the
gates, largest relative deviation over random states. \textbf{c}, Qubitized-pair representation against an explicit
gate-level emulation of the full ancilla--system state ($4^3$, Chebyshev degree 12).}\label{edfig:collocated}
\end{center}
\clearpage

\begin{center}
\edgraphic{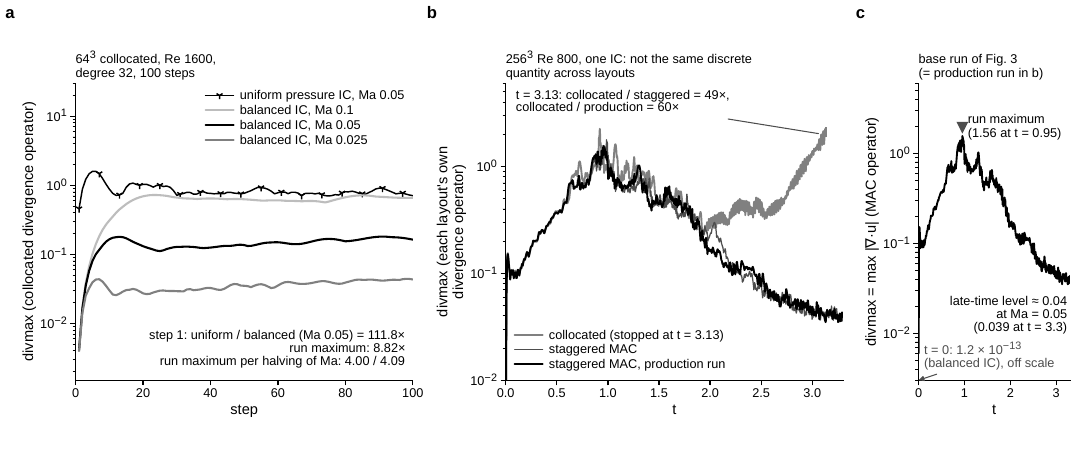}
\captionof{figure}[{Divergence control.}]{\textbf{Divergence control.} Maximum absolute discrete velocity divergence,
each layout with its own divergence operator. \textbf{a}, $64^{3}$ collocated runs ($\mathrm{Re} = 1{,}600$, Chebyshev
degree 32, 100 steps): uniform-pressure initial condition at $\mathrm{Ma} = 0.05$ against the balanced-pressure initial
condition at $\mathrm{Ma} = 0.1$, 0.05 and 0.025; the run maximum scales with $\mathrm{Ma}^{2}$. \textbf{b}, $256^{3}$,
$\mathrm{Re} = 800$: a collocated run (grey; stopped at $t = 3.13$), a second staggered run (thin) and the production
run (black). At $t = 3.13$ the collocated value is 49 and 60 times the two staggered values; the divergences are not the
same discrete quantity and the Chebyshev degrees differ (32 and 48). \textbf{c}, The production run of
Fig.~\ref{fig:turbulence} on the staggered operator: maximum 1.56 at $t = 0.95$ and 0.039 at $t = 3.3$; the $t = 0$
value, $1.2\times10^{-13}$, is off scale.}\label{edfig:divergence}
\end{center}
\clearpage

\begin{center}
\edgraphic{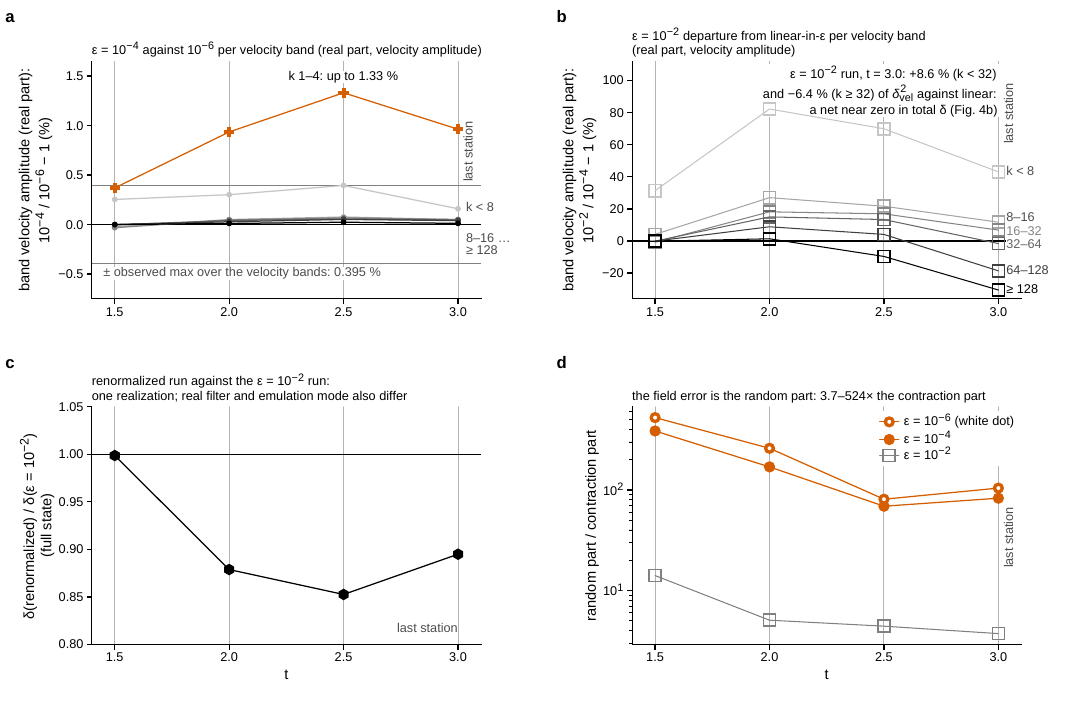}
\captionof{figure}[{Error amplitude by band and the departure at $\varepsilon = 10^{-2}$.}]{\textbf{Error amplitude by
band and the departure at $\varepsilon = 10^{-2}$.} Runs of Fig.~\ref{fig:transfer} at $\varepsilon = 10^{-6}$,
$10^{-4}$ and $10^{-2}$ (shared random draws). Bands, velocity bands $k < 8$, 8--16, 16--32, 32--64, 64--128 and
$\ge 128$ (light to dark grey); band amplitude $(E_{X}(b)/E_{U}(b))^{1/2}$ of the real part (Methods). \textbf{a},
$\varepsilon = 10^{-4}$ against $10^{-6}$ per band (dots) and for $k = 1$--4 (plus signs); lines, observed maximum
(0.4\%). \textbf{b}, $\varepsilon = 10^{-2}$ against $10^{-4}$ per band (open squares); the largest departure is at
$k < 8$. \textbf{c}, $\delta$ of a renormalized $\varepsilon = 10^{-2}$ run (per-step renormalization and a real filter)
relative to the original $\varepsilon = 10^{-2}$ run (0.85--1.00); the two runs also differ in emulation mode.
\textbf{d}, Ratio of the random part of the field error (orthogonal to $U$) to its contraction part (Methods), symbols as
in Fig.~\ref{fig:transfer}a.}\label{edfig:amplitude}
\end{center}
\clearpage

\begin{center}
\includegraphics[width=\linewidth]{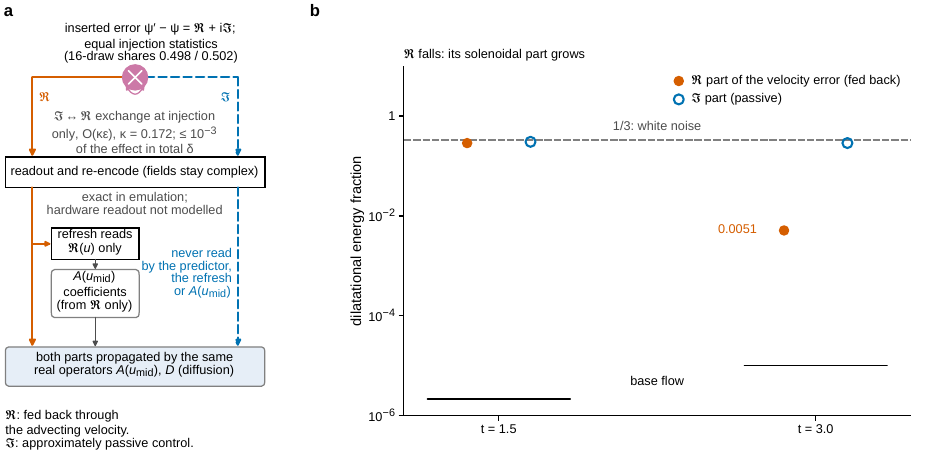}
\captionof{figure}[{The in-run passive control and the dilatational part of the error.}]{\textbf{The in-run passive
control and the dilatational part of the error.} Same run as Fig.~\ref{fig:feedback}, $\varepsilon = 10^{-4}$.
\textbf{a}, Paths of the real (fed-back; vermillion) and imaginary (passive; blue) parts of the injected error through
the step. Both parts are propagated by the same real operators; only the real part enters the refresh of
$A(u_{\mathrm{mid}})$. The injection exchanges a fraction of order $\kappa\varepsilon$ between the parts ($\kappa$,
exchange coefficient); the injected energy shares are 0.498 and 0.502 over 16 draws (Methods). \textbf{b}, Dilatational
energy fraction of the velocity error at $t = 1.5$ and 3.0 for the real (filled) and imaginary (open) parts; dashed,
white-noise value 1/3; black lines, the base flow.}\label{edfig:structure}
\end{center}
\clearpage

\begin{center}
\edgraphic{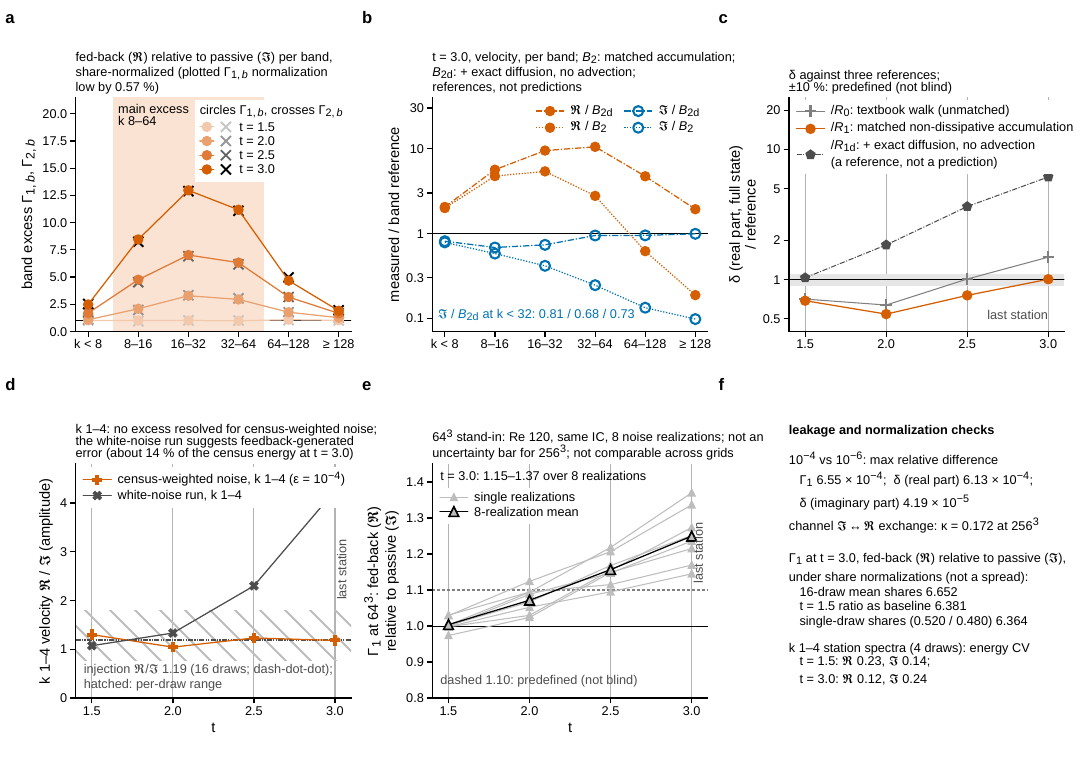}
\captionof{figure}[{Passive-control and accumulation comparisons in detail.}]{\textbf{Passive-control and accumulation
comparisons in detail.} $\varepsilon = 10^{-4}$ unless stated. \textbf{a}, Band excess of the fed-back part relative to
the passive part, normalized by injection share ($\Gamma_{1,b}$, circles) or by each part's own band reference
$B_{2\mathrm{d}}$ ($\Gamma_{2,b}$, crosses). \textbf{b}, Ratios to the band references $B_{2}$ and $B_{2\mathrm{d}}$ at
$t = 3.0$ (Methods); for $k < 32$ the passive part lies 19--32\% below $B_{2\mathrm{d}}$. \textbf{c}, $\delta$ against
$R_{1}$, $R_{1\mathrm{d}}$ and the unmatched constant-norm random walk $R_{0}$; grey band, predefined $\pm10\%$ tolerance.
\textbf{d}, Real-to-imaginary amplitude ratio at $k = 1$--4 for census-weighted and white-noise injection; dash-dotted,
injection ratio; hatched, per-draw range of the injection. \textbf{e}, $\Gamma_{1}$ of a weakly chaotic $64^{3}$ flow
($\mathrm{Re} = 120$) for eight noise realizations and their mean; context for the spread at that resolution, not an
uncertainty for $256^{3}$. \textbf{f}, Consistency checks; $\kappa$, exchange coefficient of the injection; CV,
coefficient of variation.}\label{edfig:passive}
\end{center}
\clearpage

\begin{center}
\edgraphic{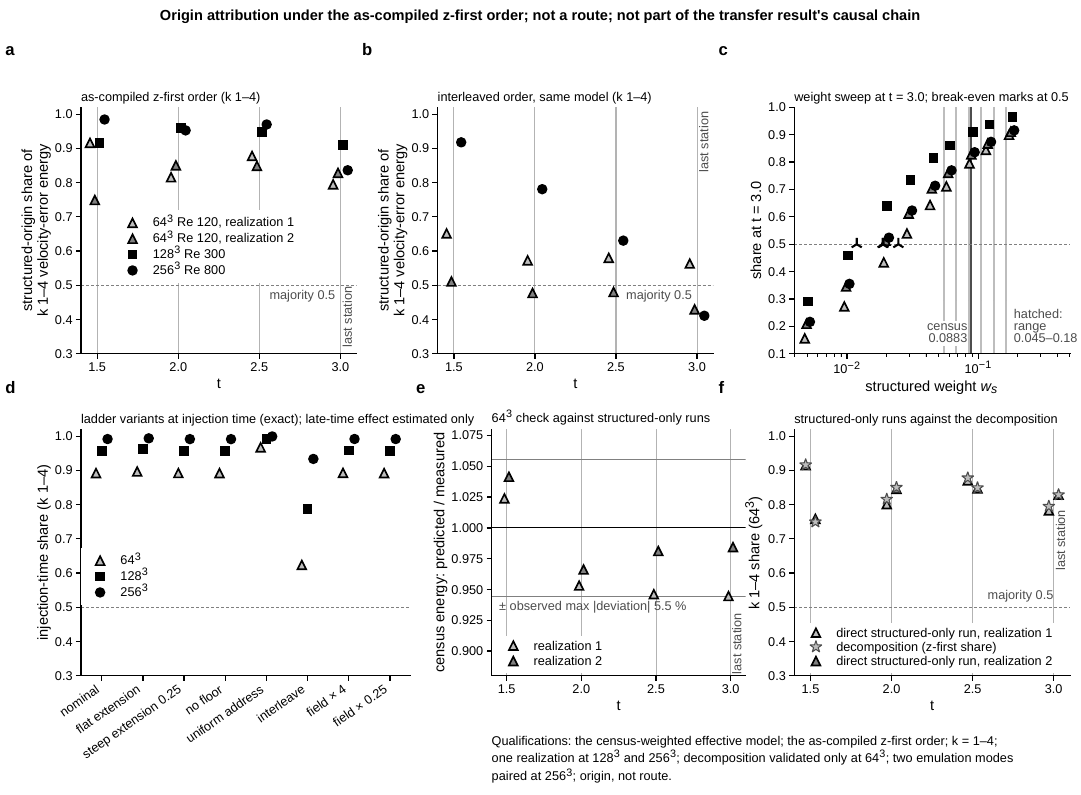}
\captionof{figure}[{Origin of the largest-scale error.}]{\textbf{Origin of the largest-scale error.} Share of the
$k = 1$--4 velocity-error energy that originates in the structured component of the census-weighted error model
($\varepsilon = 10^{-4}$), from the superposition of white and structured responses (Methods). The attribution concerns
an origin, not a route, and is not part of the transfer result of Figs.~\ref{fig:transfer} and~\ref{fig:feedback}.
\textbf{a}, As-compiled register order ($z$-index bits first) on three grids; dashed, majority. \textbf{b}, The same for
an interleaved register order ($z_0, y_0, x_0, z_1, \ldots$; an in-model permutation), where the majority becomes
marginal or disappears. \textbf{c}, Share at $t = 3.0$ against the structured weight; dark line, census weight; hatched,
range examined; marks, break-even weights. \textbf{d}, Share at injection for variants of the per-qubit ladder.
\textbf{e},\textbf{f}, Validation at $64^{3}$ against direct structured-only runs: predicted over measured energy (e;
within 5.5\%) and the $k = 1$--4 share by both routes (f).}\label{edfig:origin}
\end{center}
\clearpage

\begin{center}
\captionof{table}[{Global statistics of the production run and the reference.}]{\textbf{Global statistics of the
production run and the reference.} Kinetic energy $E$, dissipation rate $\varepsilon_{\mathrm{d}}$, Taylor-microscale
Reynolds number $\mathrm{Re}_{\lambda}$ and resolution $k_{\max}\eta$ ($k_{\max} = 128$) at six times. Shell convention,
$\varepsilon_{\mathrm{d}} = 2\nu\sum_{k}k^{2}E(k)$; scheme convention, the run's seven-point-Laplacian dissipation.
$\mathrm{Re}_{\lambda}$ of the run uses the scheme convention and that of the reference the shell convention;
$k_{\max}\eta$ uses the shell convention for both.
Reference: in-house pseudo-spectral solution from the same initial condition, 2/3-dealiased, SSP-RK3, not validated
independently.}\label{edtab:stats}
\smallskip
{\footnotesize\setlength{\tabcolsep}{3.2pt}
\begin{tabular}{ccccccccccc}
\toprule
 & \multicolumn{3}{c}{$E$} & \multicolumn{4}{c}{$\varepsilon_{\mathrm{d}}$} & \multicolumn{2}{c}{$\mathrm{Re}_{\lambda}$} & \\
\cmidrule(lr){2-4}\cmidrule(lr){5-8}\cmidrule(lr){9-10}
$t$ & ref. & run & $\Delta$ (\%) & ref. (shell) & run (scheme) & $\Delta$ (\%) & run (shell) & ref. & run & $k_{\max}\eta$ ref. / run\\
\midrule
0.5 & 1.4293 & 1.4301 & $+0.05$ & 0.2671 & 0.2650 & $-0.78$ & 0.2691 & 202.0 & 202.9 & 1.18 / 1.18\\
1.0 & 1.2007 & 1.2027 & $+0.17$ & 0.6459 & 0.6437 & $-0.35$ & 0.6735 & 109.1 & 109.5 & 0.95 / 0.94\\
1.5 & 0.8202 & 0.8298 & $+1.17$ & 0.7850 & 0.7665 & $-2.36$ & 0.8123 & 67.6 & 69.2 & 0.90 / 0.90\\
2.0 & 0.4905 & 0.5049 & $+2.94$ & 0.5024 & 0.5099 & $+1.49$ & 0.5435 & 50.5 & 51.6 & 1.01 / 0.99\\
2.5 & 0.3053 & 0.3149 & $+3.13$ & 0.2609 & 0.2723 & $+4.37$ & 0.2890 & 43.6 & 44.1 & 1.19 / 1.16\\
3.0 & 0.2086 & 0.2136 & $+2.38$ & 0.1404 & 0.1481 & $+5.49$ & 0.1558 & 40.7 & 40.5 & 1.39 / 1.35\\
\bottomrule
\end{tabular}}
\end{center}
\clearpage

\begin{center}
\captionof{table}[{Computational comparisons by task.}]{\textbf{Computational comparisons by task.} Each row compares
the integrator with another method for the same task under the stated input, output and access model; rows are not to
be combined into one ratio. Values are for the $256^{3}$ production parameters at polynomial tolerance
$\eta = 10^{-12}$ unless a grid is given; reference degrees, the proxy condition number, extrapolated gate counts and
the classical operation count are estimates. $\mathcal{N} = N^{3}$, number of grid points; $n$, number of system qubits; $\kappa_{G}$, condition number of
the discrete gradient; $d$, polynomial degree. Proposition, lemma, section and assumption labels (I1--I3, interface;
K1--K2, cost model) refer to Supplementary Note~\ref{note:C}.}\label{edtab:comparisons}
\smallskip
{\scriptsize\setlength{\tabcolsep}{3pt}
\begin{tabular}{>{\raggedright\arraybackslash}p{0.088\linewidth}>{\raggedright\arraybackslash}p{0.170\linewidth}%
>{\raggedright\arraybackslash}p{0.214\linewidth}>{\raggedright\arraybackslash}p{0.213\linewidth}%
>{\raggedright\arraybackslash}p{0.240\linewidth}}
\toprule
Task & Setting & This integrator & Comparator & Result\\
\midrule
\multicolumn{5}{l}{\textit{Structure of a step (proved for the frozen operators)}}\\
\multicolumn{5}{>{\raggedright\arraybackslash}p{0.95\linewidth}}{No projector or linear solve; $A$ anti-Hermitian and
$D$ negative semidefinite for every real frozen velocity, so every exact factor has norm at most one for any
$\Delta t$; $n = 3\log_{2}N + 2$ system qubits and a number of ancillas independent of $N$ (corrector on 32 qubits at
$256^{3}$ with ancilla reuse; the fields between steps are classical data); flow data confined to one unitary factor
(Lemmas S1--S3; Proposition E3; §5.4).}\\
\addlinespace[1pt]
\multicolumn{5}{l}{\textit{Subproblems}}\\
Pressure coupling of one step & State to state; query model; no fast diagonalizing transform (not the periodic box of
the runs); reference degrees at equal $\Delta t$ and $\eta$; unit projector constant &
Acoustic stiffness additive in the degree of the unitary factor: 31--32 calls, about 20 of them acoustic & Projected
step: 9--12 calls for transport and $\approx 3.9\times10^{3}$ for a projector by quantum singular value transformation
($\kappa_{G} = 141$, periodic-grid proxy) & 122--126 times fewer calls per step and $\approx 30$--110 per trajectory
($\approx 93$--100 per step with the diffusion factors on both sides); $\propto N$ at fixed Courant numbers. Not at
matched accuracy; assumes an acceptable splitting error where Lemma S2 is not established (Proposition E1)\\
\addlinespace[1pt]
Same, gate-weighted & Compiled circuits; arbitrary field loaded in every velocity-dependent call & Acoustic increment
of about 20 calls, each loading the field: $\approx 5.6\times10^{11}$ CNOT gates per step & Projector:
$\approx 3.9\times10^{3}$ calls without flow data, $\approx 2.5\times10^{8}$ CNOT gates per step & Fewer gates for the
projected step; the ordering changes only for loaders below $\approx 1.2\times10^{7}$ CNOT gates per call
(Proposition E2)\\
\addlinespace[1pt]
Same, with a fast transform & Periodic or rectangular box; projector by a fast transform & 31--32 velocity-dependent
calls & 9--12
velocity-dependent calls; transforms of $O(\mathrm{polylog}\,\mathcal{N})$ gates & 2.6--3.6 times fewer
velocity-dependent calls for the projected step (§8.4)\\
\addlinespace[1pt]
Frozen linear step & Integrator: state to state, query model. Comparator: field to field & $O(\tau_{A} +
\log(1/\eta))$ queries on $n = 26$ system qubits in scaling; 184--185 per step as implemented, at design success 1/64 &
Classical evaluation of the same polynomials: one stencil product of $\Theta(\mathcal{N})$ operations per degree (81 or
80 products of $A$ and $2\times55$ of $D$ in the emulation) & Same order of operator applications; a
query and a product of $\Theta(\mathcal{N})$ operations are different units and form no ratio; outputs not matched
(Lemma 2; §3)\\
\addlinespace[1pt]
Same, gates & Integrator: state to state, compiled circuits, arbitrary advecting field. Comparator: field to field &
$\Omega(\mathcal{N}/\log\mathcal{N})$
gates per convection query; staggered encoding $1.0\times10^{5}$ ($4^{3}$), $8.0\times10^{5}$ ($8^{3}$) CNOT gates
per application; diffusion walk 6,193 ($8^{3}$), 17,641 ($16^{3}$) & $\Theta(\mathcal{N})$ operations per product &
Same scaling up to a logarithm; a separation requires sublinear access to the field (Proposition 5)\\
\addlinespace[1pt]
Same, other quantum algorithms & State to state; query model & Normal factors; no quadrature or history register;
reference degrees 39--42 per step; realized splitting error $1.9$--$6.7\times10^{-6}$ per step & Linear combination of
Hamiltonian simulations (LCHS): reference degree 40--41 for one LCHS run (§9.4); no splitting error. Unamplified
series: success $\leq 1.2\times10^{-9}$ & Comparable query cost at reference degrees; the production synthesis, at design success
1/64, succeeds $1.4$--$5.3\times10^{7}$ times more often than the series, a weak baseline (Propositions S4, S5)\\
\addlinespace[1pt]
Number of time steps & Stability limits at the production parameters; not equal accuracy & 1,650 steps at acoustic
Courant number 1.63 & Explicit schemes, same weakly compressible system: $3.2$--$5.4\times10^{3}$ steps. Projection at
unit advective Courant number: 450--1,470 & 1.9--3.3 times fewer steps than explicit schemes; 1.1--3.7 times more than
projection (Proposition E3)\\
\addlinespace[1pt]
\multicolumn{5}{l}{\textit{Complete hybrid loop}}\\
Loop as designed & Field to field at every step; gates; compiled circuits, arbitrary fields, full-field readout
(assumptions I1--I3, K1--K2) & $\geq 5.3\times10^{7}$ output copies and
$\Omega(\mathcal{N}^{2}/\log^{2}\mathcal{N})$ gates per step; classical predictor of 96 stencil applications &
Same step evaluated classically: $\Theta(\mathcal{N}d)$ operations. Pseudo-spectral step of the same problem:
$\approx 2.7\times10^{10}$ floating-point operations & Floor of $\approx 1.8\times10^{15}$ CNOT gates per step at leading order; specific to full-field
output and to fields without a compact description (Propositions 5--7; Methods)\\
\bottomrule
\end{tabular}}
\end{center}
\clearpage

\begin{center}
\captionof{table}[{Three-grid ordering and a realization-spread proxy.}]{\textbf{Three-grid ordering and a
realization-spread proxy.} Transfer gain $\delta/\varepsilon$ (real part, full state) at $\varepsilon = 10^{-4}$ and
$t = 3.0$ for census-weighted and white injection, one noise realization per grid (left), and the spread over eight
noise realizations of the $64^{3}$ flow from one initial condition (right). Grid, Reynolds number, initial-condition
realization, injection placement, ladder length, resolution and noise-onset phase change together across grids: an
ordering, not a trend. The census-to-white ratio shows how the magnitude depends on the error direction; it does not
attribute the gain to the structured component. At $256^{3}$ the census-weighted and white runs used different
emulation modes.}\label{edtab:grids}
\smallskip
{\small
\begin{tabular}{lcccc}
\toprule
Grid & Re & census-weighted & white & census / white\\
\midrule
$64^{3}$ & 120 & 7.93 & 6.92 & 1.15\\
$128^{3}$ & 300 & 11.26 & 7.66 & 1.47\\
$256^{3}$ & 800 & 29.61 & 21.16 & 1.40\\
\bottomrule
\end{tabular}
\qquad
\begin{tabular}{lcc}
\toprule
$t$ & mean $\delta/\varepsilon$ & CV\\
\midrule
1.5 & 4.00 & 1.4\%\\
2.0 & 6.38 & 2.3\%\\
2.5 & 7.44 & 3.6\%\\
3.0 & 8.30 & 4.9\%\\
\bottomrule
\end{tabular}}

\smallskip
{\footnotesize Context only, not compared with $\delta/\varepsilon$: growth of the difference between noise-free twin
runs over $1.41 \le t \le 3.3$ is 2.34-fold at $64^{3}$ and 34.3-fold at $256^{3}$ (pseudo-spectral solver) and
2.19-fold for the integrator at $64^{3}$. CV, coefficient of variation.\par}
\end{center}
\clearpage

\begin{center}
\captionof{table}[{Runs.}]{\textbf{Runs.} All runs: $\Delta t = 0.002$, $\mathrm{Ma} = 0.05$, Chebyshev degree 48,
staggered grid, predictor--corrector Strang step, one initial condition per grid; errors, where present, from
$t = 1.41$. Census-weighted: the error direction of Methods; white only and structured only: its components alone.
As-compiled order: $z$-index bits first; interleaved: $z_0, y_0, x_0, z_1, \ldots$ (an in-model permutation of the
ladder weights). Real filter: the imaginary part of the decoded field is dropped after every noisy step; renormalization:
the decoded field is rescaled by $1/(1 - \varepsilon^{2}/2)$ after every noisy step. Emulation modes: both are exact
emulations of the same post-selected action (Methods). Not listed: the divergence-control runs of Extended Data
Fig.~\ref{edfig:divergence} (Supplementary Note~\ref{note:D}), the noise-free twin runs of Extended Data
Table~\ref{edtab:grids} and two further $64^{3}$ filtered runs at $\varepsilon = 10^{-2}$ (Supplementary
Note~\ref{note:E}).}\label{edtab:runs}
\smallskip
{\footnotesize\setlength{\tabcolsep}{3pt}
\begin{tabular}{>{\raggedright\arraybackslash}p{0.19\linewidth}>{\raggedright\arraybackslash}p{0.09\linewidth}%
>{\raggedright\arraybackslash}p{0.20\linewidth}>{\raggedright\arraybackslash}p{0.08\linewidth}%
>{\raggedright\arraybackslash}p{0.06\linewidth}>{\raggedright\arraybackslash}p{0.08\linewidth}%
>{\raggedright\arraybackslash}p{0.17\linewidth}}
\toprule
Run & Grid, Re & Error direction, register order & $\varepsilon$ & Realiz. & Mode & Used in\\
\midrule
Production (noise-free) & $256^{3}$, 800 & none & -- & -- & pair & Figs~\ref{fig:verification}d, \ref{fig:turbulence}; ED Fig.~\ref{edfig:divergence}b,c; ED Table~\ref{edtab:stats}; reference for all $256^{3}$ runs\\
Amplitude sweep & $256^{3}$, 800 & census-weighted, as compiled & $10^{-6}$, $10^{-4}$, $10^{-2}$ & 1 & system-only & Figs~\ref{fig:verification}d, \ref{fig:transfer}, \ref{fig:feedback}; ED Figs~\ref{edfig:amplitude}--\ref{edfig:origin}; ED Table~\ref{edtab:grids}\\
White-noise run & $256^{3}$, 800 & white only & $10^{-4}$ & 1 & pair & ED Figs~\ref{edfig:passive}d, \ref{edfig:origin}; ED Table~\ref{edtab:grids}\\
Interleaved-order run & $256^{3}$, 800 & census-weighted, interleaved & $10^{-4}$ & 1 & pair & ED Fig.~\ref{edfig:origin}b\\
Renormalized run & $256^{3}$, 800 & census-weighted, as compiled; real filter and renormalization & $10^{-2}$ & 1 & pair & ED Fig.~\ref{edfig:amplitude}c\\
Noise-free $128^{3}$, $64^{3}$ & $128^{3}$, 300; $64^{3}$, 120 & none & -- & -- & pair & references at these grids\\
$128^{3}$ noisy runs & $128^{3}$, 300 & census-weighted; white only & $10^{-4}$ & 1 & pair & ED Fig.~\ref{edfig:origin}; ED Table~\ref{edtab:grids}\\
$64^{3}$ census-weighted & $64^{3}$, 120 & census-weighted, as compiled & $10^{-4}$ & 8 & pair & ED Figs~\ref{edfig:passive}e, \ref{edfig:origin}; ED Table~\ref{edtab:grids}\\
$64^{3}$ white, structured-only, interleaved & $64^{3}$, 120 & white only; structured only; census-weighted interleaved & $10^{-4}$ & 2 & pair & ED Fig.~\ref{edfig:origin}; ED Table~\ref{edtab:grids}\\
$64^{3}$ filter tests & $64^{3}$, 120 & census-weighted; none, real filter, or real filter and renormalization & $10^{-2}$ & 1 & pair & Supplementary Note~\ref{note:E}\\
\bottomrule
\end{tabular}}

\smallskip
{\footnotesize Realiz., number of noise realizations; Mode, emulation mode: pair, qubitized-pair emulation; system-only, each factor's
post-selected polynomial applied directly to the system state; ED, Extended Data.\par}
\end{center}
\clearpage

\clearpage
\phantomsection\addcontentsline{toc}{section}{Supplementary Information}
\begin{center}{\Large Supplementary Information\par}\end{center}
\begin{SIpart}

\noindent The Supplementary Notes give the derivations, implementation details and verification procedures behind the
main text and Methods. Notes~\ref{note:A}--\ref{note:D} concern the integrator and Notes~\ref{note:E}--\ref{note:G} the
analysis of repeated per-step errors.

\begin{itemize}
\item Supplementary Note~\ref{note:A}: The time-marching scheme
\item Supplementary Note~\ref{note:B}: Block encodings, GQSP and exact emulation
\item Supplementary Note~\ref{note:C}: Computational complexity and cost structure
\item Supplementary Note~\ref{note:D}: Numerical verification procedures
\item Supplementary Note~\ref{note:E}: Effective per-step error model
\item Supplementary Note~\ref{note:F}: Contamination measures, matched accumulation and the passive control
\item Supplementary Note~\ref{note:G}: Origin attribution of the largest-scale contamination
\end{itemize}

\noindent\emph{Notation.} $N$ is the number of grid points per direction and $\mathcal{N} = N^{3}$ their total; the state
$q = (u, v, w, p_{\mathrm{s}})$ has $4\mathcal{N}$ components and occupies $n$ system qubits; $p_{\mathrm{s}} =
\mathrm{Ma}\,p$ is the scaled pressure. Re denotes the Reynolds number, and $\Re(\cdot)$ and $\Im(\cdot)$ real and
imaginary parts. $\varepsilon$ is the per-step error amplitude and $\varepsilon_{\mathrm{d}}$ the kinetic-energy
dissipation rate. $A$ and $D$ are the convection--acoustic and diffusion operators, $\bar{H}_A$ and $\bar{H}_D$ their
normalized Hermitian forms and $B_A$, $B_D$ their self-inverse block encodings. The time step is $\Delta t$; the Chebyshev degree of a factor is distinct from its number of block-encoding queries. Citations use the
single reference list at the end of the document.

\section{The time-marching scheme}\label{note:A}

This note derives the properties of the hybrid quantum–classical integrator that the Methods state without proof. The integrator is a single explicit time-marching loop in which velocity and pressure advance together through a classical midpoint predictor and a three-factor Strang corrector of GQSP matrix functions, without a per-step pressure-Poisson solve; the realization and emulation of the corrector factors are described in Supplementary Note~\ref{note:B}.

\subsection{Scaled EDAC equations}\label{si:A.1}
Entropically damped artificial compressibility (EDAC) replaces the incompressibility constraint by a pressure equation with a finite sound speed, pressure advection and pressure diffusion~\cite{Clausen2013a,Clausen2013b}. In code units, with $\nu = 1/\mathrm{Re}$, the Mach number Ma based on the initial root-mean-square velocity $u_{\mathrm{rms}}(0) = 1$ and the sound speed $c = 1/\mathrm{Ma}$, the pressure obeys
\disp{$\partial p/\partial t = -(1/\mathrm{Ma}^{2})\nabla\cdot u - u\cdot\nabla p + \nu\nabla^{2}p$,}{A1}
\noindent with the momentum viscosity as pressure diffusivity. Writing $p = p_{\mathrm{s}}/\mathrm{Ma}$ in the momentum equation and multiplying (A1) by Ma gives the integrated system
\disp{$\partial u/\partial t = -(u\cdot\nabla)u - (1/\mathrm{Ma})\nabla p_{\mathrm{s}} + \nu\nabla^{2}u$,}{A2}
\disp{$\partial p_{\mathrm{s}}/\partial t = -(u\cdot\nabla)p_{\mathrm{s}} - (1/\mathrm{Ma})\nabla\cdot u + \nu\nabla^{2}p_{\mathrm{s}}$.}{A3}
Both acoustic couplings now carry the factor $1/\mathrm{Ma}$; in terms of $p$ they carry 1 and $1/\mathrm{Ma}^{2}$. EDAC approaches the incompressible equations with $O(\mathrm{Ma}^{2})$ errors~\cite{Clausen2013b}.

\subsection{MAC operators and the frozen generator}\label{si:A.2}
On the periodic MAC grid~\cite{Harlow1965} of $N^{3}$ points ($N$ per direction, spacing $h = 2\pi/N$), $p_{\mathrm{s}}$ sits at cell centres and $u$, $v$, $w$ on the faces. With the periodic shifts $(S^{\pm}_{a}f)_{i} = f_{i \pm e_{a}}$ along axis $a$, the gradient (centre to face) and the divergence (face to centre) are
\disp{$G_{a} = (S^{+}_{a} - I)/h$,\quad $\mathrm{Div}_{a} = (I - S^{-}_{a})/h$.}{A4}
Each $S^{+}_{a}$ is a permutation with $(S^{+}_{a})^{T} = S^{-}_{a}$, so $G_{a}^{T} = -\mathrm{Div}_{a}$: $\mathrm{Div} = -G^{T}$ exactly. Advection uses the skew-symmetric form~\cite{Morinishi1998} $\tfrac{1}{2}[(\bar{u}\cdot\nabla)f + \nabla\cdot(\bar{u}f)]$ with central differences over $2h$. For a field at location $\ell \in \{u, v, w, p_{\mathrm{s}}\}$,
\disp{$(C_{\ell}f)_{i} = \sum_{a}[w^{+}_{a,\ell}(i)\,f_{i+e_{a}} + w^{-}_{a,\ell}(i)\,f_{i-e_{a}}]$,\quad $w^{\pm}_{a,\ell}(i) = \pm[U_{\ell,a}(i) + U_{\ell,a}(i \pm e_{a})]/(4h)$,}{A5}
\noindent which equals $\tfrac{1}{2}[U\cdot\nabla_{2h}f + \nabla_{2h}\cdot(Uf)]_{i}$, with $\nabla_{2h}$ the central difference over $2h$. $U_{\ell,a}$ is the $a$-component of the frozen advecting velocity $\bar{u}$ at location $\ell$: the field's own component as stored, a four-point average (two successive two-point averages) for a transverse component on a face, and a backward two-point average at the cell centre; the weight averages it once more, to the midpoint between advected nodes. The same $2h$ stencil acts at every location and on all four fields, rather than the compact Harlow–Welch flux form. With $q = (u, v, w, p_{\mathrm{s}})$,
\disp{$dq/dt = A(\bar{u})q + Dq$,\quad $A(\bar{u}) = -[A_{\mathrm{adv}}(\bar{u}) + (1/\mathrm{Ma})[[0, G], [\mathrm{Div}, 0]]] = -A_{\mathrm{adv}}(\bar{u}) + A_{\mathrm{ac}}$,\quad $A_{\mathrm{adv}}(\bar{u}) = \mathrm{diag}(C_{u}, C_{v}, C_{w}, C_{p_{\mathrm{s}}})$,\quad $A_{\mathrm{ac}} = -(1/\mathrm{Ma})[[0, G], [\mathrm{Div}, 0]]$,\quad $D = I_{4} \otimes \nu L_{h}$,}{A6}
\noindent where $L_{h} = \sum_{a}(S^{+}_{a} + S^{-}_{a} - 2I)/h^{2}$ is the periodic seven-point Laplacian.

\emph{Anti-Hermiticity of $A(\bar{u})$.} From (A5), $w^{-}_{a,\ell}(i) = -w^{+}_{a,\ell}(i - e_{a})$. Hence $C_{\ell} = F_{\ell} - F_{\ell}^{T}$ with $F_{\ell} = \sum_{a}\mathrm{diag}(w^{+}_{a,\ell})S^{+}_{a}$: each advection block is real skew-symmetric for any real coefficient field, with no divergence-free condition on $\bar{u}$. Because $\mathrm{Div} = -G^{T}$, the acoustic block $(1/\mathrm{Ma})[[0, G], [-G^{T}, 0]]$ is skew as well, so $A(\bar{u})$ is real skew-symmetric, that is anti-Hermitian, for every real $\bar{u}$. Then $e^{A\,\Delta t}$ is unitary and the frozen convection–acoustic flow conserves $\|u\|^{2} + \|p_{\mathrm{s}}\|^{2}$, since $d\|q\|^{2}/dt = \langle q, (A + A^{\dagger})q\rangle = 0$. With the physical pressure the acoustic block would be $[[0, G], [(1/\mathrm{Ma}^{2})\mathrm{Div}, 0]]$, which is not skew. For complex weights $C_{\ell}^{\dagger} = -\bar{C}_{\ell}$, so anti-Hermiticity needs a real $\bar{u}$; the coefficient refresh therefore uses only the real part $\Re(\cdot)$ of the decoded velocity, and the weights are real. Staggered gradient–divergence pairs keep acoustic operators anti-Hermitian~\cite{Bosch2025}.

\emph{Properties of $D$.} $L_{h}$ is symmetric and circulant, with eigenvalues $-\sum_{a}(4/h^{2})\sin^{2}(\theta_{a}/2) \in [-12/h^{2}, 0]$ ($\theta_{a} = k_{a}h$; the extreme value at the Nyquist corner, present for even $N$). $D$ is therefore Hermitian and negative semidefinite with $\|D\|_{2} = 12\nu/h^{2}$ (24.9007 at $256^{3}$, $\mathrm{Re} = 800$), $e^{D\,\Delta t/2}$ is a contraction for every $\Delta t$, and $D$, which does not depend on the state, is built once.

\emph{Skew pressure advection.} In continuum terms the skew operator applied to $p_{\mathrm{s}}$ is $u\cdot\nabla p_{\mathrm{s}} + \tfrac{1}{2}p_{\mathrm{s}}\nabla\cdot u$, whereas Clausen's form is $u\cdot\nabla p$. By (A1), $\nabla\cdot u = -\mathrm{Ma}^{2}(\partial_{t}p + u\cdot\nabla p - \nu\nabla^{2}p) = O(\mathrm{Ma}^{2})$ when the pressure rates are $O(1)$, so with $p_{\mathrm{s}} = \mathrm{Ma}\,p$ the extra term $\tfrac{1}{2}p_{\mathrm{s}}\nabla\cdot u$ is $O(\mathrm{Ma}^{3})$ against $O(\mathrm{Ma})$ for $u\cdot\nabla p_{\mathrm{s}}$, a relative $O(\mathrm{Ma}^{2})$; the skew-form momentum advection, $u\cdot\nabla u + \tfrac{1}{2}u\nabla\cdot u$, differs from the advective form in the same way. The statement is asymptotic, not a pointwise bound for this run, whose discrete divergence maximum reaches 1.560 at $t = 0.954$ (Extended Data Fig.~\ref{edfig:divergence}c).

\subsection{Corrector, commutators and predictor}\label{si:A.3}
The corrector is the symmetric (Strang) splitting~\cite{Strang1968}
\disp{$q_{n+1} = e^{D\,\Delta t/2}\,e^{A(u_{\mathrm{mid}})\,\Delta t}\,e^{D\,\Delta t/2}\,q_{n}$,}{A7}
\noindent with diffusion outside, the ordering also used for skew-symmetric variable-coefficient scalar transport~\cite{Akash2026}; unitary advection had earlier been split from post-selected diffusion for passive-scalar transport~\cite{Pfeffer2025}. Both kinds of factor are normal, whereas $L = A + D$ is normal only if $[A, D] = 0$, because $LL^{\dagger} - L^{\dagger}L = 2[A, D]$ (with the commutation below, only if $[A_{\mathrm{adv}}, D] = 0$); the split is what makes eigenvalue-based GQSP applicable (Supplementary Note~\ref{note:C}, §9.1–9.2). Because the factors are exponentiated, the corrector imposes no acoustic stability limit at the production acoustic Courant number $\Delta t/(\mathrm{Ma}\,h) = 1.63$; its accuracy limit there is not established.

\emph{Commutators.} On the uniform periodic MAC grid with equal momentum and pressure diffusivities, every block of $A_{\mathrm{ac}}$ and of $D$ is a polynomial in the same commuting periodic shifts, so $[D, A_{\mathrm{ac}}] = 0$ exactly (Supplementary Note~\ref{note:C}, Lemma S2; checks in Supplementary Table D3) and $[D, A] = -[D, A_{\mathrm{adv}}]$: the splitting error comes from the commutator of advection and diffusion (Supplementary Note~\ref{note:C} writes $A_{\mathrm{adv}}$ for $-\mathrm{diag}(C_{\ell})$; commutator norms are unchanged). In the commutator bound of Supplementary Note~\ref{note:C} (Proposition S3) the acoustic part enters only through the cross term $[A_{\mathrm{ac}}, [A_{\mathrm{adv}}, D]]$, which nevertheless dominates and scales as $1/\mathrm{Ma}$. On three production states the realized frozen-coefficient splitting error is $1.9\text{--}6.7 \times 10^{-6}$ of the state norm per step (evaluated with $A(u_{n})$ in place of $A(u_{\mathrm{mid}})$; freezing error excluded). Diffusion is the outer factor because it commutes with the stiff acoustic part and needs no rebuild; halving $A$ instead would need a mid-step re-linearization and a second convection factor. With a different pressure diffusivity $[D, A_{\mathrm{ac}}] \neq 0$, and with walls the commutation is not established.

\emph{Role of the predictor.} Freezing $A$ at $u_{n}$ over the whole step leaves a first-order error. In an ideal-path test ($16^{3}$ MAC grid, nonlinear, smooth test state; Supplementary Table D4), frozen-coefficient Lie and Strang steps both converge at order 0.98, whereas with one frozen linear operator ($8^{3}$) Lie gives 1.00 and Strang 2.00: the composition is second order, but the freezing error dominates. The corrector therefore takes its convection coefficient from a midpoint velocity predicted by a classical Lie half-step,
\disp{$\psi_{\mathrm{mid}} = C_{48}[e^{D\,\Delta t/2}]\,C_{48}[e^{A(u_{n})\,\Delta t/2}]\,\psi_{n}$,\quad $u_{\mathrm{mid}} = \Re(\psi_{\mathrm{mid}}^{u})\,\|q_{n}\|$,}{A8}
\noindent where $\psi_{n} = q_{n}/\|q_{n}\|$, $C_{48}$ is a degree-48 Chebyshev series of the frozen factor applied directly to the state (no ancillas, block encoding or post-selection) and $\psi_{\mathrm{mid}}^{u}$ is the velocity part of $\psi_{\mathrm{mid}}$; the predicted pressure is discarded. Because $A(u(t)) - A(u_{n}) = O(t - t_{n})$, a step frozen at $u_{n}$ has an $O(\Delta t^{2})$ local error, whereas (A8) gives $u_{\mathrm{mid}} = u(t_{n} + \Delta t/2) + O(\Delta t^{2})$ and freezing at the midpoint leaves an $O(\Delta t^{3})$ local error. Splitting and freezing errors are then both $O(\Delta t^{3})$ locally, and the order test does not separate them: the ideal-path test gives 2.03, and the production predictor–corrector path gives 2.081, 2.016 and 2.003 at $16^{3}$ (Supplementary Note~\ref{note:D}, \ref{si:D.4}). The integrator is thus a hybrid, per-step re-linearized (frozen-coefficient) exponential splitting whose second-order mechanism is classical (limitations 2 and 6 in Supplementary Note~\ref{note:B}, \ref{si:B.7}).

\subsection{EDAC-balanced initial pressure}\label{si:A.4}
The initial velocity has the energy spectrum $E(k) = 16(2/\pi)^{1/2}(u_{\mathrm{rms}}^{2}/k_{p})(k/k_{p})^{4}\exp[-2(k/k_{p})^{2}]$ with $k_{p} = 4$ and $u_{\mathrm{rms}} = 1$ (kinetic energy 1.5), usually referred to as the Passot–Pouquet spectrum~\cite{Passot1987}; it is drawn as complex Gaussian modes, projected onto the discrete solenoidal subspace with the MAC divergence symbols $m_{a} = (1 - e^{-i\theta_{a}})/h$ and rescaled to the exact $u_{\mathrm{rms}}$, leaving a MAC divergence of $1.2 \times 10^{-13}$. The pressure is chosen so that the discrete divergence does not start to change. Writing the velocity part of (A6) as $du/dt = -\mathrm{adv}(u) - (1/\mathrm{Ma})G\,p_{\mathrm{s}} + \nu L_{h}u$, where $\mathrm{adv}(u_{0}) = (C_{u}u_{0}, C_{v}v_{0}, C_{w}w_{0})$ is the advection term (the convection operator applied with zero pressure),
\disp{$d(\mathrm{Div}\,u)/dt|_{t=0} = -\mathrm{Div}\,\mathrm{adv}(u_{0}) - (1/\mathrm{Ma})\,\mathrm{Div}\,G\,p_{\mathrm{s}} + \nu L_{h}\,\mathrm{Div}\,u_{0}$.}{A9}
$\mathrm{Div}$ and $L_{h}$ commute, so the last term vanishes with $\mathrm{Div}\,u_{0}$ (to the $10^{-13}$ level), and $\mathrm{Div}\,G = -G^{T}G$ has the Fourier symbol $-|m|^{2}$, with $|m|^{2} = \sum_{a}(2\sin(\theta_{a}/2)/h)^{2}$, which on the MAC grid vanishes only at $k = 0$. Setting (A9) to zero mode by mode gives
\disp{$\hat{p}_{\mathrm{s}} = \mathrm{Ma}\,\hat{S}/|m|^{2}$,\quad $S = \mathrm{Div}\,\mathrm{adv}(u_{0})$,}{A10}
\noindent with the $k = 0$ mode set to zero (zero mean). Because $S$ and $|m|^{2}$ come from the scheme's own operators, $d(\mathrm{Div}\,u)/dt$ vanishes at $t = 0$ for the discrete system that is integrated; in the continuum limit (A10) becomes the pressure Poisson equation for $p = p_{\mathrm{s}}/\mathrm{Ma}$. At $256^{3}$, $\max|d(\mathrm{Div}\,u)/dt|$ at $t = 0$ is 241.29 with a uniform pressure and $2.28 \times 10^{-11}$ with (A10) (Supplementary Table D3). The projection and (A10) are solved by FFT at $t = 0$ (11 FFTs of size $256^{3}$) and are the only field-size linear solves of the run. The balanced pressure is designed to suppress the acoustic start-up mode that artificial-compressibility initial data excite; in the classical artificial-compressibility method even incompressible-consistent initial data excite an acoustic mode with an $O(\mathrm{Ma}^{2})$ velocity amplitude~\cite{Ohwada2010}. Its dynamic effect is examined in Supplementary Note~\ref{note:D}, \ref{si:D.5}.

\section{Block encodings, GQSP and exact emulation}\label{note:B}

“Exact emulation” means that the post-selected GQSP outputs are computed exactly for a self-inverse block encoding of the production $\bar{H}$. Circuits exist for both production factors (B.4): the diffusion encoding and the staggered convection encoding, whose top-left block is $-iA/\lambda_{A}$ by construction. In the runs the corrector factors are evaluated in exact classical emulation and no circuit is executed; the gate-level checks are in Supplementary Note~\ref{note:D}.

\subsection{Normalized operators}\label{si:B.1}
Each factor of (A7) is a function of a Hermitian operator $\bar{H}$ with $\|\bar{H}\| \leq 1$, accessed through a block encoding~\cite{LowChuang2019,Gilyen2019}:
\disp{$\bar{H}_{A} = -iA(u_{\mathrm{mid}})/\lambda_{A}$,\quad $e^{A\,\Delta t} = e^{i\tau_{A}\bar{H}_{A}}$,\quad $\tau_{A} = \lambda_{A}\,\Delta t$;\quad $\bar{H}_{D} = D/\lambda_{D}$,\quad $e^{D\,\Delta t/2} = e^{\tau_{D}\bar{H}_{D}}$,\quad $\tau_{D} = \lambda_{D}\,\Delta t/2$.}{B1}
$\bar{H}_{A}$ is Hermitian because $A$ is real skew-symmetric (Supplementary Note~\ref{note:A}). The convection normalization is a triangle-inequality bound on $\|A\|_{2}$ over the stencil terms,
\disp{$\lambda_{A} = 2(\gamma_{x} + \gamma_{y} + \gamma_{z}) + 6/(\mathrm{Ma}\,h)$,\quad $\gamma_{a} = \max_{\ell,\,i}\max(|w^{+}_{a,\ell}(i)|, |w^{-}_{a,\ell}(i)|)$,}{B2}
\noindent with acoustic part $6/(\mathrm{Ma}\,h) = 4{,}889.24$ at $256^{3}$, above $\|A_{\mathrm{ac}}\|_{2} = 2\sqrt{3}/(\mathrm{Ma}\,h)$; overall $\lambda_{A}/\|A\|_{2} \approx 1.81\text{--}1.92$ at $256^{3}$ (Supplementary Note~\ref{note:C}). For diffusion $\lambda_{D} = 12\nu/h^{2} = \|D\|_{2}$ exactly (24.9007 at $256^{3}$), so $\bar{H}_{D}$ has its spectrum in $[-1, 0]$. Production values of $\lambda_{A}$, $\tau_{A}$ and $\tau_{D}$ are in Supplementary Table B1; $\lambda_{A}$ and $\tau_{A}$ there are evaluated at $u_{n}$, because the corrector's $\lambda_{A}(u_{\mathrm{mid}})$ is not recorded.

\subsection{GQSP sequences: degree versus query count}\label{si:B.2}
GQSP~\cite{Motlagh2024} implements a polynomial in the qubitized walk $W = (2\Pi - I)B$, where $B$ is a block encoding with $\langle 0|B|0\rangle = \bar{H}$ and $\Pi$ projects onto the all-zero block-encoding register; on the two-dimensional invariant subspace of an eigenvalue $x = \cos\theta$ of $\bar{H}$, $W$ has the eigenphases $e^{\pm i\theta}$~\cite{LowChuang2019}. The production synthesis has four steps.
\begin{enumerate}
\item\relax The target $f(x)/2$, with $f = e^{i\tau_{A}x}$ or $e^{\tau_{D}x}$, is fitted by a Chebyshev series $\sum_{k=0}^{48}c_{k}T_{k}(x)$ (least squares on 784 nodes; truncation degree 48 in every block).
\item\relax With $z = e^{i\theta}$, $T_{k}(x) = (z^{k} + z^{-k})/2$, so the series is a Laurent polynomial with powers $-48$ to 48. Negative powers are realized by controlled inverse walks~\cite{Motlagh2024}, after multiplication by $z^{K}$ with $K = 48$ (the Laurent shift).
\item\relax Coefficients of the highest powers with magnitude $\leq 10^{-12}$ are trimmed; the low-power end is never trimmed.
\item\relax The sequence has $K = 48$ controlled inverse walks (on $|1\rangle$ of the GQSP signal qubit) and $k_{\mathrm{max}}$ controlled forward walks (on $|0\rangle$), $k_{\mathrm{max}}$ being the highest order with $|c_{k}|/2 > 10^{-12}$, separated by single-qubit processing rotations.
\end{enumerate}
One query is one application of controlled $W$, controlled $W^{-1}$ or a controlled choice between them, so a factor costs $K + k_{\mathrm{max}}$ queries: $53 = 48 + 5$ per diffusion half-step and 79 or $78 = 48 + 31$ or 30 for convection, that is 184–185 per step and 304,252 over the 1,650 steps. The Chebyshev degree is therefore not a query count: the counts include the 48 untrimmed inverse walks, and the $K - k_{\mathrm{max}}$ lowest powers, with coefficients $\leq 10^{-12}$, are queried anyway. The convection counts use the normalization $\lambda_{A}$, which the staggered encoding of B.4 attains exactly; the predictor's $C_{48}$ factors, applied classically, have no query count. The convection count changes from 78 to 79 between $\tau = 10.4695$ and 10.7534, where one more Laurent term passes the trimming threshold; the switches between 79 and 78 in the run (Supplementary Table B1) are consistent with $\tau_{A}(u_{\mathrm{mid}})$ crossing this point, which is an interpretation because $\tau_{A}(u_{\mathrm{mid}})$ is not recorded. The degree-48 series has a scalar truncation error of $6.76 \times 10^{-15}$ even at $\tau = 19.883$, a larger $\tau$ than that of the $256^{3}$ convection blocks; the truncation error is not monitored during the run. The angles are synthesized classically every step (Chebyshev fit, Laurent shift, a complementary polynomial from a nonlinear least-squares fit, angle recursion) at a cost independent of $N$: $\approx 0.19$ s per step against a mean emulated step time of 386.8 s (7.43 days for the run; classical emulation time, not a quantum cost). Supplementary Note~\ref{note:C} gives the query complexity, and constructions with fewer queries that are prospective and not implemented.

\subsection{Post-selection, design success and readout gain}\label{si:B.3}
The targets carry the scale 1/2, a design margin: $|P| \leq 1/2$ on the unit circle guarantees that a complementary polynomial exists, although completion also succeeds without it for the production convection factor (Supplementary Note~\ref{note:C}, §4.3). The margin sets the design success probability of 1/4 per factor. After each factor the ancillas are post-selected on $|0\rangle$; the post-selected block applies $P(\bar{H}) \approx f(\bar{H})/2$, so the success probability for a normalized input $\psi$ is $\|f(\bar{H})\psi\|^{2}/4$. For convection $f(\bar{H}_{A}) = e^{A\,\Delta t}$ is unitary and the probability is exactly 1/4 for every state (0.250000 in all 1,650 convection blocks, an in-run consistency check of the anti-Hermitian structure); for diffusion it is $\|e^{D\,\Delta t/2}\psi\|^{2}/4 \leq 1/4$ (0.249745–0.249987 over the 3,300 half-steps; means 0.249838 and 0.249837 for the two half-steps), the deficit being the physical viscous contraction. The design value per step is $(1/4)^{3} = 1/64 = 0.015625$. These probabilities are computed from exact amplitudes and never sampled; repetition for post-selection, amplitude amplification and estimation of the success probability or the norm are not modelled. The post-selected amplitudes pass unnormalized from factor to factor, so the decode is
\disp{$q_{n+1} = \psi_{\mathrm{out}}\,\|q_{n}\|\cdot 8$,\quad $8 = 1/(1/2)^{3}$,}{B3}
\noindent with 8 the deterministic design gain; because the decode does not renormalize, the viscous decay is kept. The decoded fields are complex; in the noise-free run the imaginary part stays at round-off. Post-selecting after each factor gives the same output as one joint post-selection over three disjoint ancilla registers (Supplementary Table D1).

\subsection{Implemented block encodings}\label{si:B.4}
The three implemented encodings are linear combinations of unitaries with a constant number of branches~\cite{LowChuang2019,Gilyen2019}.
\begin{itemize}
\item\relax \emph{Diffusion} (layout-independent): four branches with blocks $-I$ and $(S^{+}_{a} + S^{-}_{a})/2$ ($a = x, y, z$), with $\ell_{1}$-norm $\lambda_{D}$; three block-encoding ancillas plus the GQSP signal qubit; $\langle 0|B_{D}|0\rangle = D/\lambda_{D}$. Each branch unitary ($-I$, or $H\,V_{a}\,H$ with $V_{a} = |0\rangle\langle 1| \otimes S^{+}_{a} + |1\rangle\langle 0| \otimes S^{-}_{a}$ on one extra qubit) is a Hermitian involution, and the state-preparation conjugation preserves this, so $B_{D}^{2} = I$ by construction.
\item\relax \emph{Collocated convection} (verification only): three advective branches (weight $2\gamma_{a}$, weight diagonals loaded by uniformly controlled rotations) and six acoustic branches (field-register Pauli strings with shifts, weight $1/(2\mathrm{Ma}\,h)$ each), so $\lambda_{\mathrm{coll}} = 2\sum_{a}\gamma_{a} + 3/(\mathrm{Ma}\,h)$; six block-encoding ancillas plus the signal qubit. Each branch enters as a difference branch with top-left block $-i(F_{j} - F_{j}^{\dagger})/2$, so $\langle 0|B|0\rangle = -iA/\lambda_{\mathrm{coll}}$, and the encoding is self-inverse by construction.
\item\relax \emph{Staggered convection} (production; emulated in the runs): six difference branches on three select qubits, with the same sandwich as the collocated circuit. Advective branches (weight $2\gamma_{a}$): forward unitary $\mathrm{INC}_{a}L_{a}$, with $\mathrm{INC}_{a}$ the modular increment along $a$ and $L_{a}$ a uniformly controlled rotation of one loader qubit, conditioned on the grid address and the two field qubits, whose top-left element is $w^{+}_{a,\ell}(i)/\gamma_{a}$ for field location $\ell$. Acoustic branches (weight $2/(\mathrm{Ma}\,h)$): $\Pi_{a\leftrightarrow p}\,C_{\mathrm{field}=a}(\mathrm{INC}_{a})$, an increment applied only to velocity component $a$ followed by the field-register permutation $|a\rangle \leftrightarrow |p\rangle$, whose difference block is $(i\,\mathrm{Ma}\,h/2)K_{a}$ with $K_{a} = (1/\mathrm{Ma})(|a\rangle\langle p| \otimes G_{a} + |p\rangle\langle a| \otimes \mathrm{Div}_{a})$. Because the production weights satisfy $w^{-}_{a,\ell}(i) = -w^{+}_{a,\ell}(i - e_{a})$ exactly, the six branches give $\langle 0|B_{A}|0\rangle = -iA/\lambda_{A}$ with exactly the normalization (B2), the sum of the branch weights, and $B_{A}^{2} = I$ holds by construction for any branch unitaries. Five block-encoding ancillas (three select qubits, one loader qubit, one difference qubit) plus the signal qubit: 14 qubits for one convection factor at $4^{3}$ and 32 at $256^{3}$ (a three-factor corrector needs 32 with ancilla reuse and 40 with disjoint registers). Compiled to CNOT and one-qubit gates without optimization, one application costs 13,600, 101,476 and 803,868 CNOTs at $2^{3}$, $4^{3}$ and $8^{3}$ (about 58 CNOTs per loader angle at $4^{3}$ and $8^{3}$), so the cost grows in proportion to the number of grid points, $N^{3}$, and is 2.95 times that of the collocated circuit at $8^{3}$. A $256^{3}$ cost is only an extrapolation ($\approx 2.3 \times 10^{10}$ CNOTs per application; Supplementary Note~\ref{note:C}). Gate-level checks: Supplementary Note~\ref{note:D}, \ref{si:D.8}.
\end{itemize}
At $256^{3}$ the system register has 26 qubits ($4 \cdot 256^{3}$ amplitudes).

\subsection{The qubitized-pair representation}\label{si:B.5}
Let $B$ ($B_{A}$ or $B_{D}$) be a unitary on the block-encoding register and the system with $\langle 0|B|0\rangle = \bar{H}$, and $W = (2\Pi - I)B$. The proof needs one condition:

(S) $B^{2} = I$, which for a unitary $B$ is equivalent to $B = B^{\dagger}$.

Then $\bar{H}$ is Hermitian with $\|\bar{H}\| \leq 1$, and $W^{-1} = B(2\Pi - I)$.

\emph{Lemma 1.} For system vectors $a$ and $b$ let $\Phi(a, b) = |0\rangle a + B|0\rangle b$. Under (S),
\disp{$W\Phi(a, b) = \Phi(2\bar{H}a + b, -a)$,\quad $W^{-1}\Phi(a, b) = \Phi(-b, a + 2\bar{H}b)$.}{B4}
\emph{Proof.} By (S), $W(B|0\rangle b) = (2\Pi - I)B^{2}|0\rangle b = |0\rangle b$, and $W|0\rangle a = (2\Pi - I)B|0\rangle a = |0\rangle(2\bar{H}a) - B|0\rangle a$; adding gives the first identity. With $W^{-1} = B(2\Pi - I)$, $W^{-1}|0\rangle a = B|0\rangle a$ and $W^{-1}(B|0\rangle b) = B(2|0\rangle\bar{H}b - B|0\rangle b) = B|0\rangle(2\bar{H}b) - |0\rangle b$, which gives the second. $\blacksquare$ Without (S), $W(B|0\rangle b) = (2\Pi - I)B^{2}|0\rangle b$ is in general not of the form $\Phi(\cdot, \cdot)$, and the parametrization does not close.

\emph{Lemma 2.} At every point of a GQSP sequence the state is $|0\rangle_{g}\,\Phi(a_{0}, b_{0}) + |1\rangle_{g}\,\Phi(a_{1}, b_{1})$, where $g$ is the signal qubit. \emph{Proof} by induction: the input is $|0\rangle_{g}\,\Phi(\psi, 0) + |1\rangle_{g}\,\Phi(0, 0)$; a processing rotation acts on $g$ only and, because $\Phi$ is linear, mixes $(a_{0}, a_{1})$ and $(b_{0}, b_{1})$ component-wise; a controlled $W$ or $W^{-1}$ maps the pair of its branch by (B4). $\blacksquare$

\emph{Read-out.} Because $\langle 0|B|0\rangle = \bar{H}$ and $\|B|0\rangle b\| = \|b\|$, the post-selected system vector (all ancillas, $g$ included, in $|0\rangle$) is $a_{0} + \bar{H}b_{0}$, the success probability is $\|a_{0} + \bar{H}b_{0}\|^{2}/\|\psi\|^{2}$, and the sector norms are $\|\Phi(a, b)\|^{2} = \|a\|^{2} + \|b\|^{2} + 2\Re\langle a, \bar{H}b\rangle$.

Hence the representation is exact, not an approximation: four system-size vectors represent the full ancilla–system state, and each query costs one matrix-free application of $\bar{H}$ (plus two per factor for the read-out). The post-selected vector, the success probability and the two sector populations depend only on $\bar{H}$ and the GQSP angles, and so are the same for every self-inverse $B$ with $\langle 0|B|0\rangle = \bar{H}$; the split of the rejected population over ancilla basis states depends on the circuit and is not tracked. Because the sequence returns the scaled target polynomial on each eigenvector of $\bar{H}$, the pair and system-only backends (B.6) compute the same post-selected action up to round-off.

\emph{Existence.} A self-inverse block encoding with $\langle 0|B|0\rangle = -iA/\lambda$ exists if and only if $A$ is anti-Hermitian and $\lambda \geq \|A\|_{2}$. For Hermitian $\bar{H}$ with $\|\bar{H}\| \leq 1$ the one-qubit Hermitian dilation $B = [[\bar{H}, (I - \bar{H}^{2})^{1/2}], [(I - \bar{H}^{2})^{1/2}, -\bar{H}]]$ satisfies $B^{2} = I$, because $\bar{H}$ commutes with $(I - \bar{H}^{2})^{1/2}$; conversely, a self-inverse unitary is Hermitian, so its block is Hermitian with norm at most one. This is a statement of existence, not an efficient circuit. (S) holds by construction for the diffusion, collocated and staggered convection circuits and is verified on their gates (Supplementary Note~\ref{note:D}; for the staggered circuit on $4^{3}$ and $8^{3}$ grids). The norm drift of the recurrence in the run ($\leq 9.66 \times 10^{-15}$ over the 4,950 corrector blocks) is a necessary-condition check only and tests neither $\|\bar{H}\| \leq 1$ nor (S).

\subsection{Backends and their agreement}\label{si:B.6}
Two emulation backends (the emulation modes of the main text) compute the same post-selected action. The \emph{qubitized-pair backend} (production) evaluates (B4) on four system-size vectors; the \emph{system-only backend} applies each factor's post-selected matrix polynomial, with the GQSP-recovered coefficients and the scale 1/2, directly to the system state and returns only the post-selected output. The amplitude sweep ($\varepsilon = 10^{-6}$, $10^{-4}$ and $10^{-2}$) used the system-only backend; the base run and every other noisy run used the qubitized pair. With either backend the noise acts on the post-selected state, so no noise result involves ancilla dynamics. At $16^{3}$ and $64^{3}$ the two backends agree to round-off for the same $\bar{H}$ (Supplementary Table D2). At $256^{3}$ a five-step comparison of the scalar diagnostics agrees to a worst relative difference of $7.02 \times 10^{-12}$, in the divergence maximum; before noise starts ($t = 1.0$) the field of the amplitude-sweep run at $\varepsilon = 10^{-4}$ differs from the base run by $\delta = 1.87 \times 10^{-12}$ (real part; $2.37 \times 10^{-15}$ imaginary), whereas the white-noise run, on the base run's backend, differs from it by $1.68 \times 10^{-13}$ (Fig.~\ref{fig:verification}d). These comparisons establish emulator consistency, not circuit verification; the staggered circuit is compared with the pair representation in Supplementary Note~\ref{note:D}, \ref{si:D.8}.

\subsection{Production step accounting and scope}\label{si:B.7}
\textbf{Supplementary Table B1 | Accounting of one production step ($256^{3}$ decaying HIT, 1,650 steps).}

{\footnotesize
\setlength{\tabcolsep}{3pt}\renewcommand{\arraystretch}{1.15}\setlength{\LTleft}{0pt}\setlength{\LTright}{0pt}
\begin{longtable}{>{\raggedright\arraybackslash}p{\dimexpr 0.4376\linewidth-6pt\relax}>{\raggedright\arraybackslash}p{\dimexpr 0.5624\linewidth-6pt\relax}}
\toprule
\hspace{0pt}\textbf{Quantity} & \hspace{0pt}\textbf{Value} \\
\midrule
\endhead
\hspace{0pt}Configuration & \hspace{0pt}$256^{3}$, Re 800, Ma 0.05, $\Delta t$ 0.002, $t \leq 3.3$; MAC grid; Strang corrector with classical midpoint predictor; Chebyshev degree 48; target scale 1/2; qubitized-pair backend \\
\hspace{0pt}Classical operations per step & \hspace{0pt}1 encode; 2 coefficient builds (from $u_{n}$ and $u_{\mathrm{mid}}$); predictor with 2 Chebyshev factors (96 stencil applications); 3 phase syntheses; 1 decode; diagnostics \\
\hspace{0pt}Corrector; linear solves & \hspace{0pt}$\mathrm{GQSP}_{D}(\Delta t/2) \to \mathrm{GQSP}_{A}(u_{\mathrm{mid}})(\Delta t) \to \mathrm{GQSP}_{D}(\Delta t/2)$, each post-selected on all-zero ancillas; no linear solve (FFT solves only at $t = 0$) \\
\hspace{0pt}Queries & \hspace{0pt}per factor 53 (5 forward, 48 inverse walks), 79 or 78 (31 or 30 forward, 48 inverse) and 53; per step 184–185; per run 304,252 \\
\hspace{0pt}Convection queries over the run & \hspace{0pt}79 in 652 steps, 78 in 998 (78 in one early interval and throughout $t \geq 1.554$, preceded by an interval with 15 switches) \\
\hspace{0pt}$\lambda_{A}(u_{n})$ at $t = 0$, 0.5, …, 3.0 & \hspace{0pt}$5{,}376.7 \cdot 5{,}409.5 \cdot 5{,}460.0 \cdot 5{,}366.6 \cdot 5{,}234.9 \cdot 5{,}161.4 \cdot 5{,}121.7$ \\
\hspace{0pt}$\tau_{A} = \lambda_{A}\,\Delta t$ & \hspace{0pt}10.24–10.92 (acoustic part 9.78; advective part 0.46–1.14) \\
\hspace{0pt}$\lambda_{D}$; $\tau_{D}$ & \hspace{0pt}24.9007; 0.0249 \\
\hspace{0pt}Acoustic Courant number $\Delta t/(\mathrm{Ma}\,h)$ & \hspace{0pt}1.63 \\
\hspace{0pt}Advective Courant number $\sum_{a}|u_{a}|\,\Delta t/h$ & \hspace{0pt}0.2517–0.8885 \\
\hspace{0pt}Local Mach number & \hspace{0pt}at least about 0.26 ($\max|u_{x}| = 5.23$ at $t = 0.83$) \\
\bottomrule
\end{longtable}}

In the architecture, every step crosses the quantum–classical interface with three post-selections and one full-state readout; in the runs these are logical steps, decoded in exact classical emulation without quantum measurement or sampled post-selection events. Seven limitations apply: (1) the staggered production block-encoding circuit is constructed but not executed in the runs (the pair backend is exact for it, as for any self-inverse encoding of $\bar{H}$), and its arbitrary-field loaders cost a number of gates proportional to $N^{3}$ per query (Supplementary Note~\ref{note:C}, Proposition 5); (2) the predictor is classical, with no defined quantum realization, so the second-order mechanism is classical; (3) full readout, refresh and re-encoding are classical at every step and are not modelled for hardware (no state preparation, tomography, repeat-until-success or norm estimation); (4) end to end, the loop is subject to the per-step gate-count floor of Supplementary Note~\ref{note:C} (Proposition 7), where its costs are compared with classical computation and analysed by kind; (5) nothing was executed on hardware and no in-circuit noise is modelled; (6) second order is shown only at $16^{3}$ (smooth, non-solenoidal test state, 4–64 steps, acoustic Courant number $\leq 1.02$), not at production scale, for the turbulent initial condition or at the production acoustic Courant number of 1.63, and no time-step study of the turbulent run exists; (7) gate-level equality is shown only at $4^{3}$, for the collocated and the staggered circuits ($2^{3}$ is degenerate for convection), with sampled checks of the staggered circuit at $8^{3}$ and no GQSP check there; at $4^{3}$ the staggered GQSP factor agrees with the emulation to $3.3 \times 10^{-10}$, missing the predefined tolerance of $10^{-10}$ (Supplementary Note~\ref{note:D}, \ref{si:D.8}).

\section{Computational complexity and cost structure}\label{note:C}

\begin{NoteCnumbering}

Throughout this note $N$ is the number of grid points per direction, $\mathcal{N} = N^{3}$ the total number of grid points, $4\mathcal{N}$ the state dimension, $n = \log_{2}(4\mathcal{N})$ the number of system qubits and $M$ the number of time steps of size $\Delta t$; I1–I3 and K1–K2 label the interface and cost-model assumptions of Propositions 6–7.

\subsection{Summary}\label{si:C.0}
This note derives the cost of one step of the hybrid integrator and of the trajectory at three levels. \textbf{Level 1 (L1)} is the ideal block-encoding query complexity: calls to block encodings $B_{A}$ and $B_{D}$ of the frozen operators, assumed available. \textbf{Level 2 (L2)} is the gate complexity of the implemented encodings, including data loading, ancillas, phase synthesis and post-selection. \textbf{Level 3 (L3)} is the end-to-end hybrid cost: L2 plus the classical predictor and coefficient refresh, state preparation, full-field readout and repetition over the $M$ steps.

The results are of three kinds, which are kept apart throughout; §14 lists every comparison by kind, with its setting.

\begin{itemize}
\item\relax \textbf{(I) Structure of a step.} These properties hold for every frozen step and do not depend on a cost model. Time marching contains no linear solve (§3). The convection–acoustic factor is unitary and the diffusion factors are contractions for every real frozen velocity and every $\Delta t$ (Lemma S1; Proposition E3, part 1). The splitting error is bounded by nested commutators with explicit constants and no growth factor (Proposition S3). The state occupies $n = \log_{2}(4\mathcal{N})$ system qubits, with a number of ancillas that does not depend on $\mathcal{N}$ (§5.4). Unit-normalized constructions of the factors exist whose success probability is, up to $O(\eta)$ and the synthesis margins, the fraction of the norm that the step retains (Propositions 1–3; proved, prospective in use).
\item\relax \textbf{(II) Subproblems.} Each is compared with other methods for the same task under a stated access model. \emph{Frozen linear step.} At L1 the production configuration uses 184–185 corrector queries per step (53 / 79 or 78 / 53, against Chebyshev degree 48) at a design success of 1/4 per factor and 1/64 per step, which gives 4,852–4,868 expected corrector queries per accepted step under staged post-selection (§4). A structure-adapted synthesis would need 43–45 queries per step at $\approx 0.96$ success; it is prospective, not implemented (§4.6). At L2, loading an arbitrary advecting field costs $\Omega(\mathcal{N}/\log \mathcal{N})$ gates per query in the compiled-circuit model, so the gate count of the step has the scaling $\Theta(Q\mathcal{N})$ of classical stencil evaluation up to a logarithm; a separation requires sublinear access to the field (§5). \emph{Pressure coupling.} In the query model, and in geometries without a fast diagonalizing transform, the projection-free step needs, for transport and pressure coupling and counted with reference degrees, 122–126 times fewer block-encoding calls per step than a step with a projector applied by quantum singular value transformation, and $\approx 93\text{--}100$ times fewer with the diffusion factors of both formulations included (Proposition E1, with its conditions). With a fast-transform projector the projected step needs 2.6–3.6 times fewer data-dependent calls, and when calls are weighted by loader gates it needs fewer gates (§8.4, Proposition E2). \emph{Number of time steps.} At their stability limits explicit schemes for the same system need 1.9–3.3 times more steps than the 1650 of the production run; these counts measure refreshes and interface crossings, not total work or equal accuracy (the accuracy of the production step is not established), and a projection scheme at unit advective Courant number needs 1.1–3.7 times fewer steps than the integrator (Proposition E3).
\item\relax \textbf{(III) Complete hybrid loop.} At L3, conditional on a full-field readout every step, at least $5.3 \times 10^{7}$ output copies per step are needed at $256^{3}$, and the end-to-end floor is $\approx 1.8 \times 10^{15}$ CNOTs per step at leading order, against an estimated $\approx 2.7 \times 10^{10}$ flops for a classical pseudo-spectral RK3 step of the same problem (§6). These floors belong to the interface (full-field readout, arbitrary fields compiled into circuits) and to its cost model, not to the flow problem (§6.3).
\end{itemize}
Sections 7–10 treat accuracy, the EDAC formulation, the split and the per-step re-linearization; §11–13 collect the bounds and the comparisons with published baselines. The comparisons of kind II are statements about call, gate and step counts under the stated models, not measured run times, and a result of one kind is not carried over to another; for the complete loop as designed (kind III), the comparison with classical computation is the conditional floor of Proposition 7 (§6.3).

\subsection{Conventions}\label{si:C.1}
\textbf{Scope.} This note derives costs; it runs no simulation. ``Production'' means the exactly emulated $256^{3}$ run: the corrector factors are evaluated in exact classical emulation (qubitized-pair representation), and no circuit is executed in it. The production configuration, with 184–185 corrector queries per step, is the one reported; configurations called prospective are derived and not implemented (§4.6). The staggered convection block encoding has been constructed and verified at the gate level on small grids only (§5.5). No noise, error-correction or device model enters any count.

\textbf{Status of statements.} Propositions and lemmas are proved here (proof given) or quoted from a primary source together with its model and assumptions. A comparison holds only under the conditions stated with it, including whether its outputs and accuracy targets are matched. Numerical values are identified in words as measured in the production run, computed in the checks of Appendix A, estimates (with their basis) or prospective. A product of a lower bound and an upper estimate is never called a bound.

\textbf{Kinds of comparison and access models.} Every comparison belongs to one of the three kinds of §0: structure of a step (I), a subproblem (II) or the complete hybrid loop (III). A comparison of kind II or III states its input and output (a quantum state or a classical field), the access to the operators, the tolerance and the quantity that is counted (block-encoding queries, gates, classical arithmetic operations or time steps), with its dependence on the grid, on the number of steps and on condition numbers where they enter. Three access models are used. In the \emph{query model} (L1) the block encodings are given and every call is counted once. In the \emph{compiled-circuit model} (assumptions M1–M2 of Proposition 5) every call is a circuit of one- and two-qubit gates built from the classical coefficients, and gates are counted. With a \emph{quantum random-access memory counted as an oracle}, or with compactly described fields, the per-call bound of Proposition 5 does not apply; the comparison is then made in the query model, and the hardware of the memory, $\Theta(\mathcal{N})$ components, is stated with it. The cost of state preparation and readout is counted only in kind III and depends on the interface assumptions I1–I3 and K1–K2 of Propositions 6–7.

\textbf{Sources.} Literature is cited by its primary source, with the theorem, section or equation used; theorem numbers refer to the versions named in §12. Section numbers (§) refer to this note.

\medskip
\subsection{Notation}\label{si:C.2}
{\scriptsize
\setlength{\tabcolsep}{3pt}\renewcommand{\arraystretch}{1.15}\setlength{\LTleft}{0pt}\setlength{\LTright}{0pt}
\begin{longtable}{>{\raggedright\arraybackslash}p{\dimexpr 0.1564\linewidth-6pt\relax}>{\raggedright\arraybackslash}p{\dimexpr 0.3472\linewidth-6pt\relax}>{\raggedright\arraybackslash}p{\dimexpr 0.2786\linewidth-6pt\relax}>{\raggedright\arraybackslash}p{\dimexpr 0.2179\linewidth-6pt\relax}}
\toprule
\hspace{0pt}\textbf{Symbol} & \hspace{0pt}\textbf{Meaning} & \hspace{0pt}\textbf{Production value} & \hspace{0pt}\textbf{Source} \\
\midrule
\endhead
\hspace{0pt}$N$ & \hspace{0pt}grid points per direction & \hspace{0pt}256 & \hspace{0pt}Methods \\
\hspace{0pt}$\mathcal{N} = N^{3}$ & \hspace{0pt}total number of grid points (cells) & \hspace{0pt}$1.678 \times 10^{7}$ & \hspace{0pt}— \\
\hspace{0pt}$n_{\mathrm{dof}} = 4\mathcal{N}$ & \hspace{0pt}state components ($u$, $v$, $w$ on MAC faces; $p_{\mathrm{s}} = \mathrm{Ma}\, p$ at centres) & \hspace{0pt}$6.711 \times 10^{7} = 2^{26}$ & \hspace{0pt}Supplementary Notes~\ref{note:A} (A.2) and \ref{note:B} (B.4) \\
\hspace{0pt}$n = \log_{2}(4\mathcal{N})$ & \hspace{0pt}system qubits & \hspace{0pt}26 & \hspace{0pt}Supplementary Note~\ref{note:B}, B.4 \\
\hspace{0pt}$h = 2\pi/N$ & \hspace{0pt}mesh width & \hspace{0pt}0.0245 & \hspace{0pt}— \\
\hspace{0pt}$\nu = 1/\mathrm{Re}$ & \hspace{0pt}viscosity (momentum and EDAC pressure diffusion), code units & \hspace{0pt}$1.25 \times 10^{-3}$ & \hspace{0pt}Methods; Supplementary Note~\ref{note:A}, A.1 \\
\hspace{0pt}Ma, $c = 1/\mathrm{Ma}$ & \hspace{0pt}Mach number based on $u_{\mathrm{rms}}(0) = 1$; sound speed in these units. The local Mach number reaches $\approx 0.26$ (run maximum $\max |u_{x}| = 5.23$ at $t = 0.83$) & \hspace{0pt}0.05; 20 & \hspace{0pt}Methods; Appendix A \\
\hspace{0pt}$T$, $\Delta t$, $M = T/\Delta t$ & \hspace{0pt}physical time, time step, number of steps & \hspace{0pt}3.3, 0.002, 1650 & \hspace{0pt}Methods \\
\hspace{0pt}$A = A_{\mathrm{adv}}(\allowbreak \bar{u}) + A_{\mathrm{ac}}$ & \hspace{0pt}frozen convection–acoustic generator (anti-Hermitian for real $\bar{u}$ on the MAC grid) & \hspace{0pt}— & \hspace{0pt}Supplementary Note~\ref{note:A}, A.2 \\
\hspace{0pt}$D = I_{4} \otimes \nu L_{h}$ & \hspace{0pt}diffusion on all four fields (Hermitian, negative semidefinite, constant) & \hspace{0pt}— & \hspace{0pt}Supplementary Note~\ref{note:A}, A.2 \\
\hspace{0pt}$B_{A}$, $B_{D}$ & \hspace{0pt}self-inverse block encodings ($B^{2} = I$) of $\bar{H}_{A} = -iA/\lambda_{A}$ and $\bar{H}_{D} = D/\lambda_{D}$ & \hspace{0pt}— & \hspace{0pt}§4.1 \\
\hspace{0pt}$\alpha_{A} = \lambda_{A}$ & \hspace{0pt}LCU normalization of $A$: $2(\allowbreak \gamma_{x} + \gamma_{y} + \gamma_{z}) + 6/\allowbreak (\allowbreak \mathrm{Ma}\, h)$ & \hspace{0pt}5122–5460 on the saved $u_{n}$ & \hspace{0pt}Supplementary Note~\ref{note:B}, eq. (B2) and Table B1 \\
\hspace{0pt}$\alpha_{D} = \lambda_{D}$ & \hspace{0pt}LCU normalization of $D$: $12\nu/h^{2}$ ($= \|D\|$ exactly) & \hspace{0pt}24.90 & \hspace{0pt}Supplementary Notes~\ref{note:A} (A.2) and \ref{note:B} (B.1) \\
\hspace{0pt}$\|A_{\mathrm{ac}}\|$, $\|A\|$ & \hspace{0pt}spectral norms & \hspace{0pt}$2\sqrt{3}/\allowbreak (\allowbreak \mathrm{Ma}\, h) = 2822.8$; $\approx 2823\text{--}2855$ & \hspace{0pt}Appendix A \\
\hspace{0pt}$\tau_{A} = \lambda_{A} \Delta t$ & \hspace{0pt}convection–acoustic ``time'' of one step & \hspace{0pt}10.24–10.92 & \hspace{0pt}Supplementary Table B1 \\
\hspace{0pt}$\tau_{\mathrm{ac}} = 6\Delta t/\allowbreak (\allowbreak \mathrm{Ma}\, h)$, $\tau_{\mathrm{adv}}$ & \hspace{0pt}acoustic and advective parts of $\tau_{A}$ & \hspace{0pt}9.78; 0.46–1.14 & \hspace{0pt}Supplementary Table B1 \\
\hspace{0pt}$\tau_{D} = \lambda_{D} \Delta t/\allowbreak 2$ & \hspace{0pt}diffusion half-step & \hspace{0pt}0.0249 & \hspace{0pt}Supplementary Table B1 \\
\hspace{0pt}$\eta$ & \hspace{0pt}per-factor polynomial tolerance (sup norm on the spectral interval) & \hspace{0pt}trim $10^{-12}$; truncation $6.8 \times 10^{-15}$ & \hspace{0pt}Supplementary Note~\ref{note:B}, B.2 \\
\hspace{0pt}$\delta$ & \hspace{0pt}GQSP synthesis margin: normalization $1/(1 + \delta)$ of a factor whose target is bounded by 1 & \hspace{0pt}production $\delta = 1$ (scale 1/2) & \hspace{0pt}Supplementary Note~\ref{note:B}, B.3; Appendix A \\
\hspace{0pt}$\varepsilon_{\mathrm{acc}}$ & \hspace{0pt}global accuracy target (relative $\ell_{2}$ error at $T$, within the predictability horizon; statistical beyond it). \textbf{Not} the per-step error amplitude $\varepsilon$ of the noise study & \hspace{0pt}— & \hspace{0pt}— \\
\hspace{0pt}$d(\tau, \eta)$ & \hspace{0pt}degree of the Chebyshev (Jacobi–Anger) truncation of $e^{i\tau x}$ on $[-1, 1]$ with tail $\leq \eta$ (the reference degree) & \hspace{0pt}31–32 at $\tau_{A}$, $\eta = 10^{-12}$ & \hspace{0pt}§4.2 \\
\hspace{0pt}$d_{D}(\tau, \eta)$ & \hspace{0pt}the same for $e^{\tau x}$ & \hspace{0pt}4–5 at $\tau_{D}$ & \hspace{0pt}§4.2 \\
\hspace{0pt}$Q_{A}$, $Q_{D}$ & \hspace{0pt}block-encoding calls (queries) per factor & \hspace{0pt}production 79/78 and 53 & \hspace{0pt}Supplementary Note~\ref{note:B}, B.2 \\
\hspace{0pt}$p_{A}$, $p_{D}$, $p_{\mathrm{step}}$ & \hspace{0pt}post-selection success per factor and per step & \hspace{0pt}production 1/4, $\approx 1/4$, 1/64 (computed exactly, never sampled) & \hspace{0pt}Supplementary Note~\ref{note:B}, B.3 \\
\hspace{0pt}$C_{\mathrm{BE},A} = C_{\mathrm{shift}} + C_{\mathrm{load}}$ & \hspace{0pt}gates per call of the convection encoding; $C_{\mathrm{load}}$ is the data-dependent part & \hspace{0pt}collocated census $\approx 2.78 \times 10^{10}$ CX per walk at $256^{3}$ (upper estimate) & \hspace{0pt}§5.1 \\
\hspace{0pt}$C_{\mathrm{BE},D}$ & \hspace{0pt}gates per call of the diffusion encoding (data-independent) & \hspace{0pt}$\approx 6.3 \times 10^{4}$ CX per walk (linear fit) & \hspace{0pt}§5.1 \\
\hspace{0pt}$C_{\mathrm{prep}}$, $C_{\mathrm{read}}$ & \hspace{0pt}per-step state preparation and readout & \hspace{0pt}not modelled & \hspace{0pt}Supplementary Note~\ref{note:B}, B.3 and B.7 \\
\hspace{0pt}$C_{\mathrm{pred}}$, $C_{\mathrm{refresh}}$, $C_{\mathrm{synth}}$ & \hspace{0pt}classical predictor, coefficient build, GQSP phase synthesis & \hspace{0pt}96 stencil applications; $O(\mathcal{N}) \times 2$; $\mathrm{poly}(d) \times 3$ & \hspace{0pt}Supplementary Table B1; §3 \\
\hspace{0pt}$\kappa_{G}$, $\kappa_{\Delta}$ & \hspace{0pt}condition numbers of the MAC gradient and of the Laplacian (periodic $N^{3}$ grid; used as proxies elsewhere) & \hspace{0pt}141.1; $1.99 \times 10^{4}$ & \hspace{0pt}Appendix A \\
\hspace{0pt}$G(T)$ & \hspace{0pt}error-transfer factor: sup over $n$ of the norm of the map that transports a perturbation at step $n$ to time $T$. \textbf{Symbolic and not measured here.} The noise study's $\delta/\varepsilon$ is a cumulative transfer of one random per-step error model, not this factor & \hspace{0pt}— & \hspace{0pt}— \\
\bottomrule
\end{longtable}}

\medskip
\subsection{The implemented step and its cost inventory}\label{si:C.3}
One production step (Supplementary Table B1). The fourth column says what was actually computed in the production run; the last column says what a quantum implementation would need.

{\scriptsize
\setlength{\tabcolsep}{3pt}\renewcommand{\arraystretch}{1.15}\setlength{\LTleft}{0pt}\setlength{\LTright}{0pt}
\begin{longtable}{>{\raggedright\arraybackslash}p{\dimexpr 0.0376\linewidth-6pt\relax}>{\raggedright\arraybackslash}p{\dimexpr 0.2219\linewidth-6pt\relax}>{\raggedright\arraybackslash}p{\dimexpr 0.0756\linewidth-6pt\relax}>{\raggedright\arraybackslash}p{\dimexpr 0.2241\linewidth-6pt\relax}>{\raggedright\arraybackslash}p{\dimexpr 0.1822\linewidth-6pt\relax}>{\raggedright\arraybackslash}p{\dimexpr 0.2586\linewidth-6pt\relax}}
\toprule
\hspace{0pt}\textbf{\#} & \hspace{0pt}\textbf{Operation} & \hspace{0pt}\textbf{Count per step} & \hspace{0pt}\textbf{Implemented / emulated / classical} & \hspace{0pt}\textbf{Cost} & \hspace{0pt}\textbf{Future quantum implementation needs} \\
\midrule
\endhead
\hspace{0pt}1 & \hspace{0pt}Coefficient build from $u_{n}$ (24 weight arrays, $\gamma_{a}$, $\lambda_{A}$) & \hspace{0pt}1 & \hspace{0pt}classical & \hspace{0pt}$O(\mathcal{N})$ & \hspace{0pt}classical (or a coherent arithmetic circuit; not defined) \\
\hspace{0pt}2 & \hspace{0pt}Encode $\psi = q/\|q\|$ & \hspace{0pt}1 & \hspace{0pt}classical normalization & \hspace{0pt}$O(\mathcal{N})$ & \hspace{0pt}\textbf{state preparation}: not modelled \\
\hspace{0pt}3 & \hspace{0pt}Predictor $C_{48}[\allowbreak e^{D\Delta t/2}]\, C_{48}[\allowbreak e^{A(u_{n})\Delta t/2}]$: 48 $A$ and 48 $D$ applications & \hspace{0pt}1 & \hspace{0pt}classical & \hspace{0pt}96 stencil applications, $O(\mathcal{N})$ each & \hspace{0pt}classical by design (no quantum realization defined; Supplementary Note~\ref{note:B}, B.7) \\
\hspace{0pt}4 & \hspace{0pt}$u_{\mathrm{mid}} = \Re(\allowbreak \cdot\allowbreak )\, \|q\|$; coefficient build from $u_{\mathrm{mid}}$ & \hspace{0pt}1 & \hspace{0pt}classical & \hspace{0pt}$O(\mathcal{N})$ & \hspace{0pt}classical \\
\hspace{0pt}5 & \hspace{0pt}GQSP phase synthesis (1 convection, 2 diffusion) & \hspace{0pt}3 & \hspace{0pt}classical & \hspace{0pt}$\mathcal{N}$-independent, $\mathrm{poly}(d)$: 6 ms and 91 ms & \hspace{0pt}classical \\
\hspace{0pt}6 & \hspace{0pt}$\mathrm{GQSP}_{D}(\Delta t/2)$: 53 queries (5 forward, 48 inverse walks) & \hspace{0pt}2 & \hspace{0pt}\textbf{LCU circuit implemented} (gate-verified at $2^{3}$, $4^{3}$); production values by exact emulation & \hspace{0pt}55 $D$ applications in emulation & \hspace{0pt}block encoding exists (data-independent) \\
\hspace{0pt}7 & \hspace{0pt}$\mathrm{GQSP}_{A(u_{\mathrm{mid}})}(\allowbreak \Delta t)$: 79/78 queries (31/30 forward, 48 inverse) & \hspace{0pt}1 & \hspace{0pt}\textbf{staggered encoding constructed: gate-verified at $4^{3}$, sampled at $8^{3}$, not executed}; exact emulation & \hspace{0pt}81/80 $A$ applications in emulation & \hspace{0pt}the staggered block encoding with $O(\mathcal{N})$-gate data loading per query (constructed, §5.5) \\
\hspace{0pt}8 & \hspace{0pt}Post-selection on all-zero ancillas & \hspace{0pt}3 & \hspace{0pt}exact amplitudes (1/4 each by design; computed, never sampled) & \hspace{0pt}— & \hspace{0pt}\textbf{repetition or amplification}: not modelled \\
\hspace{0pt}9 & \hspace{0pt}Decode $\times (\|q\| \cdot 8)$ & \hspace{0pt}1 & \hspace{0pt}classical & \hspace{0pt}$O(\mathcal{N})$ & \hspace{0pt}\textbf{full-state readout and norm estimation}: not modelled \\
\hspace{0pt}10 & \hspace{0pt}Diagnostics & \hspace{0pt}1 & \hspace{0pt}classical & \hspace{0pt}$O(\mathcal{N})$ & \hspace{0pt}— \\
\bottomrule
\end{longtable}}

\begin{itemize}
\item\relax No Poisson, projection or other linear solve in the step; one FFT-based balanced-pressure solve at $t = 0$ only (11 FFTs of $256^{3}$; Supplementary Note~\ref{note:A}, A.4).
\item\relax Stencil applications per step: 286/287 ($A$ 128/129, $D$ 158; Appendix A). The code-literal arithmetic is $\approx 1.2 \times 10^{5}$ flops per cell per step ($\approx 2.0 \times 10^{12}$ per step); the hybrid's classical side proper (everything except the emulated corrector) is $\approx 2.4 \times 10^{4}$ flops per cell, dominated by the predictor (estimates).
\item\relax Times measured in the production run: 386.8 s per step for the step routine (mean) and 388.8 s per step including diagnostics and I/O (16 threads; 7.43 days for 1650 steps). These are costs of the classical emulation, not L1 or L2 quantities.
\item\relax Qubits: 26 system qubits; ancilla counts in §5.4.
\end{itemize}
\textbf{Four kinds of count} (kept separate everywhere): the Chebyshev degree 48; the production GQSP query counts 79/78 and 53 (48 untrimmed inverse walks plus the surviving positive powers); the stencil applications of the emulator (query $+ 2$ per block); and gates (L2). Supplementary Note~\ref{note:B}, B.2 and B.5.

\medskip
\subsection{Level 1: ideal block-encoding query complexity}\label{si:C.4}
\subsubsection{Access model}\label{si:C.4.1}
Assume unitaries $B_{A}$ and $B_{D}$ with $(\allowbreak \langle 0| \otimes I)\, B\, (\allowbreak |0\rangle \otimes I) = \bar{H}$, where $\bar{H}_{A} = -iA/\lambda_{A}$ (Hermitian, $\|\bar{H}_{A}\| \leq 1$) and $\bar{H}_{D} = D/\lambda_{D}$, both self-inverse ($B^{2} = I$), so that the qubitized walk $W = (\allowbreak 2|0\rangle\langle 0| \otimes I - I)\, B$ has eigenvalues $e^{\pm i\theta}$ with $\cos\theta = \lambda \in \mathrm{spec}(\allowbreak \bar{H})$ on two-dimensional invariant subspaces (Low \& Chuang~\cite{LowChuang2019}, Cor. 9). One query is one application of controlled-$W$, $W^{-1}$ or a controlled choice between them. The diffusion encoding $B_{D}$ exists as a circuit. A self-inverse staggered convection encoding (one exists for any anti-Hermitian $A$ with $\lambda \geq \|A\|$; Supplementary Note~\ref{note:B}, B.5) has been constructed with exactly $\lambda_{A}$ and verified at the gate level on small grids (§5.5); its loaders cost $O(\mathcal{N})$ gates per query, so an efficient circuit is not shown. A non-self-inverse encoding would need the standard conversion, which doubles the query counts below.

The target of the convection factor is $e^{A\Delta t} = e^{i\tau_{A}\bar{H}_{A}}$, i.e. $f(x) = e^{i\tau x}$ with $\tau = \tau_{A}$. The diffusion half-step targets $e^{D\Delta t/2} = e^{\tau_{D}\bar{H}_{D}}$.

\subsubsection{Degrees and query lower bounds}\label{si:C.4.2}
\textbf{Lemma 1 (degree).} By the Jacobi–Anger expansion $e^{i\tau x} = J_{0}(\allowbreak \tau) + 2\sum_{k \geq 1} i^{k} J_{k}(\allowbreak \tau)\, T_{k}(\allowbreak x)$, the Chebyshev truncation of degree $d$ has sup error $\leq 2\sum_{k > d}|J_{k}(\allowbreak \tau)|$ on $[-1, 1]$. With $|J_{k}(\allowbreak \tau)| \leq (\allowbreak \tau/\allowbreak 2)^{k}/\allowbreak k!$ the tail is at most a constant times $(e\tau/(2d))^{d}$, and $(\allowbreak t'/\allowbreak r)^{r} \leq e^{-\ln(1/\eta)} = \eta$ at $r = t' + \ln(1/\eta)$ (because $(1 - x/r)^{r} \leq e^{-x}$); hence $d(\allowbreak \tau,\allowbreak  \eta) \leq e\tau/\allowbreak 2 + \ln(\allowbreak 1/\allowbreak \eta) + O(\allowbreak 1)$. The sharp form is $d = \Theta(\allowbreak \tau + \log(\allowbreak 1/\allowbreak \eta)/\allowbreak \log(\allowbreak e + \log(\allowbreak 1/\allowbreak \eta)/\allowbreak \tau))$ (Gilyén, Su, Low \& Wiebe~\cite{Gilyen2019}, hereafter GSLW, Lemma 59). Numerically, $d \leq e\tau/2 + \ln(1/\eta)$ holds in all 544 cases of a grid $\tau \in [10^{-2}, 10^{3}]$, $\eta \in [10^{-16}, 10^{-2}]$ (maximum ratio 0.88; Appendix A). Values at $\eta = 10^{-12}$ (Appendix A):

{\scriptsize
\setlength{\tabcolsep}{3pt}\renewcommand{\arraystretch}{1.15}\setlength{\LTleft}{0pt}\setlength{\LTright}{0pt}
\begin{longtable}{>{\raggedright\arraybackslash}p{\dimexpr 0.1306\linewidth-6pt\relax}>{\raggedright\arraybackslash}p{\dimexpr 0.0731\linewidth-6pt\relax}>{\raggedright\arraybackslash}p{\dimexpr 0.0731\linewidth-6pt\relax}>{\raggedright\arraybackslash}p{\dimexpr 0.0731\linewidth-6pt\relax}>{\raggedright\arraybackslash}p{\dimexpr 0.2025\linewidth-6pt\relax}>{\raggedright\arraybackslash}p{\dimexpr 0.1220\linewidth-6pt\relax}>{\raggedright\arraybackslash}p{\dimexpr 0.0875\linewidth-6pt\relax}>{\raggedright\arraybackslash}p{\dimexpr 0.0875\linewidth-6pt\relax}>{\raggedright\arraybackslash}p{\dimexpr 0.1507\linewidth-6pt\relax}}
\toprule
\hspace{0pt}\textbf{$\tau$} & \hspace{0pt}\textbf{0.46} & \hspace{0pt}\textbf{0.80} & \hspace{0pt}\textbf{1.14} & \hspace{0pt}\textbf{$5.65\text{--}5.70$ ($\approx \|A\|\Delta t$)} & \hspace{0pt}\textbf{9.78 $(\tau_{\mathrm{ac}})$} & \hspace{0pt}\textbf{10.24} & \hspace{0pt}\textbf{10.92} & \hspace{0pt}\textbf{19.88 (TGV $512^{3}$)} \\
\midrule
\endhead
\hspace{0pt}$d(\tau, 10^{-12})$ & \hspace{0pt}9 & \hspace{0pt}11 & \hspace{0pt}12 & \hspace{0pt}23 & \hspace{0pt}30 & \hspace{0pt}31 & \hspace{0pt}32 & \hspace{0pt}45 \\
\bottomrule
\end{longtable}}

For the diffusion half-step, $d_{D}(0.0249, 10^{-12}) = 5$ on $[-1, 1]$ and 4 on the actual interval $[-1, 0]$ (Appendix A). The heat-kernel degree grows only as $\sqrt{\max(b, \log(1/\eta))\, \log(1/\eta)}$ for $e^{-x}$ on $[0, b]$ (Sachdeva \& Vishnoi~\cite{Sachdeva2014}, Thm 4.1; GSLW~\cite{Gilyen2019}, Cor. 64), with lower bound $\sqrt{b}/3$ (Sachdeva \& Vishnoi~\cite{Sachdeva2014}, Thm 5.3); the exact degrees are 52 at $b = 100$ and 160 at $b = 1000$, against $b + \ln(1/\eta) = 128$ and 1028 (Appendix A).

\textbf{Lemma 2 (lower bounds).} (i) For an unknown Hamiltonian given by an $(\alpha, a, 0)$ block encoding, $\Theta(\allowbreak \alpha|t| + \log(\allowbreak 1/\allowbreak \varepsilon)/\allowbreak \log(\allowbreak e + \log(\allowbreak 1/\allowbreak \varepsilon)/\allowbreak (\allowbreak \alpha|t|)))$ uses of the encoding are necessary and sufficient for $e^{itH}$ to error $\varepsilon$ (GSLW~\cite{Gilyen2019}, Cor. 60, and Thm 73 for eigenvalue transformations). (ii) With $q$ uses of the qubitized walk, the amplitude on each invariant subspace is a trigonometric polynomial of degree $\leq q$ in $\theta$ ($\cos\theta = \lambda$). For $e^{i\tau\cos\theta}$ the Jacobi–Anger series is this Fourier series, so the reference degree $d(\tau, \eta)$ is the corresponding truncation degree; it upper-bounds the minimal approximating degree and is used as the reference, while (i) is the rigorous lower bound. (iii) In the sparse-query model no-fast-forwarding gives $\Omega(\tau)$ (Berry, Ahokas, Cleve \& Sanders~\cite{BerryAhokas2007}, Thm 3; Berry, Childs \& Kothari~\cite{BerryChildsKothari2015}, Thm 2).

\textbf{Scope of the lower bounds.} They are black-box statements for an encoding of normalization $\lambda_{A}$. The integrator's operators are known stencils: the constant-coefficient parts ($A_{\mathrm{ac}}$ and $D$) are diagonalized by the QFT and can be fast-forwarded (An, Liu, Wang \& Zhao~\cite{AnLiuWangZhao2025}, Thm 11, Cor. 21); only the variable-coefficient $A_{\mathrm{adv}}(\bar{u})$ has no known fast-forwarding. ``Optimal'' below always means optimal in the black-box block-encoding model at normalization $\lambda_{A}$, which exceeds $\|A\|$ by a factor 1.81–1.92 (§4.3). It is not a lower bound for these structured operators. It would be reached by the prospective structure-adapted synthesis (Proposition 1, or directional control; §4.6), not by the emulated configuration.

\subsubsection{The convection–acoustic factor}\label{si:C.4.3}
\textbf{Proposition 1 (unit-normalized convection factor).} Let $L_{d}(z) = \sum_{|k| \leq d} i^{|k|} J_{|k|}(\tau)\, z^{k}$ (the symmetric truncated Jacobi–Anger series) with $\eta := 2\sum_{k > d}|J_{k}(\tau)|$. Then:

\begin{enumerate}
\item\relax On $|z| = 1$, $|L_{d}(z)| \leq 1 + \eta$ (because $L_{d}(e^{i\theta}) = e^{i\tau\cos\theta} - {}$tail, and $|e^{i\tau\cos\theta}| = 1$).
\item\relax $P(\allowbreak z) := z^{d} L_{d}(\allowbreak z)/\allowbreak (\allowbreak 1 + \delta)$, $\delta \geq \eta$, is a polynomial of degree $2d$ with $|P| \leq 1$ on the circle. By GQSP (Motlagh \& Wiebe~\cite{Motlagh2024}, Thms 3–4, and Thm 6 for negative powers) the Laurent polynomial $L_{d}(W)/(1 + \delta)$ is implementable with $2d$ controlled uses ($d$ of $W$, $d$ of $W^{-1}$) and one signal qubit.
\item\relax On the invariant subspace of eigenvalue $\lambda = \cos\theta$, $|0\lambda\rangle = (|+\rangle + |-\rangle)/\sqrt{2}$ with $W|\pm\rangle = e^{\pm i\theta}|\pm\rangle$. Because the coefficients are symmetric, $L_{d}(e^{i\theta}) = L_{d}(e^{-i\theta})$, so $L_{d}(W)|0\lambda\rangle = L_{d}(e^{i\theta})|0\lambda\rangle$: there is no leakage out of the encoding's ancilla state, and the block equals $e^{i\tau\lambda} + O(\eta)$.
\item\relax Hence $e^{A\Delta t}$ is implemented with $Q_{A} = 2d(\tau_{A}, \eta)$ queries and post-selection success $p_{A} \geq (1 - \eta)^{2}/(1 + \delta)^{2}$ for \textbf{every} input state, with no amplitude amplification.
\end{enumerate}
Proof: items 1–3 are shown inline; item 4 follows because the only post-selection left is the GQSP signal qubit, whose amplitude is $L_{d}(e^{i\theta})/(1 + \delta)$ on every eigencomponent. $\blacksquare$ Numerically, $\sup_{\theta}|L_{d}| = 1 + 2.1 \times 10^{-13}$ at $\tau = 10.92$, $d = 32$ (Appendix A). The proposition uses only that the target is unitary, i.e. that $A$ is anti-Hermitian. The same property holds for any formulation whose transport factor is unitary, including a projected incompressible one (§8.4); it distinguishes the integrator from structure-blind series encodings of a non-normal step (§9.4), not from projection.

Remarks. (i) Synthesis (Appendix A): the production completion and angle routines complete the symmetric Laurent polynomial ($\tau = 10.92$, degree 64) at every tested margin $\delta = 1$, $10^{-1}$, $10^{-2}$, $10^{-3}$, $10^{-6}$ and $10^{-12}$, in 3 ms each. With $P_{\mathrm{c}}$ the complementary polynomial, $\max \bigl||P|^{2} + |P_{\mathrm{c}}|^{2} - 1\bigr| \leq 4.2 \times 10^{-13}$ on the circle, and the polynomial re-evaluated from the angles is accurate to $9.3 \times 10^{-12}$ ($\leq 3.4 \times 10^{-12}$ for $\delta \leq 10^{-2}$; angle sequence of degree 63 after the production trim). Because $|L_{d}| \approx 1$ everywhere on the circle, $1 - |P|^{2}$ is nearly constant and the complement is well conditioned. The production scale 1/2 guarantees that a complementary polynomial exists (Supplementary Note~\ref{note:B}, B.3); this check shows that it is not needed for this factor. (ii) The decode needs the physical norm $\|S(\Delta t)\psi\|$ of the step. With near-unit normalization it is the square root of the success frequency, which the readout repetitions estimate with relative precision $O(R^{-1/2})$; in the emulation it is exact. Norm estimation is not modelled (Supplementary Note~\ref{note:B}, B.3).

\textbf{Counts per convection factor at $\eta = 10^{-12}$ (production $\tau_{A} = 10.24\text{--}10.92$)}

{\scriptsize
\setlength{\tabcolsep}{3pt}\renewcommand{\arraystretch}{1.15}\setlength{\LTleft}{0pt}\setlength{\LTright}{0pt}
\begin{longtable}{>{\raggedright\arraybackslash}p{\dimexpr 0.3600\linewidth-6pt\relax}>{\raggedright\arraybackslash}p{\dimexpr 0.2500\linewidth-6pt\relax}>{\raggedright\arraybackslash}p{\dimexpr 0.1700\linewidth-6pt\relax}>{\raggedright\arraybackslash}p{\dimexpr 0.2200\linewidth-6pt\relax}}
\toprule
\hspace{0pt}\textbf{Construction} & \hspace{0pt}\textbf{Queries} & \hspace{0pt}\textbf{Success} & \hspace{0pt}\textbf{Status} \\
\midrule
\endhead
\hspace{0pt}Production (emulated): $K = 48$ untrimmed inverse walks $+ 30/31$ forward; normalization 1/2 & \hspace{0pt}78/79 & \hspace{0pt}1/4 (computed exactly, never sampled) & \hspace{0pt}production, as run \\
\hspace{0pt}Same polynomial, symmetric trim ($K \to k_{\max}$) & \hspace{0pt}60–62 & \hspace{0pt}1/4 (if the scale $s = 2$ is kept) & \hspace{0pt}prospective (Appendix A) \\
\hspace{0pt}Proposition 1 (symmetric Laurent, normalization $1/(1 + \delta)$) & \hspace{0pt}62–64 (63 after the production trim) & \hspace{0pt}$\geq (\allowbreak 1 - \eta)^{2}/\allowbreak (\allowbreak 1 + \delta)^{2} \approx 1$ & \hspace{0pt}proved construction; prospective in use \\
\hspace{0pt}Directional control between $W$ and $W^{-1}$ (Berry, Motlagh, Pantaleoni \& Wiebe~\cite{Berry2024}, Thm 2): $K + 3$ controlled operations with even $K \geq d$ & \hspace{0pt}$\approx 35$ ($K = 32$) & \hspace{0pt}$\approx 1$ & \hspace{0pt}published construction (constructive upper bound); prospective in use \\
\hspace{0pt}Reference degree (black-box model) & \hspace{0pt}$d = 31\text{--}32$ & \hspace{0pt}— & \hspace{0pt}reference \\
\hspace{0pt}Spectrum-adapted polynomial (spectrum of $\bar{H}_{A}$ within $[-0.55, 0.55]$ over the run, $\approx [-0.52, 0.52]$ at $t = 1.0$) & \hspace{0pt}$d(\allowbreak \tau_{A}\|A\|/\allowbreak \lambda_{A}) = 23$ (plus the cost of keeping $|P| \leq 1$ off the interval) & \hspace{0pt}— & \hspace{0pt}prospective (an estimate; Appendix A) \\
\hspace{0pt}GSLW~\cite{Gilyen2019} Thm 58 / Cor. 62 (cos/sin LCU and robust oblivious amplitude amplification, deterministic) & \hspace{0pt}$6\tau + 9\ln(\allowbreak 12/\allowbreak \eta) \approx 337$ & \hspace{0pt}$\approx 1$ & \hspace{0pt}published construction \\
\bottomrule
\end{longtable}}

The LCU normalization exceeds the spectral norm by $\lambda_{A}/\allowbreak \|A\| \approx 1.81\text{--}1.92$: $\sqrt{3}$ on the acoustic part (triangle inequality over three axes) and $\approx 2.15$ on the advective part (Appendix A). Low \& Chuang name this ``a factor $\|\alpha\|_{1}/\|H\|$ slowdown''~\cite{LowChuang2019}; an encoding with $\alpha = \|A\|$ would cut $\tau$ by a factor $\approx 1.85$.

\subsubsection{The diffusion factor}\label{si:C.4.4}
\textbf{Proposition 2 (diffusion half-step at physical success).} Factor the identity branch of the 7-term diffusion LCU out as a scalar: $D = -(6\nu/h^{2})I + (6\nu/h^{2})\bar{H}_{D}'$ with $\bar{H}_{D}' = (D + (6\nu/h^{2})I)/(6\nu/h^{2})$, whose LCU (the six shifts, weight $\nu/h^{2}$ each) has $\ell_{1}$-norm $6\nu/h^{2}$ and spectrum in $[-1, 1]$. Then $e^{D\Delta t/2} = e^{-\beta} e^{\beta\bar{H}_{D}'}$ with $\beta = \tau_{D}/2$, and the target $g(y) = e^{-\beta}e^{\beta y}$ satisfies $0 < g \leq 1$ on $[-1, 1]$ ($= 1$ at $y = 1$, the constant mode). Its Chebyshev coefficients are $c_{0} = e^{-\beta}I_{0}(\beta)$ and $c_{k} = 2e^{-\beta}I_{k}(\beta)$ ($k \geq 1$), all positive, so the non-negative-power polynomial $P(z) = \sum_{k} c_{k} z^{k}$ obeys $|P| \leq \sum_{k} c_{k} = g(1) = 1$ on the circle and is GQSP-implementable with $d_{D}(\tau_{D}/2, \eta) \leq 4$ queries ($\eta = 10^{-12}$). The block $\langle 0\lambda|P(W)|0\lambda\rangle = \sum_{k} c_{k}\cos k\theta = g(\lambda)$; the remaining amplitude $i\sum_{k} c_{k}\sin k\theta$ leaves the encoding's ancilla state, so the post-selection is on the signal qubit and the encoding ancillas together. Its success probability is $\|e^{D\Delta t/2}\psi\|^{2}/\|\psi\|^{2}$ up to $O(\eta)$: the physical retention of the half-step. $\blacksquare$

In the production run (normalization 1/2) the mean computed success per diffusion half-step was 0.249838, i.e. a retention of $4 \times 0.249838 = 0.99935$. Phase synthesis of this polynomial (degree 5; degree 4 after the production trim) with the production routines (Appendix A) reproduces the polynomial from the angles at margins $\delta = 1$ ($7.7 \times 10^{-14}$), $\delta = 10^{-2}$ ($1.6 \times 10^{-12}$) and $\delta = 10^{-12}$ ($1.2 \times 10^{-11}$), but at $\delta = 10^{-6}$ the least-squares completion returns a complement from which the angle recursion does not reproduce $P$ (coefficient error 0.87). Here $1 - |P|^{2}$ nearly vanishes near $z = 1$ (the constant mode), so the completion is ill-conditioned and its result is not reliable at small margins. With the present routines $\delta \approx 10^{-2}$ was safe in this check; the success is then $\approx (\allowbreak 1 + \delta)^{-2} \cdot\allowbreak  0.99935 \approx 0.980$ per half-step. A better-conditioned completion (e.g. the FFT-based construction of Motlagh \& Wiebe~\cite{Motlagh2024}) may allow smaller margins; this was not tested.

\subsubsection{One step and the trajectory}\label{si:C.4.5}
\textbf{Proposition 3 (step success $=$ physical retention).} With Propositions 1–2, the success probability of one corrector step is $\|S(\Delta t)\psi\|^{2}/\|\psi\|^{2}$ up to $O(\eta)$ and the synthesis margins, $S = e^{D\Delta t/2}e^{A\Delta t}e^{D\Delta t/2}$: the fraction of the (velocity plus encoded pressure) norm retained by the frozen linear step.

\begin{itemize}
\item\relax \emph{Prospective configuration.} At production parameters this would be $\approx 0.99935^{2} \approx 0.9987$, an estimate from the measured retention (for the frozen step the checks of Appendix A give $1 - 1.07 \times 10^{-3}$, $1 - 1.85 \times 10^{-3}$ and $1 - 1.39 \times 10^{-3}$ at $t = 1.0$, 1.5 and 3.0). With the $\delta = 10^{-2}$ margin that the present synthesis routines support on both diffusion factors it is $\approx 0.96$.
\item\relax \emph{Production configuration.} The design value is 1/64: 1/4 per factor, computed exactly and never sampled.
\end{itemize}
This is consistent with the per-step failure probability ``proportional to the viscous dissipation rate'' of Akash et al.~\cite{Akash2026} (Eq. 26) and with the $\alpha = 1$ requirement of Over et al.~\cite{Over2025}.

\textbf{Proposition 4 (trajectory query count).} $Q_{\mathrm{traj}} = \sum_{n = 1}^{M}[Q_{A}(\tau_{A,n}, \eta) + 2Q_{D}(\tau_{D}, \eta)]$, with $\tau_{A,n} = \tau_{A}(u_{\mathrm{mid}})$ at step $n$. With Proposition 1 for the convection factor, Proposition 2 for diffusion and Lemma 1,

\displ{$Q_{\mathrm{traj}} \leq \sum_{n}[2d(\tau_{A,n}, \eta) + 2d_{D}] \leq e\, \bar{\lambda}_{A} T + 2M\ln(1/\eta) + 2M d_{D} + O(M)$ \quad (symmetric Laurent),}

and about half of the first two terms with directional control.

\begin{itemize}
\item\relax \textbf{What the bound covers.} It bounds the Proposition 1–2 constructions, whose use is prospective. At production parameters it is $\approx 1.52\text{--}1.57 \times 10^{5}$ ($\bar{\lambda}_{A} = 5300\text{--}5460$); the production configuration used 304,252 queries, $1.94\text{--}2.00$ times the bound (Appendix A).
\item\relax \textbf{Dependence on $\Delta t$.} The stiffness term $e\, \bar{\lambda}_{A} T$ does not depend on $\Delta t$; only the additive accuracy term grows with the number of steps. Consequently, provided accuracy permits, the largest admissible $\Delta t$ minimizes both the query count and the number $M$ of hybrid interface crossings. At production the additive term dominates: $2M\ln(10^{12}) \approx 9.1 \times 10^{4}$ against $e\, \bar{\lambda}_{A} T \approx 4.8 \times 10^{4}$ ($\bar{\lambda}_{A} \approx 5300$). Doubling $\Delta t$ would cut the total by $\approx 28$ \% and the readouts by half (Appendix A); its accuracy is not established (§9.5).
\end{itemize}
\textbf{Production consistency check} ($\eta = 10^{-12}$; Appendix A)

{\scriptsize
\setlength{\tabcolsep}{3pt}\renewcommand{\arraystretch}{1.15}\setlength{\LTleft}{0pt}\setlength{\LTright}{0pt}
\begin{longtable}{>{\raggedright\arraybackslash}p{\dimexpr 0.2938\linewidth-6pt\relax}>{\raggedright\arraybackslash}p{\dimexpr 0.1261\linewidth-6pt\relax}>{\raggedright\arraybackslash}p{\dimexpr 0.1416\linewidth-6pt\relax}>{\raggedright\arraybackslash}p{\dimexpr 0.1447\linewidth-6pt\relax}>{\raggedright\arraybackslash}p{\dimexpr 0.2938\linewidth-6pt\relax}}
\toprule
\hspace{0pt}\textbf{Configuration} & \hspace{0pt}\textbf{Status} & \hspace{0pt}\textbf{Corrector queries per step} & \hspace{0pt}\textbf{Trajectory (1650 steps)} & \hspace{0pt}\textbf{Expected corrector queries per accepted output copy} \\
\midrule
\endhead
\hspace{0pt}Production emulation, design success 1/4 per factor, 1/64 per step & \hspace{0pt}production, as run (reported) & \hspace{0pt}184–185 ($79/78 + 2 \times 53$) & \hspace{0pt}304,252 & \hspace{0pt}\textbf{4,852–4,868} under staged post-selection with early abort (assumptions S1–S4 below): $64 \cdot\allowbreak  (\allowbreak 53 + 78/\allowbreak 4 + 53/\allowbreak 16) = 4{,}852$ and $64 \cdot\allowbreak  (\allowbreak 53 + 79/\allowbreak 4 + 53/\allowbreak 16) = 4{,}868$. With the measured mean diffusion success 0.249838 in place of 1/4: 4,857–4,873. With all three post-selections read only at the end of the step: $64 \times\allowbreak  (\allowbreak 184\text{--}185) = 11{,}776\text{--}11{,}840$ \\
\hspace{0pt}Proposition 1 (Laurent) + Proposition 2 (Chebyshev form) & \hspace{0pt}prospective & \hspace{0pt}70–72 & \hspace{0pt}$\leq 1.19 \times 10^{5}$ & \hspace{0pt}$\approx 72\text{--}74$ with early abort, $\approx 73\text{--}75$ read at the end (step success $\approx 0.96$ with $\delta = 10^{-2}$ on diffusion) \\
\hspace{0pt}Symmetric Laurent for all three factors (diffusion target shifted as in Proposition 2) & \hspace{0pt}prospective & \hspace{0pt}78–84 & \hspace{0pt}$\leq 1.39 \times 10^{5}$ & \hspace{0pt}$\approx 80\text{--}86$ with early abort, $\approx 81\text{--}88$ read at the end (same margins) \\
\hspace{0pt}Directional control (convection, $K = 32$) + Proposition 2 (diffusion) & \hspace{0pt}prospective & \hspace{0pt}43–45 & \hspace{0pt}$\approx 7.1\text{--}7.4 \times\allowbreak  10^{4}$ & \hspace{0pt}$\approx 44\text{--}46$ with early abort, $\approx 45\text{--}47$ read at the end (step success $\approx 0.96$) \\
\hspace{0pt}Reference degrees (black-box model) & \hspace{0pt}reference & \hspace{0pt}$\approx 39\text{--}42$ & \hspace{0pt}$\approx 6.4\text{--}6.9 \times\allowbreak  10^{4}$ & \hspace{0pt}— \\
\bottomrule
\end{longtable}}

Assumptions of the expected-query column. The early-abort values use S1–S3. The ``read at the end'' values replace S1–S2 by deferred post-selection ($Q_{\mathrm{step}}/p_{\mathrm{step}}$). S4 applies only to the production row, the only one at success 1/64.

\begin{itemize}
\item\relax \textbf{S1 (staged post-selection).} The encoding ancillas and the GQSP signal qubit of each factor are measured right after that factor, in the step order $D$, $A$, $D$. This needs mid-circuit measurement and classical feed-forward.
\item\relax \textbf{S2 (early abort).} At the first failed post-selection the state is discarded, and the step restarts from state preparation with the same classical data. The expected number of trials per accepted copy is $1/p_{\mathrm{step}}$ (64 at the design value).
\item\relax \textbf{S3 (what is counted).} Only corrector queries are counted. Excluded: the $1/p_{\mathrm{step}}$ expected state preparations per accepted copy (§5.3), the classical predictor and refresh, phase synthesis, and the $R(\varepsilon_{r})$ accepted copies that one full readout needs (§6.3).
\item\relax \textbf{S4 (success values).} The factors succeed independently with the design values 1/4. The measured diffusion successes (0.249745–0.249987 per block) change the count by $\approx 0.1$ \%.
\end{itemize}
For success probabilities $p_{D}$, $p_{A}$, $p_{D}$ and query counts $Q_{D}$, $Q_{A}$, $Q_{D}$ in step order, the number of trials is geometric and the cost per trial is staged, so

\displ{$E = (Q_{D} + p_{D} Q_{A} + p_{D} p_{A} Q_{D})/(p_{D} p_{A} p_{D})$.}

The production counts come from a uniform, non-adaptive synthesis: $K = 48$ inverse walks are always applied, diffusion uses Chebyshev degree 48 where $\leq 5$ suffices, and every factor is normalized by 1/2. None of this is intrinsic to the architecture. The prospective rows are derived and not implemented (§4.6); they describe the integrator's own design space and are not a comparison with another algorithm.

\subsubsection{Prospective structure-adapted configuration: status and missing requirements}\label{si:C.4.6}
The 43–45-query configuration combines three pieces:

\begin{itemize}
\item\relax directional control for the convection factor: $K + 3 = 35$ controlled-$W$/$W^{-1}$ operations, with even $K = 32 \geq d = 31\text{--}32$;
\item\relax Proposition 2 for both diffusion half-steps: 4 queries each after the $10^{-12}$ trim, 5 without it;
\item\relax margins $\delta \to 0$ for convection and $\delta \approx 10^{-2}$ for diffusion, which give a step success of $\approx 0.96$.
\end{itemize}
\textbf{It is not implemented.} Using it would require each of the following; only item 1 exists, at normalization $\lambda_{A}$.

\begin{enumerate}
\item\relax \textbf{A staggered self-inverse block encoding} $B_{A}$ of $\bar{H}_{A} = -iA/\lambda_{A}$ with normalization $\lambda_{A}$ (or one closer to $\|A\|$), constructed and verified at the gate level ($\langle 0|B_{A}|0\rangle = \bar{H}_{A}$, $B_{A}^{2} = I$, gate-level equality at grids up to $16^{3}$). A non-self-inverse encoding would double every count. Status: constructed at normalization $\lambda_{A}$, verified at the gate level at $4^{3}$ and by sampling at $8^{3}$; the GQSP comparison at $4^{3}$ misses its predefined tolerance (§5.5). No encoding closer to $\|A\|$ is constructed, and no check at $16^{3}$ was run.
\item\relax \textbf{Directional control between $W$ and $W^{-1}$} with the cost of a controlled $W$. Berry, Motlagh, Pantaleoni \& Wiebe obtain it by controlling the reflections~\cite{Berry2024}. The circuit, and the GQSP phase synthesis for this sequence, have not been written; the count $K + 3$ is their Theorem 2 count, not a compiled count.
\item\relax \textbf{The diffusion encoding without its identity branch.} Proposition 2 needs the six-shift LCU with normalization $6\nu/h^{2}$, i.e. a new PREPARE, and the encoding ancillas must be post-selected together with the signal qubit.
\item\relax \textbf{Phase synthesis at near-unit normalization.} The present routines are reliable for the convection Laurent polynomial down to $\delta = 10^{-12}$ and for the diffusion polynomial at $\delta = 10^{-2}$, but fail for the latter at $\delta = 10^{-6}$ (§4.3–4.4); a better-conditioned completion is untested.
\item\relax \textbf{Mid-circuit measurement and feed-forward} for staged post-selection (S1–S2).
\item\relax \textbf{Verification} that the new sequences reproduce the emulated factors to the production tolerance, first at the gate level (up to $16^{3}$) and then in exact emulation at $256^{3}$.
\end{enumerate}
Until all six exist, every statement about 43–45 queries, $\approx 0.96$ step success or $\approx 44\text{--}47$ expected queries per accepted copy is prospective.

\medskip
\subsection{Level 2: quantum implementation complexity}\label{si:C.5}
\subsubsection{Gates per query}\label{si:C.5.1}
\begin{itemize}
\item\relax \textbf{Diffusion encoding} (implemented, data-independent): PREPARE on 2 select qubits, three controlled $\pm 1$ shift branches (incrementers built from MCX ladders) and a $-I$ phase. The census of the compiled circuits gives 6,193 CX ($8^{3}$) and 17,641 CX ($16^{3}$); a linear fit gives $\approx 6.3 \times 10^{4}$ CX per walk at $256^{3}$.
\item\relax \textbf{Convection encoding} (collocated circuit implemented; staggered circuit constructed, §5.5): every walk applies 6 uniformly controlled $R_{y}$ velocity loaders with $2^{m} = \mathcal{N}$ angles each, at $\approx 276$ CX per angle in the census compilation, so $C_{\mathrm{load}} \approx 6 \cdot 276 \cdot \mathcal{N} \approx 2.78 \times 10^{10}$ CX per walk at $256^{3}$ (a structural extrapolation from the compiled $8^{3}$ and $16^{3}$ circuits); the shifts, acoustic branches and reflection are data-independent. The 276 CX per angle comes from attaching the six outer controls to every gate of the multiplexor; a uniformly controlled rotation with $k$ controls needs $2^{k}$ CNOTs (Möttönen et al.~\cite{Mottonen2005}), i.e. $O(1)$ CX per angle, so every gate number below is an upper estimate from an unoptimized compilation (no optimized count is stated). A staggered loader must also be multiplexed over the 2 field qubits (4 weight sets per axis), i.e. $\approx 4$ times more angles; the constructed circuit has exactly $4\mathcal{N}$ angles per loader (§5.5).
\item\relax \textbf{Phase processing:} one signal qubit and $Q + 1$ single-qubit SU(2) rotations per factor.
\end{itemize}
\textbf{Proposition 5 (data loading for arbitrary coefficient fields).} The proposition holds under the following model assumptions.

\begin{itemize}
\item\relax \textbf{(M1) Compiled-circuit model.} Each block-encoding call is a circuit of one- and two-qubit gates on at most $n_{\mathrm{tot}} = \mathrm{poly}(n)$ qubits. The classical coefficients enter only through the circuit description (gate choices and parameters); no quantum memory holding the data persists between calls.
\item\relax \textbf{(M2) Gates.} Either (a) a finite gate set $\Gamma$ (the fault-tolerant setting), or (b) gates from a finite set of types, each with at most $c$ real parameters and Lipschitz in each parameter with constant $\leq 1$ in operator norm.
\item\relax \textbf{(F) Field family.} The advecting fields range over a set $F$ of $2^{b\mathcal{N}'}$ fields with $\mathcal{N}' \geq \mathcal{N}/2$ independent $b$-bit coefficients. An example is $x$-face velocities supported on alternate faces, with the other components zero. On such fields the two-point averages that define the advective weights, $w_{x}^{+}(i) = (u(i) + u(i + e_{x}))/(4h)$ (Supplementary Note~\ref{note:A}, eq. (A5)), are injective, so distinct fields give distinct operators $A(\bar{u})$.
\item\relax \textbf{(E) Accuracy.} For each $\bar{u} \in F$ the circuit block-encodes $A(\bar{u})$ with normalization $\lambda$ to accuracy $\varepsilon_{\mathrm{BE}}$, with $2\lambda\varepsilon_{\mathrm{BE}}$ below the smallest separation $\|A(\bar{u}) - A(\bar{u}')\|$ in $F$. A one-coefficient change $\delta u$ changes two weights by $\delta u/(4h)$, so $\|\Delta A\| \geq |\delta u|/(4h)$. With $\lambda \approx 6/(\mathrm{Ma}\, h)$, about $b \approx \log_{2}(\mathrm{Ma}\, u_{s}/(48\varepsilon_{\mathrm{BE}}))$ bits per coefficient are therefore resolvable (an estimate): $\approx 30$ bits at $\mathrm{Ma} = 0.05$, velocity scale $u_{s} = 1$ and $\varepsilon_{\mathrm{BE}} = 10^{-12}$.
\end{itemize}
Statement.

\begin{itemize}
\item\relax \textbf{Finite gate set (a).} A circuit with at most $G$ gates is one of at most $2(|\Gamma|\, n_{\mathrm{tot}}^{2})^{G}$ circuits. So every field in $F$, except at most a fraction $2^{1 - b\mathcal{N}'/2}$, needs $G \geq b\mathcal{N}'/(2\log_{2}(|\Gamma|\, n_{\mathrm{tot}}^{2})) = \Omega(\mathcal{N} b/\log n)$ gates.
\item\relax \textbf{Continuous parameters (b).} A covering argument gives a slightly weaker conclusion. Discretizing every parameter at spacing $\varepsilon_{\mathrm{BE}}/(cG)$ gives at most $(|\Gamma_{\mathrm{types}}|\, n_{\mathrm{tot}}^{2})^{G}\, (2\pi cG/\varepsilon_{\mathrm{BE}})^{cG}$ cells, each of operator-norm diameter $\leq 2\varepsilon_{\mathrm{BE}}$. If the separation in (E) exceeds $4\lambda\varepsilon_{\mathrm{BE}}$, each cell serves at most one field. So $G\, [\log_{2}(|\Gamma_{\mathrm{types}}|\, n_{\mathrm{tot}}^{2}) + c\log_{2}(2\pi cG/\varepsilon_{\mathrm{BE}})] \geq b\mathcal{N}'$. Since $\log G = \Theta(\log \mathcal{N})$ for $G \gtrsim \mathcal{N}'/c$, this gives $G = \Omega(\mathcal{N} b/(\log \mathcal{N} + \log(1/\varepsilon_{\mathrm{BE}})))$, i.e. $\Omega(\mathcal{N}/\log \mathcal{N})$ per call when $b \approx \log(1/\varepsilon_{\mathrm{BE}})$. No exceptional-fraction bound is derived for case (b).
\item\relax \textbf{Per step.} Hence $C_{\mathrm{BE},A} = \Omega(\mathcal{N} b/\log n)$ per call for a finite gate set, and $\Omega(\mathcal{N}/\log \mathcal{N})$ for Lipschitz-parameterized gates, for arbitrary coefficient fields. Every query re-applies the call circuit, so $G_{\mathrm{step}} \geq Q_{A}\, C_{\mathrm{BE},A}$: $\Omega(Q_{A}\mathcal{N}/\log n)$ in case (a) and $\Omega(Q_{A}\mathcal{N}/\log \mathcal{N})$ in case (b). $\blacksquare$
\end{itemize}
Scope.

\begin{itemize}
\item\relax \textbf{Persistent data register.} If the coefficients are written once into a quantum data register whose gates are counted, each call must touch every coefficient register that can change its action. A light-cone count then still gives at least $\mathcal{N}'/2$ two-qubit gates per call, provided the register holds the raw coefficients with no garbage left by earlier calls.
\item\relax \textbf{QRAM as a unit-cost oracle.} With pre-processed quantum memory, or a QRAM query counted as one oracle call, this argument gives no per-call gate bound. The comparison then reverts to the query model (L1), and the QRAM itself has $\Theta(\mathcal{N})$ hardware: its total gate volume is $\Omega(\mathcal{N})$ (Dalzell et al.~\cite{Dalzell2025}, Sec. 17.1).
\item\relax \textbf{Structured fields.} Fields with a compact description, such as a few Fourier modes or a low-degree polynomial, can be loaded with polylogarithmic circuits. A turbulent field read out from the previous step is not assumed to have one. Using the dynamics itself as the compact description (the field is determined by the initial condition and the algorithm) amounts to coherent multi-step simulation, which is outside the per-step re-linearized design.
\item\relax \textbf{Worst-case nature.} The bound is a worst-case and typical-case statement over $F$. It does not bound the circuit size for one specific field.
\item\relax \textbf{Consequence.} The classical cost of applying the same polynomial by stencil matrix–vector products is $\Theta(Q\mathcal{N})$. So \textbf{at L2 there is no per-step separation for arbitrary coefficient fields} in the compiled-circuit model, up to a logarithmic factor. The outputs are not matched: the quantum corrector produces a state and the classical product produces the field; the matched comparison is L3 (§6). A separation would need $C_{\mathrm{load}} = o(\allowbreak \mathcal{N})$, i.e. coefficient fields with a compact quantum description, or a cost model that counts QRAM depth instead of QRAM gate volume (bucket-brigade QRAM: $O(\log \mathcal{N})$ depth, $\Theta(\mathcal{N})$ hardware; Giovannetti, Lloyd \& Maccone~\cite{Giovannetti2008}).
\item\relax The census is consistent with this: 99.998 \% of the extrapolated $256^{3}$ step is loaders (a collocated structural estimate, excluding state preparation, readout, repetition for post-selection and error correction).
\end{itemize}
\subsubsection{Gate count per step (upper estimates)}\label{si:C.5.2}
Per run of the corrector, with no post-selection repetition included,

\displ{$G_{\mathrm{step}} \approx Q_{A}(C_{\mathrm{shift}} + C_{\mathrm{load}}) + 2Q_{D}\, C_{\mathrm{BE},D}$:}

\begin{itemize}
\item\relax production query counts ($79 + 2 \times 53$), collocated census: $\approx 2.2 \times 10^{12}$ CX (the census used $79 + 2 \times 48$ walks; the production counts change it by $\leq 1.3$ \%);
\item\relax the constructed staggered encoding (§5.5), extrapolated from compiled $2^{3}\text{--}8^{3}$ circuits under a different, unitary-equivalent compilation (an estimate, not a compiled $256^{3}$ count): $\approx 2.7 \times 10^{12}$ CX for the convection factor alone at 79 queries, the same order as the census;
\item\relax Proposition 1 (62–64 data calls; prospective): $\approx 1.7\text{--}1.8 \times 10^{12}$ CX;
\item\relax directional control ($\approx 35$ data calls; prospective): $\approx 9.7 \times 10^{11}$ CX.
\end{itemize}
The diffusion factors contribute $< 10^{-5}$ of every variant: the data-independent encoding is $\approx 4 \times 10^{5}$ times cheaper per call than the loader-based convection encoding. The cost of a step is set by the number of data-dependent calls, $Q_{A}$, and by $C_{\mathrm{load}}$.

\subsubsection{Post-selection and repetition}\label{si:C.5.3}
\begin{itemize}
\item\relax \textbf{Production design:} $p_{\mathrm{step}} = (1/4)^{3} = 1/64$, computed exactly in the emulation and never sampled. On hardware a failed post-selection destroys the state, so a step restarts from state preparation.
\item\relax \textbf{Expected cost per accepted output copy, with re-preparation included}, under assumptions S1–S2 of §4.5 (staged post-selection, abort at the first failure):

\displ{$E_{\mathrm{copy}} = (C_{\mathrm{prep}} + G_{D} + p_{D} G_{A} + p_{D} p_{A} G_{D})/(p_{D} p_{A} p_{D})$,}

where $G_{D} = Q_{D}\, C_{\mathrm{BE},D}$ and $G_{A} = Q_{A}\, C_{\mathrm{BE},A}$ are the gate counts of one diffusion and one convection factor.

\begin{itemize}
\item\relax Every trial starts with one state preparation, so an accepted copy costs $1/p_{\mathrm{step}} = 64$ preparations at the design value: $\approx 64 \times 3.4 \times 10^{7} \approx 2.1 \times 10^{9}$ CNOTs for generic 26-qubit states at the leading-order preparation bound (§6.3).
\item\relax In queries, $E_{\mathrm{copy}}$ counts the 4,852–4,868 corrector queries of §4.5, which exclude the preparations.
\item\relax In the loader-dominated gate model, $E_{\mathrm{copy}} \approx 64\, C_{\mathrm{prep}} + 16\, G_{A} \approx 3.5 \times 10^{13}$ CX with the collocated census ($G_{A} \approx 79 \times 2.78 \times 10^{10}$; an upper estimate, §5.1). The preparations are $\approx 10^{-4}$ of this.
\item\relax If all post-selections are read at the end, the cost is $64\, (C_{\mathrm{prep}} + G_{\mathrm{step}})$ instead.
\end{itemize}
\item\relax \textbf{What $E_{\mathrm{copy}}$ still excludes.} One full readout needs $R(\varepsilon_{r})$ accepted copies per step (§6.3), so the per-step cost is $R\, E_{\mathrm{copy}}$ plus the classical work of §6.2.
\item\relax \textbf{Amplification.} Oblivious amplitude amplification makes the convection factor deterministic in one round (amplitude $1/2 \to 1$ with 3 uses), because its target is unitary (Berry, Childs, Cleve, Kothari \& Somma~\cite{BerryChildsCleve2014}, Lemma 3.6; GSLW~\cite{Gilyen2019}, Thm 28); this alone would cut the production trials from 64 to 16. It does \textbf{not} apply to the diffusion factors: one round maps a singular value $\sigma$ to $(3\sigma - \sigma^{3})/2$ and erases the first-order damping; uniform singular-value amplification is the valid route (Fang, Lin \& Tong~\cite{Fang2023}, Lemma 2).
\item\relax In the prospective configuration (§4.6) neither is needed. With Propositions 1–3, $p_{\mathrm{step}} \approx 0.96$ at production parameters with the $\delta = 10^{-2}$ diffusion margins ($\approx 0.9987$ as the margin $\to 0$), so the expected number of trials (and state preparations) per accepted copy would be $\approx 1.04$.
\item\relax Coherent chaining without measurement would give a success of $(1/64)^{M} = 10^{-2980}$ for $M = 1650$. The integrator never chains coherently, so this is a cost the hybrid design avoids, not one it pays; what it pays instead is a re-preparation and $R(\varepsilon_{r})$ accepted copies at every step (§6.3).
\end{itemize}
\subsubsection{Ancillas and phase synthesis}\label{si:C.5.4}
\begin{itemize}
\item\relax Ancillas: block-encoding registers (3 for diffusion; 6 for the collocated convection encoding; 5 for the constructed staggered encoding: 3 select, 1 loader and 1 difference qubit) plus one GQSP signal qubit per factor, constant in $\mathcal{N}$, $T$ and the polynomial degree. With the staggered encoding the three-factor corrector needs 32 qubits with ancilla reuse after each post-selection and 40 with disjoint registers ($26 + 4 + 6 + 4$); with the collocated encoding, 33 and 41.
\item\relax Phase synthesis is classical and $\mathcal{N}$-independent: 6 ms per convection factor and 91 ms per diffusion factor, 0.19 s per step against a step time of 386.8 s (Appendix A). Reuse of the phases across steps is discussed in §9.7.
\end{itemize}
\subsubsection{The constructed staggered encoding}\label{si:C.5.5}
The staggered convection encoding of the production runs now exists as a circuit (Supplementary Note~\ref{note:D}, Table D6). It was built after the runs, which emulate it and do not execute it.

\begin{itemize}
\item\relax \textbf{Construction.} Six difference branches on 3 select qubits. Three advective branches (weight $2\gamma_{a}$) apply the modular increment after a uniformly controlled $R_{y}$ loader conditioned on the grid address and the 2 field qubits ($4\mathcal{N}$ angles per loader). Three acoustic branches (weight $2/(\mathrm{Ma}\, h)$) apply a field-controlled increment followed by the velocity–pressure field swap. $\langle 0|B_{A}|0\rangle = -iA_{\mathrm{MAC}}/\lambda_{A}$ holds by construction with exactly the production normalization $\lambda_{A}$ (the six weights sum to it, with no slack), and $B_{A}^{2} = I$ for any branch unitaries. Ancillas: 3 select, 1 loader and 1 difference qubit, plus the GQSP signal qubit.
\item\relax \textbf{Verification (gate level, small grids).}

\begin{itemize}
\item\relax $4^{3}$: the block equals $-iA/\lambda_{A}$ on every basis state to $\leq 4.2 \times 10^{-12}$ (relative Frobenius norm; $4.8 \times 10^{-12}$ on an anisotropic $8 \times 4 \times 2$ grid), and $B_{A}^{2} = I$ to $\leq 2.2 \times 10^{-11}$ uncompiled and $\leq 8.4 \times 10^{-12}$ compiled ($8 \times 4 \times 2$: $2.6 \times 10^{-11}$ and $9.4 \times 10^{-12}$).
\item\relax $4^{3}$: the GQSP factor (78 queries at $\tau_{A} = 10.5$) and the three-factor corrector equal the qubitized-pair emulation to $3.3 \times 10^{-10}$, which misses the predefined $10^{-10}$ tolerance (so do the sector-population differences, $1.06\text{--}1.07 \times 10^{-10}$). The raw difference grows by $4.2 \times 10^{-12}$ per query (floating-point accumulation); a later diagnostic of the single factor, not a predefined check, gives $9.9 \times 10^{-11}$ for its full state after removing one floating-point global phase.
\item\relax $8^{3}$: sampled checks only (no GQSP check): block actions $\leq 3.9 \times 10^{-11}$, $B_{A}^{2} = I$ to $6.6 \times 10^{-11}$ and $B_{A} = B_{A}^{\dagger}$ to $3.3 \times 10^{-12}$ on random states, single branches $\leq 1.4 \times 10^{-12}$.
\item\relax $256^{3}$: not simulated; the production numbers remain exact-emulation numbers.
\end{itemize}
\end{itemize}
\textbf{Compiled counts} (unoptimized; controls distributed to primitive gates as in the production circuit construction; compiled to \{CNOT, one-qubit\} without using idle qubits as clean ancillas, so unitary-equivalent):

{\scriptsize
\setlength{\tabcolsep}{3pt}\renewcommand{\arraystretch}{1.15}\setlength{\LTleft}{0pt}\setlength{\LTright}{0pt}
\begin{longtable}{>{\raggedright\arraybackslash}p{\dimexpr 0.0677\linewidth-6pt\relax}>{\raggedright\arraybackslash}p{\dimexpr 0.2300\linewidth-6pt\relax}>{\raggedright\arraybackslash}p{\dimexpr 0.1813\linewidth-6pt\relax}>{\raggedright\arraybackslash}p{\dimexpr 0.2912\linewidth-6pt\relax}>{\raggedright\arraybackslash}p{\dimexpr 0.2298\linewidth-6pt\relax}}
\toprule
\hspace{0pt}\textbf{Grid} & \hspace{0pt}\textbf{$B_{A}$ gates / CX / depth} & \hspace{0pt}\textbf{Controlled walk CX} & \hspace{0pt}\textbf{Collocated encoding CX (same compilation)} & \hspace{0pt}\textbf{Collocated controlled walk CX} \\
\midrule
\endhead
\hspace{0pt}$2^{3}$ & \hspace{0pt}37,556 / 13,600 / 27,022 & \hspace{0pt}19,418 & \hspace{0pt}6,048 & \hspace{0pt}8,986 \\
\hspace{0pt}$4^{3}$ & \hspace{0pt}284,260 / 101,476 / 193,222 & \hspace{0pt}145,694 & \hspace{0pt}35,316 & \hspace{0pt}46,078 \\
\hspace{0pt}$8^{3}$ & \hspace{0pt}2,253,180 / 803,868 / 1,532,118 & \hspace{0pt}1,150,422 & \hspace{0pt}272,076 & \hspace{0pt}355,558 \\
\bottomrule
\end{longtable}}

\begin{itemize}
\item\relax Per application, $B_{A}$ thus compiles to $1.0 \times 10^{5}$ CNOTs at $4^{3}$ and $8.0 \times 10^{5}$ at $8^{3}$.
\item\relax One $4^{3}$ GQSP convection factor at 78 queries, executed in simulation: $3.0 \times 10^{7}$ gates, of which $1.14 \times 10^{7}$ CX.
\item\relax The advective loaders dominate: 47.4, 58.5 and 58.3 CX per loader angle at $2^{3}$, $4^{3}$ and $8^{3}$, so $\mathrm{CX}(B_{A}) \propto \mathcal{N}$ ($\times 7.46$ and $\times 7.92$ per doubling of $N$). The data-independent acoustic branches compile to $1.4\text{--}1.7 \times 10^{3}$, $3.8\text{--}4.1 \times 10^{3}$ and $2.8\text{--}3.0 \times 10^{4}$ CX at $2^{3}$, $4^{3}$ and $8^{3}$ ($\times 2.5$ and $\times 7.4$ per doubling of $N$) in this unoptimized emission, so these counts do not show a growth as $\log \mathcal{N}$; their cost is polylogarithmic in $\mathcal{N}$ only for an ancilla-assisted synthesis of the multi-controlled increments, which was not compiled.
\item\relax These counts depend on the compilation; the census of §5.1 used $\approx 276$ CX per angle for the collocated loaders under a different compilation.
\end{itemize}
\textbf{Extrapolation to $256^{3}$} (an estimate, not a compiled circuit). At 58.3 CX per angle, $\mathrm{CX}(B_{A}) \approx 3 \times 2 \times 4\mathcal{N} \times 58.3 \approx 2.3 \times 10^{10}$ per application. This gives $\approx 3.4 \times 10^{10}$ per controlled walk ($\times 1.43$, the measured walk-to-application ratio) and $\approx 2.7 \times 10^{12}$ per convection factor at 79 queries. This is consistent with Proposition 5: for arbitrary fields the loaders cost $\Theta(\mathcal{N})$ gates per query, so the constructed encoding is not an efficient circuit, and it changes no conclusion of §5–§6.

\medskip
\subsection{Level 3: end-to-end hybrid complexity}\label{si:C.6}
\subsubsection{Cost formula}\label{si:C.6.1}
\displ{$C_{\mathrm{total}} = \sum_{n = 1}^{M}\bigl[C_{\mathrm{pred}} + 2C_{\mathrm{refresh}} + 3C_{\mathrm{synth}} + R(\varepsilon_{r})\, E_{\mathrm{copy},n}\bigr]$,}

\begin{itemize}
\item\relax $R(\varepsilon_{r})$ is the number of accepted corrector executions (output copies) needed to read out the $4\mathcal{N}$ amplitudes to accuracy $\varepsilon_{r}$.
\item\relax $E_{\mathrm{copy}}$ is the expected cost per accepted copy, including the $1/p_{\mathrm{step}}$ state preparations (§5.3). With deferred post-selection it becomes $(C_{\mathrm{prep}} + G_{\mathrm{step}})/p_{\mathrm{step}}$.
\end{itemize}
Every term inside the bracket is paid every step.

\subsubsection{Classical work per step}\label{si:C.6.2}
$C_{\mathrm{pred}} \approx 96$ stencil applications (48 $A$ and 48 $D$) $= \Theta(\allowbreak \mathcal{N} d_{\mathrm{pred}})$; refresh, encode and decode $\Theta(\mathcal{N})$; synthesis $\mathrm{poly}(d)$. The predictor needs only enough accuracy for $u_{\mathrm{mid}}$; at $\eta_{\mathrm{pred}} = 10^{-6}$ the Chebyshev degrees give 17 stencil applications instead of 96 (27 at $\eta_{\mathrm{pred}} = 10^{-12}$; Appendix A), so its present constant ($\approx 2.4 \times 10^{4}$ flops per cell) could be cut $\approx 3.5\text{--}5.5$-fold without changing the scheme. For comparison, one pseudo-spectral SSP-RK3 step of the same $256^{3}$ problem costs 27 real 3D FFTs (9 per right-hand side; Mortensen \& Langtangen~\cite{Mortensen2016}, §2), $\approx 27 \times\allowbreak  2.5\, \mathcal{N}\log_{2}\mathcal{N} \approx 2.7 \times\allowbreak  10^{10}$ flops $\approx 1.6 \times 10^{3}$ flops per cell (an estimate with the standard FFT constant). The classical predictor and refresh of a step thus cost about 1.3–15 times a complete classical pseudo-spectral step, whatever the cost of the corrector (the lower end with the reduced predictor degree and minimal arithmetic).

\subsubsection{State preparation and readout}\label{si:C.6.3}
\textbf{Preparation.} An arbitrary 26-qubit state needs $< (\allowbreak 23/\allowbreak 24)\, 2^{26} \approx 6.4 \times\allowbreak  10^{7}$ CNOTs (Plesch \& Brukner~\cite{Plesch2011}) and $\leq 2^{27} - 52$ (Shende, Bullock \& Markov~\cite{Shende2006}). For exact preparation of a generic state with CNOTs and arbitrary one-qubit gates, Plesch \& Brukner give a lower bound of $(1/2)\, 2^{n}$ CNOTs in leading order, $\approx 3.4 \times 10^{7}$ at $n = 26$, and Shende, Bullock \& Markov state that their upper bound ``lies a factor of four away from the theoretical lower bound''. Only this leading-order form is used here. A turbulent field has no efficiently integrable structure that would allow Grover–Rudolph preparation~\cite{GroverRudolph2002}.

\textbf{Proposition 6 (full-readout copy floor).} The proposition holds under the following interface assumptions.

\begin{itemize}
\item\relax \textbf{(I1) Interface.} Every step must output a classical description of the full normalized output state: each of the $4\mathcal{N}$ amplitudes to $b$ bits, or the whole vector to $\ell_{2}$ accuracy $\varepsilon_{r}$. This is the integrator's design, because its classical predictor and refresh read the full field. It is not intrinsic to the PDE problem.
\item\relax \textbf{(I2) Worst case.} The readout must succeed with probability $\geq 1 - P_{\mathrm{err}}$ for every state in a family of $K$ distinguishable output states. $b$-bit amplitude vectors give $\log_{2}K \geq (b - 1)(4\mathcal{N} - 1)$: fix $4\mathcal{N} - 1$ amplitudes on a grid of $2^{b - 1}$ values in $[0, (4\mathcal{N})^{-1/2}]$, and let normalization fix the last.
\item\relax \textbf{(I3) Access.} Either $k$ independent copies of the $n$-qubit output state (each copy is one accepted corrector run), or uses of a preparation unitary and its inverse.
\end{itemize}
Statement.

\begin{itemize}
\item\relax \textbf{Copies.} Each copy carries at most $n$ bits of accessible information (Holevo bound~\cite{Holevo1973}, in the form $I(X{:}Y) \leq \chi$ used by Haah et al.~\cite{Haah2017}). With Fano's inequality, $kn \geq (1 - P_{\mathrm{err}})\log_{2}K - 1$, i.e. $k \gtrsim (1 - P_{\mathrm{err}})\, 4\mathcal{N}(b - 1)/n$. At $256^{3}$ and $b = 32$ this is $\geq 5.3 \times 10^{7}$ copies for $P_{\mathrm{err}} = 1/3$; the leading-order value $4\mathcal{N}b/n$ is $8.3 \times 10^{7}$ at $b = 32$ ($4.1 \times 10^{7}$ at $b = 16$).
\item\relax \textbf{Tomography.} Pure-state $\ell_{2}$ tomography takes $\tilde{\Theta}(\allowbreak 4\mathcal{N}/\allowbreak \varepsilon_{r})$ uses of the preparation unitary (tight), or $\tilde{O}(\allowbreak 4\mathcal{N}/\allowbreak \varepsilon_{r}^{2})$ copies in the sampling model (van Apeldoorn, Cornelissen, Gilyén \& Nannicini~\cite{vanApeldoorn2023}; Haah et al.~\cite{Haah2017}). For $\varepsilon_{r} = 10^{-2}\text{--}10^{-4}$ that is $6.7 \times\allowbreak  10^{9}\text{--}6.7 \times\allowbreak  10^{11}$ unitary uses and $6.7 \times\allowbreak  10^{11}\text{--}6.7 \times\allowbreak  10^{15}$ copies (Appendix A). $\blacksquare$
\end{itemize}

\textbf{Proposition 7 (conditional end-to-end floor with per-step full readout).} The proposition holds under assumptions I1–I3 of Proposition 6 and the following cost-model assumptions.

\begin{itemize}
\item\relax \textbf{(K1) One run per copy.} Each output copy comes from its own run: state preparation from the classical data of the previous step, then the corrector with its post-selections. This is the integrator's hybrid loop: copies cannot be cloned, and a measured copy is consumed.
\item\relax \textbf{(K2) Gate counting.} Gates are counted, with data entering through compiled circuits (the model of Proposition 5). Preparing a worst-case (and typical) input state then costs $C_{\mathrm{prep}} = \Omega(\allowbreak \mathcal{N} b/\allowbreak \log n)$ gates for a finite gate set, by the same counting as Proposition 5, and $\Omega(\allowbreak \mathcal{N}/\allowbreak \log \mathcal{N})$ for Lipschitz-parameterized gates. The numbers below further assume exact preparation of generic states, counted in CNOTs with free one-qubit gates, and use the leading-order lower bound $C_{\mathrm{prep}} \approx (\allowbreak 1/\allowbreak 2)\, 2^{n} = 2^{n - 1}$ (Plesch \& Brukner~\cite{Plesch2011}). This numeric form does not cover approximate preparation, for which only the counting bound applies.
\end{itemize}
Statement. Per step, $C_{\mathrm{step}} \geq R\, C_{\mathrm{prep}} = \Omega(\mathcal{N} b/n) \cdot \Omega(\mathcal{N}/\log \mathcal{N}) = \Omega(\mathcal{N}^{2}/\log^{2}\mathcal{N})$ at fixed $b$. So $C_{\mathrm{total}} = \Omega(M\mathcal{N}^{2}/\log^{2}\mathcal{N})$, against the classical $\Theta(M\mathcal{N}d)$. In total gate count, a hybrid with per-step full-field readout, the integrator included, therefore cannot be faster than a classical computation of the same steps, and with arbitrary state preparation it is polynomially slower. $\blacksquare$

What the floor depends on:

\begin{itemize}
\item\relax \textbf{(I1).} With observable outputs instead of fields the floor does not apply. It is a property of the interface, not of the problem.
\item\relax \textbf{(K2).} Suppose each block-encoding call, and each preparation, takes $O(\mathrm{polylog}\, \mathcal{N})$ time on $\Theta(\mathcal{N})$ parallel hardware (for example with QRAM).

\begin{itemize}
\item\relax A run of $Q$ sequential queries still takes at least $Q$ call times, so the time per step is $\geq RQ = \Omega(\mathcal{N}Q/\log \mathcal{N})$.
\item\relax Against the sequential classical $\Theta(\mathcal{N}d)$, that leaves at most a factor $O(d\log \mathcal{N}/Q)$: a logarithmic factor when $Q \gtrsim d$, as for the same polynomial.
\item\relax Against classical hardware of the same $\Theta(\mathcal{N})$ size there is no gain (the equal-hardware objection).
\item\relax Counting only the copies (time $\geq R$ per step, each run at unit time) would leave a factor $O(d\log \mathcal{N})$. Since $d$ grows with the grid at fixed $\Delta t$ ($d \propto \mathcal{N}^{1/3} = N$), that unit-time-per-run model is not used here.
\end{itemize}
\item\relax \textbf{(K1).} Joint multi-copy circuits that share pre-processed quantum memory across copies are not the integrator's loop, and their per-copy gate floor is not established by these arguments.
\end{itemize}
Numbers at $256^{3}$, $b = 32$, generic states:

\begin{itemize}
\item\relax \textbf{Lower bound at leading order} (under I1–I3 and K1–K2, with exact preparation of generic states and the leading-order preparation bound): $R\, C_{\mathrm{prep}} \gtrsim 5.3 \times 10^{7} \times 3.4 \times 10^{7} \approx 1.8 \times 10^{15}$ CNOTs per step, with one-qubit gates free ($P_{\mathrm{err}} = 1/3$; $\approx 2.8 \times 10^{15}$ with the leading-order copy count). It holds even at step success 1, with no repetition. A pseudo-spectral RK3 step takes $\approx 2.7 \times 10^{10}$ flops (§6.2).
\item\relax \textbf{Census-based estimate for the production configuration} (an estimate, not a bound): $R\, E_{\mathrm{copy}} \approx 8.3 \times 10^{7} \times 3.5 \times 10^{13} \approx 2.9 \times 10^{21}$ CX per step. It assumes design success 1/64 and staged post-selection, with $E_{\mathrm{copy}}$ from the unoptimized collocated census (§5.3). At step success 1 it would be $\approx 8.3 \times 10^{7} \times 2.2 \times 10^{12} \approx 1.8 \times 10^{20}$. These products multiply the leading-order copy count by an upper estimate per run, so they are estimates and not bounds.
\end{itemize}
This is the common situation of every hybrid with per-step classical readout in the literature: Bharadwaj \& Sreenivasan~\cite{Bharadwaj2023}: ``O(Ng) operations that eliminate any quantum advantage''; Griffin et al.~\cite{Griffin2019}: ``an exponential speedup over a classical computer is unlikely''; Inger \& Frankel~\cite{IngerFrankel2026}: $O(\mathcal{N})$ transfer per step; Akash et al.~\cite{Akash2026} (Eq. 25): $\geq \mathcal{N}/\allowbreak \varepsilon^{2}$ copies for reconstruction.

\textbf{What would have to change for an L3 gain.} (i) Output observables instead of fields (readout $O(1/\varepsilon)$ per observable; Dalzell et al.~\cite{Dalzell2025}, Sec. 7); (ii) fewer interface crossings per physical time (larger admissible $\Delta t$; refresh every $K$ steps, whose net gain is not computed and is not claimed); (iii) sublinear state preparation and loading (structured fields, not expected for turbulence); (iv) a quantum predictor and refresh, i.e. a coherent multi-step scheme, which is not analysed here (§10.2 states only that the published coherent lower bounds do not cover the integrator's re-prepared, read-out interface). These changes are not alternatives. (i) alone leaves the data-loading floor of Proposition 5, and (iii) alone leaves the readout floor of Propositions 6–7. Moreover, the integrator's classical refresh reads the full field, so (i) would also mean replacing it.

\medskip
\subsection{Accuracy: error composition and parameter choice}\label{si:C.7}
\textbf{Local error sources per step.}

\begin{enumerate}
\item\relax Polynomial approximation: $\leq \eta$ per factor, $\leq 3\eta$ per step (Propositions 1–2).
\item\relax Strang splitting of the frozen step: $\leq \Delta t^{3}(\|[A, [A, D]]\|/12 + \|[D, [D, A]]\|/24)$ (Proposition S3, §9.3); realized $1.9\text{--}6.7 \times 10^{-6}$ per step on the production states (§9.3.1).
\item\relax Frozen-coefficient re-linearization: local $O(\Delta t^{3})$ with the midpoint predictor; second order is shown only at $16^{3}$ (§9.5).
\item\relax Hardware errors (gate errors inside the circuits, finite-sample readout) would add their own per-step errors. The effective end-of-step error model of the noise study is not a model of them (its in-circuit equivalence is untested), and no requirement follows.
\item\relax EDAC modelling error relative to incompressible NS: $O(\mathrm{Ma}^{2})$ (Clausen~\cite{Clausen2013b}, §2.3); the divergence floor scales as $\mathrm{Ma}^{2}$ in a $64^{3}$ collocated test (100 steps, Re 1600, degree 32; Extended Data Fig.~\ref{edfig:divergence}a: 4.00 / 4.09 per halving). The constant $C_{\mathrm{Ma}}$ of this error is not measured for this flow; in the $256^{3}$ run the dilatational fraction of the velocity energy (MAC projection) is $2.1 \times 10^{-6}$ at $t = 1.5$ and $1.0 \times 10^{-5}$ at $t = 3.0$.
\end{enumerate}
\textbf{Proposition 8 (linear accumulation for the frozen linear factors).} Let $S_{1}, \dots, S_{m}$ be exact factors with $\|S_{k}\| \leq 1$ and $\tilde{S}_{k}$ their implemented approximations with $\|S_{k} - \tilde{S}_{k}\| \leq \eta_{k}$, hence $\|\tilde{S}_{k}\| \leq 1 + \eta_{k}$. Then, by telescoping, $\|\prod_{k} S_{k} - \prod_{k} \tilde{S}_{k}\| \leq \sum_{k}\eta_{k}\prod_{j}(1 + \eta_{j}) \leq \bigl(\sum_{k}\eta_{k}\bigr)\exp\bigl(\sum_{k}\eta_{k}\bigr)$. All exact corrector factors are contractions (unitary or contractive) and each polynomial factor is within $\eta$ of its target, so for $m = 3M$ factors the polynomial error of the frozen problem is at most $3M\eta\,e^{3M\eta}$, linear in $M$ while $3M\eta \ll 1$ (the approximate factors are not assumed to be contractions; they are only bounded by $1 + \eta$). The bound concerns a fixed sequence of frozen operators; in the loop each operator is rebuilt from the previous output, and that feedback enters through $G(T)$ below. $\blacksquare$

\textbf{Nonlinear transport.} For the re-linearized nonlinear dynamics, an error committed at step $n$ is transported by the flow to time $T$; $G(T)$ (§2) denotes the largest such transfer and is not measured here. For a chaotic flow the pointwise $\ell_{2}$ difference saturates beyond the predictability horizon, so an $\ell_{2}$ accuracy target $\varepsilon_{\mathrm{acc}}$ is meaningful only within it; beyond it accuracy is statistical (kinetic energy, spectra, as in the base-flow comparison), and the $O(\mathrm{Ma}^{2})$ and splitting errors enter as biases of the statistics. The noise study measures something different: the cumulative transfer of one effective zero-mean per-step error model through this run, $\delta/\varepsilon = 3.4 \to 29.6$ after $46 \to 796$ injections (one realization, one initial condition, within the error model; the gains at $\varepsilon = 10^{-6}$ and $10^{-4}$ agree within 0.061\%, consistent with proportional scaling at these two amplitudes). It is not a per-error amplification factor, and whether other solvers show the same transfer is not tested. The total stayed below matched non-dissipative accumulation up to $t = 2.5$ and was comparable to it at $t = 3.0$ ($\delta/R_{1} = 1.01$, within the predefined $\pm 10$\% tolerance; full-state convention). Coherent errors (splitting, freezing) would need their own transfer estimate.

\textbf{Parameter choice (symbolic accuracy model; estimates).} If local errors were transported with a factor at most $G(T)$ and an equal share of $\varepsilon_{\mathrm{acc}}$ went to each of four sources:

\begin{itemize}
\item\relax $\eta = \varepsilon_{\mathrm{acc}}/(12 M G) \Rightarrow$ the per-factor degree grows only as $\ln(M G/\varepsilon_{\mathrm{acc}})$ (additive);
\item\relax $\Delta t$ from $C_{2}\, T\, \Delta t^{2}\, G \leq \varepsilon_{\mathrm{acc}}/4$ (second order) $\Rightarrow M = T/\Delta t \propto T^{3/2}(C_{2}G/\varepsilon_{\mathrm{acc}})^{1/2}$;
\item\relax $\varepsilon_{r} = \varepsilon_{\mathrm{acc}}/(4 M G) \Rightarrow R \propto 4\mathcal{N} M^{2}G^{2}/\varepsilon_{\mathrm{acc}}^{2}$ copies per step (sample tomography), i.e. a total readout cost $\propto \mathcal{N} M^{3} G^{2}/\varepsilon_{\mathrm{acc}}^{2}$;
\item\relax EDAC: $\mathrm{Ma} \propto \varepsilon_{\mathrm{acc}}^{1/2} \Rightarrow \tau_{\mathrm{ac}} = 6\Delta t/(\mathrm{Ma}\, h) \propto \varepsilon_{\mathrm{acc}}^{-1/2}$ at fixed $\Delta t$ and $h$ (§8.4); the leading splitting term also grows $\propto 1/\mathrm{Ma}$ (§9.3.1).
\end{itemize}
At a fixed final accuracy the cost is therefore super-linear in $T$, and grows with the flow's error transport, as for any solver of a chaotic flow. The readout term penalizes small $\Delta t$ most strongly ($\propto M^{3}$), which reinforces Proposition 4: the architecture's cost is minimized by the largest accuracy-admissible step.

\subsection{EDAC: what it removes, what it adds, and when it pays}\label{si:C.8}
\subsubsection{The two formulations}\label{si:C.8.1}
\begin{itemize}
\item\relax \textbf{EDAC as implemented} (Supplementary Note~\ref{note:A}, A.1; Clausen~\cite{Clausen2013b}, eq. (15)): $\partial u/\allowbreak \partial t = -u \cdot\allowbreak  \nabla u - (\allowbreak 1/\allowbreak \mathrm{Ma})\nabla p_{\mathrm{s}} + \nu\nabla^{2}u$ and $\partial p_{\mathrm{s}}/\allowbreak \partial t = -u \cdot\allowbreak  \nabla p_{\mathrm{s}} - (\allowbreak 1/\allowbreak \mathrm{Ma})\nabla \cdot\allowbreak  u + \nu\nabla^{2}p_{\mathrm{s}}$, with $p_{\mathrm{s}} = \mathrm{Ma}\, p$ and Ma based on $u_{\mathrm{rms}}(0)$ (§2). Pressure advection and momentum advection are in skew form, which differs from Clausen's form by $\tfrac{1}{2} p\nabla \cdot u$ and $\tfrac{1}{2} u\nabla \cdot u$, a relative $O(\mathrm{Ma}^{2})$ (Supplementary Note~\ref{note:A}, A.2). Discretized on the MAC grid: $A = A_{\mathrm{adv}}(\allowbreak \bar{u}) + A_{\mathrm{ac}}$ with $A_{\mathrm{ac}} = \mathrm{Ma}^{-1}\begin{pmatrix} 0 & -G \\ G^{T} & 0 \end{pmatrix}$ ($G$ the forward-difference gradient, $\mathrm{Div} = -G^{T}$), constant and block-circulant; $D = I_{4} \otimes \nu L_{h}$.
\item\relax \textbf{Projection (Chorin–Temam and successors)} (Guermond, Minev \& Shen~\cite{Guermond2006}, §1, Thms 3.1–3.3): each step solves decoupled elliptic problems (a pressure Poisson equation, plus Helmholtz problems if diffusion is implicit). In bounded domains the velocity $L^{2}$ error is $O(\Delta t)$ for the non-incremental (Chorin–Temam) scheme and $O(\Delta t^{2})$ for the standard and rotational incremental schemes; the rotational form improves the velocity $H^{1}$ and pressure $L^{2}$ errors to $O(\Delta t^{3/2})$; an artificial Neumann condition causes a numerical boundary layer. In a triply periodic box the projection is mode-wise and exact and adds no splitting error (pseudo-spectral: $\hat{P} = I - kk^{T}/|k|^{2}$, 9 FFTs per right-hand side; Mortensen \& Langtangen~\cite{Mortensen2016}).
\end{itemize}
\subsubsection{Operations removed and introduced}\label{si:C.8.2}
{\footnotesize
\setlength{\tabcolsep}{3pt}\renewcommand{\arraystretch}{1.15}\setlength{\LTleft}{0pt}\setlength{\LTright}{0pt}
\begin{longtable}{>{\raggedright\arraybackslash}p{\dimexpr 0.2600\linewidth-6pt\relax}>{\raggedright\arraybackslash}p{\dimexpr 0.3700\linewidth-6pt\relax}>{\raggedright\arraybackslash}p{\dimexpr 0.3700\linewidth-6pt\relax}}
\toprule
\hspace{0pt}\textbf{} & \hspace{0pt}\textbf{EDAC (this integrator)} & \hspace{0pt}\textbf{Projection} \\
\midrule
\endhead
\hspace{0pt}Nonlocal elliptic solve per step & \hspace{0pt}\textbf{none} (one FFT solve at $t = 0$ for the balanced initial condition) & \hspace{0pt}one Poisson solve: FFT, $O(\mathcal{N}\log \mathcal{N})$, in periodic boxes; fast cosine/sine transforms for MAC grids in rectangular boxes with staggered Dirichlet/Neumann conditions (Schumann \& Sweet~\cite{SchumannSweet1988}); multigrid $O(\mathcal{N})$ otherwise; in one profiled 3D DNS code (lid-driven cavity, geometric-multigrid baseline) the Poisson solve took ``no less than roughly 90\%'' of the calculation time, and much less with FFT acceleration (Costa~\cite{CostaP2022}, §4.2) \\
\hspace{0pt}Global communication (classical, parallel) & \hspace{0pt}none (local stencils) & \hspace{0pt}yes (elliptic solve; Clausen~\cite{Clausen2013b}, §1) \\
\hspace{0pt}Quantum counterpart of the solve & \hspace{0pt}none & \hspace{0pt}a Leray projector: QFT-diagonal in periodic boxes; quantum DCT/DST ($O(\log^{2}\mathcal{N})$ gates per transform; Klappenecker \& Rötteler~\cite{Klappenecker2001}) in rectangular boxes, by analogy with the classical fast-transform route (no quantum projector built this way was found); where no fast diagonalizing transform exists (complex or immersed boundaries), a QSVT projector of degree $\Theta(\kappa_{G}\log(1/\eta))$ or a QLSA with $\Theta(\allowbreak \kappa_{\Delta}\log(\allowbreak 1/\allowbreak \varepsilon))$ calls (§8.4) \\
\hspace{0pt}Unitarity of the dominant factor & \hspace{0pt}yes: $e^{A\Delta t}$ unitary for any real $\bar{u}$ ($A$ anti-Hermitian on the MAC grid) & \hspace{0pt}yes for $e^{PA_{\mathrm{adv}}P\, \Delta t}$ ($P$ Hermitian); the projector itself is non-unitary unless built into the generator \\
\hspace{0pt}Additional fields/terms & \hspace{0pt}pressure transport and pressure diffusion & \hspace{0pt}— \\
\hspace{0pt}Acoustic modes & \hspace{0pt}yes: speed $c = 1/\mathrm{Ma}$; stiffness $\tau_{\mathrm{ac}} = 6\Delta t/\allowbreak (\allowbreak \mathrm{Ma}\, h) = 9.78$ per step & \hspace{0pt}none \\
\hspace{0pt}Modelling error relative to incompressible NS & \hspace{0pt}$O(\mathrm{Ma}^{2})$ (Clausen~\cite{Clausen2013b}, §2.3), with Ma based on $u_{\mathrm{rms}}$ (local Mach number up to $\approx 0.26$ in this run); for the classical artificial-compressibility method, not EDAC, error orders with $k \sim \mathrm{Ma}^{2}$ are analysed by Ohwada \& Asinari~\cite{Ohwada2010} (2D) & \hspace{0pt}splitting error $O(\Delta t^{k})$ and boundary layers in bounded domains (Guermond, Minev \& Shen~\cite{Guermond2006}); none from a periodic pseudo-spectral projection \\
\hspace{0pt}Initial-condition requirement & \hspace{0pt}balanced pressure (one FFT solve at $t = 0$) to avoid a large acoustic transient & \hspace{0pt}— \\
\hspace{0pt}Time-step limit (explicit) & \hspace{0pt}acoustic CFL, in 2D: $\Delta t \leq \sigma[\allowbreak 1/\allowbreak \Delta x + 1/\allowbreak \Delta y + (\allowbreak 1/\allowbreak \mathrm{Ma})\sqrt{1/\Delta x^{2} + 1/\Delta y^{2}}]^{-1}$ with $\sigma = 1/2$ for the MacCormack scheme (Clausen~\cite{Clausen2013b}, eq. 21); the 3D form used in §8.4 is an extension made here, not Clausen's equation; time step ``Ma times smaller than that of INS'' (Toutant~\cite{Toutant2018}, §2) & \hspace{0pt}advective CFL \\
\hspace{0pt}Time-step limit (exponential corrector) & \hspace{0pt}\textbf{none for stability} (unitary and contractive factors). The accuracy limit is not established for HIT; the leading splitting term is dominated by $[\allowbreak A_{\mathrm{ac}},\allowbreak  [\allowbreak A_{\mathrm{adv}},\allowbreak  D]] \propto 1/\allowbreak \mathrm{Ma}$ (§9.3.1) & \hspace{0pt}— \\
\bottomrule
\end{longtable}}

\subsubsection{Classical cost comparison}\label{si:C.8.3}
\textbf{Proposition E0 (classical cost ratio).} In a periodic box, with step counts at unit Courant number, projection costs $C_{\mathrm{proj}} \approx (T|u|/h)(c_{P}\mathcal{N}\log \mathcal{N} + c_{1}\mathcal{N})$ and explicit EDAC costs $C_{\mathrm{EDAC}} \approx (T(|u| + c)/h)\, c_{2}\mathcal{N}$, so

\displ{$C_{\mathrm{EDAC}}/C_{\mathrm{proj}} \approx (1 + 1/(\mathrm{Ma}|u|))\, c_{2}/(c_{P}\log \mathcal{N} + c_{1})$.}

At production the step-count factor $1 + 1/(\mathrm{Ma}|u|)$ is 4.8–6.2 ($|u| = \max|u_{x}| = 3.84\text{--}5.23$), so the ratio is $\approx 5\text{--}6 \times c_{2}/(c_{P}\log \mathcal{N} + c_{1})$: above one unless the Poisson solve dominates the projection step, and growing as $1/\mathrm{Ma}$ at fixed $\mathcal{N}$. As $\mathcal{N} \to \infty$ at fixed Ma the ratio falls like $1/\log \mathcal{N}$ against an FFT-based projection, so asymptotically EDAC gains a factor $\log \mathcal{N}$ there, which the $1/\mathrm{Ma}$ step penalty offsets at practical Ma; against an $O(\mathcal{N})$ multigrid projection it has no asymptotic advantage. In a periodic pseudo-spectral code the projection itself is a mode-wise $O(\mathcal{N})$ operation once the FFTs needed for the right-hand side are done. $\blacksquare$ Matched targets: step counts at unit Courant number (stability limits) for two different models (EDAC against incompressible), not equal accuracy.

Practical evidence (implementation dependent; not equal-output by construction):

\begin{itemize}
\item\relax 3D decaying HIT at $256^{3}$ on one GPU (Re 4000, Ma 0.04, EDAC without pressure advection): EDAC (4th order) 0.012 s/step $\times 6000$ steps $= 72$ s at $L^{2}$ error 0.039, against a fractional-step projection method 1.107 s/step $\times 500$ steps $= 553$ s at 0.033; the Poisson solver of the projection code is not stated (Yasuda et al.~\cite{Yasuda2026}, Table 1).
\item\relax 2D: EDAC wall time 1.2 times smaller than a CG+SSOR projection code at $\mathrm{Ma} = 0.02$ (Toutant~\cite{Toutant2018}, §4.2).
\item\relax EDAC-SPH $\approx 25$ \% slower than weakly compressible SPH; no projection comparison (Ramachandran \& Puri~\cite{Ramachandran2019}).
\end{itemize}
The emulated integrator is itself a classical exponential EDAC integrator. It has no CFL limit, but it pays $\Theta(\tau_{A} + \log(1/\eta))$ stencil applications per step, i.e. $\Theta(T/(\mathrm{Ma}\, h) + M\log(1/\eta))$ in total. At production that is $\approx 4.7 \times 10^{5}$ stencil applications as run (286.5 per step $\times$ 1650), or $\approx 1 \times 10^{5}$ with Chebyshev degrees at $\eta = 10^{-12}$ and the reduced predictor ($\approx 56\text{--}69$ per step), against $\approx 2.7 \times 10^{4}$ for explicit RK4 EDAC at its imaginary-axis stability limit ($\approx 3.3 \times 10^{3}$ steps $\times$ 8 applications): $\approx 3.5\text{--}18$ times more matrix–vector products, with $\approx 2$ times fewer steps (nonlinear refreshes and interface crossings). These are estimates; the step counts are those of Proposition E3.

\subsubsection{Quantum (block-encoding) comparison}\label{si:C.8.4}
Fair baseline. The projected incompressible formulation also splits viscosity from transport and keeps the transport factor unitary: its frozen generator $P A_{\mathrm{adv}} P$ is skew-Hermitian on the full space ($P$ is an orthogonal projector), with block-encoding normalization $\lambda_{\mathrm{adv}}$ ($P$ has normalization 1 when encoded as $(I + R)/2$ with the reflection $R = 2P - I$). Two placements of the projector are considered: inside every query of the generator (exact projected dynamics), or once per step after the transport factor (Chorin-type, with its splitting error). The comparison is per step, at $\eta = 10^{-12}$, with production $\tau$ values (Appendix A). Counting conventions (estimates): Chebyshev reference degrees; the constant of $\Theta(\kappa_{G}\log(1/\eta))$ set to 1; the condition numbers of the periodic grid used as proxies for non-periodic geometries.

{\scriptsize
\setlength{\tabcolsep}{3pt}\renewcommand{\arraystretch}{1.15}\setlength{\LTleft}{0pt}\setlength{\LTright}{0pt}
\begin{longtable}{>{\raggedright\arraybackslash}p{\dimexpr 0.2302\linewidth-6pt\relax}>{\raggedright\arraybackslash}p{\dimexpr 0.1732\linewidth-6pt\relax}>{\raggedright\arraybackslash}p{\dimexpr 0.2896\linewidth-6pt\relax}>{\raggedright\arraybackslash}p{\dimexpr 0.3069\linewidth-6pt\relax}}
\toprule
\hspace{0pt}\textbf{Formulation} & \hspace{0pt}\textbf{Data-dependent calls per step} & \hspace{0pt}\textbf{Constant-coefficient calls per step} & \hspace{0pt}\textbf{Status of the projector} \\
\midrule
\endhead
\hspace{0pt}EDAC, combined generator (production) & \hspace{0pt}$d(\tau_{A}) = 31\text{--}32$ & \hspace{0pt}0 & \hspace{0pt}none needed \\
\hspace{0pt}Projected, periodic box, QFT projector & \hspace{0pt}$d(\allowbreak \tau_{\mathrm{adv}}) = 9\text{--}12$ & \hspace{0pt}1 per step (Chorin-type) or 2 per query; each $O(\allowbreak \mathrm{polylog}\, \mathcal{N})$ gates (QFT plus a mode-wise rotation computed by arithmetic) & \hspace{0pt}no 3D time-dependent implementation found; closest: steady 2D periodic Stokes (Gimenez, Cirak \& Ortiz~\cite{Gimenez2026}) \\
\hspace{0pt}Projected, rectangular box with walls, quantum DCT/DST projector & \hspace{0pt}9–12 & \hspace{0pt}as for the QFT projector, with DCT/DST transforms of $O(\log^{2}\mathcal{N})$ gates & \hspace{0pt}classical fast-transform route for MAC grids (Schumann \& Sweet~\cite{SchumannSweet1988}); quantum transforms (Klappenecker \& Rötteler~\cite{Klappenecker2001}); no quantum projector found \\
\hspace{0pt}Projected, geometry without a fast diagonalizing transform, QSVT projector & \hspace{0pt}9–12 & \hspace{0pt}$d_{P} = \Theta(\allowbreak \kappa_{G}\log(\allowbreak 1/\allowbreak \eta)) \approx 3.9 \times\allowbreak  10^{3}$ once per step; $\approx 0.9 \times 10^{5}$ inside every query & \hspace{0pt}projector degree $O(\kappa(D)\log(1/\eta_{P}))$ (Wu \& Li~\cite{WuLi2026}, Thm 4.1; GSLW~\cite{Gilyen2019}, Thm 31); $\kappa_{G} = \sqrt{3}/\allowbreak \sin(\allowbreak \pi/\allowbreak N) = 141$ on the periodic $256^{3}$ grid (proxy) \\
\hspace{0pt}Same geometry, QLSA projector & \hspace{0pt}9–12 & \hspace{0pt}$\Theta(\allowbreak \kappa_{\Delta}\log(\allowbreak 1/\allowbreak \varepsilon)) \approx 5.5 \times\allowbreak  10^{5}$ per solve & \hspace{0pt}$\kappa_{\Delta} = 3/\allowbreak \sin^{2}(\allowbreak \pi/\allowbreak N) \approx 2.0 \times\allowbreak  10^{4}$ (periodic proxy); optimal QLSA $O(\kappa\log(1/\varepsilon))$ (Costa et al.~\cite{Costa2022}) \\
\hspace{0pt}Preconditioned (BPX) quantum FEM & \hspace{0pt}— & \hspace{0pt}$\kappa = O(1)$ & \hspace{0pt}only for linear-functional outputs, not for projecting a state (Deiml \& Peterseim~\cite{Deiml2026}) \\
\bottomrule
\end{longtable}}

Grid scaling at fixed $\Delta t$ and Ma (Appendix A): $\kappa_{G} \propto N$ ($35 \to 565$ from $64^{3}$ to $1024^{3}$) and $\tau_{\mathrm{ac}} \propto N$ ($2.4 \to 39.1$).

\textbf{Proposition E1 (query-count crossover, per step).} Counting every block-encoding call equally, EDAC needs fewer calls per step than the once-per-step QSVT-projected formulation iff $d(\tau_{\mathrm{adv}} + \tau_{\mathrm{ac}}, \eta) < d(\tau_{\mathrm{adv}}, \eta) + d_{P}(\kappa_{G}, \eta)$. At production $31\text{--}32 < 9\text{--}12 + 3.9 \times 10^{3}$, counted with the reference degrees $d(\tau, \eta)$ of §4 and the reference projector degree $d_{P}$ at equal $\Delta t$ and $\eta$ (not with the 184--185 production queries, which include the Laurent synthesis): $122\text{--}126\times$ fewer calls per step ($\approx 2.3\text{--}2.9 \times 10^{3}$ times fewer against the projector-in-every-query form, $\approx 1.7 \times 10^{4}$ times fewer against the QLSA projector). The diffusion factors, which both formulations contain (4–5 calls each at the reference degrees), cancel in the inequality and are left out of these ratios; with them on both sides the per-step ratio against the once-per-step QSVT projector is $\approx 93\text{--}100$ (Appendix A).

\begin{itemize}
\item\relax \textbf{Per trajectory.} The ratio is per step. A projection scheme at unit advective Courant number would need $1.1\text{--}3.7$ times fewer steps than the integrator (Proposition E3), so per trajectory the ratio is $\approx 30\text{--}110$.
\item\relax \textbf{Under refinement at fixed $\Delta t$ and Ma} the per-step ratio grows:

\begin{itemize}
\item\relax with $\tau_{\mathrm{adv}}$ held at 0.8 (Appendix A): from $\approx 55$ ($64^{3}$) through 126 ($256^{3}$) to $\approx 220$ ($1024^{3}$);
\item\relax with $\tau_{\mathrm{adv}}$ scaled $\propto N$, as it is at fixed $\Delta t$: $\approx 61 \to 89 \to 126 \to 166 \to 211$ (Bessel-tail degrees; Appendix A).
\end{itemize}
It grows because $\ln(1/\eta) \approx 28$ still dominates $d(\tau, \eta)$ at these $\tau$, and it tends to a constant only when $\tau_{\mathrm{ac}} \gg \ln(1/\eta)$.
\item\relax \textbf{At fixed Courant numbers} ($\Delta t$ shrinking with $h$), $\tau_{\mathrm{ac}}$ stays constant while $\kappa_{G}$ grows like $N$, so the ratio grows like $N$. $\blacksquare$
\end{itemize}
Conditions:

\begin{itemize}
\item\relax \textbf{(a) Query model.} Every call costs the same.
\item\relax \textbf{(b) Geometry.} No fast diagonalizing transform exists; periodic and rectangular boxes are excluded.
\item\relax \textbf{(c) Splitting error.} It must stay acceptable in that geometry. Lemma S2 ($[D, A_{\mathrm{ac}}] = 0$) is proved only on the uniform periodic MAC grid with equal diffusivities. With walls it would require a compatible (mimetic) vector Laplacian with consistent boundary conditions, which is not established, so the acoustic part may enter the splitting error directly.
\item\relax \textbf{(d) Accuracy target.} It must not lie below the $O(\mathrm{Ma}^{2})$ EDAC error (Proposition E4).
\end{itemize}
Matched targets. The comparison is per step, at equal $\Delta t$ and equal polynomial tolerance $\eta$. It is \textbf{not} at matched solution accuracy or matched output: EDAC solves a weakly compressible model with an $O(\mathrm{Ma}^{2})$ difference from incompressible NS, plus the acoustic splitting term of §9.3; the projected step carries its own splitting error; and both output a state, but of different models. E1 is therefore a statement about the call count of one step of each formulation, not a cost comparison at equal error.

\textbf{Proposition E2 (gate-weighted crossover).} If a data-dependent call costs $C_{\mathrm{load}} \gg C_{G}$ (the cost of a constant-coefficient call), the combined EDAC generator is cheaper than the projected formulation iff $[d(\tau_{\mathrm{adv}} + \tau_{\mathrm{ac}}) - d(\tau_{\mathrm{adv}})]\, C_{\mathrm{load}} < d_{P}\, C_{G}$ (projector once per step). With the collocated census $C_{\mathrm{load}} \approx 2.78 \times 10^{10}$ CX and $C_{G}$ taken as the cost of the implemented data-free stencil walk, $\approx 6.3 \times 10^{4}$ CX at $256^{3}$ (the diffusion walk, §5.1), the left side is $\approx 20 \times 2.78 \times 10^{10} \approx 5.6 \times 10^{11}$ and the right side $\approx 3.9 \times 10^{3} \times 6.3 \times 10^{4} \approx 2.5 \times 10^{8}$ (estimates): \textbf{the condition fails} at $256^{3}$, in periodic and in general geometry. The threshold $C_{\mathrm{load}} < d_{P}\, C_{G}/\Delta d \approx 1.2 \times 10^{7}$ CX lies below $6\mathcal{N} \approx 1.0 \times 10^{8}$, the number of loader angles per walk, so the condition fails for any loader that spends at least one two-qubit gate per angle; the conclusion does not depend on the census value of 276 CX per angle, which is an unoptimized upper estimate (§5.1). In the loader-dominated gate model, EDAC as implemented pays its acoustic stiffness on the expensive data-loading calls: about 20 of its 31–32 data calls per step with trimmed polynomials ($2.6\text{--}3.6$ times more than a projected formulation, $\approx 65$ \% of the step's data-loading gates), or about 21 of the 79 calls of the production polynomial (the acoustic part alone needs 77 of them; Appendix A). $\blacksquare$

\textbf{Proposition E3 (acoustic stiffness: no stability limit, an additive degree cost).}

\begin{enumerate}
\item\relax \emph{Stability.} Every corrector factor is a contraction for any $\Delta t$ (a unitary convection–acoustic factor and contractive diffusion factors), so the acoustic stiffness imposes no stability limit on the step. Whether the accuracy-admissible $\Delta t$ is free of the acoustic scale is not shown: the leading Strang term is dominated by $[A_{\mathrm{ac}}, [A_{\mathrm{adv}}, D]] \propto 1/\mathrm{Ma}$ ($\geq 99$ \% of the worst case, $\geq 99.8$ \% of the leading term on the production states, §9.3.1), and second order is shown only at $16^{3}$ with acoustic Courant number $\leq 1.02$ (production: 1.63).
\item\relax \emph{Degree.} The stiffness enters the per-step degree additively, through $\tau_{\mathrm{ac}}$ in $d(\tau_{\mathrm{adv}} + \tau_{\mathrm{ac}}, \eta)$; over the trajectory its total is at most $\approx e\, \lambda_{\mathrm{ac}} T/2$ per convection factor, independent of $\Delta t$ (Proposition 4). This is the optimal black-box scaling of Hamiltonian simulation at the encoding normalization (Lemma 2, with its scope in §4.2); it is the scaling of the reference degree, which the production synthesis does not reach.
\item\relax \emph{Step counts} (recomputed from the production run's per-step metrics, Appendix A; definition-dependent). The integrator uses 1650 steps (acoustic Courant number $c\, \Delta t/h = 1.63$; logged 3D advective CFL at most 0.889). An explicit acoustically limited scheme needs $\approx 3.2\text{--}5.4 \times 10^{3}$ steps ($1.9\text{--}3.3$ times more): 3205 with a 1D Courant number on $c + \max|u_{x}|$ and $\max|u_{x}|(0) = 3.84$; 3392 with the run maximum $\max|u_{x}| = 5.23$ ($t = 0.83$); 5061 with a 3D extension, made here, of Clausen's 2D limit (eq. 21) at $\sigma = 1$ (the 2D MacCormack scheme uses $\sigma = 1/2$); 3294–3332 for RK4 and 5379–5440 for RK3 at their imaginary-axis limits with $\|A\| = 2823\text{--}2855$. A projection scheme at unit advective Courant number needs $\approx 450\text{--}1470$ steps ($1.1\text{--}3.7$ times fewer than the integrator): 703 with a fixed $\Delta t$ and a 1D Courant number at the velocity peak; 447 with an adaptive $\Delta t$ (1D); 1466 with a fixed $\Delta t$ and the code's 3D-sum CFL at its logged peak; 871 with an adaptive $\Delta t$ (3D sum).
\end{enumerate}
So the step-count benefit holds against acoustically limited hybrids (explicit EDAC or artificial compressibility, lattice-Boltzmann-type marching), not against projection, and only if the production $\Delta t$ is accuracy-admissible. $\blacksquare$

Matched targets. Parts 1 and 2 are proved (part 1 is a stability statement only). Part 3 compares stability limits at unit Courant number, not equal accuracy; the integrator's own count (1650) uses the production $\Delta t$, whose accuracy is not established (§9.5); and the counts measure interface crossings and nonlinear refreshes, not total work.

\textbf{Proposition E4 (accuracy dependence).} If the target accuracy relative to incompressible NS is $\varepsilon_{\mathrm{acc}}$ and the EDAC error is $C_{\mathrm{Ma}}\mathrm{Ma}^{2}$ (Ma based on $u_{\mathrm{rms}}$, as here), then $\mathrm{Ma} \leq (\varepsilon_{\mathrm{acc}}/C_{\mathrm{Ma}})^{1/2}$ and $\tau_{\mathrm{ac}} = 6\Delta t/(\mathrm{Ma}\, h) \geq 6\Delta t\, (C_{\mathrm{Ma}}/\varepsilon_{\mathrm{acc}})^{1/2}/h$: the per-step degree (at fixed $\Delta t$ and $h$) and the trajectory's stiffness term $e\, \lambda_{\mathrm{ac}} T$ (at fixed $h$) grow as $\varepsilon_{\mathrm{acc}}^{-1/2}$, up to the logarithmic tolerance term of the degree. At fixed $\Delta t$ the leading splitting term, $\propto 1/\mathrm{Ma}$ (§9.3.1), grows in the same way. The projected formulation has neither term, so EDAC is most attractive at moderate accuracy (for turbulence the $\ell_{2}$ target applies only within the predictability horizon; §7). $\blacksquare$ $C_{\mathrm{Ma}}$ is not measured for this flow (§7, item 5).

\subsubsection{Summary for EDAC}\label{si:C.8.5}
\begin{itemize}
\item\relax \textbf{Its benefit is the elimination of a difficult quantum subroutine} (a statement about the pressure-coupling subproblem, not about the total cost of the loop, §6). It removes the only nonlocal operator of the step. Where no fast diagonalizing transform exists (complex or immersed boundaries; periodic boxes are diagonalized by the QFT and rectangular boxes by DCT/DST), the projector costs $\Theta(\kappa_{G}\log(1/\eta))$ calls per application ($\kappa_{G} \propto N$), or $\Theta(\kappa_{\Delta}\log(1/\varepsilon))$ with a QLSA ($\kappa_{\Delta} \propto N^{2}$). EDAC replaces it by an additive acoustic increment of the per-step degree. By Proposition E1, counted with reference degrees at equal $\Delta t$ and $\eta$ (not with the production queries), that means $122\text{--}126\times$ fewer calls per step at $256^{3}$ than a once-per-step QSVT projection, $\approx 30\text{--}110\times$ per trajectory, and the per-step ratio grows from $\approx 55\text{--}61$ at $64^{3}$ to $\approx 211\text{--}220$ at $1024^{3}$ at fixed $\Delta t$ and Ma. This holds in the query-count model only, under conditions (a)–(d) of Proposition E1 (the geometry, an acceptable splitting error there and an accuracy target above the $O(\mathrm{Ma}^{2})$ error), and it is not a comparison at matched solution accuracy.
\item\relax \textbf{It keeps the whole step inside the class that GQSP executes directly:} products of bounded matrix functions of local sparse operators, with no non-unitary projector. The unitary dominant factor (Proposition 1) is shared with the transport factor of a projected formulation; what EDAC removes is the projector.
\item\relax \textbf{Other settings:}

\begin{itemize}
\item\relax (i) classically: by Proposition E0 the cost ratio to an FFT projection falls like $1/\log \mathcal{N}$ at fixed Ma and carries the step-count factor $1 + 1/(\mathrm{Ma}|u|)$, 4.8–6.2 at production, and the ratio to an $O(\mathcal{N})$ multigrid projection has no $\log \mathcal{N}$ factor; practical GPU speed-ups are implementation dependent;
\item\relax (ii) in periodic or rectangular boxes, where fast-transform projectors exist and a projected formulation needs $2.6\text{--}3.6$ times fewer data-dependent calls (§8.4);
\item\relax (iii) in the loader-dominated gate model, in any geometry: by Proposition E2 the additional data-dependent calls of the combined generator cost $\approx 5.6 \times 10^{11}$ CX per step at $256^{3}$, against $\approx 2.5 \times 10^{8}$ CX for the projector once per step (estimates).
\end{itemize}
\item\relax \textbf{It removes the acoustic stability limit} (Proposition E3). Whether it also removes the acoustic scale from the accuracy-admissible step is not shown: the leading splitting term is $\propto 1/\mathrm{Ma}$, and the degree term grows as $\varepsilon_{\mathrm{acc}}^{-1/2}$ (Proposition E4). At unit advective Courant number a projection scheme needs $1.1\text{--}3.7$ times fewer steps than the integrator (Proposition E3, part 3).
\end{itemize}
\medskip
\subsection{The structure-preserving split}\label{si:C.9}
\subsubsection{Structure}\label{si:C.9.1}
\textbf{Lemma S1.} For any real frozen $\bar{u}$ on the MAC grid, $A$ is real skew-symmetric ($A^{\dagger} = -A$; skew residual $\leq 4.4 \times 10^{-18}$ for random coefficient fields, Supplementary Table D3; $\|A_{\mathrm{adv}} + A_{\mathrm{adv}}^{T}\|_{F} = 0$ at $8^{3}\text{--}32^{3}$, Appendix A), so $e^{A\Delta t}$ is unitary. $D = D^{\dagger} \leq 0$ with $\|D\| = \lambda_{D}$, so $\|e^{D\Delta t/2}\| \leq 1$ ($\|e^{D\Delta t/2}\|_{2} - 1 = 2 \times 10^{-16}$, Supplementary Table D3).

\textbf{Lemma S2 ($[D, A_{\mathrm{ac}}] = 0$ on the periodic MAC grid).} On a uniform periodic MAC grid, with $D = I_{4} \otimes \nu L_{h}$ using the same index-space periodic Laplacian on every field (equal diffusivity $\nu$ for momentum and for $p_{\mathrm{s}}$), every block of $A_{\mathrm{ac}}$ is a polynomial in the same commuting periodic shifts. Circulant matrices commute, so $[D, A_{\mathrm{ac}}] = 0$ exactly. Hence $[D, A] = [D, A_{\mathrm{adv}}]$, $[D, [D, A]] = [D, [D, A_{\mathrm{adv}}]]$, and $[A, [A, D]] = [A_{\mathrm{adv}}, [A_{\mathrm{adv}}, D]] + [A_{\mathrm{ac}}, [A_{\mathrm{adv}}, D]]$. For uniform $\bar{u}$, $A_{\mathrm{adv}}$ is also circulant and the split is exact. $\blacksquare$

Scope. If the pressure diffusivity $\nu_{p}$ differs from $\nu$, the velocity–pressure blocks of $[D, A_{\mathrm{ac}}]$ are $\pm(\nu_{p} - \nu)L_{h}G/\mathrm{Ma} \neq 0$. With walls the commutation requires a compatible (mimetic) vector Laplacian with consistent boundary conditions and is not established; component-wise Laplacians and cut-cell or immersed-boundary treatments break it. Numerically, $\|[D, A_{\mathrm{ac}}]\|_{F} = 0$ at $8^{3}\text{--}32^{3}$ (Supplementary Table D3), and $\|[D, A_{\mathrm{ac}}]q\|$ is at rounding level ($5.0 \times 10^{-15}\text{--}1.3 \times 10^{-14}$ relative) on the three saved production states (§9.3.1).

\textbf{Lemma S3 (non-normality is the same commutator).} For $L = A + D$ with $A^{\dagger} = -A$ and $D^{\dagger} = D$, $LL^{\dagger} - L^{\dagger}L = 2[A, D] = 2[A_{\mathrm{adv}}, D]$ (the second equality by Lemma S2). So the unsplit frozen generator is normal iff $[A_{\mathrm{adv}}, D] = 0$. $\blacksquare$ The commutator that makes the unsplit operator non-normal, and therefore outside eigenvalue-based QSP, is the same one that controls the Strang error.

\subsubsection{Why the split is what makes GQSP applicable}\label{si:C.9.2}
QSP and GQSP transform eigenvalues of a Hermitian operator or of a unitary (Low \& Chuang~\cite{LowChuang2019}; Motlagh \& Wiebe~\cite{Motlagh2024}); QSVT transforms singular values (GSLW~\cite{Gilyen2019}). For non-normal $L$, $e^{L\Delta t}$ is neither a function of the eigenvalues in a unitarily diagonalizable basis nor a function of the singular values of $L$. Hence, by Lemma S3, the unsplit step cannot be executed by a single QSP/QSVT polynomial of a block encoding of $L$. The split reduces it to an anti-Hermitian factor (eigenvalue polynomial of $\bar{H}_{A}$) and a Hermitian factor (eigenvalue polynomial of $\bar{H}_{D}$), both normal.

\subsubsection{A factor-free Strang bound with explicit commutator constants}\label{si:C.9.3}
Prior art. Factor-free second-order bounds for contractive splittings are classical. Jahnke \& Lubich~\cite{JahnkeLubich2000} (Thm 2.1(b)) assume $\|e^{tA}\|$, $\|e^{tB}\|$, $\|e^{t(A + B)}\| \leq 1$ ($A$ generating a strongly continuous semigroup, $B$ bounded) and bound the Strang error $e^{\tau B/2}e^{\tau A}e^{\tau B/2}v - e^{\tau(A + B)}v$ by $C_{2}\tau^{3}$ (times a norm of $v$). The principal term is $(1/12)c_{2}\tau^{3}$, with $c_{2}$ a bound on $[A, [A, B]]$ (their eq. 2.7); the constant $C_{2}$ also depends on a bound $c_{1}$ on $[A, B]$ and on $\|B\|$, through remainder terms of order $\tau^{3}\|B\|^{3}$ and $\tau^{3}c_{1}(c_{1} + \|B\|)$ with unspecified constants. In the commutator-scaling Trotter theory a factor-free bound with explicit commutator constants is proved for anti-Hermitian summands; for general summands it carries a factor $e^{2t\Upsilon\sum_{\gamma}\|H_{\gamma}\|}$ (Childs, Su, Tran, Wiebe \& Zhu~\cite{Childs2021}, Thm 11, Prop. 16). With $\Upsilon = 2$ for Strang and the measured spectral norms this is $\approx e^{23}$ at production ($e^{22.8}\text{--}e^{23.0}$), so vacuous.

The following proposition gives, for unitary-plus-dissipative splitting, a bound in nested commutators only, with the explicit constants 1/12 and 1/24 of the unitary case. In this setting it sharpens Jahnke–Lubich~\cite{JahnkeLubich2000} Thm 2.1(b), since no $\|B\|$-dependent remainder appears. The splitting literature was not searched exhaustively for this exact form; nothing beyond the explicit constants is claimed as new.

\textbf{Proposition S3 (Strang bound with contractive outer factors).} Let $A$ be anti-Hermitian and $D$ Hermitian negative semidefinite (bounded). Then for $t \geq 0$

\displ{$\|e^{tD/2}\, e^{tA}\, e^{tD/2} - e^{t(A + D)}\| \leq (t^{3}/12)\|[A, [A, D]]\| + (t^{3}/24)\|[D, [D, A]]\|$.}

\emph{Proof.} Interaction picture with respect to $A$: let $D(s) := e^{-sA} D e^{sA}$ (Hermitian, $\leq 0$, unitarily equivalent to $D$). Then $e^{t(A + D)} = e^{tA}W(t)$ with $W' = D(t)W$, $W(0) = I$; the propagator $W(t, s)$ of $W' = D(t)W$ is a contraction because $\frac{d}{dt}\|Wv\|^{2} = 2\langle Wv, D(t)Wv\rangle \leq 0$. Also $e^{tD/2}e^{tA}e^{tD/2} = e^{tA}\, e^{(t/2)D(t)}\, e^{(t/2)D} =: e^{tA}P(t)$, because $e^{-tA}e^{(t/2)D}e^{tA} = e^{(t/2)D(t)}$. Since $e^{tA}$ is unitary, the error equals $\|P(t) - W(t)\|$. Let $F = P - W$, $F(0) = 0$, and $\Delta(s) := P'(s) - D(s)P(s)$. Then $F' = D(s)F + \Delta$, so $F(t) = \int_{0}^{t} W(t, s)\Delta(s)\, ds$ and $\|F(t)\| \leq \int_{0}^{t}\|\Delta(s)\|\, ds$. With $X(s) = (s/2)D(s)$, $X' = D(s)/2 + (s/2)D'(s)$, and $\frac{d}{ds}e^{X} = \int_{0}^{1} e^{\sigma X}X'e^{(1 - \sigma)X}d\sigma$. Because $X$ commutes with $D(s)$, the $D(s)/2$ part gives $e^{X}D(s)/2$, and $D(s)e^{X} = e^{X}D(s)$. Collecting terms,

\displ{$\Delta(s) = -\tfrac{1}{2}\, e^{X}\bigl[\int_{0}^{s} D'(r)\, dr\bigr]\, e^{(s/2)D} + (s/2)\int_{0}^{1} e^{\sigma X}D'(s)e^{(1 - \sigma)X}d\sigma \cdot e^{(s/2)D}$.}

Add and subtract $(s/2)e^{X}D'(s)e^{(s/2)D}$. The first group becomes $\tfrac{1}{2} e^{X}\bigl[\int_{0}^{s}\int_{r}^{s} D''(\rho)\, d\rho\, dr\bigr]e^{(s/2)D}$ with $D''(\rho) = e^{-\rho A}[A, [A, D]]e^{\rho A}$, of norm $\leq (s^{2}/4)\|[A, [A, D]]\|$ (all exponentials of $X$ and $D$ are contractions, and conjugation by $e^{\rho A}$ is unitary). The second group is $(s/2)\int_{0}^{1}e^{\sigma X}[D'(s), e^{(1 - \sigma)X}]d\sigma \cdot e^{(s/2)D}$, and $[D', e^{\tau X}] = \int_{0}^{\tau} e^{(\tau - \rho)X}[D', X]e^{\rho X}d\rho$ with $[D'(s), X(s)] = (s/2)e^{-sA}[[D, A], D]e^{sA}$ up to sign, so its norm is $\leq (s/2) \cdot \tfrac{1}{2} \cdot (s/2)\|[D, [D, A]]\| = (s^{2}/8)\|[D, [D, A]]\|$. Hence $\|\Delta(s)\| \leq (s^{2}/4)\|[A, [A, D]]\| + (s^{2}/8)\|[D, [D, A]]\|$, and integrating over $[0, t]$ gives the bound. $\blacksquare$

Remarks.

\begin{enumerate}
\item\relax Only forward exponentials of the dissipative operator appear, so no growth factor arises; unitarity of $e^{sA}$ is used for the conjugations.
\item\relax The constants coincide with the Hermitian case (Childs et al.~\cite{Childs2021}, eq. (150)) and with the leading term of the symmetric BCH expansion $E_{3} = (1/12)[A, [A, D]] - (1/24)[D, [D, A]]$ (a fit at step size $10^{-2}$ reproduces the coefficients $-1/24$ and $+1/12$; Appendix A).
\item\relax With Lemma S2 (periodic MAC grid, equal diffusivities), $[A, [A, D]] = [A_{\mathrm{adv}}, [A_{\mathrm{adv}}, D]] + [A_{\mathrm{ac}}, [A_{\mathrm{adv}}, D]]$: the acoustic part enters only through the cross term with $[A_{\mathrm{adv}}, D]$. This cross term nevertheless dominates. It carries $\geq 99$ \% of $\|[A, [A, D]]\|$ at $8^{3}\text{--}32^{3}$ (Frobenius norm; Appendix A) and $\geq 99.8$ \% of the leading term on the production states (§9.3.1), and it scales as $1/\mathrm{Ma}$, because $A_{\mathrm{ac}} \propto 1/\mathrm{Ma}$ while $[A_{\mathrm{adv}}, D]$ does not depend on Ma. A projected formulation has no acoustic operator, so against projection this term is an additional error. It is small only in comparison with splitting the acoustic part off (§9.6).
\item\relax Combined with Proposition 8, the frozen-problem splitting error over $M$ steps is $\leq M\Delta t^{3}(\|[A, [A, D]]\|/12 + \|[D, [D, A]]\|/24)$.
\item\relax The leading BCH term is not an asymptotic statement at production: $\Delta t\|A\| \approx 5.7$, so $\Delta t^{3}\|E_{3}q\|$ is an estimate, not a bound. Proposition S3 is a bound for every $t \geq 0$.
\end{enumerate}
\textbf{Domain of Proposition S3.}

\begin{itemize}
\item\relax \textbf{Operators.} $A$ and $D$ are bounded operators on one Hilbert space with one inner product, in which $A$ is anti-Hermitian and $D$ is Hermitian and negative semidefinite. In the integrator this is the Euclidean inner product on the $4\mathcal{N}$ MAC unknowns of a uniform grid ($u$, $v$, $w$ on faces, $p_{\mathrm{s}}$ at centres), with $A = A(\bar{u})$ frozen for a real $\bar{u}$ (Lemma S1) and $D = I_{4} \otimes \nu L_{h}$. A non-uniform grid needs the weighted inner product in which $A$ is skew; unbounded PDE operators need domain conditions of the Jahnke–Lubich~\cite{JahnkeLubich2000} type.
\item\relax \textbf{Ordering.} Only the symmetric $D/2$–$A$–$D/2$ ordering, with the dissipative factor outside. The proof does not cover $A/2$–$D$–$A/2$.
\item\relax \textbf{Norm.} The operator norm induced by that inner product. Vector-dependent (realized) errors are smaller (§9.3.1).
\item\relax \textbf{Exponentials.} Exact exponentials of the frozen operators. The polynomial errors ($\leq \eta$ per factor) add separately (Proposition 8). The bound says nothing about the freezing (re-linearization) error of the nonlinear step, the predictor, the post-selection normalization or hardware errors.
\item\relax \textbf{Time.} $t \geq 0$, with no step-size restriction.
\end{itemize}
The bound controls the splitting error of one frozen step for any $\Delta t$. It is not a statement that the full nonlinear scheme is accurate at a given $\Delta t$ (§9.5).

\textbf{Numbers.} Worst case (operator norm, $256^{3}$, production field at $t = 1.0$): $\|[D, A]\| = 565$, $\|[D, [D, A]]\| = 4.60 \times 10^{3}$ and $\|[A, [A, D]]\| = 6.50 \times 10^{5}$ (Appendix A). These are converged Lanczos values and formally lower bounds on the norms, so the evaluated bound is an estimate of Proposition S3 from below: $\Delta t^{3}(5.41 \times 10^{4} + 1.9 \times 10^{2}) \approx 4.3 \times 10^{-4}$ per step. At least 99 \% of $[A, [A, D]]$ is the cross term $[A_{\mathrm{ac}}, [A_{\mathrm{adv}}, D]]$ at $8^{3}\text{--}32^{3}$. These worst-state values are set by grid-scale acoustic modes; realized values on the production states follow.

\paragraph*{9.3.1 Realized splitting error on the production states}
Method (Appendix A). The check uses the saved $256^{3}$ production states $q$ at $t = 1.0$, 1.5 and 3.0 (steps 500, 750 and 1500) with the frozen operators $A(u_{n})$ and $D$. $A(u_{n})$ stands in for $A(u_{\mathrm{mid}})$, which is not saved; their coefficients differ by $\approx 1\text{--}2$ \%. On each state it evaluates the exact frozen step $e^{(A + D)\Delta t}q$, the Strang step $Sq$ and the hypothetical sub-split step by high-order Taylor sub-stepping (truncation $\approx 10^{-26}$ per sub-step), together with the leading-order terms. The measured quantity is the frozen-coefficient splitting error only; the re-linearization (freezing) error is not included.

{\scriptsize
\setlength{\tabcolsep}{3pt}\renewcommand{\arraystretch}{1.15}\setlength{\LTleft}{0pt}\setlength{\LTright}{0pt}
\begin{longtable}{>{\raggedright\arraybackslash}p{\dimexpr 0.5540\linewidth-6pt\relax}>{\raggedright\arraybackslash}p{\dimexpr 0.1470\linewidth-6pt\relax}>{\raggedright\arraybackslash}p{\dimexpr 0.1470\linewidth-6pt\relax}>{\raggedright\arraybackslash}p{\dimexpr 0.1519\linewidth-6pt\relax}}
\toprule
\hspace{0pt}\textbf{Per step, relative to $\|q\|$} & \hspace{0pt}\textbf{$t = 1.0$} & \hspace{0pt}\textbf{$t = 1.5$} & \hspace{0pt}\textbf{$t = 3.0$} \\
\midrule
\endhead
\hspace{0pt}$\|[D, A_{\mathrm{ac}}]q\|$ (Lemma S2 check) & \hspace{0pt}$5.0 \times 10^{-15}$ & \hspace{0pt}$5.2 \times 10^{-15}$ & \hspace{0pt}$1.3 \times 10^{-14}$ \\
\hspace{0pt}Lie term $(\Delta t^{2}/2)\|[D, A]q\|$ & \hspace{0pt}$2.81 \times 10^{-5}$ & \hspace{0pt}$3.68 \times 10^{-5}$ & \hspace{0pt}$1.15 \times 10^{-5}$ \\
\hspace{0pt}Strang leading order $\Delta t^{3}\|E_{3}q\|$ (estimate) & \hspace{0pt}$7.33 \times 10^{-6}$ & \hspace{0pt}$9.58 \times 10^{-6}$ & \hspace{0pt}$2.51 \times 10^{-6}$ \\
\hspace{0pt}of which $(\allowbreak \Delta t^{3}/\allowbreak 12)\|[\allowbreak A_{\mathrm{ac}},\allowbreak  [\allowbreak A_{\mathrm{adv}},\allowbreak  D]]q\|$ (ratio to the full $[A, [A, D]]$ term) & \hspace{0pt}$7.32 \times 10^{-6}$ (99.8 \%) & \hspace{0pt}$9.57 \times 10^{-6}$ (99.9 \%) & \hspace{0pt}$2.51 \times 10^{-6}$ (99.97 \%) \\
\hspace{0pt}$(\allowbreak \Delta t^{3}/\allowbreak 12)\|[\allowbreak A_{\mathrm{adv}},\allowbreak  [\allowbreak A_{\mathrm{adv}},\allowbreak  D]]q\|$ & \hspace{0pt}$5.0 \times 10^{-7}$ & \hspace{0pt}$5.1 \times 10^{-7}$ & \hspace{0pt}$6.6 \times 10^{-8}$ \\
\hspace{0pt}$(\allowbreak \Delta t^{3}/\allowbreak 24)\|[\allowbreak D,\allowbreak  [\allowbreak D,\allowbreak  A]]q\|$ & \hspace{0pt}$3.9 \times 10^{-8}$ & \hspace{0pt}$4.9 \times 10^{-8}$ & \hspace{0pt}$1.1 \times 10^{-8}$ \\
\hspace{0pt}\textbf{Realized Strang ($D$/$A$) error $\|Sq - e^{(A + D)\Delta t}q\|$} & \hspace{0pt}\textbf{$5.10 \times 10^{-6}$} & \hspace{0pt}\textbf{$6.70 \times 10^{-6}$} & \hspace{0pt}\textbf{$1.93 \times 10^{-6}$} \\
\hspace{0pt}Hypothetical acoustic/advective sub-split, leading order & \hspace{0pt}$1.44 \times 10^{-3}$ & \hspace{0pt}$1.88 \times 10^{-3}$ & \hspace{0pt}$5.92 \times 10^{-4}$ \\
\hspace{0pt}\textbf{Hypothetical sub-split, realized} & \hspace{0pt}\textbf{$1.31 \times 10^{-3}$} & \hspace{0pt}\textbf{$1.71 \times 10^{-3}$} & \hspace{0pt}\textbf{$5.52 \times 10^{-4}$} \\
\hspace{0pt}Frozen-step norm change $\|e^{(A + D)\Delta t}q\|/\allowbreak \|q\| - 1$ & \hspace{0pt}$-5.36 \times 10^{-4}$ & \hspace{0pt}$-9.23 \times 10^{-4}$ & \hspace{0pt}$-6.93 \times 10^{-4}$ \\
\hspace{0pt}Unitarity check $\|e^{A\Delta t}q\|/\|q\| - 1$ & \hspace{0pt}$1.3 \times 10^{-14}$ & \hspace{0pt}$6.7 \times 10^{-16}$ & \hspace{0pt}$-2.2 \times 10^{-14}$ \\
\bottomrule
\end{longtable}}

Findings.

\begin{enumerate}
\item\relax The realized Strang error is $1.9\text{--}6.7 \times 10^{-6}$ per step. That is $\approx 85$ times below the operator-norm bound at $t = 1.0$ ($4.3 \times 10^{-4}$) and $\approx 0.70\text{--}0.77$ times the leading-order estimate.
\item\relax At production $\Delta t$, Strang is only $\approx 5.5\text{--}6.0$ times below the Lie term on these states (2.6 times in the worst case), because the remaining $O(\Delta t^{3})$ term is dominated by the acoustic cross term.
\item\relax The cross term $[A_{\mathrm{ac}}, [A_{\mathrm{adv}}, D]]$ carries 99.8–99.97 \% of the leading term. It scales as $1/\mathrm{Ma}$ by the algebra of Remark 3; no other Ma was tested.
\item\relax Structure of the leading term at $t = 1.0$ (Appendix A): 99.1 \% of $\|\Delta t^{3}E_{3}q\|^{2}$ lies in the $p_{\mathrm{s}}$ block and 0.9 \% in the velocity block, of which 64 \% is dilatational (MAC projection). The solenoidal velocity part is therefore $\approx 0.3$ \% of the error energy, i.e. $\approx 6$ \% of its amplitude, or $\approx 4 \times 10^{-7}$ of $\|q\|$ per step (an estimate from the leading term). The local error is acoustic in character; errors in $p_{\mathrm{s}}$ and in the dilatational velocity are coupled back into the velocity by later steps.
\item\relax The hypothetical acoustic/advective sub-split (§9.6) has a realized error of $0.55\text{--}1.7 \times 10^{-3}$ per step, $\approx 255\text{--}286$ times the $D$/$A$ split. It is not viable at production parameters.
\item\relax If the realized per-step value were representative of every step, the linear accumulation of Proposition 8 would give at most $1650 \times (1.9\text{--}6.7) \times 10^{-6} \approx 0.3\text{--}1.1 \times 10^{-2}$ relative over the run for the frozen splitting part (an upper estimate). This is in the full state norm, dominated by $p_{\mathrm{s}}$. It is not a measured trajectory error and says nothing directly about kinetic energy or spectra.
\end{enumerate}
The three states come from one run, and the leading-order value is an estimate, not a bound, at $\Delta t\|A\| \approx 5.7$ (Remark 5).

\subsubsection{Valid unsplit baselines and their costs}\label{si:C.9.4}
{\scriptsize
\setlength{\tabcolsep}{3pt}\renewcommand{\arraystretch}{1.15}\setlength{\LTleft}{0pt}\setlength{\LTright}{0pt}
\begin{longtable}{>{\raggedright\arraybackslash}p{\dimexpr 0.2377\linewidth-6pt\relax}>{\raggedright\arraybackslash}p{\dimexpr 0.2377\linewidth-6pt\relax}>{\raggedright\arraybackslash}p{\dimexpr 0.2377\linewidth-6pt\relax}>{\raggedright\arraybackslash}p{\dimexpr 0.1579\linewidth-6pt\relax}>{\raggedright\arraybackslash}p{\dimexpr 0.1290\linewidth-6pt\relax}}
\toprule
\hspace{0pt}\textbf{Unsplit method for one frozen step $e^{(A + D)\Delta t}$} & \hspace{0pt}\textbf{Normalization / success per step} & \hspace{0pt}\textbf{Query complexity (as stated)} & \hspace{0pt}\textbf{Extra registers} & \hspace{0pt}\textbf{Splitting error} \\
\midrule
\endhead
\hspace{0pt}Direct QSP / QSVT / GQSP of $L = A + D$ & \hspace{0pt}not applicable ($L$ non-normal unless $[A_{\mathrm{adv}}, D] = 0$, Lemma S3) & \hspace{0pt}— & \hspace{0pt}— & \hspace{0pt}— \\
\hspace{0pt}Truncated Taylor or Dyson LCU of $L$ & \hspace{0pt}$\ell_{1}$-norm $\sum_{k}(\allowbreak \alpha_{L}\Delta t)^{k}/\allowbreak k! = e^{\alpha_{L}\Delta t}$; success at most $\approx 10^{-9}$ ($\leq e^{-2\alpha_{L}\Delta t} = 3.0 \times\allowbreak  10^{-10}\text{--}1.2 \times\allowbreak  10^{-9}$, $\alpha_{L} = \lambda_{A} + \lambda_{D}$; Proposition S4, Appendix A) & \hspace{0pt}$O(\allowbreak \alpha_{L}\Delta t \cdot\allowbreak  \log(\allowbreak 1/\allowbreak \eta)/\allowbreak \mathrm{loglog})$ per step & \hspace{0pt}$O(\log$ of the truncation order) & \hspace{0pt}none \\
\hspace{0pt}… with oblivious amplitude amplification & \hspace{0pt}not applicable (non-unitary target; Fang, Lin \& Tong~\cite{Fang2023}, Sec. 1.3; Berry et al.~\cite{BerryChildsCleve2014}) & \hspace{0pt}— & \hspace{0pt}— & \hspace{0pt}— \\
\hspace{0pt}… with uniform singular-value amplification & \hspace{0pt}success $\Omega(1)$ per step & \hspace{0pt}$\times\allowbreak \, O(\allowbreak (\allowbreak \alpha/\allowbreak (\allowbreak \delta\|\Xi\|))\log(\allowbreak \alpha/\allowbreak (\allowbreak \|\Xi\|\varepsilon)))$ (Fang, Lin \& Tong~\cite{Fang2023}, Lemma 2) & \hspace{0pt}+1 & \hspace{0pt}none \\
\hspace{0pt}LCHS (An, Liu \& Lin~\cite{AnLiuLin2023}; An, Childs \& Lin~\cite{AnChildsLin2026}) & \hspace{0pt}$\sum|c| = O(1)$; success $\approx$ physical retention$/O(1)$ & \hspace{0pt}stated: $\tilde{O}(\allowbreak (\allowbreak \|u_{0}\|/\allowbreak \|u(\allowbreak T)\|)\, \alpha T(\allowbreak \log 1/\allowbreak \varepsilon)^{1/\beta})$, $\beta \in (0, 1)$ (An, Childs \& Lin~\cite{AnChildsLin2026}, Cor. 17); in the proof each quadrature node $k$ costs $O(\allowbreak \|kL + H\|T + \log(\allowbreak 1/\allowbreak \varepsilon))$, giving the uniform per-run cost $(\allowbreak \alpha_{L}K + \alpha_{H})T + \log(\allowbreak 1/\allowbreak \varepsilon_{1})$ (arXiv:2312.03916v2, §4.2.1) & \hspace{0pt}quadrature register; LCU over $O(\allowbreak T\|L\|(\allowbreak \log 1/\allowbreak \varepsilon)^{1 + 1/\beta})$ nodes (Lemma 11) & \hspace{0pt}none \\
\hspace{0pt}Schrödingerisation & \hspace{0pt}$\approx$ physical retention after projection onto $p > 0$ & \hspace{0pt}$\tilde{O}(\allowbreak (\allowbreak \|u(\allowbreak 0)\|/\allowbreak \|u(\allowbreak t)\|)\, s\, t\, \|H_{\mathrm{total}}\|_{\max})$ with $\|H_{\mathrm{total}}\|_{\max} \sim \max(\allowbreak \|H\|_{\max}/\allowbreak \varepsilon,\allowbreak  \|\bar{H}\|_{\max})$ (Jin, Liu \& Yu~\cite{JinLiuYu2024}, Thm 4, general case; Thm 3 is the heat equation) & \hspace{0pt}$p$-register ($\Delta p \sim \varepsilon$) & \hspace{0pt}none \\
\hspace{0pt}QLSA-based ODE solvers (Berry, Childs, Ostrander \& Wang~\cite{BerryChildsOstrander2017}; Krovi~\cite{Krovi2023}; Berry \& Costa~\cite{BerryCosta2024}) & \hspace{0pt}$\Omega(1)$ with flag & \hspace{0pt}$O(\allowbreak \kappa_{V}\, s\, g\, T\|A\|\, \mathrm{polylog})$; $O(gT\|A\|C(A)\, \mathrm{poly})$; $O(\allowbreak R\lambda_{A}T\log(\allowbreak 1/\allowbreak \varepsilon)\log(\allowbreak \lambda T/\allowbreak \varepsilon))$, respectively & \hspace{0pt}history-state registers & \hspace{0pt}none \\
\hspace{0pt}\textbf{Split, production configuration} & \hspace{0pt}design success 1/64 per step (1/4 per factor; computed exactly, never sampled) & \hspace{0pt}184–185 & \hspace{0pt}none beyond the block encodings and one signal qubit & \hspace{0pt}Proposition S3; realized $1.9\text{--}6.7 \times 10^{-6}$ per step \\
\hspace{0pt}\textbf{Split, structure-adapted (Propositions 1–3; prospective, not implemented)} & \hspace{0pt}success $=$ physical retention up to $O(\eta)$ and the synthesis margins ($\approx 0.9987$; $\approx 0.96$ with $\delta = 10^{-2}$ on diffusion) & \hspace{0pt}70–72 (Proposition 1 + Proposition 2) to 78–84 (all Laurent); 43–45 with directional control & \hspace{0pt}as above & \hspace{0pt}as above \\
\bottomrule
\end{longtable}}

\textbf{Proposition S4 (exponential normalization gap to unstructured series).} A truncated Taylor or Dyson LCU of $e^{L\Delta t}$ with an LCU of $L$ of normalization $\alpha_{L}$ has coefficient $\ell_{1}$-norm $\sum_{k \leq K}(\alpha_{L}\Delta t)^{k}/k! \to e^{\alpha_{L}\Delta t}$; splitting the step into $r$ sub-steps multiplies $r$ normalizations $e^{\alpha_{L}\Delta t/r}$, again $e^{\alpha_{L}\Delta t}$. Without amplification the success probability is $\leq \|e^{L\Delta t}\psi\|^{2}e^{-2\alpha_{L}\Delta t}$, and oblivious amplification is unavailable because $e^{L\Delta t}$ is not unitary. The split's normalization is 1 per factor for the constructions of Propositions 1–2, whose use is prospective. At production the ratio of success probabilities is then $\approx e^{2\alpha_{L}\Delta t} \approx 0.9\text{--}3.4 \times 10^{9}$ (and $1.4\text{--}5.3 \times 10^{7}$ against the emulated design value 1/64; Appendix A). $\blacksquare$ This is a provable comparative property against a weak baseline; modern non-unitary methods (LCHS, uniform amplification) close the normalization gap. Matched targets: the same frozen step, output (post-selected state) and polynomial tolerance $\eta$. The series encoding has no splitting error while the split carries the Strang error, so S4 compares normalization (success probability), not total error.

\textbf{Proposition S5 (one frozen step: split versus LCHS, at the level of the published per-run cost).} With the Proposition 1 construction (prospective), the split admits the additive form $\Theta(\allowbreak \tau_{A} + \log(\allowbreak 1/\allowbreak \eta))$ of optimal Hamiltonian simulation for its dominant factor (Low \& Chuang~\cite{LowChuang2019}, Thm 1; GSLW~\cite{Gilyen2019}, Cor. 60), plus $O(\log(1/\eta))$ for the diffusion factors. For LCHS, An, Childs \& Lin~\cite{AnChildsLin2026} state the bound $\tilde{O}(\allowbreak (\allowbreak \|u_{0}\|/\allowbreak \|u(\allowbreak T)\|)\alpha T(\allowbreak \log 1/\allowbreak \varepsilon)^{1/\beta})$, but the proof of their Cor. 17 uses the uniform per-run cost $(\allowbreak \alpha_{L}K + \alpha_{H})T + \log(\allowbreak 1/\allowbreak \varepsilon_{1})$, in which $K = O(\allowbreak (\allowbreak \log 1/\allowbreak \varepsilon)^{1/\beta})$ multiplies only the dissipative normalization. For one step of the integrator $\alpha_{L}\Delta t = \lambda_{D}\Delta t = 0.05$ and $\alpha_{H}\Delta t = \tau_{A} = 10.24\text{--}10.92$. The truncation $K$ carries a constant that depends on the kernel. For the kernel of An, Childs \& Lin, $g(k) = e^{2^{\beta} - (1 + ik)^{\beta}}/(2\pi(1 - ik))$ with $0 < \beta < 1$, $|g(k)|$ decays as $\exp(-\cos(\beta\pi/2)|k|^{\beta})/|k|$, so the constant grows without bound as $\beta \to 1$, and $\beta = 1$ is not admissible. Requiring the $\ell_{1}$ tail $2\int_{K}^{\infty}|g(k)|\, dk$ of the truncated kernel to be at most $\eta = 10^{-12}$ gives $K \approx 1282$, 264, 239 and 423 for $\beta = 0.5$, 0.75, 0.82 and 0.95 (Appendix A), that is 1.7 to 13 times the value $(\ln 1/\eta)^{1/\beta}$ obtained by setting the constant to one. The normalized evolution time is smallest at $\beta = 0.82$. With the dissipative part at its full LCU normalization $\lambda_{D} = 12\nu/h^{2}$ it is $(\allowbreak K\lambda_{D} + \lambda_{A})\Delta t = 22.2\text{--}22.8$. The identity branch of the diffusion LCU can be factored out as in Proposition 2 (for LCHS it becomes a scalar phase of each quadrature node, which changes neither $|g|$ nor $K$); the dissipative normalization is then $6\nu/h^{2} = \lambda_{D}/2$, and the normalized evolution time is $(\allowbreak K\lambda_{D}/2 + \lambda_{A})\Delta t = 16.2\text{--}16.9$, about 1.6 times the $\tau_{A}$ of the split's dominant factor. Because the degree is additive in the normalized evolution time and in the tolerance term (Lemma 1), the reference degree of one LCHS run is 40–41 with this factorization and 48–49 without it, against 31–32 for the unitary factor of the split plus 4–5 for each of its diffusion factors (39–42 per step): a ratio of about 1.0 with the same factorization on both sides, and of about 1.2 otherwise (Appendix A). The query costs are therefore comparable within this kernel family; the constants hidden in $\tilde{O}$ beyond the truncation are not known, and later variants of LCHS were not assessed. LCHS adds an LCU over quadrature nodes, amplitude amplification and a quadrature register, but it has no splitting error. No query-count advantage of the split over LCHS is claimed. $\blacksquare$ What the split does differently is implementation: no quadrature register or LCU over nodes, eigenvalue-based QSP of two normal factors, and a constant, data-free diffusion factor (§9.7). Matched targets: the same frozen step and output (normalized state) at tolerance $\varepsilon$. LCHS has no splitting error, so the two reach the same accuracy only when the Strang error (realized $1.9\text{--}6.7 \times 10^{-6}$ per step) is within the budget.

\subsubsection{Accuracy trade-off and the role of the predictor}\label{si:C.9.5}
\begin{itemize}
\item\relax Lie splitting has local error $(\Delta t^{2}/2)\|[D, A]\|$: worst case $1.13 \times 10^{-3}$ per step at $256^{3}$ (Appendix A); realized $(\Delta t^{2}/2)\|[D, A]q\|/\|q\| = 1.15\text{--}3.68 \times 10^{-5}$ on the production states. Strang removes the $O(\Delta t^{2})$ term, but at production $\Delta t$ the remaining $O(\Delta t^{3})$ term is only $\approx 2.6$ times (worst case) to $\approx 5.5\text{--}6.0$ times (realized) smaller per step, because it is dominated by the acoustic cross term (§9.3.1).
\item\relax The $16^{3}$ order test (smooth non-solenoidal test initial condition, 4–64 steps, acoustic Courant number $\leq 1.02$, against 1.63 in production) shows that without the predictor the frozen-coefficient step is first order (Lie 0.98, Strang 0.98, on the ideal classical path): the first-order freezing error dominates. With the classical midpoint predictor the order is 2.03 on the ideal path and 2.081 / 2.016 / 2.003 through the production GQSP/pair path (Supplementary Table D4; Fig.~\ref{fig:verification}c). Splitting and freezing errors are then both $O(\Delta t^{3})$ locally, and the order test does not separate them.
\item\relax No time-step study exists for the production HIT case, so the $\Delta t$ optimization of §4.5 and §11 is conditional on one; the realized frozen splitting error on the production states is in §9.3.1.
\end{itemize}
\subsubsection{Hypothetical acoustic/advective sub-split (a design variant, not implemented)}\label{si:C.9.6}
Splitting $e^{A\Delta t}$ further into $e^{A_{\mathrm{ac}}\Delta t/2}e^{A_{\mathrm{adv}}\Delta t}e^{A_{\mathrm{ac}}\Delta t/2}$ would have three effects:

\begin{itemize}
\item\relax (i) it would confine the data-dependent calls to $d(\tau_{\mathrm{adv}}) = 9\text{--}12$ per step instead of 31–32;
\item\relax (ii) it would allow the constant-coefficient acoustic factor to be executed by a QFT diagonalization (its $4 \times 4$ symbol has eigenvalues 0, 0, $\pm i|g(\theta)|$) or by a cheap constant-coefficient encoding;
\item\relax (iii) it would remove the acoustic stiffness from the loader-dominated gate count (Proposition E2).
\end{itemize}
But $[\allowbreak A_{\mathrm{ac}},\allowbreak  A_{\mathrm{adv}}] \neq 0$. The worst-case Strang error of the sub-split is $\approx 0.08$ per step at $256^{3}$, $\approx 185$ times the production $D$/$A$ split ($\|[\allowbreak A_{\mathrm{ac}},\allowbreak  [\allowbreak A_{\mathrm{ac}},\allowbreak  A_{\mathrm{adv}}]]\| = 2.33 \times\allowbreak  10^{8}$; Appendix A). Its realized error on the production states is $0.55\text{--}1.7 \times\allowbreak  10^{-3}$ per step, $\approx 255\text{--}286$ times the realized $D$/$A$ split error (§9.3.1). An interaction-picture algorithm (Low \& Wiebe~\cite{LowWiebe2018}) would avoid the sub-split error but carries a multiplicative log factor. At the per-step $\tau_{\mathrm{adv}} \approx 1$ of the production configuration, where the additive $\log(1/\eta)$ term already dominates, that gives no gain. The sub-split is therefore not supported at production parameters; it is a candidate only for larger per-step $\tau_{\mathrm{ac}}$.

\subsubsection{Reuse and isolation of the data dependence}\label{si:C.9.7}
\begin{itemize}
\item\relax $D$ is constant, so its block encoding is data-free: $\approx 6.3 \times 10^{4}$ CX per walk at $256^{3}$, against $\approx 2.78 \times 10^{10}$ (upper estimate) for the loader-based convection walk. Its phases depend only on $\tau_{D}$, so they could be computed once (the production code recomputes them every step, at 91 ms each).
\item\relax The convection phases depend only on the scalar $\tau_{A}(u_{\mathrm{mid}})$, so they form a one-parameter family that could be tabulated; the step's whole data dependence sits in one unitary factor's block encoding.
\item\relax For the split diffusion factor the heat-kernel degree grows as $\sqrt{\tau_{D}\log(1/\eta)}$: degree 160 at $b = 1000$ (Appendix A), against the linear bound $b + \ln(1/\eta) \approx 1028$. That 1028 is a formula value, not the degree of any unsplit method. This comparison is conditional and not at matched accuracy, because the splitting error grows with $\tau_{D}$, and it is immaterial at production ($\tau_{D} = 0.025$). The per-run cost of LCHS-class methods contains $K\alpha_{L}T$, linear in $\alpha_{L}$ (Proposition S5), and no matched-accuracy comparison with them was made in this regime.
\end{itemize}
\subsubsection{Summary for the split (tests 1–5)}\label{si:C.9.8}
{\footnotesize
\setlength{\tabcolsep}{3pt}\renewcommand{\arraystretch}{1.15}\setlength{\LTleft}{0pt}\setlength{\LTright}{0pt}
\begin{longtable}{>{\raggedright\arraybackslash}p{\dimexpr 0.4116\linewidth-6pt\relax}>{\raggedright\arraybackslash}p{\dimexpr 0.5884\linewidth-6pt\relax}}
\toprule
\hspace{0pt}\textbf{Test} & \hspace{0pt}\textbf{Result} \\
\midrule
\endhead
\hspace{0pt}1. Provable asymptotic gain & \hspace{0pt}Against unstructured Taylor/Dyson series without amplification, a weak baseline: exponential in $\alpha_{L}\Delta t$ (Proposition S4). Against LCHS-class methods: comparable per-run cost (Proposition S5) \\
\hspace{0pt}2. Provable conditional query-count gain & \hspace{0pt}Against LCHS: the published per-run cost $(\allowbreak \alpha_{L}K + \alpha_{H})T + \log(\allowbreak 1/\allowbreak \varepsilon)$ is comparable for one step (Proposition S5), and LCHS has no splitting error. Against uniform singular-value amplification of a series encoding the split avoids the amplification factor; this was not quantified \\
\hspace{0pt}3. Normalization or polynomial approximation & \hspace{0pt}Proved for the constructions, relative to non-unitary encodings of the unsplit step; their use in the integrator is prospective (§4.6). The dominant unitary factor has unit normalization and state-independent success (Proposition 1; shared with the transport factor of a projected formulation). Diffusion succeeds with its physical retention (Proposition 2). Stiff diffusion has square-root degree (§9.7; immaterial at production) \\
\hspace{0pt}4. Implementation difference & \hspace{0pt}No quadrature or history register, constant ancillas, data dependence confined to one unitary factor, constant diffusion encoding and phases (§9.7); a difference of implementation, not of query count \\
\hspace{0pt}5. Stability or modularity (not a complexity statement) & \hspace{0pt}Every exact factor of a frozen step is a contraction for any $\Delta t$ (no stability limit for the frozen factors; not a stability proof of the nonlinear scheme); errors accumulate linearly (Proposition 8); the second-order splitting bound has no growth factor and explicit constants (Proposition S3; realized $1.9\text{--}6.7 \times 10^{-6}$ per step) \\
\bottomrule
\end{longtable}}

Exploiting the structure of $A$ and $D$ therefore makes eigenvalue-based QSP applicable at all (Lemma S3, §9.2) and gives unit-normalized and physical-retention constructions whose success probability is provably higher than that of unamplified series encodings of the unsplit step, a weak baseline (Proposition S4; their use is prospective); its implementation differs from LCHS-class methods (no quadrature register). Its query cost for one step is comparable with that of the unsplit method assessed here (LCHS; Proposition S5), and it adds a commutator-controlled error dominated by the acoustic cross term ($\propto 1/\mathrm{Ma}$; realized $1.9\text{--}6.7 \times 10^{-6}$ per step on the production states).

\medskip
\subsection{Per-step re-linearization}\label{si:C.10}
\subsubsection{Comparison with approaches that keep the nonlinearity on the quantum side}\label{si:C.10.1}
Symbols inside the rows of published methods follow their sources (e.g. $N$ the Carleman truncation order, $n$ the ODE dimension, $D$ the spatial dimension).

{\scriptsize
\setlength{\tabcolsep}{3pt}\renewcommand{\arraystretch}{1.15}\setlength{\LTleft}{0pt}\setlength{\LTright}{0pt}
\begin{longtable}{>{\raggedright\arraybackslash}p{\dimexpr 0.2276\linewidth-6pt\relax}>{\raggedright\arraybackslash}p{\dimexpr 0.1325\linewidth-6pt\relax}>{\raggedright\arraybackslash}p{\dimexpr 0.1844\linewidth-6pt\relax}>{\raggedright\arraybackslash}p{\dimexpr 0.2276\linewidth-6pt\relax}>{\raggedright\arraybackslash}p{\dimexpr 0.1177\linewidth-6pt\relax}>{\raggedright\arraybackslash}p{\dimexpr 0.1102\linewidth-6pt\relax}}
\toprule
\hspace{0pt}\textbf{Approach} & \hspace{0pt}\textbf{Quantum state dimension} & \hspace{0pt}\textbf{Condition for convergence / efficiency} & \hspace{0pt}\textbf{Cost (as stated)} & \hspace{0pt}\textbf{Output} & \hspace{0pt}\textbf{Per-step readout} \\
\midrule
\endhead
\hspace{0pt}\textbf{Per-step re-linearization (this integrator)} & \hspace{0pt}$4\mathcal{N}$ ($n = 26$ qubits at $256^{3}$) for a given grid; constant ancillas (implemented encodings) & \hspace{0pt}none on nonlinearity strength; second order in $\Delta t$ ($16^{3}$ test) & \hspace{0pt}per step at fixed $\Delta t$ and $\eta$: $\Theta(\allowbreak \tau_{A} + \log(\allowbreak 1/\allowbreak \eta))$ queries (L1); $\Omega(\mathcal{N}/\log n)$ loading gates per query (L2) & \hspace{0pt}classical field every step & \hspace{0pt}\textbf{yes} \\
\hspace{0pt}Carleman (Liu et al.~\cite{Liu2021}, Thm 1) & \hspace{0pt}$\Delta = O(n^{N})$ at truncation order $N$ (Eq. 3.6) & \hspace{0pt}$R < 1$ and $\|F_{0}\| \leq \|F_{2}\|$ (arXiv v4), sufficient for efficiency (Thm 1); worst case exponential in $T$ for $R \geq \sqrt{2}$ (Thm 2) and for $R \geq 1$ (Lewis et al.~\cite{Lewis2024}, Thm 3) & \hspace{0pt}$sT^{2}q[\allowbreak \dots]/\allowbreak (\allowbreak g\varepsilon) \cdot\allowbreak  \mathrm{poly}(\allowbreak \log)$ & \hspace{0pt}normalized state at $T$ & \hspace{0pt}no (readout not addressed; their §7) \\
\hspace{0pt}Carleman, improved (Krovi~\cite{Krovi2023}, Thm 8; Costa et al.~\cite{CostaSchleich2025}) & \hspace{0pt}$O(d^{N})$ & \hspace{0pt}$R < 1$ with log-norm; time-independent $F_{0}$, $F_{1}$, $F_{2}$ & \hspace{0pt}$O(\allowbreak gT\|A\|\, \mathrm{poly}(\allowbreak N,\allowbreak  \dots))$ & \hspace{0pt}state & \hspace{0pt}no \\
\hspace{0pt}LBM–Carleman, end-to-end (Jennings et al.~\cite{Jennings2026}) & \hspace{0pt}Carleman of the LBE & \hspace{0pt}$\mathrm{Re} \leq \mathrm{Re}_{T}$ (numerical threshold) & \hspace{0pt}advantage bounded by $O(\mathrm{Re}^{3D/8})$ (Eq. 183); $O(\allowbreak \mathrm{Re}^{\beta(D - 1)/2})$ with drag readout & \hspace{0pt}scalar observable (drag) & \hspace{0pt}no \\
\hspace{0pt}LBM–Carleman (Li et al.~\cite{LiYin2025}) & \hspace{0pt}$n^{3}Q^{3} + n^{2}Q^{2} + nQ$ ($k = 3$) & \hspace{0pt}$\mathrm{Ma} \ll 1$; $\kappa_{J} \leq 1.152 \times\allowbreak  10^{5}$ & \hspace{0pt}$O(\allowbreak t_{c}^{-1}\tilde{T}\, \mathrm{polylog})$ (contested: $\|C\| \propto \Delta x^{-1}$; Jennings et al.~\cite{Jennings2026}, §IV) & \hspace{0pt}state & \hspace{0pt}no \\
\hspace{0pt}KvN / Liouville (Joseph~\cite{Joseph2020}; Jin \& Liu~\cite{JinLiu2024}) & \hspace{0pt}phase-space grid (exponential in the number of variables for a field) & \hspace{0pt}— & \hspace{0pt}— & \hspace{0pt}observables & \hspace{0pt}no \\
\hspace{0pt}Mean-field / copies (Lloyd et al.~\cite{Lloyd2020}; Leyton \& Osborne~\cite{Leyton2008}) & \hspace{0pt}$n$ per copy, many copies & \hspace{0pt}stability-limited times & \hspace{0pt}copies $(16/\varepsilon^{2})^{m}$ for $m$ steps (Leyton \& Osborne~\cite{Leyton2008}) & \hspace{0pt}state & \hspace{0pt}no \\
\hspace{0pt}Hybrid per-step (Gaitan~\cite{Gaitan2020}; Lapworth~\cite{Lapworth2022}; Bharadwaj \& Sreenivasan~\cite{Bharadwaj2023}; Inger \& Frankel~\cite{IngerFrankel2026}; Akash et al.~\cite{Akash2026}; Kim et al.~\cite{Kim2025}) & \hspace{0pt}$n$ & \hspace{0pt}— & \hspace{0pt}per step $O(\mathcal{N})$-type readout (§6.3) & \hspace{0pt}classical field & \hspace{0pt}yes \\
\bottomrule
\end{longtable}}

\subsubsection{What re-linearization avoids and what it pays}\label{si:C.10.2}
\textbf{Avoids} (each statement is from the cited paper; its application to the integrator is our reading).

\begin{enumerate}
\item\relax \textbf{State-space enlargement:} $O(n^{N})$ Carleman dimension; for a given grid the integrator stays at $4\mathcal{N}$ amplitudes (26 qubits at $256^{3}$), independent of nonlinearity strength.
\item\relax \textbf{Truncation order and $R$-conditions:} $R < 1$ is a sufficient condition for Carleman efficiency (Liu et al.~\cite{Liu2021}, Thm 1), with worst-case hardness for $R \geq \sqrt{2}$ (Thm 2) and for $R \geq 1$ (Lewis et al.~\cite{Lewis2024}, Thm 3). Turbulence has $R \gg 1$, and for a chaotic initial condition the Carleman error ``does not decrease by increasing Carleman truncation number'' (Lewis et al.~\cite{Lewis2024}, §4). The integrator has no truncation order.
\item\relax \textbf{Time dependence:} Carleman/QLSA solvers assume fixed $F_{1}$, $F_{2}$ (``not obvious how to implement'' time-dependent coefficients; Liu et al.~\cite{Liu2021}, §7). The integrator freezes the coefficients per step, so every step is a time-independent linear problem (an exponential-midpoint/Magnus-type step; Hochbruck \& Ostermann~\cite{Hochbruck2010}, §2.8, §3.1), with no Dyson series or time-ordering register.
\item\relax \textbf{The scope of the coherent lower bounds.} Lewis et al.~\cite{Lewis2024} (Thm 5) prove that producing the normalized solution state of a chaotic system has complexity exponential in time; they state that this ``does not preclude potential for polynomial speedup'', and they exclude ``reducing the dynamics to linear with classical preprocessing'' (§5). Ameri et al.~\cite{Ameri2026} (Thms 24–25) prove that $e^{\Omega(T)}$ copies of the initial state are needed for 2D Euler at error $\leq e^{-CT}$, with no bound in $\mathcal{N}$ or Re. These bounds do not apply to the integrator because it re-prepares the state from classical data every step and reads it out in full, at the price of Proposition 7. The reason is not its $T$-dependence. At fixed $\Delta t$ and polynomial tolerance the query count grows linearly in $T$, but at a fixed final accuracy $\Delta t$ and the tolerances must shrink with the flow's error transfer (§7), as for any solver of a chaotic flow.
\end{enumerate}
\textbf{Pays.}

\begin{enumerate}
\item\relax A full-state readout and a state preparation every step: at least $\approx (1 - P_{\mathrm{err}})\, 4\mathcal{N}(b - 1)/n$ copies ($\approx 4\mathcal{N}b/n$ at leading order; Proposition 6), each with a fresh preparation of $\Omega(\mathcal{N}/\log \mathcal{N})$ gates for worst-case or typical states, i.e. $\Omega(\mathcal{N}^{2}/\log^{2}\mathcal{N})$ gates per step (Proposition 7).
\item\relax A classical refresh and predictor: $\Theta(\mathcal{N})$ and $\Theta(\mathcal{N} d_{\mathrm{pred}})$ per step (§6.2). In the words of Lewis et al.~\cite{Lewis2024} (§5), ``All the complexity of the problem would therefore be contained in the map.''
\item\relax Data loading inside every query for an arbitrary advecting field: $\Omega(\mathcal{N} b/\log n)$ gates for a finite gate set, $\Omega(\mathcal{N}/\log \mathcal{N})$ for Lipschitz-parameterized gates (Proposition 5, compiled-circuit model).
\item\relax A re-linearization error, $O(\Delta t^{2})$ globally with the midpoint predictor (shown at $16^{3}$ only), transported by the flow (§7).
\end{enumerate}
\subsubsection{Error control}\label{si:C.10.3}
Carleman truncation errors grow like $R^{N + 1 - j}$ (Liu et al.~\cite{Liu2021}, Cor. 1; $N$ the truncation order) and have no convergence guarantee for $R \geq 1$. The integrator's per-step errors are polynomial ($\eta$), splitting (Proposition S3; realized $1.9\text{--}6.7 \times 10^{-6}$ per step) and re-linearization ($O(\Delta t^{3})$ local). It has no truncation order, so there is no method-specific non-convergence of the Carleman kind. These errors are transported by the flow; their transport, and what the noise study measures of it, are discussed in §7.

\subsubsection{Summary}\label{si:C.10.4}
For a given grid, per-step classical re-linearization keeps the quantum object at fixed size ($\log_{2}(4\mathcal{N})$ system qubits and a constant number of ancillas for the implemented encodings), turns every step into a time-independent linear problem whose query cost is set by the frozen operator's normalization alone ($\Theta(\tau_{A} + \log(1/\eta))$ per convection factor), and needs no truncation order and no condition on nonlinearity strength. At fixed $\Delta t$ and $\eta$ the query count grows linearly in $T$; at fixed final accuracy it grows faster, with the flow's error transfer, as for any solver of a chaotic flow.

The alternatives have known limits. Carleman-type embeddings have efficiency guarantees only for $R < 1$ (or below an empirical $\mathrm{Re}_{T}$), with a truncation-dependent dimension. Coherent multi-step algorithms face exponential-in-$T$ lower bounds for chaotic or Euler dynamics, and those bounds do not cover the integrator because it re-prepares from classical data and reads out every step. The price is paid every step: a classical refresh, a classical predictor and a full-state readout. With arbitrary state preparation the per-step gate count is $\Omega(\mathcal{N}^{2}/\log^{2}\mathcal{N})$ (Proposition 7; §10.2).

\subsection{Overall complexity bounds}\label{si:C.11}
All quantities are at $\eta = 10^{-12}$ unless stated; levels L1–L3 as in §0. Production values are those of the reported configuration; prospective values are derived and not implemented (§4.6).

{\scriptsize
\setlength{\tabcolsep}{3pt}\renewcommand{\arraystretch}{1.15}\setlength{\LTleft}{0pt}\setlength{\LTright}{0pt}
\begin{longtable}{>{\raggedright\arraybackslash}p{\dimexpr 0.0400\linewidth-6pt\relax}>{\raggedright\arraybackslash}p{\dimexpr 0.0700\linewidth-6pt\relax}>{\raggedright\arraybackslash}p{\dimexpr 0.1700\linewidth-6pt\relax}>{\raggedright\arraybackslash}p{\dimexpr 0.3600\linewidth-6pt\relax}>{\raggedright\arraybackslash}p{\dimexpr 0.3600\linewidth-6pt\relax}}
\toprule
\hspace{0pt}\textbf{\#} & \hspace{0pt}\textbf{Level} & \hspace{0pt}\textbf{Quantity} & \hspace{0pt}\textbf{Formula} & \hspace{0pt}\textbf{Values} \\
\midrule
\endhead
\hspace{0pt}1 & \hspace{0pt}L1 & \hspace{0pt}Queries, convection factor & \hspace{0pt}black-box: $\Theta(\allowbreak \tau_{A} + \log(\allowbreak 1/\allowbreak \eta)/\allowbreak \log(\allowbreak e + \log(\allowbreak 1/\allowbreak \eta)/\allowbreak \tau_{A}))$ (GSLW~\cite{Gilyen2019}, Cor. 60); Proposition 1: $\leq 2d(\tau_{A}, \eta)$; directional control $K + 3$, $K \geq d$ even & \hspace{0pt}production 78/79; prospective 62–64 (Laurent) or $\approx 35$ (directional, $K = 32$); reference degree $d = 31\text{--}32$ \\
\hspace{0pt}2 & \hspace{0pt}L1 & \hspace{0pt}Queries, diffusion half-step & \hspace{0pt}$d_{D}(\tau_{D}/2, \eta)$ (Chebyshev form, Proposition 2) to $2d_{D}$ (Laurent); $\sqrt{\max(\tau_{D}, \log)\log}$ for stiff diffusion & \hspace{0pt}production 53; prospective 4–5 (Chebyshev form) or 8–10 (Laurent) \\
\hspace{0pt}3 & \hspace{0pt}L1 & \hspace{0pt}Queries per step & \hspace{0pt}$Q_{\mathrm{step}} = Q_{A} + 2Q_{D}$ & \hspace{0pt}\textbf{production 184–185 (reported)}; prospective 43–45 (directional + Proposition 2), 70–72 (Proposition 1 + Proposition 2), 78–84 (all Laurent) \\
\hspace{0pt}4 & \hspace{0pt}L1 & \hspace{0pt}Queries per trajectory & \hspace{0pt}bound for the Proposition 1–2 constructions: $Q_{\mathrm{traj}} \leq e\, \bar{\lambda}_{A} T + 2M\ln(\allowbreak 1/\allowbreak \eta) + 2M d_{D} + O(\allowbreak M) \approx 1.5\text{--}1.6 \times\allowbreak  10^{5}$ (Laurent); about half of the first two terms with directional control & \hspace{0pt}production 304,252, $\approx 2$ times the bound; prospective $\approx 7.1\text{--}7.4 \times\allowbreak  10^{4}$ (directional) to $\leq 1.39 \times 10^{5}$ (all Laurent) \\
\hspace{0pt}5 & \hspace{0pt}L1 & \hspace{0pt}Dependence on $\Delta t$ at fixed $T$ & \hspace{0pt}stiffness term $\Delta t$-independent; additive term $\propto M = T/\Delta t$ & \hspace{0pt}prospective, $m = 2$, 4, 8 times $\Delta t$: $4.95 \times 10^{4}$, $3.59 \times 10^{4}$, $2.87 \times 10^{4}$ queries (directional, convention $K = d + 3$; Appendix A) and $M/m$ readouts. Context: $\Delta t/\tau_{\eta} = 0.10$, 0.20, 0.40 ($\tau_{\eta} = 0.0399$ at the dissipation peak); code CFL 1.78, 3.55, 7.11; acoustic Courant number 3.3, 6.5, 13.0. Only $m = 2$ is a plausible candidate without a time-step study; accuracy not checked \\
\hspace{0pt}6 & \hspace{0pt}L1/L2 & \hspace{0pt}Post-selection success per step & \hspace{0pt}prospective: $\|S(\allowbreak \Delta t)\psi\|^{2}/\allowbreak \|\psi\|^{2}$ up to $O(\eta)$ and the synthesis margins; design: $(1/s)^{6}$ with scale $s$ per factor & \hspace{0pt}production 1/64 (design; 1/4 per factor; computed exactly, never sampled); prospective $\approx 0.9987$ (margins $\to 0$) or $\approx 0.96$ ($\delta = 10^{-2}$ on diffusion) \\
\hspace{0pt}7 & \hspace{0pt}L1/L2 & \hspace{0pt}Expected corrector queries per accepted output copy & \hspace{0pt}$E = (\allowbreak Q_{D} + p_{D} Q_{A} + p_{D} p_{A} Q_{D})/\allowbreak (\allowbreak p_{D} p_{A} p_{D})$ under S1–S4 of §4.5 (staged post-selection, early abort, restart from state preparation; corrector queries only); $Q_{\mathrm{step}}/\allowbreak p_{\mathrm{step}}$ if all post-selections are read at the end & \hspace{0pt}production \textbf{4,852–4,868} (design 1/4 per factor; 4,857–4,873 with the measured mean diffusion success) or 11,776–11,840 (read at the end); prospective $\approx 44\text{--}46$ (early abort) or $\approx 45\text{--}47$ (read at the end) \\
\hspace{0pt}8 & \hspace{0pt}L2 & \hspace{0pt}Gates per query & \hspace{0pt}$C_{\mathrm{BE},D} = O(\allowbreak \mathrm{polylog}\, \mathcal{N})$, data-free, for an ancilla-assisted synthesis of the multi-controlled shifts (the compiled emission grows irregularly between $8^{3}$ and $16^{3}$); $C_{\mathrm{BE},A} = C_{\mathrm{shift}} + C_{\mathrm{load}}$, with $C_{\mathrm{load}} = \Omega(\allowbreak \mathcal{N} b/\allowbreak \log n)$ for a finite gate set and $\Omega(\allowbreak \mathcal{N}/\allowbreak \log \mathcal{N})$ for Lipschitz-parameterized gates, for arbitrary fields in the compiled-circuit model (Proposition 5) & \hspace{0pt}$\approx 6.3 \times 10^{4}$ CX (diffusion walk); $\approx 2.78 \times 10^{10}$ CX (collocated convection walk, $6 \cdot 276 \cdot \mathcal{N}$; unoptimized upper estimate) \\
\hspace{0pt}9 & \hspace{0pt}L2 & \hspace{0pt}Gates per corrector run & \hspace{0pt}$G_{\mathrm{step}} \approx Q_{A}\, C_{\mathrm{BE},A} + 2Q_{D}\, C_{\mathrm{BE},D} = \Omega(\allowbreak Q_{A}\mathcal{N}/\allowbreak \log n)$ & \hspace{0pt}upper estimates: $\approx 2.2 \times 10^{12}$ CX at production counts (collocated census; excludes state preparation, readout, repetition for post-selection and error correction); $\approx 2.7 \times 10^{12}$ for the convection factor alone with the constructed staggered encoding, extrapolated from compiled $2^{3}\text{--}8^{3}$ circuits (§5.5); $\approx 9.7 \times 10^{11}$ (directional, prospective) \\
\hspace{0pt}10 & \hspace{0pt}L2 & \hspace{0pt}Expected gates per accepted copy, re-preparation included & \hspace{0pt}$E_{\mathrm{copy}} = (\allowbreak C_{\mathrm{prep}} + G_{D} + p_{D} G_{A} + p_{D} p_{A} G_{D})/\allowbreak (\allowbreak p_{D} p_{A} p_{D})$ & \hspace{0pt}$\approx 64\, C_{\mathrm{prep}} + 16\, G_{A} \approx 3.5 \times\allowbreak  10^{13}$ CX at production (census estimate; the preparations, $\approx 64 \times\allowbreak  3.4 \times\allowbreak  10^{7}$ CNOTs at the leading-order bound, are $\approx 10^{-4}$ of it) \\
\hspace{0pt}11 & \hspace{0pt}L3 & \hspace{0pt}Classical work per step & \hspace{0pt}$C_{\mathrm{pred}} + 2C_{\mathrm{refresh}} + 3C_{\mathrm{synth}} + O(\allowbreak \mathcal{N}) = \Theta(\allowbreak \mathcal{N} d_{\mathrm{pred}})$ & \hspace{0pt}96 stencil applications ($\approx 2.4 \times 10^{4}$ flops per cell), reducible to 17–27; $\approx 1.3\text{--}15$ times a pseudo-spectral RK3 step ($\approx 1.6 \times 10^{3}$ flops per cell) \\
\hspace{0pt}12 & \hspace{0pt}L3 & \hspace{0pt}State preparation per copy & \hspace{0pt}$< (23/24)\, 2^{n}$ CNOTs (upper); $(1/2)\, 2^{n} = 2^{n - 1}$ in leading order (lower bound, exact preparation of generic states) & \hspace{0pt}$< 6.4 \times 10^{7}$; $\approx 3.4 \times 10^{7}$ (leading order) \\
\hspace{0pt}13 & \hspace{0pt}L3 & \hspace{0pt}Readout per step (interface I1) & \hspace{0pt}$kn \geq (\allowbreak 1 - P_{\mathrm{err}})\log_{2}K - 1$, $\log_{2}K \geq (\allowbreak b - 1)(\allowbreak 4\mathcal{N} - 1)$ (Holevo and Fano); $\tilde{\Theta}(\allowbreak 4\mathcal{N}/\allowbreak \varepsilon_{r})$ uses of the preparation unitary; $\tilde{O}(\allowbreak 4\mathcal{N}/\allowbreak \varepsilon_{r}^{2})$ copies (sampling) & \hspace{0pt}conditional on I1–I3: $\geq 5.3 \times 10^{7}$ copies ($b = 32$, $P_{\mathrm{err}} = 1/3$; leading order $4\mathcal{N}b/\allowbreak n = 8.3 \times\allowbreak  10^{7}$); $6.7 \times\allowbreak  10^{9}\text{--}6.7 \times\allowbreak  10^{11}$ uses; $6.7 \times\allowbreak  10^{11}\text{--}6.7 \times\allowbreak  10^{15}$ copies ($\varepsilon_{r} = 10^{-2}\text{--}10^{-4}$) \\
\hspace{0pt}14 & \hspace{0pt}L3 & \hspace{0pt}End-to-end per step & \hspace{0pt}$C_{\mathrm{cl}} + R(\allowbreak \varepsilon_{r})\, E_{\mathrm{copy}} \geq R\, C_{\mathrm{prep}} = \Omega(\allowbreak \mathcal{N}^{2}/\allowbreak \log^{2}\mathcal{N})$ gates (fixed $b$) under I1–I3 and K1–K2 (Proposition 7) & \hspace{0pt}lower bound at leading order, with exact preparation of generic states: $\approx 1.8 \times 10^{15}$ CNOTs (one-qubit gates free); census estimate $\approx 2.9 \times 10^{21}$ CX at the production design (not a bound); against $\approx 2.7 \times 10^{10}$ flops for a pseudo-spectral RK3 step (an estimate) \\
\hspace{0pt}15 & \hspace{0pt}all & \hspace{0pt}Accuracy composition (within the predictability horizon) & \hspace{0pt}relative error at $T \lesssim G(\allowbreak T)\sum_{n}(\allowbreak 3\eta + \varepsilon_{\mathrm{split},n} + \varepsilon_{\mathrm{lin},n} + \varepsilon_{r,n}) + O(\allowbreak \mathrm{Ma}^{2})$, with $G(T)$ the symbolic error-transfer factor of §2 (not measured; not the noise study's $\delta/\varepsilon$) & \hspace{0pt}$\varepsilon_{\mathrm{split}}$: worst case $\approx 4.3 \times 10^{-4}$ per step (Lanczos estimate of the Proposition S3 bound); realized $1.9\text{--}6.7 \times 10^{-6}$ per step (§9.3.1); $\varepsilon_{\mathrm{lin}}$, $\varepsilon_{r}$ and $G(T)$ not measured \\
\hspace{0pt}16 & \hspace{0pt}all & \hspace{0pt}Parameter choice (symbolic accuracy model) & \hspace{0pt}$\eta \propto \varepsilon_{\mathrm{acc}}/\allowbreak (\allowbreak MG)$; $\Delta t \propto (\allowbreak \varepsilon_{\mathrm{acc}}/\allowbreak (\allowbreak C_{2}TG))^{1/2}$; readout cost $\propto \mathcal{N} M^{3}G^{2}/\allowbreak \varepsilon_{\mathrm{acc}}^{2}$; $\mathrm{Ma} \propto \varepsilon_{\mathrm{acc}}^{1/2}$ & \hspace{0pt}— \\
\bottomrule
\end{longtable}}

\textbf{Optimization over the algorithmic parameters} (symbolic; §7 and Proposition 4). The query count and every interface term are minimized by the largest accuracy-admissible $\Delta t$; $\eta$ enters only logarithmically; and the normalization should be 1 per factor, up to a synthesis margin, which would turn 64 expected trials per step into $\approx 1.04$ (prospective). This optimization is conditional on a time-step study at production scale, which does not exist (§9.5). The production configuration is a non-optimized point of this design space: fixed $\Delta t$, fixed Chebyshev degree 48, $K = 48$, scale 2.

\medskip
\subsection{Published baselines}\label{si:C.12}
Each entry gives the result used, the access and output models, and the cost level (§0) at which it can be compared with the integrator. Symbols inside the entries follow the sources. Theorem, lemma, corollary, section and equation numbers refer to the version of each work that was read. GSLW numbers are those of arXiv:1806.01838v1, whose numbering differs from the STOC proceedings, and Krovi's those of arXiv:2202.01054v5, a 2025 revision (the journal version corresponds to v3). For Motlagh \& Wiebe (arXiv:2308.01501v2), Jin, Liu \& Yu (arXiv:2212.13969v1), Costa et al. 2022 (arXiv:2111.08152v1) and Linden, Montanaro \& Shao (arXiv:2004.06516v2) the numbering of the published version was not compared.

{\scriptsize
\setlength{\tabcolsep}{3pt}\renewcommand{\arraystretch}{1.15}\setlength{\LTleft}{0pt}\setlength{\LTright}{0pt}
\begin{longtable}{>{\raggedright\arraybackslash}p{\dimexpr 0.0417\linewidth-6pt\relax}>{\raggedright\arraybackslash}p{\dimexpr 0.2116\linewidth-6pt\relax}>{\raggedright\arraybackslash}p{\dimexpr 0.1944\linewidth-6pt\relax}>{\raggedright\arraybackslash}p{\dimexpr 0.2116\linewidth-6pt\relax}>{\raggedright\arraybackslash}p{\dimexpr 0.2095\linewidth-6pt\relax}>{\raggedright\arraybackslash}p{\dimexpr 0.1311\linewidth-6pt\relax}}
\toprule
\hspace{0pt}\textbf{ID} & \hspace{0pt}\textbf{Paper} & \hspace{0pt}\textbf{Problem / access / output} & \hspace{0pt}\textbf{Result used (as stated)} & \hspace{0pt}\textbf{Conditions} & \hspace{0pt}\textbf{Comparable level} \\
\midrule
\endhead
\hspace{0pt}B1 & \hspace{0pt}Low \& Chuang, Quantum 3, 163 (2019)~\cite{LowChuang2019} & \hspace{0pt}Hamiltonian simulation; block encoding; unitary & \hspace{0pt}$\Theta(Q)$, $Q = O(\allowbreak \alpha t + \log(\allowbreak 1/\allowbreak \varepsilon))$, Thm 1, Cor. 16; a self-inverse block encoding qubitizes, Cor. 9 & \hspace{0pt}block encoding with normalization $\alpha$ & \hspace{0pt}L1, convection factor \\
\hspace{0pt}B2 & \hspace{0pt}Gilyén, Su, Low, Wiebe~\cite{Gilyen2019}, STOC 2019 (arXiv:1806.01838v1) & \hspace{0pt}QSVT & \hspace{0pt}$\Theta(\allowbreak \alpha|t| + \log(\allowbreak 1/\allowbreak \varepsilon)/\allowbreak \log(\allowbreak e + \log(\allowbreak 1/\allowbreak \varepsilon)/\allowbreak (\allowbreak \alpha|t|)))$ necessary and sufficient (Cor. 60); $6\alpha|t| + 9\ln(\allowbreak 12/\allowbreak \varepsilon)$ explicit (Cor. 62); heat-kernel degree (Cor. 64); threshold projectors $O(\log(1/\varepsilon)/\delta)$ (Thm 31) & \hspace{0pt}unknown $H$ & \hspace{0pt}L1 (both factors; projector of §8) \\
\hspace{0pt}B3 & \hspace{0pt}Motlagh \& Wiebe, PRX Quantum 5, 020368 (2024)~\cite{Motlagh2024}; Berry, Motlagh, Pantaleoni, Wiebe, PRA 110, 012612 (2024)~\cite{Berry2024} & \hspace{0pt}GQSP & \hspace{0pt}any $P$ with $|P| \leq 1$ on the circle, degree $=$ queries (Thms 3–4); Laurent polynomials via the inverse (Thm 6); directional control halves the controlled operations & \hspace{0pt}— & \hspace{0pt}L1, Propositions 1–2 \\
\hspace{0pt}B4 & \hspace{0pt}An, Liu, Lin, PRL 131, 150603 (2023)~\cite{AnLiuLin2023}; An, Childs, Lin, CMP 407, 19 (2026)~\cite{AnChildsLin2026} & \hspace{0pt}non-unitary linear ODE, $L \succeq 0$; state output & \hspace{0pt}An, Childs \& Lin: $\tilde{O}(\allowbreak (\allowbreak \|u_{0}\|/\allowbreak \|u(\allowbreak T)\|)\alpha T(\allowbreak \log 1/\allowbreak \varepsilon)^{1/\beta})$ (Cor. 17; the original LCHS of An, Liu \& Lin has a polynomial dependence on $1/\varepsilon$); per-run cost $(\allowbreak \alpha_{L}K + \alpha_{H})T + \log(\allowbreak 1/\allowbreak \varepsilon)$ in its proof (arXiv:2312.03916v2, §4.2.1); $O(\|u_{0}\|/\|u(T)\|)$ state preparations & \hspace{0pt}linear, oracle-given generator & \hspace{0pt}L1, one frozen step (Proposition S5: comparable) \\
\hspace{0pt}B5 & \hspace{0pt}Jin, Liu, Yu, PRL 133, 230602 (2024)~\cite{JinLiuYu2024} & \hspace{0pt}Schrödingerisation; state & \hspace{0pt}Thm 4 (general $A = H + i\bar{H}$): $\tilde{O}(\allowbreak (\allowbreak \|u(\allowbreak 0)\|/\allowbreak \|u(\allowbreak t)\|)\, s\, t\, \|H_{\mathrm{total}}\|_{\max})$, $\|H_{\mathrm{total}}\|_{\max} \sim \max(\allowbreak \|H\|_{\max}/\allowbreak \varepsilon,\allowbreak  \|\bar{H}\|_{\max})$; Thm 3 (heat) $\tilde{O}(\dots/\varepsilon)$ & \hspace{0pt}first order in $p$ & \hspace{0pt}L1 \\
\hspace{0pt}B6 & \hspace{0pt}Fang, Lin, Tong, Quantum 7, 955 (2023)~\cite{Fang2023} & \hspace{0pt}time-marching with per-step block encodings & \hspace{0pt}$O(\allowbreak \alpha^{2}T^{2}Q\log(\allowbreak \alpha TQ)\log(\allowbreak \alpha TQ/\allowbreak \varepsilon)/\allowbreak \mathrm{loglog})$ (Thm 8); $4^{L}$ decay with a factor-2 subnormalization (Sec. 1.3); uniform amplification (Lemma 2) & \hspace{0pt}linear & \hspace{0pt}L1/L2: per-step normalization (§5.3) \\
\hspace{0pt}B7 & \hspace{0pt}Berry, Childs, Ostrander, Wang, CMP 356, 1057 (2017)~\cite{BerryChildsOstrander2017}; Krovi, Quantum 7, 913 (2023)~\cite{Krovi2023}; Berry \& Costa, Quantum 8, 1369 (2024)~\cite{BerryCosta2024} & \hspace{0pt}linear ODE via QLSA; state & \hspace{0pt}$\kappa_{V}$, $g$, $C(A)$, $R$ factors (Thm 9; Thm 7; Thm 4.1) & \hspace{0pt}linear, oracle-given & \hspace{0pt}L1 (one step: overkill) \\
\hspace{0pt}B8 & \hspace{0pt}Costa et al., PRX Quantum 3, 040303 (2022)~\cite{Costa2022} & \hspace{0pt}QLSP & \hspace{0pt}$O(\kappa\log(1/\varepsilon))$ (Thm 19) & \hspace{0pt}$\|A\| = 1$ & \hspace{0pt}projector baseline (§8.4) \\
\hspace{0pt}B9 & \hspace{0pt}Liu et al., PNAS 118, e2026805118 (2021)~\cite{Liu2021}; Krovi~\cite{Krovi2023} 2023; Costa et al., npj QI 11, 141 (2025)~\cite{CostaSchleich2025} & \hspace{0pt}dissipative quadratic ODE, Carleman; state & \hspace{0pt}$sT^{2}q[\allowbreak \dots]/\allowbreak (\allowbreak g\varepsilon) \cdot\allowbreak  \mathrm{poly}(\allowbreak \log)$ (Thm 1); dimension $O(n^{N})$ at truncation order $N$ & \hspace{0pt}$R < 1$ sufficient; worst-case hard for $R \geq \sqrt{2}$ (Thm 2) & \hspace{0pt}L3 (coherent variant only) \\
\hspace{0pt}B10 & \hspace{0pt}Lewis, Eidenbenz, Nadiga, Subaşı, Quantum 8, 1509 (2024)~\cite{Lewis2024} & \hspace{0pt}lower bound, chaos; normalized-state output & \hspace{0pt}``complexity that scales exponentially in time'' (Thm 5); $R \geq 1$ hardness (Thm 3); ``does not preclude potential for polynomial speedup'' (§5) & \hspace{0pt}excludes classical-preprocessing linearizations (§5) & \hspace{0pt}scope statement (§10) \\
\hspace{0pt}B11 & \hspace{0pt}Ameri, Carolan, Childs, Krovi~\cite{Ameri2026}, arXiv:2603.12161 & \hspace{0pt}lower bound, KdV and 2D Euler; copies of the initial state & \hspace{0pt}$\Omega(T^{2})$ (KdV); $e^{\Omega(T)}$ at error $\leq e^{-CT}$ (Euler, Thms 24–25) & \hspace{0pt}no bound in $\mathcal{N}$ or Re & \hspace{0pt}scope statement \\
\hspace{0pt}B12 & \hspace{0pt}Jennings et al., PRX Quantum 7, 033060 (2026)~\cite{Jennings2026} & \hspace{0pt}LBM–Carleman, end-to-end, drag & \hspace{0pt}advantage $\leq O(\mathrm{Re}^{3D/8})$ (Eq. 183) & \hspace{0pt}$\mathrm{Re} \leq \mathrm{Re}_{T}$; steady target & \hspace{0pt}L3 (observable output) \\
\hspace{0pt}B13 & \hspace{0pt}Li et al., PRR 7, 013036 (2025)~\cite{LiYin2025} & \hspace{0pt}LBM–Carleman & \hspace{0pt}$O(\allowbreak t_{c}^{-1}\tilde{T}\, \mathrm{polylog}(\allowbreak n/\allowbreak \varepsilon))$ (Eq. 120), contested & \hspace{0pt}$\mathrm{Ma} \ll 1$ & \hspace{0pt}L3 (coherent) \\
\hspace{0pt}B14 & \hspace{0pt}Over et al., PRA 112, L010401 (2025)~\cite{Over2025} & \hspace{0pt}advection–diffusion, $\alpha = 1$ direct encoding; state & \hspace{0pt}$\tilde{O}(\allowbreak T\mathcal{N}^{2/d}d/\allowbreak \varepsilon)$ gates ($d$ the spatial dimension); $O(\allowbreak \|\varphi(\allowbreak 0)\|^{2}/\allowbreak \|\varphi(\allowbreak T)\|^{2})$ attempts & \hspace{0pt}prescribed velocity; forward Euler & \hspace{0pt}L1/L2 per step ($\alpha = 1$ principle) \\
\hspace{0pt}B15 & \hspace{0pt}Brearley \& Laizet, PRA 110, 012430 (2024)~\cite{Brearley2024} & \hspace{0pt}advection by Hamiltonian simulation & \hspace{0pt}depth $\tilde{O}(\allowbreak T\mathcal{N}^{1/D}Dk/\allowbreak \varepsilon)$ ($D$ the spatial dimension) & \hspace{0pt}prescribed velocity & \hspace{0pt}L1/L2 per step \\
\hspace{0pt}B16 & \hspace{0pt}Wu \& Li~\cite{WuLi2026}, arXiv:2605.00794 & \hspace{0pt}Stokes on MAC, Zeno/QSVT projector & \hspace{0pt}$\tilde{O}(t h^{-3})$ direct; projector degree $O(\kappa(D)\log(1/\eta))$ (Thm 4.1) & \hspace{0pt}linear Stokes & \hspace{0pt}§8.4 projector baseline \\
\hspace{0pt}B17 & \hspace{0pt}Jin, Tang, Zhai, Zhang~\cite{Jin2026}, arXiv:2607.00281 & \hspace{0pt}Stokes + penalty AC, Schrödingerisation & \hspace{0pt}$\tilde{O}(\allowbreak d^{3}T^{2}\|u(\allowbreak 0)\|^{4}/\allowbreak (\allowbreak h^{8}\delta^{4}))$ (Thm 4.1) & \hspace{0pt}linear; parabolic pressure & \hspace{0pt}contrast with EDAC \\
\hspace{0pt}B18 & \hspace{0pt}Akash et al.~\cite{Akash2026}, arXiv:2608.27712 & \hspace{0pt}1D skew-symmetric variable-coefficient transport, Strang $D/2$–$A$–$D/2$, per-step re-preparation & \hspace{0pt}$O(\mathcal{N})$ kernel (Eq. 23); end-to-end model with $\geq \mathcal{N}/\allowbreak \varepsilon^{2}$ copies (Eq. 25); failure $\propto$ dissipation (Eq. 26) & \hspace{0pt}1D, static velocity & \hspace{0pt}closest per-step analogue (L2/L3) \\
\hspace{0pt}B19 & \hspace{0pt}Lapworth~\cite{Lapworth2022} 2022; S. S. Bharadwaj \& Sreenivasan~\cite{Bharadwaj2023} 2023; Inger \& Frankel~\cite{IngerFrankel2026} 2026 & \hspace{0pt}hybrid per-step readout & \hspace{0pt}$O(\mathcal{N})$-type readout per step; hybrid 10 times slower than classical (Lapworth~\cite{Lapworth2022}, §6) & \hspace{0pt}— & \hspace{0pt}L3 \\
\hspace{0pt}B20 & \hspace{0pt}Montanaro \& Pallister, PRA 93, 032324 (2016)~\cite{Montanaro2016}; Linden, Montanaro, Shao, CMP 395, 601 (2022)~\cite{Linden2022}; Dalzell et al., CUP (2025)~\cite{Dalzell2025} & \hspace{0pt}equal-output limits in fixed dimension & \hspace{0pt}at most polynomial ($\approx$ quadratic) speedup; the QLSA route is never faster than the best classical method for the heat equation & \hspace{0pt}scalar outputs & \hspace{0pt}L3 context \\
\hspace{0pt}B21 & \hspace{0pt}Classical: projection (Guermond, Minev \& Shen~\cite{Guermond2006} 2006), multigrid (Brandt~\cite{Brandt1977} 1977), FFT pseudo-spectral (Mortensen \& Langtangen~\cite{Mortensen2016} 2016), explicit EDAC (Clausen~\cite{Clausen2013b} SAND2013-2608; Yasuda et al.~\cite{Yasuda2026} ICCFD13), exponential integrators (Hochbruck \& Ostermann~\cite{Hochbruck2010} 2010) & \hspace{0pt}classical fields & \hspace{0pt}§8.3 & \hspace{0pt}— & \hspace{0pt}classical baselines \\
\hspace{0pt}B22 & \hspace{0pt}Jahnke \& Lubich, BIT 40, 735 (2000)~\cite{JahnkeLubich2000} & \hspace{0pt}Strang splitting error for contractive semigroups & \hspace{0pt}$C_{2}\tau^{3}$ local error, principal term $(1/12)c_{2}\tau^{3}$, $C_{2}$ depending on $c_{1}$, $c_{2}$ and $\|B\|$ (Thm 2.1(b), eq. 2.7) & \hspace{0pt}$\|e^{tA}\|$, $\|e^{tB}\|$, $\|e^{t(A + B)}\| \leq 1$ & \hspace{0pt}prior art for Proposition S3 \\
\hspace{0pt}B23 & \hspace{0pt}Schumann \& Sweet, JCP 75, 123 (1988)~\cite{SchumannSweet1988}; Klappenecker \& Rötteler~\cite{Klappenecker2001}, quant-ph/0111038 (ISPA 2001, pp. 464–468) & \hspace{0pt}fast Poisson solvers on staggered grids in rectangular boxes; quantum DCT/DST & \hspace{0pt}DCT/DST diagonalization of the staggered-grid Poisson operator; quantum DCT/DST in $O(\log^{2}\mathcal{N})$ gates & \hspace{0pt}rectangular boxes, Dirichlet/Neumann conditions & \hspace{0pt}§8.2, §8.4 (geometry condition of Proposition E1) \\
\bottomrule
\end{longtable}}

\medskip
\subsection{Comparison and crossover conditions}\label{si:C.13}
\textbf{Comparison table} (one frozen step, then the trajectory). Every row states the kind of the comparison (II, subproblem; III, complete loop; §0), its level, and whether the output and the accuracy target are matched; a comparison that is not matched is not a cost comparison at equal error. The settings of the rows are given in §14.

{\scriptsize
\setlength{\tabcolsep}{3pt}\renewcommand{\arraystretch}{1.15}\setlength{\LTleft}{0pt}\setlength{\LTright}{0pt}
\begin{longtable}{>{\raggedright\arraybackslash}p{\dimexpr 0.1488\linewidth-6pt\relax}>{\raggedright\arraybackslash}p{\dimexpr 0.0986\linewidth-6pt\relax}>{\raggedright\arraybackslash}p{\dimexpr 0.1321\linewidth-6pt\relax}>{\raggedright\arraybackslash}p{\dimexpr 0.1405\linewidth-6pt\relax}>{\raggedright\arraybackslash}p{\dimexpr 0.1321\linewidth-6pt\relax}>{\raggedright\arraybackslash}p{\dimexpr 0.1739\linewidth-6pt\relax}>{\raggedright\arraybackslash}p{\dimexpr 0.1739\linewidth-6pt\relax}}
\toprule
\hspace{0pt}\textbf{Alternative} & \hspace{0pt}\textbf{Kind; level} & \hspace{0pt}\textbf{Matched output?} & \hspace{0pt}\textbf{Matched accuracy target?} & \hspace{0pt}\textbf{Where the integrator needs fewer resources} & \hspace{0pt}\textbf{Where it needs more} & \hspace{0pt}\textbf{Result} \\
\midrule
\endhead
\hspace{0pt}Unstructured Taylor/Dyson LCU of the unsplit step (no amplification) & \hspace{0pt}II; L1 & \hspace{0pt}yes: same frozen step, post-selected state & \hspace{0pt}no: the series has no splitting error; normalization is compared at equal $\eta$ & \hspace{0pt}success probability $1.4\text{--}5.3 \times 10^{7}$ times higher at the design value 1/64, and $\approx 10^{9}$ times for the unit-normalized constructions (prospective) (Proposition S4) & \hspace{0pt}— & \hspace{0pt}proved, against a weak baseline \\
\hspace{0pt}LCHS (one step) & \hspace{0pt}II; L1 & \hspace{0pt}yes & \hspace{0pt}only if the Strang error (realized $1.9\text{--}6.7 \times 10^{-6}$ per step) is within the budget & \hspace{0pt}no quadrature register or LCU over nodes & \hspace{0pt}LCHS has no splitting error & \hspace{0pt}comparable query cost: reference degrees 39–42 per step against 40–41 for one LCHS run (48–49 with the dissipative part at its full LCU normalization; Proposition S5) \\
\hspace{0pt}Projection-based quantum marcher with a fast-transform projector (periodic: QFT; rectangular box: DCT/DST) & \hspace{0pt}II; L1 & \hspace{0pt}no: EDAC against incompressible ($O(\mathrm{Ma}^{2})$) & \hspace{0pt}no & \hspace{0pt}— & \hspace{0pt}the projected step needs $2.6\text{--}3.6$ times fewer data-dependent calls per step and, at unit advective Courant number, $1.1\text{--}3.7$ times fewer steps; it has no acoustic splitting term & \hspace{0pt}$2.6\text{--}3.6$ times fewer data-dependent calls for the projected step in these geometries; not at equal error (§8.4; Proposition E3, part 3) \\
\hspace{0pt}QSVT- or QLSA-projected incompressible step, geometry without a fast diagonalizing transform & \hspace{0pt}II; L1 (queries), L2 (gates) & \hspace{0pt}no & \hspace{0pt}no: per step at equal $\Delta t$ and $\eta$ (§8.4) & \hspace{0pt}L1: $122\text{--}126\times$ fewer calls per step ($\approx 30\text{--}110\times$ per trajectory) (Proposition E1) & \hspace{0pt}L2: more gates with $\Theta(\mathcal{N})$ loaders, in every geometry (Proposition E2); accuracy targets below $C_{\mathrm{Ma}}\mathrm{Ma}^{2}$ (Proposition E4); a larger splitting error if Lemma S2 fails with walls & \hspace{0pt}$122\text{--}126$ times fewer calls per step for the integrator at L1 for transport and pressure coupling, counted with reference degrees and conditional on the query model, the geometry and the splitting error (Proposition E1, conditions (a)–(d)); more gates at L2 (Proposition E2) \\
\hspace{0pt}Explicit acoustically limited hybrid (explicit EDAC or artificial compressibility, LBM-type marching) & \hspace{0pt}II; step counts (interface crossings, L3) & \hspace{0pt}similar model (exactly the same for explicit EDAC) & \hspace{0pt}no: stability limits at unit Courant number & \hspace{0pt}$1.9\text{--}3.3$ times fewer steps at production, depending on the Courant definition; the ratio $M/\allowbreak M_{\mathrm{explicit}} \approx C\, \mathrm{Ma}\, h/\allowbreak (\allowbreak \Delta t\, (\allowbreak 1 + \mathrm{Ma}\max|u|))$ is $\propto \mathrm{Ma}$ (Proposition E3) & \hspace{0pt}if the accuracy-admissible $\Delta t$ lies below the acoustic CFL step (not shown; leading splitting term $\propto 1/\mathrm{Ma}$) & \hspace{0pt}$1.9\text{--}3.3$ times fewer steps for the integrator, conditional on an accuracy-admissible $\Delta t$; counts interface crossings only (Proposition E3) \\
\hspace{0pt}Carleman-type coherent algorithms & \hspace{0pt}scope; L1/L3 & \hspace{0pt}no: normalized state at $T$ against fields every step & \hspace{0pt}no & \hspace{0pt}applicability at $R \gg 1$ and for chaotic flows, without a truncation order (§10) & \hspace{0pt}when $R < 1$ and only a final state or observable is needed (no per-step readout) & \hspace{0pt}scope statement, not a cost comparison \\
\hspace{0pt}Classical exponential/Chebyshev integration of the same step & \hspace{0pt}II: L1, L2; III: L3 & \hspace{0pt}L1, L2: no (state against field); L3: yes (field) & \hspace{0pt}yes (same polynomial) & \hspace{0pt}L1: every operator application is one query on $n$ qubits in place of a $\Theta(\mathcal{N})$ stencil product, with the same order of applications; for the flow-independent diffusion factor a query is a data-free circuit (6,193 and 17,641 CX at $8^{3}$ and $16^{3}$, §5.1) & \hspace{0pt}L2: equal $\Theta(Q\mathcal{N})$ up to a log factor for arbitrary advecting fields in the compiled-circuit model (Proposition 5); L3: $\Omega(\allowbreak \mathcal{N}^{2}/\allowbreak \log^{2}\mathcal{N})$ gates per step (fixed $b$) against the classical $\Theta(\mathcal{N}d)$, under I1–I3 and K1–K2 (Proposition 7) & \hspace{0pt}same order of operator applications at L1; same gate scaling up to a logarithm in the compiled-circuit model, a separation being conditional on sublinear access to the field (X-L2); complete loop in §6.3 \\
\hspace{0pt}Classical pseudo-spectral or projection DNS & \hspace{0pt}III; L3 & \hspace{0pt}yes (fields) & \hspace{0pt}no: different discretizations and step sizes & \hspace{0pt}— & \hspace{0pt}the hybrid's classical side alone costs $\approx 1.3\text{--}15$ pseudo-spectral RK3 steps (§6.2); the quantum floor is $\approx 1.8 \times 10^{15}$ CNOTs per step at leading order against an estimated $\approx 2.7 \times 10^{10}$ flops (Proposition 7) & \hspace{0pt}Proposition 7 applies under I1–I3 and K1–K2 (§6.3) \\
\bottomrule
\end{longtable}}

\textbf{Crossover inequalities (derived).}

\begin{itemize}
\item\relax \textbf{(X-E1)} EDAC vs once-per-step QSVT projection, query count, per step: $d(\tau_{\mathrm{adv}} + \tau_{\mathrm{ac}}, \eta) < d(\tau_{\mathrm{adv}}, \eta) + c\, \kappa_{G}\ln(1/\eta)$. At $256^{3}$ it holds by $122\text{--}126\times$ with reference degrees on both sides (not the production queries), $c = 1$ and the periodic $\kappa_{G} = \sqrt{3}/\sin(\pi/N)$ as a proxy. Conditions (a)–(d) of Proposition E1: the query model, a geometry without a fast diagonalizing transform, an acceptable splitting error there, and an accuracy target above the $O(\mathrm{Ma}^{2})$ error; outputs and solution accuracy are not matched.
\item\relax \textbf{(X-E2)} EDAC vs projected formulation, loader-weighted gates: $[d(\tau_{\mathrm{adv}} + \tau_{\mathrm{ac}}) - d(\tau_{\mathrm{adv}})]\, C_{\mathrm{load}} < d_{P}\, C_{G}$. At $256^{3}$ it requires $C_{\mathrm{load}} \lesssim d_{P}\, C_{G}/\Delta d \approx 1.2 \times 10^{7}$ CX, which is below the $6\mathcal{N} \approx 1.0 \times 10^{8}$ loader angles per walk (Proposition E2). It could hold only if the advecting field had a compact (sublinear) quantum description, or if the acoustic increment were not paid on data-loading calls.
\item\relax \textbf{(X-S5)} Split vs LCHS, one step: $2d(\tau_{A}, \eta) + 2d_{D}$ queries (Propositions 1–2; reference degrees $d(\tau_{A}, \eta) + 2d_{D}$) against $O((\lambda_{D}K + \lambda_{A})\Delta t + \log(1/\eta))$ per LCHS run, plus the LCU over quadrature nodes. With the truncation that the kernel requires at $\eta = 10^{-12}$ and the identity branch of the diffusion LCU factored out, the normalized evolution time of an LCHS run is 16.2–16.9 against $\tau_{A} = 10.24\text{--}10.92$ for the split, and the reference degrees are 40–41 against 39–42 (39–40 when the diffusion factors of the split are factored in the same way); with the dissipative part at its full LCU normalization the LCHS values are 22.2–22.8 and 48–49 (Proposition S5). The costs are comparable; the split also needs its splitting error within budget (realized $1.9\text{--}6.7 \times 10^{-6}$ per step, §9.3.1).
\item\relax \textbf{(X-M)} Interface crossings vs explicit acoustic hybrids: $T/\Delta t_{\mathrm{acc}} < T(c + \max|u|)/(Ch)$, i.e. $\Delta t_{\mathrm{acc}} > C\, \mathrm{Ma}\, h/(1 + \mathrm{Ma}\max|u|)$. At production, $\Delta t = 0.002$ exceeds $1.03 \times 10^{-3}$ ($C = 1$, $\max|u_{x}| = 3.84$) and $9.7 \times 10^{-4}$ ($\max|u_{x}| = 5.23$). Whether $\Delta t = 0.002$ is accuracy-admissible is not established.
\item\relax \textbf{(X-L2)} Per-step gate separation from classical: requires $C_{\mathrm{load}} = o(\mathcal{N})$ (Proposition 5), which is not available for a turbulent field re-read every step.
\item\relax \textbf{(X-L3)} End-to-end gate count with per-step full-field readout: $\Omega(\mathcal{N}^{2}/\log^{2}\mathcal{N})$ per step (fixed $b$) against the classical $\Theta(\mathcal{N}d)$, under the interface and cost model of Proposition 7 (I1–I3, K1–K2). A crossover would require observable outputs or $o(\mathcal{N})$ extractable information per step, together with compactly described advecting fields. The dependence on the time model is given in §6.3.
\end{itemize}
\medskip
\subsection{Comparisons by kind and their settings}\label{si:C.14}
This section restates the comparisons of §4–§13 by kind (§0) and gives the setting of each: input and output, access to the operators, tolerance, the quantity counted, and the dependence on the grid, on the number of steps and on condition numbers. It contains no new result; every value is taken from the section named with it.

\subsubsection{Structure of a step (kind I)}\label{si:C.14.1}
{\scriptsize
\setlength{\tabcolsep}{3pt}\renewcommand{\arraystretch}{1.15}\setlength{\LTleft}{0pt}\setlength{\LTright}{0pt}
\begin{longtable}{>{\raggedright\arraybackslash}p{\dimexpr 0.1700\linewidth-6pt\relax}>{\raggedright\arraybackslash}p{\dimexpr 0.4100\linewidth-6pt\relax}>{\raggedright\arraybackslash}p{\dimexpr 0.2900\linewidth-6pt\relax}>{\raggedright\arraybackslash}p{\dimexpr 0.1300\linewidth-6pt\relax}}
\toprule
\hspace{0pt}\textbf{Property} & \hspace{0pt}\textbf{Statement} & \hspace{0pt}\textbf{Scope} & \hspace{0pt}\textbf{Where} \\
\midrule
\endhead
\hspace{0pt}Projection-free pressure evolution & \hspace{0pt}after one balanced-pressure solve at $t = 0$, a step applies only matrix functions of local stencil operators; no Poisson, projection or other linear solve, so no condition number enters the query count or the success probability & \hspace{0pt}every step; the model differs from incompressible Navier–Stokes by $O(\mathrm{Ma}^{2})$ & \hspace{0pt}§3, §8.2 \\
\hspace{0pt}Unitary transport factor & \hspace{0pt}$A(\bar{u})$ is anti-Hermitian for every real frozen $\bar{u}$, without a divergence-free condition, so $e^{A\Delta t}$ is unitary & \hspace{0pt}uniform staggered grid, skew-symmetric advection & \hspace{0pt}Lemma S1 \\
\hspace{0pt}Contractive diffusion factors & \hspace{0pt}$D$ is Hermitian, negative semidefinite, constant and independent of the flow & \hspace{0pt}equal diffusivities for momentum and pressure & \hspace{0pt}Lemma S1 \\
\hspace{0pt}Bounded factors for any $\Delta t$ & \hspace{0pt}every exact corrector factor is a contraction, so the acoustic stiffness imposes no stability limit on the step; the polynomial factors exceed norm one by at most $\eta$, and their errors accumulate linearly & \hspace{0pt}a stability statement for the frozen factors; not a statement about the accuracy of the nonlinear scheme at a given $\Delta t$ & \hspace{0pt}Proposition E3, part 1; Proposition 8 \\
\hspace{0pt}Normal factors & \hspace{0pt}the unsplit generator is normal only if $[A_{\mathrm{adv}}, D] = 0$; after the split each factor is an eigenvalue polynomial of one normal operator with a known spectral interval & \hspace{0pt}other non-unitary algorithms apply the unsplit step by other means (§14.2) & \hspace{0pt}Lemma S3, §9.2 \\
\hspace{0pt}Controlled splitting error & \hspace{0pt}the Strang error is bounded by $(t^{3}/12)\|[A, [A, D]]\| + (t^{3}/24)\|[D, [D, A]]\|$ for any $t \geq 0$, with no growth factor; $[D, A_{\mathrm{ac}}] = 0$, so the split is exact for uniform $\bar{u}$; realized error $1.9\text{--}6.7 \times 10^{-6}$ per step & \hspace{0pt}Lemma S2 holds on the uniform periodic grid with equal diffusivities and is not established with walls; the leading term is the acoustic cross term, $\propto 1/\mathrm{Ma}$ & \hspace{0pt}Lemma S2, Proposition S3, §9.3.1 \\
\hspace{0pt}Success set by normalization & \hspace{0pt}production: 1/4 per factor by design (diffusion: 1/4 times the retained norm fraction); unit-normalized constructions: the step succeeds with the fraction of the norm that it retains & \hspace{0pt}unit-normalized constructions are proved and prospective in use (§4.6) & \hspace{0pt}Supplementary Note~\ref{note:B}, B.3; Propositions 1–3 \\
\hspace{0pt}Classical re-linearization & \hspace{0pt}the state keeps the size of the discretized field; no truncation order and no condition on the strength of the nonlinearity; every step is a time-independent linear problem; the midpoint predictor gives second order & \hspace{0pt}second order measured at $16^{3}$; the loop pays a classical refresh and an interface crossing every step (§10.2) & \hspace{0pt}§9.5, §10 \\
\hspace{0pt}State representation & \hspace{0pt}$n = \log_{2}(4\mathcal{N})$ system qubits (26 at $256^{3}$); 5 encoding ancillas for convection and 3 for diffusion, plus one signal qubit per factor, independent of $\mathcal{N}$, $T$ and the degree; 32 qubits for the corrector with ancilla reuse, 40 without & \hspace{0pt}the size of the quantum register during a step; the advecting field and the field between steps are classical data & \hspace{0pt}§5.4 \\
\hspace{0pt}Confined data dependence & \hspace{0pt}the flow enters one unitary factor only; the diffusion encoding and its phases are constant, and the convection phases depend on the scalar $\tau_{A}$ & \hspace{0pt}— & \hspace{0pt}§9.7 \\
\bottomrule
\end{longtable}}

The number of qubits in the table is the size of the register that holds the state during a step; the field is stored classically between steps, and a quantum random-access memory for the advecting field would have $\Theta(\mathcal{N})$ components (Proposition 5, scope).

\subsubsection{The frozen linear step (kind II)}\label{si:C.14.2}
The task is to apply $S(\Delta t) = e^{D\Delta t/2}e^{A(u_{\mathrm{mid}})\Delta t}e^{D\Delta t/2}$, or the unsplit $e^{(A + D)\Delta t}$, to a given state, with tolerance $\eta$ per factor ($\eta = 10^{-12}$ in the values below).

{\scriptsize
\setlength{\tabcolsep}{3pt}\renewcommand{\arraystretch}{1.15}\setlength{\LTleft}{0pt}\setlength{\LTright}{0pt}
\begin{longtable}{>{\raggedright\arraybackslash}p{\dimexpr 0.1500\linewidth-6pt\relax}>{\raggedright\arraybackslash}p{\dimexpr 0.1600\linewidth-6pt\relax}>{\raggedright\arraybackslash}p{\dimexpr 0.2900\linewidth-6pt\relax}>{\raggedright\arraybackslash}p{\dimexpr 0.1500\linewidth-6pt\relax}>{\raggedright\arraybackslash}p{\dimexpr 0.2500\linewidth-6pt\relax}}
\toprule
\hspace{0pt}\textbf{Method} & \hspace{0pt}\textbf{Input and output; access} & \hspace{0pt}\textbf{Cost, as counted} & \hspace{0pt}\textbf{Dependence on grid, steps and condition numbers} & \hspace{0pt}\textbf{Relation to the integrator} \\
\midrule
\endhead
\hspace{0pt}This integrator: split step, GQSP of two normal operators & \hspace{0pt}state to post-selected state $\propto S\psi$; block encodings $B_{A}$, $B_{D}$ (L1) or their compiled circuits (L2) & \hspace{0pt}queries: $O(\tau_{A} + \log(1/\eta))$ for the unitary factor (optimal black-box scaling in Lemma 2) and $O(\log(1/\eta))$ for diffusion; production 184–185 per step at design success 1/64; reference degrees $\approx 39\text{--}42$; prospective 43–45 at success $\approx 0.96$ (§4.5). Gates per query: diffusion, data-free (6,193 and 17,641 CX at $8^{3}$ and $16^{3}$); convection, $\Omega(\mathcal{N}/\log \mathcal{N})$ for arbitrary fields in the compiled-circuit model (§5) & \hspace{0pt}$\tau_{A}$ independent of $N$ at fixed Courant numbers; $\tau_{\mathrm{ac}} \propto N$ at fixed $\Delta t$; no condition number; per trajectory a stiffness term independent of $\Delta t$ and an additive term $\propto M$ (Proposition 4) & \hspace{0pt}— \\
\hspace{0pt}Classical exponential integration: polynomial evaluation of the product of the same matrix functions with a vector (Hochbruck \& Ostermann~\cite{Hochbruck2010}, Section 4, on implementation) & \hspace{0pt}field to field; stencil matrix–vector products & \hspace{0pt}$d(\tau_{A}, \eta) + 2d_{D} \approx 39\text{--}42$ products per step at the reference degrees, each of $\Theta(\mathcal{N})$ arithmetic operations; the emulation applies 81/80 products of $A$ and $2 \times 55$ of $D$ for the corrector (§3). No count was made for a Krylov-subspace evaluation with the same products & \hspace{0pt}number of products as for the queries above; cost per product $\propto \mathcal{N} = N^{3}$ & \hspace{0pt}same polynomials. One product corresponds to one query, so the numbers of operator applications are of the same order. In the compiled-circuit model with arbitrary fields the gate count has the same $\Theta(Q\mathcal{N})$ scaling up to a logarithm (Proposition 5); the outputs are not matched (state against field) \\
\hspace{0pt}Hamiltonian simulation by qubitization and QSP or QSVT (Low \& Chuang~\cite{LowChuang2019}; GSLW~\cite{Gilyen2019}) & \hspace{0pt}state to state, unitary $e^{-iHt}$; block encoding of $H$ with normalization $\alpha$ & \hspace{0pt}$\Theta(\alpha t + \log(1/\varepsilon)/\log(e + \log(1/\varepsilon)/(\alpha t)))$ queries, necessary and sufficient in the black-box model (Lemma 2) & \hspace{0pt}linear in $\alpha t$, additive in $\log(1/\varepsilon)$; no condition number & \hspace{0pt}this is the task of the convection–acoustic factor, with $\alpha t = \tau_{A}$: the split places the dominant factor in this class. The production synthesis uses 78–79 queries against the reference degree 31–32; Proposition 1 gives $2d$ and directional control $\approx d + 3$ (prospective) \\
\hspace{0pt}One eigenvalue or singular-value polynomial of the unsplit generator $L = A + D$ & \hspace{0pt}— & \hspace{0pt}not applicable: $L$ is non-normal unless $[A_{\mathrm{adv}}, D] = 0$ (Lemma S3) & \hspace{0pt}— & \hspace{0pt}the split is what makes QSP, QSVT and GQSP applicable to the step (§9.2) \\
\hspace{0pt}LCHS (An, Liu \& Lin~\cite{AnLiuLin2023}; An, Childs \& Lin~\cite{AnChildsLin2026}) & \hspace{0pt}state to state $\propto e^{L\Delta t}\psi$; block encodings of the Hermitian and anti-Hermitian parts of $L$ & \hspace{0pt}per run $(\alpha_{L}K + \alpha_{H})T + \log(1/\varepsilon)$ with $K = O((\log 1/\varepsilon)^{1/\beta})$; for one step, with the truncation that the kernel requires at $\eta = 10^{-12}$ ($\beta = 0.82$), a normalized evolution time of 16.2–16.9 and a reference degree of 40–41 when the identity branch of the diffusion LCU is factored out, 22.2–22.8 and 48–49 otherwise, against $\tau_{A} = 10.24\text{--}10.92$ (Proposition S5); quadrature register, LCU over nodes and amplitude amplification & \hspace{0pt}linear in the normalizations and in $T$; success $\approx$ physical retention$/O(1)$ & \hspace{0pt}comparable query cost within this kernel family (Proposition S5); LCHS has no splitting error; the split needs no quadrature register and keeps the diffusion factor data-free \\
\hspace{0pt}Schrödingerisation (Jin, Liu \& Yu~\cite{JinLiuYu2024}) & \hspace{0pt}state to state after projection onto $p > 0$; sparse access; auxiliary $p$ register & \hspace{0pt}$\tilde{O}(\allowbreak (\allowbreak \|u(\allowbreak 0)\|/\allowbreak \|u(\allowbreak t)\|)\, s\, t\, \allowbreak \|H_{\mathrm{total}}\|_{\max})$ with $\|H_{\mathrm{total}}\|_{\max} \sim \max(\allowbreak \|H\|_{\max}/\allowbreak \varepsilon,\allowbreak  \|\bar{H}\|_{\max})$ (Thm 4) & \hspace{0pt}linear in $t$ and, in the theorem read, in $1/\varepsilon$ & \hspace{0pt}no splitting error; the precision enters polynomially in the theorem read, against $\log(1/\eta)$ for the factors of the split. No count was made for the production operators, and later variants were not assessed \\
\hspace{0pt}Truncated Taylor or Dyson series of the unsplit step (LCU) & \hspace{0pt}state to post-selected state; LCU of $L$ with normalization $\alpha_{L}$ & \hspace{0pt}$O(\alpha_{L}\Delta t \cdot \log(1/\eta)/\mathrm{loglog})$ per step; success $\leq 3.0 \times 10^{-10}\text{--}1.2 \times 10^{-9}$ without amplification, $\Omega(1)$ with uniform singular-value amplification at an additional factor (Fang, Lin \& Tong~\cite{Fang2023}, Lemma 2) & \hspace{0pt}success decays as $e^{-2\alpha_{L}\Delta t}$ & \hspace{0pt}the design success 1/64 of the split is $1.4\text{--}5.3 \times 10^{7}$ times higher, and that of its unit-normalized constructions (prospective) $0.9\text{--}3.4 \times 10^{9}$ times (Proposition S4); a weak baseline, and a comparison of normalization, not of total error \\
\hspace{0pt}Linear-system-based ODE solvers (Berry, Childs, Ostrander \& Wang~\cite{BerryChildsOstrander2017}; Krovi~\cite{Krovi2023}; Berry \& Costa~\cite{BerryCosta2024}) & \hspace{0pt}state to history or final state; oracle access to the generator; a quantum linear-systems algorithm inside & \hspace{0pt}$O(\kappa_{V}\, s\, g\, T\|A\|\, \mathrm{polylog})$; $O(gT\|A\|C(A)\, \mathrm{poly})$; $O(R\lambda_{A}T\log(1/\varepsilon)\log(\lambda T/\varepsilon))$ (§9.4) & \hspace{0pt}linear in $T\|A\|$, with the additional factors $\kappa_{V}$, $g$, $C(A)$ and $R$ of the respective theorems & \hspace{0pt}designed for coherent evolution over long times; for one frozen step they add history-state registers and a linear solve, and were not counted (§12, B7) \\
\bottomrule
\end{longtable}}

\textbf{Linear-systems algorithms are not comparators for this task.} The algorithm of Harrow, Hassidim \& Lloyd~\cite{Harrow2009} and its successors, up to the optimal scaling $O(\kappa\log(1/\varepsilon))$ (Costa et al.~\cite{Costa2022}), prepare a state proportional to the solution of a linear system, at a cost that grows with its condition number $\kappa$. Applying a matrix exponential is a different task, and the step of the integrator contains no linear system, so no ratio between the cost of its factors and that of a linear-systems solver is formed. Such solvers enter the comparisons only through formulations that contain a linear solve: the pressure projection of a projected incompressible step (§14.3) and the history-state ODE solvers of the last row.

\subsubsection{The pressure coupling (kind II)}\label{si:C.14.3}
The task is to advance velocity and pressure over one step, with the incompressibility constraint treated by pressure evolution (this integrator) or by projection (§8.4).

\begin{itemize}
\item\relax \textbf{Setting.} State to state. Query model: every block-encoding call is counted once. Chebyshev reference degrees at equal $\Delta t$ and $\eta = 10^{-12}$; the constant of $\Theta(\kappa_{G}\log(1/\eta))$ set to 1; the condition numbers of the periodic grid, $\kappa_{G} = 141$ and $\kappa_{\Delta} \approx 2.0 \times 10^{4}$, used as proxies for geometries without a fast diagonalizing transform. The two formulations solve different models (weakly compressible against incompressible, $O(\mathrm{Ma}^{2})$ apart) and carry different splitting errors, so the comparison is of call counts per step, not of cost at equal error.
\item\relax \textbf{Projector by quantum singular value transformation, once per step.} 31–32 calls against $9\text{--}12 + \approx 3.9 \times 10^{3}$: 122–126 times fewer calls per step and $\approx 30\text{--}110$ times per trajectory. \emph{Grid:} the per-step ratio grows from $\approx 55\text{--}61$ at $64^{3}$ to $\approx 211\text{--}220$ at $1024^{3}$ at fixed $\Delta t$, and $\propto N$ at fixed Courant numbers. \emph{Condition number:} $\kappa_{G} \propto N$ enters the projector only. Conditions (a)–(d) of Proposition E1.
\item\relax \textbf{Projector by a quantum linear-systems algorithm.} $\approx 5.5 \times 10^{5}$ calls per solve at the optimal scaling ($\kappa_{\Delta} \propto N^{2}$). This is the cost of the projector inside the projected formulation (Proposition E1); it is not a comparison of a linear-systems solver with the exponential factors of the integrator.
\item\relax \textbf{Projector by a fast transform} (periodic box: QFT; rectangular box: quantum cosine and sine transforms). The projected step needs $2.6\text{--}3.6$ times fewer data-dependent calls per step (9–12 against 31–32; §8.4, Proposition E2), and it has no acoustic splitting term (§9.3).
\item\relax \textbf{Gate-weighted count in the compiled-circuit model.} The acoustic increment of the integrator, about 20 calls per step, is paid on calls that load the field, and the projector calls are data-free: $\approx 5.6 \times 10^{11}$ against $\approx 2.5 \times 10^{8}$ CX per step at $256^{3}$ (estimates), so the projected step needs fewer gates, and the crossover would require loaders below $\approx 1.2 \times 10^{7}$ CX per call, fewer than the loader angles per walk (Proposition E2).
\item\relax \textbf{Accuracy target.} The acoustic degree term and the leading splitting term grow as $\varepsilon_{\mathrm{acc}}^{-1/2}$ when the target relative to incompressible Navier–Stokes is tightened (Proposition E4); $C_{\mathrm{Ma}}$ is not measured.
\end{itemize}
\subsubsection{The number of time steps (kind II)}\label{si:C.14.4}
\begin{itemize}
\item\relax \textbf{Stability of the frozen factors (proved).} Every exact corrector factor is a contraction for any $\Delta t$, so the acoustic stiffness imposes no stability limit on the step (Proposition E3, part 1); this is not a stability proof for the nonlinear integration.
\item\relax \textbf{Step counts at stability limits} (same weakly compressible system; production parameters). 1650 steps at acoustic Courant number 1.63 and advective CFL at most 0.889, against $\approx 3.2\text{--}5.4 \times 10^{3}$ for explicit acoustically limited schemes ($1.9\text{--}3.3$ times more; 3294–3332 for RK4 and 5379–5440 for RK3). An incompressible projection scheme at unit advective Courant number needs $\approx 450\text{--}1470$ steps, $1.1\text{--}3.7$ times fewer than the integrator (Proposition E3, part 3).
\item\relax \textbf{Query count against step size.} The stiffness term of the trajectory query count does not depend on $\Delta t$, and only the additive tolerance term grows with the number of steps (Proposition 4). For the constructions of Propositions 1–2 (prospective), a larger step therefore lowers both the query count and the number of interface crossings, provided its accuracy permits, which is not established (§9.5).
\item\relax \textbf{Classical arithmetic.} As a classical exponential integrator, the scheme applies $\approx 18$ times more matrix–vector products than explicit RK4 at its stability limit as run, and $\approx 3.5$ times more with Chebyshev degrees at $\eta = 10^{-12}$ and the reduced predictor, with $\approx 2$ times fewer steps (§8.3).
\item\relax \textbf{Accuracy evidence.} Second-order convergence is measured at $16^{3}$ for acoustic Courant numbers up to 1.02 (§9.5); the realized frozen splitting error at the production step is $1.9\text{--}6.7 \times 10^{-6}$ per step (§9.3.1); and the production run at $\Delta t = 0.002$ is compared with a reference solution in the main text, a comparison that combines spatial, modelling and temporal differences.
\item\relax \textbf{Not established.} Whether the production step is accuracy-admissible, the largest accuracy-admissible step at production scale, and therefore whether the step counts above hold at equal accuracy. The comparison is one of stability limits; establishing more requires a time-step study at production scale (§9.5).
\end{itemize}
\subsubsection{Access to the advecting field}\label{si:C.14.5}
The cost of one convection query depends on how the classical field enters the circuit (Proposition 5 and its scope).

{\scriptsize
\setlength{\tabcolsep}{3pt}\renewcommand{\arraystretch}{1.15}\setlength{\LTleft}{0pt}\setlength{\LTright}{0pt}
\begin{longtable}{>{\raggedright\arraybackslash}p{\dimexpr 0.2200\linewidth-6pt\relax}>{\raggedright\arraybackslash}p{\dimexpr 0.3000\linewidth-6pt\relax}>{\raggedright\arraybackslash}p{\dimexpr 0.1700\linewidth-6pt\relax}>{\raggedright\arraybackslash}p{\dimexpr 0.3100\linewidth-6pt\relax}}
\toprule
\hspace{0pt}\textbf{Access model} & \hspace{0pt}\textbf{Cost of one convection query} & \hspace{0pt}\textbf{Hardware} & \hspace{0pt}\textbf{Consequence for the frozen step} \\
\midrule
\endhead
\hspace{0pt}Compiled circuit, arbitrary field (M1–M2) & \hspace{0pt}$\Omega(\mathcal{N} b/\log n)$ gates for a finite gate set, $\Omega(\mathcal{N}/\log \mathcal{N})$ for Lipschitz-parameterized gates; the constructed staggered encoding compiles to $1.0 \times 10^{5}$ and $8.0 \times 10^{5}$ CNOTs per application at $4^{3}$ and $8^{3}$ and extrapolates to $\approx 2.3 \times 10^{10}$ per application ($\approx 3.4 \times 10^{10}$ per controlled walk) at $256^{3}$ (§5.5) & \hspace{0pt}system and ancilla qubits only & \hspace{0pt}gate count $\Theta(Q\mathcal{N})$ up to a logarithm, the scaling of classical stencil evaluation \\
\hspace{0pt}Persistent data register with counted gates & \hspace{0pt}at least $\mathcal{N}'/2$ two-qubit gates per call (light-cone count) & \hspace{0pt}a register holding the coefficients & \hspace{0pt}as for compiled circuits \\
\hspace{0pt}Compactly described field (a few Fourier modes, a low-degree polynomial) & \hspace{0pt}polylogarithmic in $\mathcal{N}$ & \hspace{0pt}system and ancilla qubits only & \hspace{0pt}the query counts carry over to gates; a turbulent field read out from the previous step is not assumed to have such a description \\
\hspace{0pt}Quantum random-access memory counted as an oracle (bucket brigade: depth $O(\log \mathcal{N})$; Giovannetti, Lloyd \& Maccone~\cite{Giovannetti2008}) & \hspace{0pt}one oracle call & \hspace{0pt}$\Theta(\mathcal{N})$ memory components; total gate volume $\Omega(\mathcal{N})$ (Dalzell et al.~\cite{Dalzell2025}, Sec. 17.1) & \hspace{0pt}the query model applies; for the complete loop see the time model of §6.3 \\
\bottomrule
\end{longtable}}

\subsubsection{The complete hybrid loop (kind III)}\label{si:C.14.6}
The task is to map the classical field at one step to the classical field at the next, for $M$ steps. The setting is that of Propositions 6–7: a full-field output at every step (I1–I3), one run per output copy (K1) and gate counting with data entering through compiled circuits (K2).

{\scriptsize
\setlength{\tabcolsep}{3pt}\renewcommand{\arraystretch}{1.15}\setlength{\LTleft}{0pt}\setlength{\LTright}{0pt}
\begin{longtable}{>{\raggedright\arraybackslash}p{\dimexpr 0.2200\linewidth-6pt\relax}>{\raggedright\arraybackslash}p{\dimexpr 0.5000\linewidth-6pt\relax}>{\raggedright\arraybackslash}p{\dimexpr 0.2800\linewidth-6pt\relax}}
\toprule
\hspace{0pt}\textbf{Component, per step} & \hspace{0pt}\textbf{Cost at $256^{3}$} & \hspace{0pt}\textbf{Scaling; where} \\
\midrule
\endhead
\hspace{0pt}Classical predictor and refresh & \hspace{0pt}96 stencil applications, reducible to 17–27; $\approx 1.3\text{--}15$ times a pseudo-spectral RK3 step (estimate) & \hspace{0pt}$\Theta(\mathcal{N} d_{\mathrm{pred}})$; §6.2 \\
\hspace{0pt}State preparation, per copy & \hspace{0pt}$\approx 3.4 \times 10^{7}$ CNOTs at the leading-order bound for exact preparation of generic states & \hspace{0pt}$\Omega(\mathcal{N}/\log \mathcal{N})$ gates; §6.3 \\
\hspace{0pt}Corrector, per run & \hspace{0pt}184–185 queries; 4,852–4,868 expected per accepted copy at the design success 1/64; $\approx 2.2 \times 10^{12}$ CX per run (census, upper estimate) & \hspace{0pt}$\Omega(Q_{A}\mathcal{N}/\log n)$ gates for arbitrary fields (Proposition 5, finite gate set); §4.5, §5.1–5.2 \\
\hspace{0pt}Readout & \hspace{0pt}at least $5.3 \times 10^{7}$ copies ($b = 32$, $P_{\mathrm{err}} = 1/3$) & \hspace{0pt}$\gtrsim (1 - P_{\mathrm{err}})\, 4\mathcal{N}(b - 1)/n$ copies; Proposition 6 \\
\hspace{0pt}Total & \hspace{0pt}at least $\approx 1.8 \times 10^{15}$ CNOTs at leading order, against $\approx 2.7 \times 10^{10}$ flops for a pseudo-spectral RK3 step & \hspace{0pt}$\Omega(\mathcal{N}^{2}/\log^{2}\mathcal{N})$ against $\Theta(\mathcal{N}d)$; Proposition 7 \\
\bottomrule
\end{longtable}}

The totals are those of Proposition 7 and hold in this setting. They depend on the assumptions as follows (§6.3): with observable outputs in place of fields (I1) the readout floor does not apply; with a quantum memory in a time model (K2) at most a logarithmic factor remains against sequential classical computation; and joint multi-copy circuits (K1) are outside the loop analysed. The changes that would lift these floors are listed in §6.3; they have to be made together, since observable outputs alone leave the loading floor of Proposition 5 and sublinear loading alone leaves the readout floor.

\subsubsection{Relation between the kinds}\label{si:C.14.7}
A result of kind II is a property of a subproblem under its access model. Whether it becomes a gain for the complete computation is decided by the interface of kind III. In the loop as designed the interface terms of §6 dominate, so the results of kind II do not change the result of kind III. Conversely, the result of kind III does not remove the properties of kinds I and II: they are properties of the step and hold for any loop that uses the same step behind a different interface.

\medskip
\subsection*{Appendix A. Computational checks}\phantomsection\addcontentsline{toc}{subsection}{Appendix A. Computational checks}
The values attributed to Appendix A in the text were computed in the following checks.

{\footnotesize
\setlength{\tabcolsep}{3pt}\renewcommand{\arraystretch}{1.15}\setlength{\LTleft}{0pt}\setlength{\LTright}{0pt}
\begin{longtable}{>{\raggedright\arraybackslash}p{\dimexpr 0.0912\linewidth-6pt\relax}>{\raggedright\arraybackslash}p{\dimexpr 0.6021\linewidth-6pt\relax}>{\raggedright\arraybackslash}p{\dimexpr 0.3068\linewidth-6pt\relax}}
\toprule
\hspace{0pt}\textbf{Check} & \hspace{0pt}\textbf{What} & \hspace{0pt}\textbf{Used in} \\
\midrule
\endhead
\hspace{0pt}C01 & \hspace{0pt}Chebyshev degrees of $e^{i\tau x}$ and $e^{\tau x}$; query tables; Laurent sup norm; Taylor-LCU normalization; $\kappa_{G}$, $\kappa_{\Delta}$ and grid scaling; heat-kernel degree & \hspace{0pt}§2, §4.2–4.5, §8.4, §9.4 \\
\hspace{0pt}C02 & \hspace{0pt}Numerical support for $d(\allowbreak \tau,\allowbreak  \eta) \leq e\tau/\allowbreak 2 + \ln(\allowbreak 1/\allowbreak \eta)$ (544 cases, maximum ratio 0.88); heat-kernel degree; readout floors; predictor degrees & \hspace{0pt}§4.2, §6.2, §6.3, §9.7 \\
\hspace{0pt}C03 & \hspace{0pt}Leading-order and realized frozen splitting errors on saved production states ($t = 1.0$, 1.5, 3.0), $D$/$A$ split and hypothetical sub-split; frozen-step retention & \hspace{0pt}§4.5, §9.1, §9.3.1, §9.5, §9.6 \\
\hspace{0pt}C04 & \hspace{0pt}Spectrum-adapted degrees (23); trajectory counts for $\Delta t \times 1$, 2, 4, 8 & \hspace{0pt}§4.3, §4.5, §11 \\
\hspace{0pt}C05 & \hspace{0pt}GQSP phase synthesis at near-unit normalization with the production routines (convection Laurent, diffusion Proposition 2); block and projection structure of the leading splitting term at $t = 1.0$ & \hspace{0pt}§4.3, §4.4, §9.3.1 \\
\hspace{0pt}C06 & \hspace{0pt}Cost model of the production code: stencil counts, step and synthesis times, gate census of the compiled encodings, trimmed query counts; operator norms and commutators at $8^{3}\text{--}32^{3}$ and $256^{3}$ (Lanczos); BCH coefficients; commutators of the acoustic sub-split & \hspace{0pt}§2, §3, §4.3, §5, §8.4, §9.1, §9.3, §9.5, §9.6 \\
\hspace{0pt}C07 & \hspace{0pt}Kernel of An, Childs \& Lin by one-dimensional quadrature: normalization, $\ell_{1}$ norm, truncation $K(\beta, \eta)$ from the $\ell_{1}$ tail and its minimum over $\beta$ at $\eta = 10^{-12}$; normalized evolution time and reference degree of the resulting LCHS run with the dissipative part at its full and at its factored LCU normalization; call ratio of Proposition E1 with the diffusion factors included & \hspace{0pt}§8.4, §9.4 \\
\bottomrule
\end{longtable}}

The step counts of Proposition E3 and the run maxima ($\max|u_{x}| = 5.23$ at $t = 0.83$; logged CFL sum 871.3, maximum 0.889) were recomputed from the production run's per-step metrics and run log.

\end{NoteCnumbering}

\section{Numerical verification procedures}\label{note:D}

This note gives the procedures and results behind Fig.~\ref{fig:verification}, Fig.~\ref{fig:turbulence}, Extended Data Figs~\ref{edfig:collocated} and~\ref{edfig:divergence} and Extended Data Table~\ref{edtab:stats}: the gate-level checks of the collocated circuits (D.1--D.2; Extended Data Fig.~\ref{edfig:collocated}) and of the staggered convection circuit used by the integrator (\ref{si:D.8}; Fig.~\ref{fig:verification}a,b). The checks used Python 3.12.13, NumPy 2.4.4, SciPy 1.17.1, Numba 0.65.1, Qiskit~\cite{Qiskit2024} 2.2.3 and Qiskit Aer 0.17.2 with the production code; every tolerance was fixed before the comparison.

\subsection{Gate-level checks of the collocated circuits}\label{si:D.1}
The explicit gate-level emulator, which propagates the full ancilla–system state, is compared with exact Qiskit statevector evolution of the same collocated circuits (Supplementary Table D1; Extended Data Fig.~\ref{edfig:collocated}a). The grid is $4^{3}$ because on $2^{3}$ the collocated convection generator vanishes. The Strang step is evolved block by block, each block on its own register with the post-selected vector passed on; because a joint 23-qubit circuit at $4^{3}$ is infeasible, sequential and joint post-selection are compared on $2^{3}$, where the identity is structural. Tolerances: full-state relative L2 error $\leq 10^{-9}$; maximum absolute amplitude error $\leq 10^{-10}$; $1 -$ fidelity $\leq 10^{-12}$; post-selected relative L2 error $\leq 10^{-9}$; $|\Delta P(\mathrm{success})| \leq 10^{-10}$. All cases pass.

\textbf{Supplementary Table D1 | Explicit gate-level emulator against Qiskit statevector evolution (collocated circuits; Chebyshev degree 6; logical-gate level, not transpiled).}

{\scriptsize
\setlength{\tabcolsep}{3pt}\renewcommand{\arraystretch}{1.15}\setlength{\LTleft}{0pt}\setlength{\LTright}{0pt}
\begin{longtable}{>{\raggedright\arraybackslash}p{\dimexpr 0.3113\linewidth-6pt\relax}>{\raggedright\arraybackslash}p{\dimexpr 0.0763\linewidth-6pt\relax}>{\raggedright\arraybackslash}p{\dimexpr 0.1154\linewidth-6pt\relax}>{\raggedright\arraybackslash}p{\dimexpr 0.1112\linewidth-6pt\relax}>{\raggedright\arraybackslash}p{\dimexpr 0.0654\linewidth-6pt\relax}>{\raggedright\arraybackslash}p{\dimexpr 0.1967\linewidth-6pt\relax}>{\raggedright\arraybackslash}p{\dimexpr 0.1236\linewidth-6pt\relax}}
\toprule
\hspace{0pt}\textbf{Block and parameters} & \hspace{0pt}\textbf{Qubits} & \hspace{0pt}\textbf{Full-state rel. L2} & \hspace{0pt}\textbf{Max abs. amplitude error} & \hspace{0pt}\textbf{$1 - F$} & \hspace{0pt}\textbf{Post-selected rel. L2} & \hspace{0pt}\textbf{$|\Delta P|$} \\
\midrule
\endhead
\hspace{0pt}Convection, pressure-dominated: $4^{3}$, Ma 0.05, $|u| \sim 0.3$, $\Delta t$ 0.01 (advective part 1.4\% of $\|A\|_{F}$); 10,561 gates & \hspace{0pt}15 & \hspace{0pt}$6.05 \times 10^{-11}$ & \hspace{0pt}$6.34 \times 10^{-12}$ & \hspace{0pt}$\approx 2 \times 10^{-16}$ & \hspace{0pt}$6.16 \times 10^{-11}$ & \hspace{0pt}$3.08 \times 10^{-11}$ \\
\hspace{0pt}Convection, advection-dominated: $4^{3}$, Ma 5, $|u| \sim 3$, $\Delta t$ 0.02 (advective part 99.8\% of $\|A\|_{F}$) & \hspace{0pt}15 & \hspace{0pt}$6.37 \times 10^{-11}$ & \hspace{0pt}$8.32 \times 10^{-12}$ & \hspace{0pt}$3.3 \times 10^{-16}$ & \hspace{0pt}$6.46 \times 10^{-11}$ & \hspace{0pt}$3.23 \times 10^{-11}$ \\
\hspace{0pt}Diffusion: $4^{3}$, Re 800, $\Delta t$ 0.01; 437 gates & \hspace{0pt}12 & \hspace{0pt}$3.96 \times 10^{-13}$ & \hspace{0pt}$3.15 \times 10^{-14}$ & \hspace{0pt}$3.3 \times 10^{-16}$ & \hspace{0pt}$3.99 \times 10^{-13}$ & \hspace{0pt}$1.87 \times 10^{-13}$ \\
\hspace{0pt}One Strang step: diffusion ($\Delta t/2$) $\to$ convection ($\Delta t$) $\to$ diffusion ($\Delta t/2$), $4^{3}$, block by block & \hspace{0pt}12 / 15 / 12 & \hspace{0pt}— & \hspace{0pt}— & \hspace{0pt}— & \hspace{0pt}$6.24 \times 10^{-11}$ & \hspace{0pt}$1.95 \times 10^{-12}$ ($P_{\mathrm{step}}$ 0.0156241) \\
\hspace{0pt}Joint post-selection over three disjoint ancilla registers, $2^{3}$ & \hspace{0pt}20 & \hspace{0pt}— & \hspace{0pt}— & \hspace{0pt}— & \hspace{0pt}joint vs sequential 0.0 (bit-identical); joint vs emulator $1.23 \times 10^{-11}$ & \hspace{0pt}0.0 ($P_{\mathrm{step}}$ 0.0156248) \\
\bottomrule
\end{longtable}}

The $\approx 6 \times 10^{-11}$ residual of the convection cases is mostly norm loss in Qiskit's numerical synthesis of the state preparation and the uniformly controlled rotations (norm deficit $6.05$ and $6.37 \times 10^{-11}$; component orthogonal to the emulator state $1.6$ and $1.1 \times 10^{-12}$, pressure- and advection-dominated case). The reference is the emulated degree-6 polynomial action, not the exponential, from which the Qiskit post-selected convection block differs by the truncation ($1.8 \times 10^{-9}$ and $2.6 \times 10^{-10}$).

\subsection{\texorpdfstring{Self-inverse encodings, pair exactness and the $256^{3}$ backends}{Self-inverse encodings, pair exactness and the 256³ backends}}\label{si:D.2}
The checks of D.1 do not test (S): the diffusion circuit and its emulator both use $W^{-1} = B(2\Pi - I)$, and both convection implementations build $B^{\dagger}$ explicitly. (S) is therefore tested directly on the Qiskit block-encoding gates (tolerance $10^{-10}$) and on the gate-level emulator (Supplementary Table D2; Extended Data Fig.~\ref{edfig:collocated}b). The $B^{2} - I$ residual on the gates, larger than that of $B - B^{\dagger}$, is about twice the gate's own unitarity error, the precision of Qiskit's synthesis, not a failure of (S). The qubitized pair, which relies on (S), equals the explicit emulator, which does not, to $1.91 \times 10^{-13}$ (Extended Data Fig.~\ref{edfig:collocated}c), an independent consistency check of (S) for the collocated encoding. The two emulation backends are compared at $16^{3}$ and $64^{3}$ in Supplementary Table D2 and at $256^{3}$ in Supplementary Note~\ref{note:B}, B.6 (Fig.~\ref{fig:verification}d).

\textbf{Supplementary Table D2 | Self-inverse property of the collocated encodings and exactness of the qubitized pair.}

{\footnotesize
\setlength{\tabcolsep}{3pt}\renewcommand{\arraystretch}{1.15}\setlength{\LTleft}{0pt}\setlength{\LTright}{0pt}
\begin{longtable}{>{\raggedright\arraybackslash}p{\dimexpr 0.3657\linewidth-6pt\relax}>{\raggedright\arraybackslash}p{\dimexpr 0.2978\linewidth-6pt\relax}>{\raggedright\arraybackslash}p{\dimexpr 0.3365\linewidth-6pt\relax}}
\toprule
\hspace{0pt}\textbf{Object} & \hspace{0pt}\textbf{Test} & \hspace{0pt}\textbf{Result} \\
\midrule
\endhead
\hspace{0pt}Convection gate, $4 \times 1 \times 1$ (3 of 9 branches active), 10 qubits & \hspace{0pt}dense, maximum absolute error & \hspace{0pt}$B^{2} - I\ 4.4 \times 10^{-13}$; $B - B^{\dagger}\ 1.1 \times\allowbreak  10^{-14}$; unitarity $4.4 \times 10^{-13}$ \\
\hspace{0pt}Convection gate, $2^{3}$ (9 of 9 branches), 11 qubits & \hspace{0pt}dense, maximum absolute error & \hspace{0pt}$B^{2} - I\ 1.3 \times 10^{-12}$; $B - B^{\dagger}\ 2.7 \times\allowbreak  10^{-14}$; unitarity $1.3 \times 10^{-12}$ \\
\hspace{0pt}Diffusion gate, $2^{3}$ / $4^{3}$, 8 / 11 qubits & \hspace{0pt}dense, maximum absolute error & \hspace{0pt}$B^{2} - I\ 1.8 \times 10^{-14}$ / $4.3 \times 10^{-14}$; $B - B^{\dagger}\ 9.5 \times\allowbreak  10^{-15}$ / $7.7 \times 10^{-15}$; unitarity $4.0 \times 10^{-15}$ / $4.0 \times 10^{-14}$ \\
\hspace{0pt}Convection gate, $4^{3}$, the two convection cases of Supplementary Table D1, 14 qubits & \hspace{0pt}3 random full states each, relative L2 & \hspace{0pt}$B^{2} - I \leq 6.0 \times\allowbreak  10^{-12}$ (maximum $6.03 \times 10^{-12}$); $B - B^{\dagger} \leq 8.8 \times\allowbreak  10^{-14}$; unitarity $\approx 3.0 \times 10^{-12}$ \\
\hspace{0pt}Diffusion gate, $4^{3}$, 11 qubits & \hspace{0pt}3 random full states, relative L2 & \hspace{0pt}$B^{2} - I \leq 4.2 \times\allowbreak  10^{-14}$; $B - B^{\dagger} \leq 4.5 \times\allowbreak  10^{-15}$; unitarity $\approx 2.0 \times 10^{-14}$ \\
\hspace{0pt}Gate-level emulator, $4^{3}$ & \hspace{0pt}$B^{2} = I$ on random full ancilla states & \hspace{0pt}convection $4.30 \times 10^{-14}$; diffusion $6.33 \times 10^{-16}$ \\
\hspace{0pt}Gate-level emulator, $4^{3}$, degree 8 & \hspace{0pt}top-left block against target & \hspace{0pt}$-iA/\lambda$ to $4.88 \times 10^{-15}$; $D/\lambda_{D}$ to $4.41 \times 10^{-16}$ \\
\hspace{0pt}Qubitized pair against explicit gate-level emulator, collocated, $4^{3}$, Chebyshev degree 12 & \hspace{0pt}post-selected output; success probability; sector populations & \hspace{0pt}$1.91 \times 10^{-13}$; $|\Delta P|\ 9.49 \times\allowbreak  10^{-14}$; $2.38 \times 10^{-13}$ \\
\hspace{0pt}Qubitized pair against system-only backend (same $\bar{H}$) & \hspace{0pt}$16^{3}$, degree 24; $64^{3}$, degree 32 & \hspace{0pt}$16^{3}$: convection block $4.37 \times 10^{-16}$ (staggered) and $3.80 \times 10^{-16}$ (collocated); Strang step $5.29 \times 10^{-16}$ (both layouts). $64^{3}$ staggered Strang step: $7.11 \times 10^{-16}$ \\
\bottomrule
\end{longtable}}

\subsection{Operator-structure checks}\label{si:D.3}
The structural properties used in Supplementary Note~\ref{note:A} were checked on the production operators (Supplementary Table D3).

\textbf{Supplementary Table D3 | Operator-structure checks on the production code.}

{\footnotesize
\setlength{\tabcolsep}{3pt}\renewcommand{\arraystretch}{1.15}\setlength{\LTleft}{0pt}\setlength{\LTright}{0pt}
\begin{longtable}{>{\raggedright\arraybackslash}p{\dimexpr 0.3659\linewidth-6pt\relax}>{\raggedright\arraybackslash}p{\dimexpr 0.2779\linewidth-6pt\relax}>{\raggedright\arraybackslash}p{\dimexpr 0.3562\linewidth-6pt\relax}}
\toprule
\hspace{0pt}\textbf{Check} & \hspace{0pt}\textbf{Grid and inputs} & \hspace{0pt}\textbf{Result} \\
\midrule
\endhead
\hspace{0pt}Skewness of the MAC generator, $\langle x, Ay\rangle + \langle Ax, y\rangle$ (worst relative residual) & \hspace{0pt}dense $8^{3}$; bilinear $16^{3}$, $32^{3}$; 3 random coefficient fields & \hspace{0pt}combined $A$ $4.4 \times 10^{-18}$; advection blocks (face velocities and $p_{\mathrm{s}}$) $2.4 \times 10^{-17}$; acoustic coupling $6.3 \times 10^{-18}$ \\
\hspace{0pt}Norm preservation of the ideal $e^{A\,\Delta t}$ & \hspace{0pt}same & \hspace{0pt}$1.1 \times 10^{-15}$ \\
\hspace{0pt}$\lambda_{A}/\|A\|_{2}$ (power iteration) & \hspace{0pt}$16^{3}$, 3 random fields of amplitude 1.5 & \hspace{0pt}1.897–1.908 ($\|\bar{H}_{A}\| \approx 0.53$) \\
\hspace{0pt}MAC advection: max$|L + L^{T}|$; consistency error & \hspace{0pt}$16^{3}$, $32^{3}$ & \hspace{0pt}0 at both grids; errors fall at second order ($2.94 \times\allowbreak  10^{-2} \to 7.86 \times\allowbreak  10^{-3}$; $4.63 \times\allowbreak  10^{-2} \to 1.25 \times\allowbreak  10^{-2}$) \\
\hspace{0pt}Production operator against an independent mini-solver & \hspace{0pt}$16^{3}$, collocated / staggered & \hspace{0pt}$2.27 \times 10^{-16}$ / $2.24 \times 10^{-16}$; anti-Hermiticity $1.74 \times 10^{-15}$ / $1.32 \times 10^{-15}$; LCU bound $\geq$ spectral norm: $172.66 \geq 86.43$ and $325.45 \geq 172.03$ \\
\hspace{0pt}Diffusion $D$ & \hspace{0pt}dense and $16^{3}$, $32^{3}$ & \hspace{0pt}$\|D - D^{H}\| = 0$; largest eigenvalue $2 \times 10^{-17}$ (constant mode); largest Rayleigh quotient $-0.049$; $\|e^{D\,\Delta t/2}\|_{2} - 1 = 2 \times\allowbreak  10^{-16}$; production polynomial half-step, max $\|P(D)x\|/\|x\| - 1 = -4.9 \times\allowbreak  10^{-5}$ \\
\hspace{0pt}MAC adjoint pair, $\mathrm{Div} + \mathrm{Grad}^{T}$; inner-product form $\langle \mathrm{Div}\,u, p\rangle = -\langle u, \mathrm{Grad}\,p\rangle$ & \hspace{0pt}dense $8^{3}$; bilinear $16^{3}$, $32^{3}$; inner product $16^{3}$ & \hspace{0pt}0 exactly ($8^{3}$); $1.7 \times 10^{-17}$ and $1.3 \times 10^{-17}$; inner product $1.62 \times 10^{-16}$ \\
\hspace{0pt}Null space of $\mathrm{Div}\cdot\mathrm{Grad}$ & \hspace{0pt}grids of the adjoint-pair check; $16^{3}$ (mode count) & \hspace{0pt}MAC: the constant mode only (no null mode among the 4,095 nonzero modes at $16^{3}$); collocated: the constant mode plus 7 Nyquist modes \\
\hspace{0pt}$[D, A_{\mathrm{ac}}]$ & \hspace{0pt}$8^{3}\text{--}32^{3}$; production states & \hspace{0pt}$\|[D, A_{\mathrm{ac}}]\|_{F} = 0.0$; $\|[D, A_{\mathrm{ac}}]q\|$ $5.0 \times\allowbreak  10^{-15}\text{--}1.3 \times\allowbreak  10^{-14}$ relative (three production states) \\
\hspace{0pt}Balanced-pressure static condition, max$|d(\mathrm{Div}\,u)/dt|$ at $t = 0$ from $A(u_{0})q_{0} + Dq_{0}$, uniform $\to$ balanced pressure & \hspace{0pt}MAC; production initial-condition generator ($k_{p}$ 4, $u_{\mathrm{rms}}$ 1, Ma 0.05); collocated formula at $64^{3}$ & \hspace{0pt}$64^{3}$: $236.45 \to 9.6 \times 10^{-13}$; $256^{3}$: $241.29 \to 2.28 \times\allowbreak  10^{-11}$; collocated $231.32 \to 2.49 \times 10^{-13}$; $\mathrm{divmax}(u_{0})$ $2.3 \times\allowbreak  10^{-14}$ ($64^{3}$) and $1.1987 \times\allowbreak  10^{-13}$ ($256^{3}$; run value $1.1990 \times 10^{-13}$, and $\max|p_{\mathrm{s}}| = 0.32888$ as in the run) \\
\bottomrule
\end{longtable}}

\subsection{\texorpdfstring{Temporal order at $16^{3}$}{Temporal order at 16³}}\label{si:D.4}
The order test runs the production predictor–corrector step on a $16^{3}$ MAC grid (Ma 0.05, Re 800) to $T = 0.08$ with $\Delta t = 0.02$, 0.01, 0.005, 0.0025 and 0.00125 (4–64 steps; acoustic Courant number 1.02 to 0.064, against 1.63 in production). The test state is smooth but neither solenoidal nor pressure-balanced (four random sine modes per field, the pressure 0.1 times such a field; maximum divergence after the first step 3.41 at $\Delta t = 0.00125$, against $1.2 \times 10^{-13}$ for the production initial condition). Orders are Richardson self-convergence orders, $\log_{2}$ of the ratio of successive differences $|s(\Delta t) - s(\Delta t/2)|$. Four paths share the state: the \emph{ideal path} (classical ideal Chebyshev action, degrees 44 and 48), the \emph{GQSP path} (production step, system-only backend), the \emph{ideal-corrector path} (production step with an ideal degree-48 corrector) and the \emph{production path} (production step on the qubitized-pair backend: predictor, coefficient refresh, three post-selected GQSP factors and design-gain decoding, in exact emulation). Results: Supplementary Table D4; Fig.~\ref{fig:verification}c.

\textbf{Supplementary Table D4 | Temporal order at $16^{3}$ (MAC grid; Richardson self-convergence orders from five time steps) and the separation of error sources.}

{\scriptsize
\setlength{\tabcolsep}{3pt}\renewcommand{\arraystretch}{1.15}\setlength{\LTleft}{0pt}\setlength{\LTright}{0pt}
\begin{longtable}{>{\raggedright\arraybackslash}p{\dimexpr 0.3588\linewidth-6pt\relax}>{\raggedright\arraybackslash}p{\dimexpr 0.3588\linewidth-6pt\relax}>{\raggedright\arraybackslash}p{\dimexpr 0.1498\linewidth-6pt\relax}>{\raggedright\arraybackslash}p{\dimexpr 0.1325\linewidth-6pt\relax}}
\toprule
\hspace{0pt}\textbf{Paths or quantity} & \hspace{0pt}\textbf{Full state} & \hspace{0pt}\textbf{Velocity} & \hspace{0pt}\textbf{$p_{\mathrm{s}}$} \\
\midrule
\endhead
\hspace{0pt}All four paths: ideal (degrees 44 and 48), GQSP, ideal-corrector and production & \hspace{0pt}2.081 / 2.016 / 2.003 & \hspace{0pt}2.032 / 2.007 / 2.001 & \hspace{0pt}2.147 / 2.029 / 2.007 \\
\hspace{0pt}Separate ideal-path test at degree 44: Lie / Strang / Strang with the classical midpoint predictor; with one frozen linear operator ($8^{3}$), Lie / Strang & \hspace{0pt}0.98 / 0.98 / 2.03 (local orders of the last: 2.08 / 2.02 / 2.00); 1.00 / 2.00 & \hspace{0pt}— & \hspace{0pt}— \\
\hspace{0pt}Absolute Richardson differences (ideal path at degree 44 and production path, equal to four digits) & \hspace{0pt}$7.867 \times 10^{-2}$ / $1.859 \times 10^{-2}$ / $4.598 \times 10^{-3}$ / $1.147 \times 10^{-3}$ & \hspace{0pt}— & \hspace{0pt}— \\
\hspace{0pt}Error separation at equal $\Delta t$ (relative, maximum over the five $\Delta t$) & \hspace{0pt}production $-$ GQSP $\leq 1.4 \times 10^{-14}$ (backend); ideal-corrector $-$ ideal(48) $\leq 3.3 \times 10^{-15}$ (encode, predictor, refresh, decode); GQSP $-$ ideal-corrector $\leq 4.2 \times 10^{-10}$ (GQSP polynomial and angles); ideal(48) $-$ ideal(44) $\leq 4.9 \times 10^{-14}$ & \hspace{0pt}— & \hspace{0pt}— \\
\hspace{0pt}Temporal Richardson difference (relative) & \hspace{0pt}$1.4 \times 10^{-3}$ ($\Delta t$ 0.02 against 0.01) to $2.0 \times 10^{-5}$ (0.0025 against 0.00125); imaginary parts $\leq 4.1 \times 10^{-14}$ & \hspace{0pt}— & \hspace{0pt}— \\
\bottomrule
\end{longtable}}

The non-temporal sources are more than four orders of magnitude below the smallest temporal difference, so the observed order is that of the time integrator; on the production path at $\Delta t = 0.00125$ the success probabilities are 0.250000 (convection) and 0.249997 (diffusion) in all 64 steps. The test establishes second order at $16^{3}$ for this state and these step sizes only (limitation 6 in Supplementary Note~\ref{note:B}, B.7).

\subsection{Divergence control}\label{si:D.5}
\textbf{Supplementary Table D5 | Divergence-control evidence (divmax $=$ maximum absolute discrete divergence, each layout with its own operator).}

{\scriptsize
\setlength{\tabcolsep}{3pt}\renewcommand{\arraystretch}{1.15}\setlength{\LTleft}{0pt}\setlength{\LTright}{0pt}
\begin{longtable}{>{\raggedright\arraybackslash}p{\dimexpr 0.2030\linewidth-6pt\relax}>{\raggedright\arraybackslash}p{\dimexpr 0.3158\linewidth-6pt\relax}>{\raggedright\arraybackslash}p{\dimexpr 0.2077\linewidth-6pt\relax}>{\raggedright\arraybackslash}p{\dimexpr 0.2735\linewidth-6pt\relax}}
\toprule
\hspace{0pt}\textbf{Test} & \hspace{0pt}\textbf{Configuration} & \hspace{0pt}\textbf{Quantity} & \hspace{0pt}\textbf{Values} \\
\midrule
\endhead
\hspace{0pt}Balanced against uniform pressure (dynamic) & \hspace{0pt}$64^{3}$ collocated, Ma 0.05, Re 1600 (default setting, not logged), Chebyshev degree 32, $\Delta t$ 0.002, 100 steps, system-only backend & \hspace{0pt}divmax after the first step; run maximum & \hspace{0pt}$0.4566 \to 0.004083$ ($111.8 \times$ lower); $1.592 \to 0.1805$ ($8.82 \times$ lower) \\
\hspace{0pt}Ma scaling (balanced) & \hspace{0pt}same, $\mathrm{Ma} = 0.10$ / 0.05 / 0.025 & \hspace{0pt}run maximum & \hspace{0pt}0.7230 / 0.1805 / 0.04412; 4.00 and 4.09 per halving of Ma \\
\hspace{0pt}Chebyshev degree & \hspace{0pt}same, Ma 0.025, degree 32 against 48 & \hspace{0pt}divmax after the first step; run maximum & \hspace{0pt}identical \\
\hspace{0pt}Collocated against MAC & \hspace{0pt}$256^{3}$, Re 800, Ma 0.05, one initial condition; collocated: degree 32, system-only, stopped at $t = 3.13$; MAC: degree 48, Lie splitting, system-only; MAC production run: degree 48, Strang with predictor, qubitized pair & \hspace{0pt}collocated divmax over MAC divmax at $t = 3.13$ & \hspace{0pt}$49.06 \times$ (MAC) and $59.99 \times$ (MAC production run) \\
\bottomrule
\end{longtable}}

The dynamic effect of the balanced pressure is shown only at $64^{3}$ on the collocated grid (Extended Data Fig.~\ref{edfig:divergence}a); at $256^{3}$ on the MAC grid only the static check exists (Supplementary Table D3). In the $64^{3}$ test the remaining run maximum scales as $\mathrm{Ma}^{2}$ over three Ma values, consistent with the $O(\mathrm{Ma}^{2})$ acoustic residual of artificial compressibility~\cite{Ohwada2010}; no incompressible-limit convergence was shown. At $256^{3}$ the collocated divergence maximum rose from 0.2288 at $t = 1.926$ to 2.097 at $t = 3.13$, while both MAC runs decreased over the same window (Extended Data Fig.~\ref{edfig:divergence}b). The ratios $49.06 \times$ and $59.99 \times$ are not a like-for-like comparison: each layout reports its own discrete divergence (a central difference on the collocated grid, the backward face-to-centre difference on the MAC grid); the Chebyshev degrees differ (32 against 48), and degree independence is shown only at $64^{3}$ for 100 steps; the two MAC runs differ in splitting and backend, and their ordering alternates between steps; single-instant ratios are imprecise (over $t = 3.09\text{--}3.14$ the divergence maximum varies by about $\pm 10$\% around its mean); and one initial condition is used. Attributing the collocated growth to the seven Nyquist null modes of the collocated layout (Supplementary Table D3) is an interpretation; no experiment isolates those modes. In the production run the divergence maximum reaches its run maximum of 1.560 at $t = 0.954$, to which no cause is assigned, and is 0.0389 at $t = 3.3$ (Extended Data Fig.~\ref{edfig:divergence}c).

\subsection{Reference solution and flow diagnostics}\label{si:D.6}
The reference is an in-house pseudo-spectral solution of the incompressible Navier–Stokes equations (Fourier–Galerkin, exact spectral projection, SSP-RK3 time stepping~\cite{ShuOsher1988}, 2/3 dealiasing~\cite{Orszag1971} with the cube mask $|k_{i}| < 85.3$) at $N = 256$, $\mathrm{Re} = 800$ and $\Delta t = 0.002$ to $t = 3.3$, started from the same Gaussian draws as the run but projected spectrally. It has not been independently validated. Its differences from the run combine spatial and temporal error and are not attributed to the time integrator; no external HIT data and no HIT time-step study enter, and one initial condition is used.
\begin{itemize}
\item\relax \emph{Kinetic energy} (Fig.~\ref{fig:turbulence}b): $\mathrm{KE}_{\mathrm{run}}/\mathrm{KE}_{\mathrm{ref}} - 1$ at the 414 common output times, with KE the domain mean of $\tfrac{1}{2}|u|^{2}$ of the real fields as reported by each solver; maximum +3.248\% at $t = 2.312$.
\item\relax \emph{Spectra} (Fig.~\ref{fig:turbulence}c), at $t = 0.5$, 1.0, …, 3.0: for the run, $E(n) = \sum\tfrac{1}{2}|\mathrm{FFT}(f)/N^{3}|^{2}$ over the modes with $\mathrm{rint}(|k|) = n$, summed over $f = \Re(u), \Re(v), \Re(w)$ as stored on the faces (no interpolation or re-centring); shells 0–128 are kept and cube-corner modes with $|k| > 128.5$ dropped. The reference uses the same normalization and binning with real-FFT half-spectrum weights; shells up to 85 lie wholly inside its retained cube.
\item\relax \emph{Bands}: half-open shell ranges 1–4 (shells 1–3), 4–8, 8–16, 16–32, 32–64 and 64–85 (shells 64–84); a band ratio is $\sum E_{\mathrm{run}}/\sum E_{\mathrm{ref}}$ over the band. The large scales of the main text are the two lowest bands (shells 1–7), which contain the spectral peak of both solutions at all six times (shell 4 up to $t = 1.5$, shell 2 from $t = 2.0$); there the band ratios differ from one by at most 1.56\% (band 1–4, $t = 2.0$). Band ratios are averages: single-shell ratios in shells 1–7 differ from one by up to 6.3\% (shell 6, $t = 2.0$). In the three higher bands the ratios depart further from one, most in the 32–64 band (0.971 at $t = 1.5$ to 1.099 at $t = 3.0$), and single-shell ratios there range over 0.93–1.23 (largest at the top of the 32–64 band at $t = 3.0$). The 64–85 band, which reaches 1.23, lies next to the reference's dealiasing edge, where the cutoff changes the dynamics, and is reported separately, not counted in the comparison.
\item\relax \emph{Dissipation} (Extended Data Table~\ref{edtab:stats}): the shell convention $\varepsilon_{\mathrm{d}} = 2\nu\sum_{n}n^{2}E(n)$ (primary, and applied to the reference), the mode-sum $2\nu\sum_{k}|k|^{2}P(k)$ over all modes, and the scheme convention $2\nu\sum_{k}k_{\mathrm{eff}}^{2}P(k)$ with $k_{\mathrm{eff}}^{2} = (4/h^{2})\sum_{a}\sin^{2}(k_{a}h/2)$, the dissipation of the run's seven-point Laplacian. The run's scheme value differs from the reference's shell value by $-2.36$ … +5.49\%. The kinetic energies of Extended Data Table~\ref{edtab:stats} are shell sums, $\mathrm{KE} = \sum_{n=0}^{128}E(n)$; this spectral comparison and the solver-level one of Fig.~\ref{fig:turbulence}b differ by up to 0.11 percentage points (+1.17\% against +1.28\% at $t = 1.5$).
\item\relax \emph{Resolution}: $k_{\mathrm{max}}\eta$ with $k_{\mathrm{max}} = N/2 = 128$ for both codes and $\eta = (\nu^{3}/\varepsilon_{\mathrm{d}})^{1/4}$ in the shell convention; the run's minimum is 0.896 at $t = 1.5$ (mode-sum 0.884, scheme 0.909), below 1 at $t = 1.0$, 1.5 and 2.0. No value on a dealiased ($N/3$) basis is quoted.
\item\relax \emph{Taylor-microscale Reynolds number}: $\mathrm{Re}_{\lambda} = u_{\mathrm{rms}}\lambda/\nu$, with $u_{\mathrm{rms}} = (2\mathrm{KE}/3)^{1/2}$ and $\lambda = (15\nu u_{\mathrm{rms}}^{2}/\varepsilon_{\mathrm{d}})^{1/2}$; this formula reproduces the run's shell-convention $\mathrm{Re}_{\lambda}$ to $\leq 1.7 \times 10^{-12}$.
\item\relax \emph{Divergence and dilatation}: divmax $= \max|\mathrm{Div}\,u|$, with the MAC backward-difference divergence of the real face fields. The dilatational fraction is $\sum_{k}|m(k)\cdot F(k)|^{2}/|m(k)|^{2}$ over the total energy ($F$ the velocity transform, $m$ the MAC divergence symbol); because $\mathrm{Div} = -G^{T}$ this is an exact orthogonal discrete Helmholtz split, and white noise gives 1/3. For the base flow it is $2.14 \times 10^{-6}$ at $t = 1.5$ and $1.02 \times 10^{-5}$ at $t = 3.0$. The solver's per-step enstrophy uses collocated central differences of the face fields, is not MAC-consistent and is used only for the relative changes quoted in Supplementary Note~\ref{note:E}.7.
\end{itemize}

\subsection{Visualization of the vortex field}\label{si:D.7}
Fig.~\ref{fig:turbulence}a is a representative instantaneous visualization; it is not validation evidence, and no
quantitative claim depends on it. It shows the noise-free production run at $t = 1.5$ (step 750) from the saved
production snapshot, without further simulation; the physical field is the real part of the complex decoded fields
(max$|\Im|$/max$|\Re|$ $\approx 3.6 \times 10^{-13}$ at this step). The MAC face velocities are averaged to the cell
centres with the production cell-centre average, $u_{c}[i] = \tfrac{1}{2}(u[i] + u[i - 1])$ and likewise along $y$ and
$z$, and the velocity-gradient tensor $G_{ij} = \partial u_{i}/\partial x_{j}$ is formed by periodic second-order centred
differences with $h = 2\pi/256$; $S$ and $\Omega$ are its symmetric and antisymmetric parts,
$Q = \tfrac{1}{2}(\|\Omega\|^{2}_{F} - \|S\|^{2}_{F})$ and $\omega = \nabla \times u$. The off-diagonal gradient terms,
and hence $\omega$, equal the edge-to-centre average of the compact MAC differences; the diagonal terms are smoothed by
the averaging, so $\mathrm{tr}\,G$ is not the MAC divergence. On analytic MAC-sampled periodic fields ($N = 32$--128) this
procedure converges at second order for $Q$ and $|\omega|$, and it reproduces the exact periodic identity
$\langle Q\rangle = -\tfrac{1}{2}\langle(\mathrm{tr}\,G)^{2}\rangle$. Every point with $Q > 0$ is rotation-dominated
($\|\Omega\|_{F} > \|S\|_{F}$); 39.7\% of the cells are. The isosurface is $Q = 672$, the 98th percentile of the
positive values of $Q$ in this snapshot, which equals $3.0\langle\|\Omega\|^{2}_{F}\rangle$ ($\approx 4.0\ Q_{\mathrm{rms}}$
in code units): it selects the most strongly rotation-dominated 2\% of the rotation-dominated cells, a display level
anchored to the volume mean of $\|\Omega\|^{2}_{F}$ rather than a vortex-identification threshold. Isosurfaces at the 95th, 97th
and 99th percentiles were rendered as well, and the major tubes visible at the 98th persist at the 97th and 99th. The
surface is coloured by $|\omega|$ over one fixed range, 44.6--91.9 (the 2nd--98th percentiles of $|\omega|$ where $Q$
exceeds its 95th positive percentile; $\approx 2.1\text{--}4.3\ \omega_{\mathrm{rms}}$). On the surface
$|\omega|^{2} = 4Q + 2\|S\|^{2}_{F}$, so $|\omega| \gtrsim 2\sqrt{Q} \approx 51.9$ and the lowest part of the colour bar is
essentially unused (exact in the continuum; interpolated vertices can fall slightly below); values above 91.9 take the top
colour. The full periodic cube $[0, 2\pi]^{3}$ is viewed in perspective along the $(1, 1, 1)$ body diagonal with $z$ up,
and tubes are cut at the box faces. Rendering used off-screen PyVista~\cite{PyVista2019} 0.48.4 and VTK 9.6.2: a contour
of $Q$ on the periodically extended grid, $|\omega|$ interpolated linearly, the viridis colour map restricted to
0.55--0.90, smooth shading without specular highlight, no smoothing or size filtering of the isosurface, and Lanczos
resampling to 72 mm at 600 dpi (shown at 62 mm in Fig.~\ref{fig:turbulence}a). No quantity is read from the image: tube
numbers, sizes, intensities and alignment are not results, and the projected tube density does not measure spatial
inhomogeneity, because the line of sight is longest ($\sqrt{3}\,L$) at the image centre. The finest tubes are a few
grid cells wide at this resolution ($k_{\max}\eta = 0.90$ at $t = 1.5$). The pseudo-spectral reference saved spectra and
time series only, not velocity fields, so no structural comparison with it is possible.

\subsection{Gate-level checks of the staggered convection circuit}\label{si:D.8}
The staggered circuit of Supplementary Note~\ref{note:B}, B.4 uses the components of the collocated circuit with its own six-branch table and two new forward gates; no run executed it. Block, random-state and sampled $8^{3}$ checks use two references, the production operator and an independent MAC operator assembled from the definitions by explicit index loops without production code; the two agree to $\leq 6.0 \times 10^{-17}$ (relative Frobenius norm) and both are exactly skew-symmetric. Per-branch checks use the independent per-axis pieces, and GQSP checks the production qubitized-pair emulator, the dense $e^{A\,\Delta t}/2$ and the Chebyshev polynomial. Circuits were evolved with Qiskit statevector simulation of the uncompiled gates and with Qiskit Aer on a unitary-equivalent compilation to CNOT and one-qubit gates (idle qubits not treated as clean ancillas; the global phase applied explicitly); the two agree to $\leq 1.34 \times 10^{-11}$ on six full output columns per case and to $4.61 \times 10^{-11}$ on a whole degree-4 GQSP circuit. Predefined tolerances: $10^{-10}$ for every block, unitarity and GQSP comparison, and $10^{-9}$ against the matrix function. Results: Supplementary Table D6; Fig.~\ref{fig:verification}a,b.

\textbf{Supplementary Table D6 | Gate-level checks of the staggered convection circuit.}

{\footnotesize
\setlength{\tabcolsep}{3pt}\renewcommand{\arraystretch}{1.15}\setlength{\LTleft}{0pt}\setlength{\LTright}{0pt}
\begin{longtable}{>{\raggedright\arraybackslash}p{\dimexpr 0.2943\linewidth-6pt\relax}>{\raggedright\arraybackslash}p{\dimexpr 0.3626\linewidth-6pt\relax}>{\raggedright\arraybackslash}p{\dimexpr 0.3431\linewidth-6pt\relax}}
\toprule
\hspace{0pt}\textbf{Check} & \hspace{0pt}\textbf{Grids and cases} & \hspace{0pt}\textbf{Result} \\
\midrule
\endhead
\hspace{0pt}Normalization & \hspace{0pt}seven cases (five velocity fields at $4^{3}$, anisotropic $8 \times 4 \times 2$, $2^{3}$); $\lambda_{A}$ also at $8^{3}$ & \hspace{0pt}circuit $\lambda_{A}$ equal to the production $\lambda_{A}$ bitwise; independent $\lambda_{A}$ equal bitwise on the cubic grids and to $1.4 \times 10^{-16}$ on $8 \times 4 \times 2$; every emitted loader satisfies $|w/\gamma| \leq 1$ with maximum 1 (six cases; $\bar{u} = 0$ has no advective weights) \\
\hspace{0pt}Top-left block against $-iA/\lambda_{A}$ on every basis state, both references & \hspace{0pt}the seven cases: $4^{3}$ at Ma 0.05 (production) and Ma 5 (advection-dominated), divergence-free and non-solenoidal curl fields, $\bar{u} = 0$ (pure acoustic); anisotropic $8 \times 4 \times 2$; $2^{3}$ & \hspace{0pt}max abs $\leq 1.3 \times 10^{-12}$; relative Frobenius $\leq 4.8 \times 10^{-12}$ ($\leq 4.2 \times 10^{-12}$ at $4^{3}$) \\
\hspace{0pt}Each branch, per pair of fields & \hspace{0pt}same & \hspace{0pt}advective $\leq 9.3 \times 10^{-14}$, acoustic $\leq 7.3 \times 10^{-14}$; acoustic branches couple only $u$ with $p_{\mathrm{s}}$, $v$ with $p_{\mathrm{s}}$ and $w$ with $p_{\mathrm{s}}$; impulse responses reproduce the production forward gradient and backward divergence \\
\hspace{0pt}$B_{A}^{2} = I$ and $B_{A} = B_{A}^{\dagger}$ on random full ancilla–system states & \hspace{0pt}the seven cases (compiled and uncompiled); $8^{3}$ & \hspace{0pt}$B_{A}^{2} = I$ at $4^{3}$ $\leq 8.4 \times 10^{-12}$ compiled and $\leq 2.2 \times 10^{-11}$ uncompiled, on $8 \times 4 \times 2$ $9.4 \times 10^{-12}$ and $2.6 \times 10^{-11}$; $B_{A} = B_{A}^{\dagger}$ $\leq 4.3 \times 10^{-13}$ at $4^{3}$ and $4.8 \times 10^{-13}$ on $8 \times 4 \times 2$; $8^{3}$: $6.6 \times 10^{-11}$ and $3.3 \times 10^{-12}$ \\
\hspace{0pt}One GQSP convection factor against the qubitized pair (production angle chain, Chebyshev degree 48; post-selected output, success probability, sector populations, full ancilla–system state) & \hspace{0pt}$4^{3}$; $\tau_{A} = 0.155$, 3.0 and 10.5 (55, 65 and 78 queries) & \hspace{0pt}$2.3\text{--}3.3 \times 10^{-10}$ (predefined tolerance $10^{-10}$ not met); success-probability differences $\leq 4.4 \times 10^{-11}$; sector-population differences $\leq 1.1 \times 10^{-10}$ ($1.06 \times 10^{-10}$ at $\tau_{A} = 10.5$, also above $10^{-10}$); full state $9.9 \times 10^{-11}$ after removal of one global phase ($\tau_{A} = 10.5$; later diagnostic, not part of the predefined comparison); against $e^{A\,\Delta t}/2$ $\leq 3.3 \times 10^{-10}$ \\
\hspace{0pt}Three-factor corrector D/2–A–D/2 against the qubitized pair & \hspace{0pt}$4^{3}$; Re 800 and Re 1; $\tau_{A} = 10.5$ & \hspace{0pt}$3.3 \times 10^{-10}$ (tolerance not met); success-probability differences per factor $\leq 4.4 \times 10^{-11}$; sector-population differences of the convection factor $1.06\text{--}1.07 \times 10^{-10}$ (also above $10^{-10}$); no full-state or phase-aligned comparison \\
\hspace{0pt}Sampled checks & \hspace{0pt}$8^{3}$ (block action on four random states and eight basis columns; branches on two random states; one output column compiled against uncompiled) & \hspace{0pt}block action $\leq 3.9 \times 10^{-11}$; branches $\leq 1.4 \times 10^{-12}$; compiled against uncompiled $8.9 \times 10^{-11}$; no GQSP check at $8^{3}$ \\
\bottomrule
\end{longtable}}

The excess over the GQSP tolerance is consistent with floating-point accumulation in the double-precision simulation of about $3 \times 10^{7}$ gates. Followed query by query at $\tau_{A} = 10.5$, the difference is $4.6 \times 10^{-12}$ after one query and grows linearly by $4.18 \times 10^{-12}$ per query, equal to the compiled circuit's own error per application ($4.15 \times 10^{-12}$, from $\|B_{A}^{2}\psi - \psi\|/2$), and it is mostly a global phase (after 78 queries, $-3.15 \times 10^{-10}$ rad for the full state and $-3.13 \times 10^{-10}$ rad for the post-selected output); with that phase removed, the full-state difference is $9.9 \times 10^{-11}$. No identity, branch or index condition fails; a wrong sign, shift or field would give differences of order one. On $2^{3}$ the advective part vanishes identically ($S^{+} = S^{-}$), so that case tests only the acoustic branches. The circuit was not simulated at $256^{3}$.

\section{Effective per-step error model}\label{note:E}

This note defines the effective per-step error model (E.1–E.5) and two consequences of its form that enter the Results: the built-in contraction and the random part of the field error (E.6), and the departure from linearity at $\varepsilon = 10^{-2}$, examined with a renormalized run (E.7). The runs are listed in Extended Data Table~\ref{edtab:runs}.

\subsection{The channel}\label{si:E.1}
Once per step, after the third corrector factor and its post-selection and before decoding, the post-selected system state $\psi$ (complex, of dimension $4N^{3}$) is replaced by

\disp{$\psi' = (1 - \varepsilon^{2}/2)\,\psi + \varepsilon\,(1 - \varepsilon^{2}/4)^{1/2}\,\|\psi\|\,n_{\perp}$,}{E1}
\noindent where $n_{\perp}$ is a unit vector orthogonal to $\psi$ in the complex inner product (E.2). $\psi$ is not renormalized after post-selection; decoding then multiplies it by the positive real constant $\|q_{n}\| \cdot 8$ (Methods; Supplementary Note~\ref{note:B}). Since $\langle \psi, n_{\perp}\rangle = 0$,

\disp{$\|\psi'\|^{2} = \bigl[(1 - \varepsilon^{2}/2)^{2} + \varepsilon^{2}(1 - \varepsilon^{2}/4)\bigr]\,\|\psi\|^{2} = \|\psi\|^{2}$,\quad $\|\psi' - \psi\|^{2} = \bigl[\varepsilon^{4}/4 + \varepsilon^{2}(1 - \varepsilon^{2}/4)\bigr]\,\|\psi\|^{2} = \varepsilon^{2}\,\|\psi\|^{2}$}{E2}
The channel therefore preserves the norm and sets $\varepsilon = \|\psi' - \psi\|/\|\psi\|$ exactly, by construction rather than by iteration; because decoding is a uniform real scaling, $\varepsilon$ is also the per-step relative L2 change of the field ($u$, $v$, $w$, $p_{\mathrm{s}}$), and the realized values agree with the set $\varepsilon$ to about $10^{-14}$ relative. The per-step overlap infidelity, $1 - |\langle \psi, \psi'\rangle|/\|\psi\|^{2}$, is $\varepsilon^{2}/2$, and the squared-overlap infidelity is $\approx \varepsilon^{2}$. $\varepsilon$ is set directly: it is a per-step relative state error, not a gate error rate or an infidelity, and no device error rate enters the channel.

\subsection{The direction}\label{si:E.2}
A white complex Gaussian vector $g$ of dimension $4N^{3}$ (independent standard normal real and imaginary parts) gives $\hat{g} = g/\|g\|$, which does not depend on the state. The structured part does: for each system qubit $b$, a random traceless complex $2 \times 2$ matrix $G_{b}$ (complex Gaussian entries, trace removed) acts on the pairs of amplitudes of $\psi$ whose flat indices differ in bit $b$, and

\disp{$\hat{d}_{\mathrm{sys}} = \sum_{b} \sqrt{w_{b}}\,G_{b}\psi/\|G_{b}\psi\| \,\big/\, \bigl\|\sum_{b} \sqrt{w_{b}}\,G_{b}\psi/\|G_{b}\psi\|\bigr\|$,}{E3}
\noindent so that each bit is normalized before weighting. The two unit parts are mixed without renormalization,

\disp{$d = \sqrt{0.91}\,\hat{g} + \sqrt{0.09}\,\hat{d}_{\mathrm{sys}}$,}{E4}
\noindent then projected off $\psi$, $d_{\perp} = d - (\langle \psi, d\rangle/\|\psi\|^{2})\,\psi$, and normalized, $n_{\perp} = d_{\perp}/\|d_{\perp}\|$. Over 16 draws at $256^{3}$ (four per station), the projected structured part $s_{\perp}$ has amplitude $\rho = \|s_{\perp}\| = 0.853$ (per draw 0.64–0.97), and the projected mixture has norm $c = \|d_{\perp}\| = 0.9882$ (per draw 0.973–0.998), which enters Supplementary Note~\ref{note:G}. White-only runs use $d = g$, and the structured-only runs (at $64^{3}$ only) use white weight 0. The system register holds the address bits of the C-order flat grid index, z first (bit 0 is the least significant z bit), then y, then x, followed by two field bits that select among ($u$, $v$, $w$, $p_{\mathrm{s}}$): 20, 23 and 26 system qubits at $64^{3}$, $128^{3}$ and $256^{3}$ (at $256^{3}$, bits 0–7 are z, 8–15 y, 16–23 x and 24–25 the field bits).

\subsection{Gate census and per-qubit ladder}\label{si:E.3}
The weights come from the two-qubit gate census of a compiled collocated $16^{3}$ convection walk with 7 walk ancillas, 12 address qubits and 2 field qubits (Supplementary Table E1). The walk ancillas carry 0.9117 of its two-qubit gate touches and the system qubits 0.0883 (0.91 and 0.09 in the model); the ancilla share is modelled as the white part, on the rationale that the post-selection back-action of the walk ancillas is not local in the system register. The ladder $w_{b}$ is proportional to the measured address-qubit touch counts, embedded least-significant bit first to match the control ladder of the uniformly controlled loaders, and the two field qubits take their measured count. Beyond the 12 measured address bits the ladder continues from the last measured entry with ratio 0.5, floored at the field-qubit count because the measured deep-address tail flattens instead of halving, and the whole ladder is normalized to unit sum; hence 6 of 18, 9 of 21 and 12 of 24 address bits are extrapolated at $64^{3}$, $128^{3}$ and $256^{3}$. Only two-qubit touches are counted; Supplementary Note~\ref{note:G} examines the sensitivity to the weight, the ladder construction and the register order.

\textbf{Supplementary Table E1 | Two-qubit gate touches of the compiled collocated $16^{3}$ convection walk and the ladder built from them.}

{\footnotesize
\setlength{\tabcolsep}{3pt}\renewcommand{\arraystretch}{1.15}\setlength{\LTleft}{0pt}\setlength{\LTright}{0pt}
\begin{longtable}{>{\raggedright\arraybackslash}p{\dimexpr 0.4489\linewidth-6pt\relax}>{\raggedright\arraybackslash}p{\dimexpr 0.5511\linewidth-6pt\relax}}
\toprule
\hspace{0pt}\textbf{Item} & \hspace{0pt}\textbf{Value} \\
\midrule
\endhead
\hspace{0pt}7 walk ancillas & \hspace{0pt}1,516,776; 1,616,606; 1,765,738; 1,663,150; 2,011,390; 1,573,096; 2,510,944 \\
\hspace{0pt}12 address qubits, bit 0 (least significant) to bit 11 & \hspace{0pt}593,568; 298,368; 152,064; 77,184; 40,608; 21,888; 13,824; 8,064; 6,048; 4,608; 5,184; 4,032 \\
\hspace{0pt}2 field qubits & \hspace{0pt}320; 320 \\
\hspace{0pt}Extension beyond 12 address bits & \hspace{0pt}ratio 0.5 from the last measured entry, floor 320: 2,016; 1,008; 504; 320; 320; … \\
\bottomrule
\end{longtable}}

\subsection{Random numbers, onset and stations}\label{si:E.4}
Each step draws from its own generator, keyed by the noise realization and the step, so any step can be reproduced or restarted. The white block is drawn before the per-qubit matrices $G_{b}$, which follow in bit order whatever the weights. Runs of the same realization therefore share the white draw at every step, and runs that use the structured part (census-weighted in either register order, or structured only) also share the $G_{b}$, which act on each run's own pre-injection state. All $256^{3}$ noisy runs use one noise realization and the same initial condition, and the amplitude sweep ($\varepsilon = 10^{-6}$, $10^{-4}$ and $10^{-2}$) scales one draw sequence by $\varepsilon$: a linearity test, not a spread estimate. Noise acts from $t = 1.41$ (near the dissipation maximum) to $t = 3.3$, 946 noisy steps. Snapshots, taken after the step's injection, decoding and any field processing, give the stations $t = 1.5$, 2.0, 2.5 and 3.0 after 46, 296, 546 and 796 noisy steps; there is no snapshot at $t = 3.3$.

\subsection{What the model leaves out}\label{si:E.5}
The model is a lumped end-of-step perturbation of the post-selected system register; it acts on the system register only, with either backend, so no result involves ancilla dynamics. It has no one-qubit, idle or readout errors, no coherent or biased errors, no time correlation and no drift; readout, shot and measurement noise are not modelled, and post-selection is not sampled. The census split and ladder come from a collocated $16^{3}$ circuit and are untested at $256^{3}$, where 12 of the 24 address bits are extrapolated, and for the staggered layout, whose block-encoding circuit has no census. Whether end-of-step injection represents errors inside the circuit is untested. Folding gate errors into one effective per-step channel is itself an assumption: noise on quantum processors is layer-local and structured~\cite{vandenBerg2023}, folding it into white noise can fail for structured circuits~\cite{Foldager2024}, and a stochastic effective channel is justified only on average under randomized compiling~\cite{Wallman2016}, which is assumed here, not performed. The census-weighted model is a defined effective model, not a device model; every result of this note and of Supplementary Notes~\ref{note:F} and~\ref{note:G} is obtained within it, in exact classical emulation.

\subsection{Built-in contraction and the random part of the field error}\label{si:E.6}
Because (E1) preserves the norm, each injection shortens the component of the state along its pre-injection direction by exactly $1 - \varepsilon^{2}/2$, a property of the channel formula, not a measured property of any noise. Ignoring the dynamics between injections, this compounds to $(1 - \varepsilon^{2}/2)^{2N_{n}} \approx \exp(-N_{n}\varepsilon^{2})$ in energy over $N_{n}$ injections. At $\varepsilon = 10^{-2}$ the measured kinetic-energy deficit of the $256^{3}$ run is 84–91 \% of this prediction ($t = 1.5\text{--}3.3$), and at $64^{3}$ ($t = 3.3$) the deficit is 8.3 \% against 9.0 \% predicted. The field error is split by projecting the noisy field $V$ on the noise-free field $U$ over the four real fields,

\disp{$V = \hat{a}U + R$,\quad $\hat{a} = \langle V, U\rangle/\langle U, U\rangle$,\quad $\langle R, U\rangle = 0$,}{E5}
\noindent into a contraction part $|1 - \hat{a}|$ and a random part $\delta_{R} = \|R\|/\|U\|$, with $\|V - U\|^{2}/\|U\|^{2} = (1 - \hat{a})^{2} + \delta_{R}^{2}$ and $\|V\|^{2}/\|U\|^{2} = \hat{a}^{2} + \delta_{R}^{2}$ exactly ($\hat{a} - 1$ is evaluated as $\langle V - U, U\rangle/\langle U, U\rangle$, without cancellation). The random part exceeds the contraction part at every amplitude and station, by factors from 3.7 ($\varepsilon = 10^{-2}$, $t = 3.0$) to 524 ($\varepsilon = 10^{-6}$, $t = 1.5$) (Supplementary Table E2; Extended Data Fig.~\ref{edfig:amplitude}d), so the contamination is not merely the energy removed by the contraction. The measured $1 - \hat{a}$ is not a pure contraction: at small $\varepsilon$ it also contains a term of order $\varepsilon$, and it exceeds the predicted shrink 3.1–9.0-fold at $\varepsilon = 10^{-4}$ and 1.05–1.99-fold at $\varepsilon = 10^{-2}$.

\subsection{\texorpdfstring{The departure at $\varepsilon = 10^{-2}$ and the renormalized run}{The departure at ε = 10⁻² and the renormalized run}}\label{si:E.7}
In total $\delta$ (all bands, full state) the $\varepsilon = 10^{-2}$ run departs from the linear response by +13 \% at $t = 2.0$ and 2.5 and by +1.1 \% at $t = 3.0$ (Supplementary Table E2). Band by band (Extended Data Fig.~\ref{edfig:amplitude}b), the excess at $k < 8$ persists to $t = 3.0$ (+31, +82, +70 and +43 \%), while the response becomes sub-linear at high wavenumbers, from $t = 2.5$ in the band $k \geq 128$ and at $t = 3.0$ in all three bands at $k \geq 32$; the small $t = 3.0$ total is therefore a net (relative to linear, $\delta_{\mathrm{vel}}^{2}$ is +8.6 \% at $k < 32$ and $-6.4$ \% at $k \geq 32$). At $t = 1.5$ the $k < 8$ excess is a coherent band-wise shrink of the noise-free field equal to 0.97 of the built-in contraction; from $t = 2.0$ a perpendicular excess appears as well (+20, +27 and +18 \% at $k < 8$).

The renormalized run repeats the $\varepsilon = 10^{-2}$ run with two operations on the decoded field after every noisy step, both of which feed back because the field is re-encoded every step: renormalization by $1/(1 - \varepsilon^{2}/2)$, which undoes the contraction, and a real filter, which drops the imaginary part (exact for the noise-free dynamics, whose imaginary energy is about $10^{-25}$ of the total). It used the qubitized-pair backend, whereas the amplitude sweep used the system-only backend. With the contraction cancelled, every band stays within about 1 \% of linear up to $t = 2.0$, the $k < 8$ band stays within 1.5 \% of linear to $t = 3.0$ (0.997, 0.996, 0.993 and 0.985), and the kinetic-energy deficit at $t = 3.3$ falls from 7.57 \% to 0.46 \% (94 \% removed); a sub-linear response at $k \geq 32$ remains (0.89, 0.77 and 0.71 of linear in the bands 32–64, 64–128 and $\geq 128$ at $t = 3.0$), for which no mechanism is identified (Extended Data Fig.~\ref{edfig:amplitude}c). No $256^{3}$ run applies the real filter alone. At $64^{3}$ ($\varepsilon = 10^{-2}$, one backend) the real filter alone changes $\delta(3.0)$ by $-0.21$ \% and the $k < 8$ band ratio by $\leq 0.6$ \%, while filter plus renormalization returns the $k < 8$ ratio to 0.997 of linear; the low-wavenumber excess is therefore attributed mainly to the built-in contraction, with the real filter excluded as the cause only at $64^{3}$. At $64^{3}$, real-filtered and renormalized $\varepsilon = 10^{-2}$ runs with three noise realizations (qubitized pair) end at $-0.034$ \%, $-0.006$ \% and $-0.90$ \% in kinetic energy at $t = 3.3$, with $\delta(3.0) = 0.079$, 0.085 and 0.087 (0.095 for the unprocessed run of the first realization) and a raised enstrophy (+2.07 \%, +1.31 \% and +1.05 \%; the solver's finite-difference diagnostic of Supplementary Note~\ref{note:D}.6, quoted only as a relative change): model-level consequences that require a known $\varepsilon$, not a technique.

\textbf{Supplementary Table E2 | Amplitude dependence at $256^{3}$.} Rows 1–3, $\delta_{R}/|1 - \hat{a}|$ (E.6); other rows, ratio of $\delta/\varepsilon$, or of the velocity band amplitude per unit $\varepsilon$, between the runs named (1 is the linear response; percentages are the ratio minus 1; ``renormalized'', the run of E.7).

{\scriptsize
\setlength{\tabcolsep}{3pt}\renewcommand{\arraystretch}{1.15}\setlength{\LTleft}{0pt}\setlength{\LTright}{0pt}
\begin{longtable}{>{\raggedright\arraybackslash}p{\dimexpr 0.4000\linewidth-6pt\relax}>{\raggedright\arraybackslash}p{\dimexpr 0.1500\linewidth-6pt\relax}>{\raggedright\arraybackslash}p{\dimexpr 0.1500\linewidth-6pt\relax}>{\raggedright\arraybackslash}p{\dimexpr 0.1500\linewidth-6pt\relax}>{\raggedright\arraybackslash}p{\dimexpr 0.1500\linewidth-6pt\relax}}
\toprule
\hspace{0pt}\textbf{Quantity} & \hspace{0pt}\textbf{$t = 1.5$} & \hspace{0pt}\textbf{$t = 2.0$} & \hspace{0pt}\textbf{$t = 2.5$} & \hspace{0pt}\textbf{$t = 3.0$} \\
\midrule
\endhead
\hspace{0pt}$\delta_{R}/|1 - \hat{a}|$, $\varepsilon = 10^{-6}$ & \hspace{0pt}524.3 & \hspace{0pt}259.8 & \hspace{0pt}80.9 & \hspace{0pt}104.4 \\
\hspace{0pt}$\delta_{R}/|1 - \hat{a}|$, $\varepsilon = 10^{-4}$ & \hspace{0pt}385.9 & \hspace{0pt}169.6 & \hspace{0pt}68.9 & \hspace{0pt}82.8 \\
\hspace{0pt}$\delta_{R}/|1 - \hat{a}|$, $\varepsilon = 10^{-2}$ & \hspace{0pt}14.09 & \hspace{0pt}5.07 & \hspace{0pt}4.42 & \hspace{0pt}3.73 \\
\midrule
\hspace{0pt}$\varepsilon = 10^{-2}$ over $10^{-4}$, total $\delta$ & \hspace{0pt}+0.22 \% & \hspace{0pt}+13.05 \% & \hspace{0pt}+13.12 \% & \hspace{0pt}+1.07 \% \\
\hspace{0pt}$\varepsilon = 10^{-2}$ over $10^{-4}$, band $k < 8$ & \hspace{0pt}1.314 & \hspace{0pt}1.822 & \hspace{0pt}1.699 & \hspace{0pt}1.431 \\
\hspace{0pt}$\varepsilon = 10^{-2}$ over $10^{-4}$, band 8–16 & \hspace{0pt}1.041 & \hspace{0pt}1.270 & \hspace{0pt}1.217 & \hspace{0pt}1.117 \\
\hspace{0pt}$\varepsilon = 10^{-2}$ over $10^{-4}$, band 16–32 & \hspace{0pt}0.992 & \hspace{0pt}1.182 & \hspace{0pt}1.169 & \hspace{0pt}1.068 \\
\hspace{0pt}$\varepsilon = 10^{-2}$ over $10^{-4}$, band 32–64 & \hspace{0pt}0.999 & \hspace{0pt}1.150 & \hspace{0pt}1.134 & \hspace{0pt}0.984 \\
\hspace{0pt}$\varepsilon = 10^{-2}$ over $10^{-4}$, band 64–128 & \hspace{0pt}0.999 & \hspace{0pt}1.089 & \hspace{0pt}1.041 & \hspace{0pt}0.814 \\
\hspace{0pt}$\varepsilon = 10^{-2}$ over $10^{-4}$, band $\geq 128$ & \hspace{0pt}0.999 & \hspace{0pt}1.014 & \hspace{0pt}0.904 & \hspace{0pt}0.695 \\
\hspace{0pt}Renormalized over $\varepsilon = 10^{-2}$, total $\delta$ & \hspace{0pt}0.999 & \hspace{0pt}0.879 & \hspace{0pt}0.853 & \hspace{0pt}0.895 \\
\hspace{0pt}Renormalized over $\varepsilon = 10^{-4}$, total $\delta$ & \hspace{0pt}+0.09 \% & \hspace{0pt}$-0.65$ \% & \hspace{0pt}$-3.56$ \% & \hspace{0pt}$-9.55$ \% \\
\hspace{0pt}Renormalized over $\varepsilon = 10^{-4}$, band $k < 8$ & \hspace{0pt}0.997 & \hspace{0pt}0.996 & \hspace{0pt}0.993 & \hspace{0pt}0.985 \\
\bottomrule
\end{longtable}}

\section{Contamination measures, matched accumulation and the passive control}\label{note:F}

This note defines the contamination measures (F.1), the matched-accumulation references (F.2), the passive control (F.3), the $\Gamma$ statistics (F.4) and the decomposition identity (F.5), states when the criteria were fixed (F.6), and gives the $64^{3}$ stand-in (F.7) and the spectral and solenoidal diagnostics of Fig.~\ref{fig:feedback}d and Extended Data Fig.~\ref{edfig:structure}b (F.8); Supplementary Tables F1–F4 hold the values behind Fig.~\ref{fig:feedback} and Extended Data Fig.~\ref{edfig:passive}. Unless stated otherwise, values are for the census-weighted $\varepsilon = 10^{-4}$ run at $256^{3}$, per unit $\varepsilon$, up to $t = 3.0$, the last measured station.

\subsection{Contamination measures}\label{si:F.1}
For the noisy state $V$ and the noise-free state $U$ at equal step, $\Delta = V - U$ over the decoded complex fields ($u$, $v$, $w$, $p_{\mathrm{s}}$), with $p_{\mathrm{s}} = \mathrm{Ma} \cdot p$, and

\disp{$\delta = \|\Re\Delta\|/\|U\|$,\quad $\delta_{\Im} = \|\Im\Delta\|/\|U\|$}{F1}
\noindent over all four fields; $\delta$ measures the fed-back part and $\delta_{\Im}$ the passive part (F.3), and the complex contamination is $(\delta^{2} + \delta_{\Im}^{2})^{1/2}$. Velocity-only versions ($\delta_{\mathrm{vel}}$) use ($u$, $v$, $w$) and the velocity norm of $U$. Bands are those of Methods ($k < 8$, 8–16, 16–32, 32–64, 64–128 and $\geq 128$; ``$k < 8$'' is shells 1–7, 1,790 modes; ``$k = 1\text{--}4$'' is shells 1–3, 178 modes), and band amplitudes $(E_{X}(b)/E_{U}(b))^{1/2}$ are quoted per unit $\varepsilon$. The imaginary part of $U$, a uniform global phase of $2\text{--}6 \times 10^{-13}$, lies more than six orders of magnitude below the smallest $\delta_{\Im}$ used ($3.27 \times 10^{-6}$). The values depend on the convention: from $t = 1.5$ to 3.0, $\delta/\varepsilon$ grows 8.73-fold, its velocity-only version 9.86-fold and the complex contamination 6.35-fold (Supplementary Table F1). The encoded-pressure contamination per unit $\varepsilon$, $(p_{\mathrm{s}}\ \text{share of the contamination energy})^{1/2} \cdot \delta/\varepsilon$, is nearly equal in the two parts at every station (2.15 against 2.11 at $t = 3.0$), so the feedback adds no encoded-pressure (acoustic) error.

\subsection{Matched accumulation references}\label{si:F.2}
The references are per unit $\varepsilon$, for the real or the imaginary injection component $x$, for the full state or the velocity only. Their inputs are: (i) the per-mode spectra of the unit injected vector $n_{\perp}$ for 16 draws (the four steps from one before to two after each station), regenerated with the runs' own random sequence on the noise-free state, which stands in for the noisy state because the direction depends on the state only through its direction (the two differ by $\leq 3 \times 10^{-3}$ relative at $\varepsilon = 10^{-4}$); per draw the spectra over $x$ (real or imaginary velocity or $p_{\mathrm{s}}$) and $k$ sum to one, $S_{x}(k; j)$ is the four-draw mean of the station nearest step $j$, $s_{x}(j) = \sum_{k} S_{x}(k; j)$, and $s_{x}$ is the 16-draw mean; (ii) the noise-free norm $\|U(j)\|$ at every noisy step $j$, from the kinetic energy with the $p_{\mathrm{s}}$ share interpolated between snapshots (error $\leq 4.5 \times 10^{-4}$ of $\|U\|^{2}$), with $N_{n}$ the number of noisy steps up to step $n$ and $\|U_{\mathrm{set}}(n)\|$ the noise-free norm at the observation step; and (iii) $\lambda(k) = -(1/\mathrm{Re})(4/h^{2})\sum_{a}\sin^{2}(\pi k_{a}/N)$, the exact seven-point-Laplacian eigenvalue, applied as $e^{\lambda\,\Delta t}$ per step. Then

\disp{$R_{0}$ $= (N_{n}\,s_{x})^{1/2}$,\quad $R_{1}$ $= \bigl[\sum_{j} \|U(j)\|^{2}\,s_{x}(j)\bigr]^{1/2}/\|U_{\mathrm{set}}(n)\|$,\quad $R_{1\mathrm{d}}$ $= \bigl[\sum_{k}\sum_{j} \|U(j)\|^{2}\,S_{x}(k; j)\,e^{2\lambda(k)(n - j)\Delta t}\bigr]^{1/2}/\|U_{\mathrm{set}}(n)\|$.}{F2}
Full-state versions sum the velocity and $p_{\mathrm{s}}$ parts and use the full-state norm; velocity-only versions use the velocity part and norm. $R_{0}$, the textbook constant-norm walk, ignores the decay of the noise-free norm and is not a matched reference (Extended Data Fig.~\ref{edfig:passive}c). $R_{1}$ is matched non-dissipative accumulation (every injection keeps its norm), $R_{1\mathrm{d}}$ adds the exact decay of each injected mode, and $B_{2}$ and $B_{2\mathrm{d}}$ are the same per velocity band, relative to the noise-free band energy. Nothing is fitted; $R_{1\mathrm{d}}$ and $B_{2\mathrm{d}}$ contain no advection, acoustic exchange or feedback and are references, not predictions of passive transport. Real and imaginary versions differ by at most 1.8 \%, an independent recomputation of $R_{1\mathrm{d}}$ agrees to within 0.8 \%, and at $k = 1\text{--}4$ the four-draw band references carry about $\pm 6\text{--}12$ \% amplitude sampling uncertainty. Ratios are classified BELOW ($< 0.90$), CONSISTENT (0.90–1.10) or ABOVE ($> 1.10$); this predefined $\pm 10$ \% tolerance applies to the all-band ratios against $R_{0}$, $R_{1}$ and $R_{1\mathrm{d}}$ (full state and velocity only), not to the band ratios against $B_{2}$ and $B_{2\mathrm{d}}$, which have no criterion (Extended Data Fig.~\ref{edfig:passive}b).

\subsection{The passive control}\label{si:F.3}
A code-path audit of the production step (Supplementary Note~\ref{note:B}) shows that the refresh reads only real parts, to rounding: the predictor uses $u_{\mathrm{mid}} = \Re(\psi_{\mathrm{mid}}^{u}) \cdot \|q_{n}\|$; the coefficient kernel takes only the real part of the decoded velocity; the corrector weights and $\lambda_{A}$ come from the real $u_{\mathrm{mid}}$; and the GQSP polynomials (real-matrix-valued to $\leq 2.5 \times 10^{-15}$), the encoding norm and the post-selection gain (8) are real. In a direct test ($32^{3}$, production initial condition), a purely imaginary perturbation of relative size $10^{-4}\text{--}10^{-2}$ changed the real part after ten production steps by $\leq 1.6 \times 10^{-15}$, independently of its size. Both parts are therefore propagated by the same real operator (advection by the actual flow, acoustic coupling and diffusion), but only the real velocity contamination $u'$ changes the advection operator at the next refresh, acting on $U$ through the linearized advective term $(u' \cdot \nabla)U$, plus $\tfrac{1}{2}(\nabla \cdot u')U$ in the skew-symmetric form, the standard coupling of the Navier–Stokes equations linearized about a base state~\cite{Moulin2019}. The imaginary part is thus an approximately passive control, approximate in three measured respects (Supplementary Table F3).

(i) The channel uses the full complex state ($G_{b}\psi$, $\langle \psi, d\rangle$ and $\|\psi\|$), so each injection exchanges a fraction of order $\kappa\varepsilon$ between the parts, with the channel's real–imaginary exchange coefficient $\kappa = \|\Delta\Re(\text{injection})\|/(\varepsilon\|\Im\psi\|) = 0.172$ at $256^{3}$; per step this is $\kappa\delta_{\Im}/\sqrt{s_{\Re}}$ of the real injection, 0.8–1.1 \% at $\varepsilon = 10^{-2}$ and 0.008–0.011 \% at $10^{-4}$. It is the only route back from the passive part: at $64^{3}$ a run whose channel sees only the real part reproduces the real-filtered run to $1.3 \times 10^{-13}$ of the contamination. (ii) The built-in contraction (Supplementary Note~\ref{note:E}) acts on the essentially real state at order $\varepsilon^{2}$ per step, which shows as a real-only $k = 1\text{--}4$ departure between $\varepsilon = 10^{-4}$ and $10^{-6}$ of up to 1.33 \%. Both asymmetries vanish with $\varepsilon$ relative to the linear response: $\varepsilon = 10^{-4}$ and $10^{-6}$, a factor 100 apart, give the same $\Gamma_{1}$ within $6.6 \times 10^{-4}$, which bounds both below $10^{-3}$ of the effect in total $\delta$ (Extended Data Fig.~\ref{edfig:passive}f). (iii) The two parts of each injection are different random fields with equal statistics (0.498 and 0.502 of the injected energy over 16 draws). At $k = 1\text{--}4$ the match holds only in expectation (injection real-to-imaginary ratio 1.19 in amplitude; per draw 0.78–1.80), so the census-weighted ratio there (1.30, 1.05, 1.23 and 1.18 at the four stations) resolves no feedback excess. The band does not appear free of feedback: in the white-only run, which injects little there, the ratio grows from 1.08 at $t = 1.5$ to 4.26 at $t = 3.0$, and by the superposition of Supplementary Note~\ref{note:G} about 14 \% of the census-weighted $k = 1\text{--}4$ energy at $t = 3.0$ is feedback-generated (Extended Data Fig.~\ref{edfig:passive}d).

\subsection{\texorpdfstring{$\Gamma$ statistics relative to the passive control}{Γ statistics relative to the passive control}}\label{si:F.4}
With $s_{\Re}$ and $s_{\Im}$ the 16-draw mean shares of the full-state injection energy, the two statistics relative to the passive control are

\disp{$\Gamma_{1} = (\delta_{\Re}/\sqrt{s_{\Re}})/(\delta_{\Im}/\sqrt{s_{\Im}})$,\quad $\Gamma_{2} = (\delta_{\Re}/R_{1\mathrm{d},\Re})/(\delta_{\Im}/R_{1\mathrm{d},\Im})$,}{F3}
\noindent with $\delta_{\Re} \equiv \delta$. $\Gamma_{1}$ equals 1 for identical dynamics of the two parts and removes the number of injections, the decaying noise-free norm and the initial real-to-imaginary weight; $\Gamma_{2}$ further removes any real–imaginary difference of the injection spectra under exact diffusion. Velocity-only versions use the velocity shares (0.3813 and 0.3857). The band version $\Gamma_{1,b}$ uses the 16-draw band fractions of the velocity injection; because the exact normalizer would also carry $(s_{\Im,\mathrm{vel}}/s_{\Re,\mathrm{vel}})^{1/2}$, the reported $\Gamma_{1,b}$ is low by 0.57 \% (12.97 would be 13.05 at $k = 16\text{--}32$, $t = 3.0$).

$\Gamma_{1} = 1.04$, 1.91, 3.87 and 6.65 and $\Gamma_{2} = 1.03$, 1.92, 3.84 and 6.74 at the four stations, while $\delta/R_{1} = 0.689$, 0.544, 0.757 and 1.007. $\Gamma_{1}$ exceeds the guide value 1.10 (F.6) from $t = 2.0$ on. At $t = 3.0$ two alternative share normalizations give 6.36 (the single draw at $t = 1.5$, shares 0.520 and 0.480) and 6.38 (the measured $t = 1.5$ ratio as the baseline) against 6.65 for the predefined 16-draw mean, physically the best choice but also the largest value; 6.4–6.7 is therefore a share-normalization range, not a statistical spread. $\Gamma_{2} = 6.74$ needs no share normalization and lies just outside that range. Per band at $t = 3.0$, $\Gamma_{1,b}$ is 8.5–13.0 in the three bands spanning $k = 8\text{--}64$, against 1.9–4.7 at $k < 8$, 64–128 and $\geq 128$ (Supplementary Table F2; Extended Data Fig.~\ref{edfig:passive}a).

\subsection{Decomposition identity}\label{si:F.5}
By construction,

\disp{$\delta/R_{1} = \Gamma_{1} \times (\delta_{\Im}/R_{1,\Im}) \times f$,\quad $f = (s_{\Re}/s_{\Im})^{1/2} \cdot R_{1,\Im}/R_{1,\Re}$}{F4}
Here $f = 0.984$, 0.989, 0.992 and 0.994 at the four stations. At $t = 3.0$, $6.652 \times 0.1523 \times 0.994 = 1.007$, and on the logarithmic axis of Fig.~\ref{fig:feedback}c the gap between $\delta/R_{1}$ and $\delta_{\Im}/R_{1,\Im}$ equals $\Gamma_{1}$ within 1.6 \%. The identity restates the measurements and is not an independent check. It shows that the near-equality with matched accumulation at $t = 3.0$ combines a passive part that keeps only 15 \% of non-dissipative accumulation ($\delta_{\Im}/R_{1,\Im} = 0.152$), because its grid-scale input dissipates, with a fed-back part 6.65 times the passive part after share normalization: the near-equality is not evidence of passivity. $\delta/R_{1}$ rises from 0.544 at $t = 2.0$ to 0.757 and 1.007, so $t = 3.0$ is a crossing at the last measured station, and nothing is known after it; in the velocity-only convention the $t = 3.0$ ratio is 1.147 (ABOVE). The passive part itself stays near accumulation plus diffusion: within 9 \% of $R_{1\mathrm{d}}$ in the full state (0.874 of it for the velocity only at $t = 3.0$) and, per band at $t = 3.0$, within about 5 \% of $B_{2\mathrm{d}}$ for $k \geq 32$ (0.947–0.991) but 19–32 \% below it for $k < 32$, as expected for a field that advection carries toward small scales, where it dissipates (Extended Data Fig.~\ref{edfig:passive}b). Band ratios of the fed-back part to $B_{2}$ or $B_{2\mathrm{d}}$ are not attributed to feedback by themselves; that attribution rests on the passive control.

\subsection{Predefinition and timing}\label{si:F.6}
The definitions of $\Gamma_{1}$ and $\Gamma_{2}$, the guide value $\Gamma_{1} > 1.10$ and the $\pm 10$ \% tolerance were fixed before $\Gamma_{1}$ or any reference value was computed, but not blind: running logs had already shown $\delta/\varepsilon$ and $\delta_{\Im}/\varepsilon$ at two stations, and the raw real-to-imaginary ratio ($1.04 \to 6.6$) was known. The class does not depend on that choice ($\Gamma_{1} = 6.65$ against 1.10 at $t = 3.0$, with $\delta/R_{1} = 1.007$); the rationale of the guide is that the imaginary-to-real exchange and the per-draw spread of the shares lie far below 10 \%.

\subsection{\texorpdfstring{The $64^{3}$ stand-in (context only)}{The 64³ stand-in (context only)}}\label{si:F.7}
Eight noise realizations of a weakly chaotic $64^{3}$ Re 120 stand-in (same initial condition, census-weighted model, z-first order, $\varepsilon = 10^{-4}$) give a coefficient of variation of $\delta/\varepsilon$ of 1.36–4.91 \% over the four stations and 16.6–19.3 \% in the band $k = 1\text{--}4$ (Extended Data Table~\ref{edtab:grids}; Supplementary Table F4). Their individual $\Gamma_{1}$ exceeds 1 in all eight from $t = 2.0$ but 1.10 in all eight only at $t = 3.0$ (1.15–1.37; Extended Data Fig.~\ref{edfig:passive}e), far weaker than in the single $256^{3}$ realization (6.65). The configurations differ beyond the realization (a same-initial-condition twin perturbation in the spectral solver grows $\times 2.34$ at $64^{3}$ against $\times 34.3$ at $256^{3}$; $N$, Re and further factors co-vary), so these statistics are context, not an uncertainty bar for $256^{3}$.

\subsection{Spectral and solenoidal diagnostics}\label{si:F.8}
Velocity shell spectra use the real-to-complex Fourier transform with half-space weights (2; 1 on $k_{z} = 0$ and $k_{z} = N/2$), shells $n$ given by $|k|$ rounded to the nearest integer, and bands that include their lower and exclude their upper shell bound. The velocity-error spectra of Fig.~\ref{fig:feedback}d are normalized as $E_{\Delta}(k)/(\varepsilon^{2}\sum_{k}E_{U})$, so that each sums to $(\delta_{\mathrm{vel}}/\varepsilon)^{2}$: $3.00^{2}$ at $t = 1.5$ and $29.53^{2}$ at $t = 3.0$ for the fed-back (real) part, $2.87^{2}$ and $3.94^{2}$ for the passive (imaginary) part. The noise-free dissipation spectrum $k^{2}E_{U}(k)$ is drawn on an independent right axis; the two peak heights are not comparable. The injection is grid-scale: 90.8 \% of the velocity energy of its real part and 90.6 \% of that of its imaginary part lie above $k = 64$ (16 draws). At $t = 3.0$, 0.74 of the fed-back velocity-error energy and 0.71 of the noise-free dissipation lie in shells 16–63 (dissipation 0.57 at $t = 1.5$); the fed-back error peaks at shell 21 at $t = 3.0$, where the dissipation spectrum also peaks at $t = 1.5$ and 3.0. Both parts lose grid-scale energy: the passive part's share above $k = 64$ falls from 0.91 at injection to 0.81 at $t = 1.5$ and 0.53 at $t = 3.0$. Changes shared by both parts are not attributed to the feedback; their differences ($\Gamma_{1,b}$) are consistent with it but are not attributed to it alone, because the two parts are different random fields weakly coupled by the channel (F.3). No inter-scale flux was measured. In the MAC projection (Methods; Extended Data Fig.~\ref{edfig:structure}b), the dilatational energy fraction of the fed-back velocity error falls from 0.288 at $t = 1.5$ to 0.0051 at $t = 3.0$, whereas that of the passive part stays near the white-noise value of 1/3 (sampled 0.333): 0.305 at $t = 1.5$ and 0.289 at $t = 3.0$; that of the noise-free flow is $2.1 \times 10^{-6}$ and $1.0 \times 10^{-5}$ at the same times. Because the encoded-pressure contamination is nearly the same in both parts (F.1), the falling dilatational fraction of the fed-back part reflects the growth of its solenoidal part, not a reduction of pressure error.

\textbf{Supplementary Table F1 | Contamination, references and statistics per station (census-weighted $\varepsilon = 10^{-4}$ run; per unit $\varepsilon$ where applicable).} Upper block, full state; lower block, velocity only (velocity part of each component and velocity norm of $U$; the velocity version of $R_{0}$ carries the factor $\|U_{\mathrm{full}}\|^{2}/\|U_{\mathrm{vel}}\|^{2}$ at noise onset). Classes use the predefined $\pm 10$ \% tolerance (B, BELOW; C, CONSISTENT; A, ABOVE).

{\scriptsize
\setlength{\tabcolsep}{3pt}\renewcommand{\arraystretch}{1.15}\setlength{\LTleft}{0pt}\setlength{\LTright}{0pt}
\begin{longtable}{>{\raggedright\arraybackslash}p{\dimexpr 0.4707\linewidth-6pt\relax}>{\raggedright\arraybackslash}p{\dimexpr 0.1245\linewidth-6pt\relax}>{\raggedright\arraybackslash}p{\dimexpr 0.1349\linewidth-6pt\relax}>{\raggedright\arraybackslash}p{\dimexpr 0.1349\linewidth-6pt\relax}>{\raggedright\arraybackslash}p{\dimexpr 0.1349\linewidth-6pt\relax}}
\toprule
\hspace{0pt}\textbf{Quantity} & \hspace{0pt}\textbf{$t = 1.5$} & \hspace{0pt}\textbf{$t = 2.0$} & \hspace{0pt}\textbf{$t = 2.5$} & \hspace{0pt}\textbf{$t = 3.0$} \\
\midrule
\endhead
\hspace{0pt}Noisy steps & \hspace{0pt}46 & \hspace{0pt}296 & \hspace{0pt}546 & \hspace{0pt}796 \\
\hspace{0pt}$\delta/\varepsilon$ (fed-back part) & \hspace{0pt}3.39 & \hspace{0pt}7.75 & \hspace{0pt}16.67 & \hspace{0pt}29.61 \\
\hspace{0pt}$\delta_{\Im}/\varepsilon$ (passive part) & \hspace{0pt}3.27 & \hspace{0pt}4.08 & \hspace{0pt}4.33 & \hspace{0pt}4.47 \\
\hspace{0pt}$(\delta^{2} + \delta_{\Im}^{2})^{1/2}/\varepsilon$ & \hspace{0pt}4.71 & \hspace{0pt}8.76 & \hspace{0pt}17.22 & \hspace{0pt}29.95 \\
\hspace{0pt}$R_{0}$, real / imaginary & \hspace{0pt}4.79 / 4.81 & \hspace{0pt}12.14 / 12.19 & \hspace{0pt}16.49 / 16.56 & \hspace{0pt}19.91 / 19.99 \\
\hspace{0pt}$R_{1}$, real / imaginary & \hspace{0pt}4.92 / 4.87 & \hspace{0pt}14.25 / 14.16 & \hspace{0pt}22.01 / 21.92 & \hspace{0pt}29.40 / 29.36 \\
\hspace{0pt}$R_{1\mathrm{d}}$, real / imaginary & \hspace{0pt}3.28 / 3.25 & \hspace{0pt}4.20 / 4.24 & \hspace{0pt}4.58 / 4.57 & \hspace{0pt}4.82 / 4.91 \\
\hspace{0pt}$\delta/R_{0}$ & \hspace{0pt}0.709 (B) & \hspace{0pt}0.638 (B) & \hspace{0pt}1.011 (C) & \hspace{0pt}1.487 (A) \\
\hspace{0pt}$\delta/R_{1}$ & \hspace{0pt}0.689 (B) & \hspace{0pt}0.544 (B) & \hspace{0pt}0.757 (B) & \hspace{0pt}1.007 (C) \\
\hspace{0pt}$\delta/R_{1\mathrm{d}}$ & \hspace{0pt}1.034 (C) & \hspace{0pt}1.847 (A) & \hspace{0pt}3.637 (A) & \hspace{0pt}6.141 (A) \\
\hspace{0pt}$\delta_{\Im}/R_{0,\Im}$ & \hspace{0pt}0.680 (B) & \hspace{0pt}0.335 (B) & \hspace{0pt}0.262 (B) & \hspace{0pt}0.224 (B) \\
\hspace{0pt}$\delta_{\Im}/R_{1,\Im}$ & \hspace{0pt}0.672 (B) & \hspace{0pt}0.288 (B) & \hspace{0pt}0.198 (B) & \hspace{0pt}0.152 (B) \\
\hspace{0pt}$\delta_{\Im}/R_{1\mathrm{d},\mathrm{im}}$ & \hspace{0pt}1.007 (C) & \hspace{0pt}0.963 (C) & \hspace{0pt}0.948 (C) & \hspace{0pt}0.911 (C) \\
\hspace{0pt}$\delta/R_{1}$ at $\varepsilon = 10^{-2}$ & \hspace{0pt}0.691 (B) & \hspace{0pt}0.615 (B) & \hspace{0pt}0.857 (B) & \hspace{0pt}1.018 (C) \\
\hspace{0pt}$\Gamma_{1}$ & \hspace{0pt}1.042 & \hspace{0pt}1.907 & \hspace{0pt}3.865 & \hspace{0pt}6.652 \\
\hspace{0pt}$\Gamma_{1}$ at $\varepsilon = 10^{-6}$ & \hspace{0pt}1.042 & \hspace{0pt}1.907 & \hspace{0pt}3.863 & \hspace{0pt}6.648 \\
\hspace{0pt}$\Gamma_{2}$ & \hspace{0pt}1.026 & \hspace{0pt}1.918 & \hspace{0pt}3.838 & \hspace{0pt}6.741 \\
\hspace{0pt}f in (F4) & \hspace{0pt}0.984 & \hspace{0pt}0.989 & \hspace{0pt}0.992 & \hspace{0pt}0.994 \\
\hspace{0pt}Encoded-pressure contamination per unit $\varepsilon$, fed-back / passive & \hspace{0pt}1.59 / 1.58 & \hspace{0pt}1.96 / 1.97 & \hspace{0pt}2.07 / 2.06 & \hspace{0pt}2.15 / 2.11 \\
\midrule
\hspace{0pt}$\delta_{\mathrm{vel}}/\varepsilon$, real part & \hspace{0pt}3.00 & \hspace{0pt}7.50 & \hspace{0pt}16.54 & \hspace{0pt}29.53 \\
\hspace{0pt}$\delta_{\mathrm{vel}}/\varepsilon$, imaginary part & \hspace{0pt}2.87 & \hspace{0pt}3.58 & \hspace{0pt}3.81 & \hspace{0pt}3.94 \\
\hspace{0pt}$R_{0}$ (velocity), real / imaginary & \hspace{0pt}4.19 / 4.21 & \hspace{0pt}10.63 / 10.69 & \hspace{0pt}14.43 / 14.51 & \hspace{0pt}17.43 / 17.53 \\
\hspace{0pt}$R_{1}$ (velocity), real / imaginary & \hspace{0pt}4.32 / 4.25 & \hspace{0pt}12.50 / 12.39 & \hspace{0pt}19.29 / 19.19 & \hspace{0pt}25.75 / 25.70 \\
\hspace{0pt}$R_{1\mathrm{d}}$ (velocity), real / imaginary & \hspace{0pt}2.89 / 2.85 & \hspace{0pt}3.76 / 3.81 & \hspace{0pt}4.15 / 4.14 & \hspace{0pt}4.41 / 4.51 \\
\hspace{0pt}Velocity, real/$R_{0}$ & \hspace{0pt}0.715 (B) & \hspace{0pt}0.706 (B) & \hspace{0pt}1.146 (A) & \hspace{0pt}1.695 (A) \\
\hspace{0pt}Velocity, real/$R_{1}$ & \hspace{0pt}0.694 (B) & \hspace{0pt}0.600 (B) & \hspace{0pt}0.858 (B) & \hspace{0pt}1.147 (A) \\
\hspace{0pt}Velocity, real/$R_{1\mathrm{d}}$ & \hspace{0pt}1.035 (C) & \hspace{0pt}1.994 (A) & \hspace{0pt}3.984 (A) & \hspace{0pt}6.691 (A) \\
\hspace{0pt}Velocity, imaginary/$R_{0}$ & \hspace{0pt}0.680 (B) & \hspace{0pt}0.335 (B) & \hspace{0pt}0.262 (B) & \hspace{0pt}0.225 (B) \\
\hspace{0pt}Velocity, imaginary/$R_{1}$ & \hspace{0pt}0.674 (B) & \hspace{0pt}0.289 (B) & \hspace{0pt}0.199 (B) & \hspace{0pt}0.153 (B) \\
\hspace{0pt}Velocity, imaginary/$R_{1\mathrm{d}}$ & \hspace{0pt}1.004 (C) & \hspace{0pt}0.940 (C) & \hspace{0pt}0.920 (C) & \hspace{0pt}0.874 (B) \\
\hspace{0pt}$\Gamma_{1}$, velocity & \hspace{0pt}1.052 & \hspace{0pt}2.109 & \hspace{0pt}4.367 & \hspace{0pt}7.536 \\
\hspace{0pt}$\Gamma_{2}$, velocity & \hspace{0pt}1.031 & \hspace{0pt}2.122 & \hspace{0pt}4.328 & \hspace{0pt}7.653 \\
\bottomrule
\end{longtable}}

\textbf{Supplementary Table F2 | Band-resolved values at $t = 3.0$ (velocity, census-weighted $\varepsilon = 10^{-4}$ run; band amplitudes per unit $\varepsilon$).}

{\scriptsize
\setlength{\tabcolsep}{3pt}\renewcommand{\arraystretch}{1.15}\setlength{\LTleft}{0pt}\setlength{\LTright}{0pt}
\begin{longtable}{>{\raggedright\arraybackslash}p{\dimexpr 0.2069\linewidth-6pt\relax}>{\raggedright\arraybackslash}p{\dimexpr 0.1067\linewidth-6pt\relax}>{\raggedright\arraybackslash}p{\dimexpr 0.1089\linewidth-6pt\relax}>{\raggedright\arraybackslash}p{\dimexpr 0.1089\linewidth-6pt\relax}>{\raggedright\arraybackslash}p{\dimexpr 0.1111\linewidth-6pt\relax}>{\raggedright\arraybackslash}p{\dimexpr 0.1177\linewidth-6pt\relax}>{\raggedright\arraybackslash}p{\dimexpr 0.1415\linewidth-6pt\relax}>{\raggedright\arraybackslash}p{\dimexpr 0.0981\linewidth-6pt\relax}}
\toprule
\hspace{0pt}\textbf{Band} & \hspace{0pt}\textbf{$k < 8$} & \hspace{0pt}\textbf{8–16} & \hspace{0pt}\textbf{16–32} & \hspace{0pt}\textbf{32–64} & \hspace{0pt}\textbf{64–128} & \hspace{0pt}\textbf{$\geq 128$} & \hspace{0pt}\textbf{$k = 1\text{--}4$} \\
\midrule
\endhead
\hspace{0pt}Real part: measured / $B_{2}$ / $B_{2\mathrm{d}}$ & \hspace{0pt}4.44 / 2.23 / 2.16 & \hspace{0pt}18.02 / 3.80 / 3.20 & \hspace{0pt}38.91 / 7.22 / 4.11 & \hspace{0pt}73.43 / 26.46 / 6.99 & \hspace{0pt}136.55 / 220.57 / 28.91 & \hspace{0pt}269.26 / 1,433.27 / 139.42 & \hspace{0pt}1.65 / 1.77 / 1.76 \\
\hspace{0pt}Imaginary part: measured / $B_{2}$ / $B_{2\mathrm{d}}$ & \hspace{0pt}1.67 / 2.13 / 2.06 & \hspace{0pt}2.21 / 3.81 / 3.23 & \hspace{0pt}3.08 / 7.41 / 4.20 & \hspace{0pt}6.64 / 26.99 / 7.01 & \hspace{0pt}29.27 / 219.29 / 30.81 & \hspace{0pt}140.29 / 1,431.36 / 141.57 & \hspace{0pt}1.40 / 1.55 / 1.54 \\
\hspace{0pt}Imaginary part / $B_{2\mathrm{d}}$ & \hspace{0pt}0.811 & \hspace{0pt}0.684 & \hspace{0pt}0.735 & \hspace{0pt}0.947 & \hspace{0pt}0.950 & \hspace{0pt}0.991 & \hspace{0pt}0.905 \\
\hspace{0pt}Real / imaginary, measured & \hspace{0pt}2.66 & \hspace{0pt}8.15 & \hspace{0pt}12.62 & \hspace{0pt}11.06 & \hspace{0pt}4.67 & \hspace{0pt}1.92 & \hspace{0pt}1.18 \\
\hspace{0pt}$\Gamma_{1,b}$ & \hspace{0pt}2.48 & \hspace{0pt}8.46 & \hspace{0pt}12.97 & \hspace{0pt}11.18 & \hspace{0pt}4.67 & \hspace{0pt}1.91 & \hspace{0pt}— \\
\hspace{0pt}$\Gamma_{2,b}$ & \hspace{0pt}2.53 & \hspace{0pt}8.25 & \hspace{0pt}12.90 & \hspace{0pt}11.09 & \hspace{0pt}4.97 & \hspace{0pt}1.95 & \hspace{0pt}— \\
\hspace{0pt}Injected velocity energy share, real / imaginary & \hspace{0pt}0.0034 / 0.0030 & \hspace{0pt}0.0057 / 0.0061 & \hspace{0pt}0.0161 / 0.0170 & \hspace{0pt}0.0668 / 0.0683 & \hspace{0pt}0.4600 / 0.4618 & \hspace{0pt}0.4480 / 0.4437 & \hspace{0pt}— \\
\bottomrule
\end{longtable}}

\textbf{Supplementary Table F3 | Approximations of the passive control and their measured size.}

{\footnotesize
\setlength{\tabcolsep}{3pt}\renewcommand{\arraystretch}{1.15}\setlength{\LTleft}{0pt}\setlength{\LTright}{0pt}
\begin{longtable}{>{\raggedright\arraybackslash}p{\dimexpr 0.4707\linewidth-6pt\relax}>{\raggedright\arraybackslash}p{\dimexpr 0.5293\linewidth-6pt\relax}}
\toprule
\hspace{0pt}\textbf{Item} & \hspace{0pt}\textbf{Measured size} \\
\midrule
\endhead
\hspace{0pt}Deterministic route from the imaginary to the real part ($32^{3}$, ten production steps, imaginary perturbation $10^{-4}\text{--}10^{-2}$) & \hspace{0pt}$\leq 1.6 \times 10^{-15}$, independent of size; the imaginary part keeps 0.9966 of its norm \\
\hspace{0pt}$64^{3}$ run with the channel fed the real part only (46 noisy steps, $\varepsilon = 10^{-2}$) & \hspace{0pt}real part equals the real-filtered run to $1.3 \times 10^{-13}$ of the contamination (unfiltered run: 1.216 \%) \\
\hspace{0pt}Channel real–imaginary exchange coefficient $\kappa$ & \hspace{0pt}0.172 at $256^{3}$ (reverse 0.171; measured on the $\varepsilon = 10^{-2}$ state at $t = 1.5$, identical at $\varepsilon = 10^{-4}$; 0.172 from the state-dependent direction alone, 0.018 from projection and norm alone); 0.283 at $64^{3}$ \\
\hspace{0pt}Per-step exchange relative to the real injection, $\kappa\delta_{\Im}/\sqrt{s_{\Re}}$ & \hspace{0pt}0.8–1.1 \% ($\varepsilon = 10^{-2}$), 0.008–0.011 \% ($10^{-4}$), $8\text{--}11 \times 10^{-5}$ \% ($10^{-6}$) \\
\hspace{0pt}Built-in contraction (real part only), $k = 1\text{--}4$ departure between $\varepsilon = 10^{-4}$ and $10^{-6}$ & \hspace{0pt}real +0.37 / +0.94 / +1.33 / +0.96 \%; imaginary $\leq 0.024$ \% \\
\hspace{0pt}$\varepsilon = 10^{-4}$ against $10^{-6}$, maximum relative difference & \hspace{0pt}$\Gamma_{1}$ $6.6 \times 10^{-4}$; $\delta$ $6.2 \times 10^{-4}$; $\delta_{\Im}$ $4.2 \times 10^{-5}$ \\
\hspace{0pt}Injection energy shares, 16 draws & \hspace{0pt}real 0.498 (0.474–0.525); imaginary 0.502 (0.475–0.526) \\
\hspace{0pt}Real against imaginary injection, by component & \hspace{0pt}white part 50/50 to $10^{-4}$; $p_{\mathrm{s}}$ shares equal (0.1165 each); velocity band shares within 0.005 in every band (largest 0.0043, $k \geq 128$); $k < 8$ shares differ by 14 \% relative \\
\hspace{0pt}$k = 1\text{--}4$ injection, real/imaginary & \hspace{0pt}1.19 in amplitude (1.42 in energy); per draw 0.78–1.80 in amplitude, 0.61–3.25 in energy \\
\hspace{0pt}Injections regenerated on the noise-free state, against the injections of the runs & \hspace{0pt}reproduced to $\leq 1.4 \times 10^{-13}$ \\
\bottomrule
\end{longtable}}

\textbf{Supplementary Table F4 | The $64^{3}$ Re 120 stand-in: eight noise realizations (same initial condition, census-weighted model, z-first order, $\varepsilon = 10^{-4}$).} CV, coefficient of variation (sample standard deviation over the mean); SD, sample standard deviation.

{\scriptsize
\setlength{\tabcolsep}{3pt}\renewcommand{\arraystretch}{1.15}\setlength{\LTleft}{0pt}\setlength{\LTright}{0pt}
\begin{longtable}{>{\raggedright\arraybackslash}p{\dimexpr 0.0691\linewidth-6pt\relax}>{\raggedright\arraybackslash}p{\dimexpr 0.1209\linewidth-6pt\relax}>{\raggedright\arraybackslash}p{\dimexpr 0.1606\linewidth-6pt\relax}>{\raggedright\arraybackslash}p{\dimexpr 0.1140\linewidth-6pt\relax}>{\raggedright\arraybackslash}p{\dimexpr 0.1623\linewidth-6pt\relax}>{\raggedright\arraybackslash}p{\dimexpr 0.1866\linewidth-6pt\relax}>{\raggedright\arraybackslash}p{\dimexpr 0.0742\linewidth-6pt\relax}>{\raggedright\arraybackslash}p{\dimexpr 0.1123\linewidth-6pt\relax}}
\toprule
\hspace{0pt}\textbf{t} & \hspace{0pt}\textbf{$\delta/\varepsilon$, mean} & \hspace{0pt}\textbf{$\delta/\varepsilon$, CV} & \hspace{0pt}\textbf{$\Gamma_{1}$, mean} & \hspace{0pt}\textbf{$\Gamma_{1}$, SD} & \hspace{0pt}\textbf{$\Gamma_{1}$, range} & \hspace{0pt}\textbf{$\Gamma_{1} > 1$} & \hspace{0pt}\textbf{$\Gamma_{1} > 1.10$} \\
\midrule
\endhead
\hspace{0pt}1.5 & \hspace{0pt}3.998 & \hspace{0pt}1.36 \% & \hspace{0pt}1.004 & \hspace{0pt}0.018 & \hspace{0pt}0.974–1.030 & \hspace{0pt}4/8 & \hspace{0pt}0/8 \\
\hspace{0pt}2.0 & \hspace{0pt}6.378 & \hspace{0pt}2.29 \% & \hspace{0pt}1.072 & \hspace{0pt}0.035 & \hspace{0pt}1.024–1.124 & \hspace{0pt}8/8 & \hspace{0pt}1/8 \\
\hspace{0pt}2.5 & \hspace{0pt}7.444 & \hspace{0pt}3.64 \% & \hspace{0pt}1.157 & \hspace{0pt}0.042 & \hspace{0pt}1.095–1.219 & \hspace{0pt}8/8 & \hspace{0pt}7/8 \\
\hspace{0pt}3.0 & \hspace{0pt}8.298 & \hspace{0pt}4.91 \% & \hspace{0pt}1.250 & \hspace{0pt}0.077 & \hspace{0pt}1.145–1.370 & \hspace{0pt}8/8 & \hspace{0pt}8/8 \\
\bottomrule
\end{longtable}}

\section{Origin attribution of the largest-scale contamination}\label{note:G}

This note gives the construction behind the origin attribution of Methods and Extended Data Fig.~\ref{edfig:origin} (G.1), its validation (G.2), the register orders (G.3), the shares (G.4), their sensitivity to the structured weight (G.5) and to the ladder construction (G.6), the placement rule (G.7) and the scope of the result (G.8). The attribution says from which component of the census-weighted model the $k = 1\text{--}4$ velocity-error energy originates, not by which route that energy arrives. It is not part of the transfer measured in Figs.~\ref{fig:transfer} and~\ref{fig:feedback}: the growth of the gain and the feedback excess lie mainly in $k = 8\text{--}64$, and at $t = 3.0$ the band $k < 8$ holds about 1 \% of the velocity contamination energy.

\subsection{Superposition and share}\label{si:G.1}
In this note $\Delta_{X}$ is the real part of the velocity contamination of run $X$ at $\varepsilon = 10^{-4}$, and $\langle \cdot, \cdot\rangle_{b}$ and $\|\cdot\|_{b}$ are restricted to velocity band $b$. Write $\hat{g}_{\perp}$ and $s_{\perp}$ for the white and the structured parts of the direction after projection off $\psi$ (Supplementary Note~\ref{note:E}). With the same white draws as the white-only run, the census-weighted injection direction is $(\sqrt{w}\,\hat{g}_{\perp} + \sqrt{w_{s}}\,s_{\perp})/c$, with $w = 0.91$, $w_{s} = 0.09$ and $c = \|d_{\perp}\|$ the draw-mean norm of the projected direction (0.9882 at $256^{3}$). In the linear regime the census-weighted contamination is therefore

\disp{$\Delta_{C} = (\sqrt{w}\,\Delta_{W} + \sqrt{w_{s}}\,\Delta_{\mathrm{sp}})/c$,}{G1}
\noindent where $\Delta_{W}$ is the white-only run's contamination and $\Delta_{\mathrm{sp}}$ the response to the unnormalized structured part $s_{\perp}$. With $E_{X} = \|\Delta_{X}\|^{2}_{b}$ ($X = C$, $W$, sp) and $X_{\mathrm{CW}} = \langle \Delta_{C}, \Delta_{W}\rangle_{b}$, solving (G1) for $\Delta_{\mathrm{sp}}$ gives

\disp{$E_{\mathrm{sp}} = (c^{2}E_{C} - 2c\sqrt{w}\,X_{\mathrm{CW}} + wE_{W})/w_{s}$,\quad $X_{\mathrm{Wsp}} = (cX_{\mathrm{CW}} - \sqrt{w}\,E_{W})/\sqrt{w_{s}}$,}{G2}
\noindent and, for any structured weight $w_{s}'$ ($w' = 1 - w_{s}'$),

\disp{$\mathrm{share}(w_{s}') = w_{s}'E_{\mathrm{sp}}/\bigl[w'E_{W} + w_{s}'E_{\mathrm{sp}} + 2(w'w_{s}')^{1/2}\,X_{\mathrm{Wsp}}\bigr]$.}{G3}
The normalization $c(w_{s}')$ of the reweighted direction $(\sqrt{w'}\,\hat{g}_{\perp} + \sqrt{w_{s}'}\,s_{\perp})/c(w_{s}')$ scales numerator and denominator alike and cancels; the nominal $c$ enters only through $E_{\mathrm{sp}}$ and $X_{\mathrm{Wsp}}$. The structured and white shares need not sum to one; the remainder is their cross term (at $256^{3}$ and $t = 3.0$ the white-origin fraction $(w/c^{2})E_{W}/E_{C}$ is 0.151, against a structured share of 0.836). The break-even weight $w_{s}^{*}$ is the $w_{s}'$ at which (G3) equals $\tfrac{1}{2}$, found by bisection on $(0, 0.5]$. The z-first superposition shares at the model weight (Supplementary Table G1; Extended Data Fig.~\ref{edfig:origin}a) use $c = 1$, that is, $\Delta_{S} = (\Delta_{C} - \sqrt{0.91}\,\Delta_{W})/\sqrt{0.09}$ and share $= 0.09\,E_{S}/E_{C}$; (G3) with the draw-mean $c$ at $w_{s}' = 0.09$ agrees with them to within one unit in the third decimal. Using the measured effective white coefficient (about 0.963 rather than $\sqrt{0.91}$) changes the shares by $\leq 1.2 \times 10^{-3}$.

\subsection{Validation and status per grid}\label{si:G.2}
At $64^{3}$ (Re 120; two noise realizations), direct structured-only runs (white weight 0, identical per-qubit draws) give the structured response directly, $\Delta_{\mathrm{sp}} = \bar{\rho}\,\Delta_{\mathrm{str}}$, where $\Delta_{\mathrm{str}}$ is the structured-only run's contamination and $\bar{\rho} = 0.876$ the pooled draw-mean structured amplitude (0.865 and 0.887 for the two realizations, so $\bar{\rho}^{2}$ is uncertain by about $\pm 2.5$ \%). The structured response estimated by the superposition correlates with the direct runs at 0.993–0.996 (residual 0.040–0.095; two realizations, four stations), and the census-weighted $k = 1\text{--}4$ energy predicted from the white-only and structured-only runs matches the measured one within 2–6 \% (ratios 1.024, 0.953, 0.946 and 0.945 at the four stations, first realization), the size of that normalization uncertainty (Extended Data Fig.~\ref{edfig:origin}e,f).

At $128^{3}$ (Re 300) and $256^{3}$ (Re 800) no structured-only run exists: the decomposition is applied, not independently validated, with one realization, and at $128^{3}$ linearity in $\varepsilon$ was not tested. At $256^{3}$ the census-weighted run (system-only backend) is paired with the white-only and interleaved-order runs (qubitized-pair backend). Before noise the two backends differ by $\delta = 1.87 \times 10^{-12}$ (Methods), against contamination of order $10^{-4}\text{--}10^{-3}$; the cross-backend floor is $\leq 1 \times 10^{-7}$ of the smallest band error used at $t = 1.5$ and, within the noise window, is bounded by the agreement between $\varepsilon = 10^{-4}$ and $10^{-6}$ at $\leq 3.9 \times 10^{-4}$ ($k < 8$) and $\leq 1.1 \times 10^{-3}$ ($k = 1\text{--}4$). That agreement is within 0.061 \% in total $\delta$ and within 0.4 \% in every band, but the real part at $k = 1\text{--}4$ departs by up to 1.33 \% (the real-only contraction; Supplementary Note~\ref{note:F}).

\subsection{Register orders}\label{si:G.3}
The as-compiled order is z-first: ladder rank $r$ ($r = 0$ carries the largest weight) sits on bit $r$ of the flat index, so the ranks run through the z digits, then y, then x, least significant first, followed by the field bits; in the compiled $16^{3}$ circuit the loader touch counts follow this flat-index bit order, which fixes the z-first placement. At $256^{3}$ ($L = 8$ digits per axis) the extrapolated ranks 12–15 fall on y4–y7, and the remaining extrapolated ranks, at the floor value, on x0–x7. The interleaved order places rank $r$ on bit $(r \bmod 3) \cdot L + \lfloor r/3\rfloor$, that is, on z0, y0, x0, z1, y1, x1, \ldots. It is an in-model alternative, implemented only as a permutation of the census ladder weights over the grid digits; no interleaved circuit was compiled. The interleaved runs ($64^{3}$, both realizations; $256^{3}$) are decomposed with the z-first $c$; using the interleaved ladder's own $c$ (0.9834 at $256^{3}$, 0.9893 at $64^{3}$) moves no share by more than $3.3 \times 10^{-4}$.

\subsection{Shares}\label{si:G.4}
In the as-compiled order the structured component is the majority origin of the $k = 1\text{--}4$ energy at every station on all three grids: 0.75–0.92 at $64^{3}$ (two realizations), 0.91–0.96 at $128^{3}$ and 0.84–0.98 at $256^{3}$ over $t = 1.5\text{--}3.0$, and 0.80–0.91 at $t = 3.0$ (0.795 and 0.829 at $64^{3}$, 0.910 at $128^{3}$, 0.836 at $256^{3}$; Supplementary Table G1; Extended Data Fig.~\ref{edfig:origin}a). Under the interleaved order the share at $t = 3.0$ is 0.564 at $64^{3}$ for the first realization (marginal) and 0.430 for the second (below one half from $t = 2.0$), and 0.411 at $256^{3}$, where it falls from 0.917 at $t = 1.5$ (Extended Data Fig.~\ref{edfig:origin}b). The majority is therefore a property of the as-compiled order, not of every register layout.

The $k < 8$ and total velocity shares are weight-sensitive values, not majority results. The structured share of the total velocity-error energy is $\leq 0.34$ at $64^{3}$ and, at $128^{3}$ and $256^{3}$, below one half to $t = 2.0$ and 0.52–0.58 after; several z-first $k < 8$ break-even weights lie inside the weight range of G.5 ($256^{3}$, $t = 3.0$: 0.0639; $64^{3}$ first realization, $t = 2.0$, 2.5 and 3.0: 0.0498, 0.0456 and 0.0551; $64^{3}$ second realization, $t = 1.5$: 0.0772). Separately, the census-weighted direction gives a total $\delta$ 14–47 \% above that of a white direction on the three grids (Extended Data Table~\ref{edtab:grids}), a model dependence of the magnitude; the $k = 1\text{--}4$ origin share does not attribute the gain to either component.

\textbf{Supplementary Table G1 | Structured-origin share of the $k = 1\text{--}4$ velocity-error energy by register order (real part, $\varepsilon = 10^{-4}$).} ``Share'' is at the model weight 0.09; z-first superposition shares use $c = 1$ (G.1), as plotted in Extended Data Fig.~\ref{edfig:origin}a, and direct rows come from the structured-only runs (G.2). The minimum over $w_{s}' \in [0.045, 0.18]$ and the break-even weight $w_{s}^{*}$ use (G3) with the draw-mean $c$. For the interleaved order each break-even weight is followed by its multiple of the census weight 0.0883 in parentheses, and * marks a break-even within $\pm 20$ \% of the census weight.

{\scriptsize
\setlength{\tabcolsep}{3pt}\renewcommand{\arraystretch}{1.15}\setlength{\LTleft}{0pt}\setlength{\LTright}{0pt}
\begin{longtable}{>{\raggedright\arraybackslash}p{\dimexpr 0.1400\linewidth-6pt\relax}>{\raggedright\arraybackslash}p{\dimexpr 0.2000\linewidth-6pt\relax}>{\raggedright\arraybackslash}p{\dimexpr 0.1600\linewidth-6pt\relax}>{\raggedright\arraybackslash}p{\dimexpr 0.1250\linewidth-6pt\relax}>{\raggedright\arraybackslash}p{\dimexpr 0.1250\linewidth-6pt\relax}>{\raggedright\arraybackslash}p{\dimexpr 0.1250\linewidth-6pt\relax}>{\raggedright\arraybackslash}p{\dimexpr 0.1250\linewidth-6pt\relax}}
\toprule
\hspace{0pt}\textbf{Grid (realization)} & \hspace{0pt}\textbf{Order, route} & \hspace{0pt}\textbf{Quantity} & \hspace{0pt}\textbf{$t = 1.5$} & \hspace{0pt}\textbf{$t = 2.0$} & \hspace{0pt}\textbf{$t = 2.5$} & \hspace{0pt}\textbf{$t = 3.0$} \\
\midrule
\endhead
\hspace{0pt}$64^{3}$ Re 120 (first) & \hspace{0pt}z-first, superposition & \hspace{0pt}share & \hspace{0pt}0.916 & \hspace{0pt}0.816 & \hspace{0pt}0.878 & \hspace{0pt}0.795 \\
\hspace{0pt} & \hspace{0pt} & \hspace{0pt}minimum in range & \hspace{0pt}0.821 & \hspace{0pt}0.679 & \hspace{0pt}0.763 & \hspace{0pt}0.643 \\
\hspace{0pt} & \hspace{0pt} & \hspace{0pt}break-even $w_{s}^{*}$ & \hspace{0pt}0.0115 & \hspace{0pt}0.0218 & \hspace{0pt}0.0153 & \hspace{0pt}0.0258 \\
\hspace{0pt}$64^{3}$ Re 120 (first) & \hspace{0pt}z-first, direct (structured-only run) & \hspace{0pt}share & \hspace{0pt}0.915 & \hspace{0pt}0.801 & \hspace{0pt}0.871 & \hspace{0pt}0.783 \\
\hspace{0pt} & \hspace{0pt} & \hspace{0pt}minimum in range & \hspace{0pt}0.822 & \hspace{0pt}0.660 & \hspace{0pt}0.751 & \hspace{0pt}0.626 \\
\hspace{0pt} & \hspace{0pt} & \hspace{0pt}break-even $w_{s}^{*}$ & \hspace{0pt}0.0113 & \hspace{0pt}0.0235 & \hspace{0pt}0.0163 & \hspace{0pt}0.0277 \\
\hspace{0pt}$64^{3}$ Re 120 (second) & \hspace{0pt}z-first, superposition & \hspace{0pt}share & \hspace{0pt}0.749 & \hspace{0pt}0.850 & \hspace{0pt}0.849 & \hspace{0pt}0.829 \\
\hspace{0pt}$64^{3}$ Re 120 (second) & \hspace{0pt}z-first, direct (structured-only run) & \hspace{0pt}share & \hspace{0pt}0.758 & \hspace{0pt}0.846 & \hspace{0pt}0.847 & \hspace{0pt}0.829 \\
\hspace{0pt}$128^{3}$ Re 300 & \hspace{0pt}z-first, superposition & \hspace{0pt}share & \hspace{0pt}0.916 & \hspace{0pt}0.959 & \hspace{0pt}0.947 & \hspace{0pt}0.910 \\
\hspace{0pt} & \hspace{0pt} & \hspace{0pt}minimum in range & \hspace{0pt}0.845 & \hspace{0pt}0.916 & \hspace{0pt}0.882 & \hspace{0pt}0.814 \\
\hspace{0pt} & \hspace{0pt} & \hspace{0pt}break-even $w_{s}^{*}$ & \hspace{0pt}0.0080 & \hspace{0pt}0.0045 & \hspace{0pt}0.0074 & \hspace{0pt}0.0117 \\
\hspace{0pt}$256^{3}$ Re 800 & \hspace{0pt}z-first, superposition & \hspace{0pt}share & \hspace{0pt}0.984 & \hspace{0pt}0.953 & \hspace{0pt}0.970 & \hspace{0pt}0.836 \\
\hspace{0pt} & \hspace{0pt} & \hspace{0pt}minimum in range & \hspace{0pt}0.973 & \hspace{0pt}0.915 & \hspace{0pt}0.928 & \hspace{0pt}0.714 \\
\hspace{0pt} & \hspace{0pt} & \hspace{0pt}break-even $w_{s}^{*}$ & \hspace{0pt}0.0008 & \hspace{0pt}0.0034 & \hspace{0pt}0.0046 & \hspace{0pt}0.0182 \\
\midrule
\hspace{0pt}$64^{3}$ Re 120 (first) & \hspace{0pt}interleaved, superposition & \hspace{0pt}share & \hspace{0pt}0.652 & \hspace{0pt}0.574 & \hspace{0pt}0.581 & \hspace{0pt}0.564 \\
\hspace{0pt} & \hspace{0pt} & \hspace{0pt}break-even $w_{s}^{*}$ & \hspace{0pt}0.0515 (0.58) & \hspace{0pt}0.0698 (0.79) & \hspace{0pt}0.0686 (0.78) & \hspace{0pt}0.0717* (0.81) \\
\hspace{0pt}$64^{3}$ Re 120 (second) & \hspace{0pt}interleaved, superposition & \hspace{0pt}share & \hspace{0pt}0.511 & \hspace{0pt}0.477 & \hspace{0pt}0.480 & \hspace{0pt}0.430 \\
\hspace{0pt} & \hspace{0pt} & \hspace{0pt}break-even $w_{s}^{*}$ & \hspace{0pt}0.0864* (0.98) & \hspace{0pt}0.0981* (1.11) & \hspace{0pt}0.0969* (1.10) & \hspace{0pt}0.1161 (1.31) \\
\hspace{0pt}$256^{3}$ Re 800 & \hspace{0pt}interleaved, superposition & \hspace{0pt}share & \hspace{0pt}0.917 & \hspace{0pt}0.780 & \hspace{0pt}0.631 & \hspace{0pt}0.411 \\
\hspace{0pt} & \hspace{0pt} & \hspace{0pt}break-even $w_{s}^{*}$ & \hspace{0pt}0.0072 (0.08) & \hspace{0pt}0.0229 (0.26) & \hspace{0pt}0.0558 (0.63) & \hspace{0pt}0.1245 (1.41) \\
\bottomrule
\end{longtable}}

\subsection{Sensitivity to the structured weight}\label{si:G.5}
The census counts two-qubit touches only, so one-qubit, idle and readout errors could move the structured weight either way. The predefined range is therefore $w_{s}' \in [0.045, 0.18]$, a factor 2 either way of the census value $0.0883 \approx 0.09$. In the z-first order the $k = 1\text{--}4$ majority holds over the whole range at every station on all three grids, for both $64^{3}$ realizations and both $64^{3}$ routes; the smallest share inside the range is 0.591 ($64^{3}$ second realization, $t = 1.5$), and the break-even weight is 0.075–3.1 \%, a factor $\geq 2.8$ below the census value of 8.83 \% (Supplementary Table G1; Extended Data Fig.~\ref{edfig:origin}c). Under the interleaved order the $64^{3}$ break-even weights are 0.58–1.31 times the census weight, four of them within $\pm 20$ \% of it, and at $256^{3}$ they rise from 0.08 times ($t = 1.5$) to 1.41 times ($t = 3.0$).

\subsection{Ladder construction variants}\label{si:G.6}
Variants of the ladder construction were evaluated exactly at injection time, with the same per-qubit draws (Supplementary Table G2): flat extension (ratio 1.0, floor kept), steep extension (ratio 0.25, floor kept), no floor (ratio 0.5), field-qubit weight $\times 4$ and $\times 0.25$, a uniform address ladder (every address bit at the mean of the extended ladder) and the interleaved order. The construction variants move the injection-time $k = 1\text{--}4$ structured share by $\leq 0.006$, the largest move being +0.0051 ($128^{3}$, flat extension); the uniform address ladder raises the share (by 0.075 at $64^{3}$), and the interleaved order lowers it. Late-time shares exist only for the two orders that were run, z-first and interleaved. For the other variants there is only an illustrative proportional-transfer estimate, which scales the nominal late-time $E_{\mathrm{sp}}$ by the variant-to-nominal ratio of the injection-time structured energy in $k < 8$; it puts the construction variants within 0.03 of nominal and the uniform ladder higher, but its error against the measured interleaved shares reaches 0.10, so it is not a measured robustness result (Extended Data Fig.~\ref{edfig:origin}d).

\textbf{Supplementary Table G2 | Injection-time $k = 1\text{--}4$ structured share at the census weight, by ladder variant (exact; same per-qubit draws).}

{\scriptsize
\setlength{\tabcolsep}{3pt}\renewcommand{\arraystretch}{1.15}\setlength{\LTleft}{0pt}\setlength{\LTright}{0pt}
\begin{longtable}{>{\raggedright\arraybackslash}p{\dimexpr 0.4026\linewidth-6pt\relax}>{\raggedright\arraybackslash}p{\dimexpr 0.1043\linewidth-6pt\relax}>{\raggedright\arraybackslash}p{\dimexpr 0.1043\linewidth-6pt\relax}>{\raggedright\arraybackslash}p{\dimexpr 0.1043\linewidth-6pt\relax}>{\raggedright\arraybackslash}p{\dimexpr 0.2844\linewidth-6pt\relax}}
\toprule
\hspace{0pt}\textbf{Variant} & \hspace{0pt}\textbf{$64^{3}$} & \hspace{0pt}\textbf{$128^{3}$} & \hspace{0pt}\textbf{$256^{3}$} & \hspace{0pt}\textbf{$256^{3}$ break-even $w_{s}^{*}$} \\
\midrule
\endhead
\hspace{0pt}Nominal & \hspace{0pt}0.892 & \hspace{0pt}0.957 & \hspace{0pt}0.992 & \hspace{0pt}0.00092 \\
\hspace{0pt}Flat extension, ratio 1.0 & \hspace{0pt}0.897 & \hspace{0pt}0.962 & \hspace{0pt}0.994 & \hspace{0pt}0.00072 \\
\hspace{0pt}Steep extension, ratio 0.25 & \hspace{0pt}0.892 & \hspace{0pt}0.957 & \hspace{0pt}0.991 & \hspace{0pt}0.00094 \\
\hspace{0pt}No floor & \hspace{0pt}0.892 & \hspace{0pt}0.957 & \hspace{0pt}0.991 & \hspace{0pt}0.00093 \\
\hspace{0pt}Field-qubit weight $\times 4$ & \hspace{0pt}0.893 & \hspace{0pt}0.958 & \hspace{0pt}0.992 & \hspace{0pt}0.00089 \\
\hspace{0pt}Field-qubit weight $\times 0.25$ & \hspace{0pt}0.892 & \hspace{0pt}0.957 & \hspace{0pt}0.992 & \hspace{0pt}0.00092 \\
\hspace{0pt}Uniform address ladder & \hspace{0pt}0.967 & \hspace{0pt}0.992 & \hspace{0pt}0.999 & \hspace{0pt}0.00013 \\
\hspace{0pt}Interleaved order & \hspace{0pt}0.624 & \hspace{0pt}0.788 & \hspace{0pt}0.933 & \hspace{0pt}0.00711 \\
\bottomrule
\end{longtable}}

\subsection{Placement rule, timing and the post-hoc scope resolution}\label{si:G.7}
A placement rule for this attribution ($k = 1\text{--}4$; majority $=$ share $> 0.5$) was fixed before any share was computed and before the $64^{3}$ and $128^{3}$ inputs existed, but after the $256^{3}$ inputs (band energies and cross spectra) had been computed. Its four grades were ``robust'' (late-time majority at every station on all three grids, over the whole weight range and for both orders, and injection-time majority for all eight ladder variants), ``main text with strong qualification'' (late-time majority for the nominal order over the whole range on all grids, but failing for the interleaved order or for some variant at injection time, or holding only with a break-even margin below a factor 2 somewhere), ``supporting only'' (late-time majority failing somewhere in the range for the nominal order) and ``not robust'' (a break-even weight within 20 \% of the nominal weight anywhere). Two grades apply, and the rule fixed no precedence between them: the second, and, read literally, the fourth, through four interleaved $64^{3}$ break-even weights within $\pm 20$ \% of nominal (Supplementary Table G1); no z-first break-even comes within a factor 2.8 of nominal. Confining the fourth clause to the order whose majority is claimed, the as-compiled z-first order, is a post-hoc scope resolution: that order is a fact of the compiled circuit, whereas the interleaved order tests generalization beyond it, and the literal reading is not monotone (an interleaved break-even just above nominal would grade ``not robust'', a deeper failure with break-even above 1.2 times nominal, as at $256^{3}$ and $t = 3.0$, would not). For the as-compiled z-first order and the census-weighted model the $k = 1\text{--}4$ majority therefore holds with strong qualification; read as layout-general, it is not robust.

\subsection{What the attribution says and does not say}\label{si:G.8}
Within the scope set out in G.2–G.4 and G.7, the attribution is a quantitative, in-model origin share and its sensitivity, not a mechanism, and it implies no choice of register layout. It does not say whether the $k = 1\text{--}4$ energy arrives by direct injection into long-period address digits or by transfer between scales: at $t = 3.0$ about 15 \% of the census-weighted $k = 1\text{--}4$ energy is white-origin, and almost all of that is dynamically generated (Supplementary Note~\ref{note:F}). That the significance of an address qubit corresponds to the length scale it addresses, and that an error on one register qubit acts on the whole encoded field, are known~\cite{Jaksch2023,Meng2024}.

\end{SIpart}

\clearpage
\phantomsection\addcontentsline{toc}{section}{References}
\bibliographystyle{unsrt}
\bibliography{references}

\end{document}